\documentclass[aps,amsmath,amssymb, preprintnumbers, reprint, 
longbibliography,superscriptaddress,nofootinbib]{revtex4-2}

\usepackage{bm,graphicx,wasysym}
\usepackage{ulem,hyperref, cleveref}
\usepackage{slashed}
\usepackage{amsmath, amsfonts, amsthm, amssymb, xcolor, mathrsfs}
\usepackage{color}

\allowdisplaybreaks[3]

\def\({\left(}
\def\){\right)}
\def\[{\left[}
\def\]{\right]}
\def\<{\left\langle}
\def\>{\right\rangle}
\def\d{\mathrm{d}}

\newcommand{\f}[2]{\frac{#1}{#2}}
\def \bal#1\eal  {\begin{align} #1 \end{align}}
\newcommand{\be} {\begin{equation}}
\newcommand{\ee} {\end{equation}}
\newcommand{\bc}{\begin{center}}
\newcommand{\ec}{\end{center}}
\newcommand{\bim} {\begin{itemize}[noitemsep]}

\newcommand{\eim} {\end{itemize}}

\newcommand{\nn} {\nonumber\\}
\newcommand{\eref}[1]{Eq.~(\ref{#1})}

\newcommand{\nd} {\nabla}
\newcommand{\pd} {\partial}
\newcommand{\mc} {\mathcal}
\newcommand{\bsb}{\boldsymbol}
   \newcommand{\bfk} {{\bf k}}   
      
   \newcommand{\bfq} {{\bf q}}   
      
      \newcommand{\bfx} {{\bf x}}
   
   \newcommand{\bfB} {{\bf B}}

      \newcommand{\bfO} {{\bf O}}
\newcommand{\bfP} {{\bf P}}      \newcommand{\bfR} {{\bf R}}
      
      \newcommand{\bfX} {{\bf X}}
   
\newcommand{\mn} {{\mu\nu}}

\newcommand{\ai}{{\alpha}}
\newcommand{\bi}{{\beta}}
\newcommand{\gi}{{\gamma}}

\newcommand{\ri}{{\rho}}
\newcommand{\si}{{\sigma}}
\newcommand{\li}{{\lambda}}
\newcommand{\ti}{{\tau}}

\newcommand{\oi}{\omega}
\newcommand{\epi}{\epsilon}
\newcommand{\thi}{\theta}

\begin{document}

\hfill {{\footnotesize USTC-ICTS/PCFT-24-51}}

\title{Non-topological solitons and quasi-solitons}

\author{Shuang-Yong Zhou}
\email{zhoushy@ustc.edu.cn}
\affiliation{Interdisciplinary Center for Theoretical Study, University of
Science and Technology of China, Hefei, Anhui 230026, China}
\affiliation{Peng Huanwu Center for Fundamental Theory, Hefei, Anhui 230026, China}

\begin{abstract}

Solitons in relativistic field theories are not necessarily topologically charged. In particular, non-topological solitons—known as Q-balls—arise naturally in nonlinear field theories endowed with attractive interactions and internal symmetries. Even without stabilizing internal symmetries, quasi-solitons known as oscillons, which are long-lived, can also exist. Both Q-balls and oscillons have significant applications in cosmology and particle physics. This review is an updated account of the intriguing properties and dynamics of these non-topological solitons and quasi-solitons, as well as their important roles in early-universe scenarios and particle physics models.

\end{abstract}

\maketitle

\tableofcontents

\section{Introduction and summary}

Following extensive and fruitful applications of the perturbative approach to quantum field theory, the quest for nonperturbative field-theoretical structures has gained significant traction since the 1960s. A multitude of solitons have been discovered in various relativistic field theories. These are typically localized, smooth solutions to the fully nonlinear field equations, exhibiting particle-like characteristics.

The study of solitons dates back at least to the investigations of solitary waves on shallow water surfaces by naval architect John Scott Russell in the early 1800s, who famously chased a solitary wave along a canal in Scotland. Despite being rebuked by the scientific establishment at the time, solitary waves were later shown by Korteweg and de Vries to be described by the celebrated KdV equation, building on earlier work by Boussinesq. General multi-solitary wave solutions were subsequently obtained using the inverse scattering method in the 1960s \cite{Gardner:1967wc}, coinciding with the rising popularity of solitons in quantum field theory. Solitons like the solitary waves differ from those in relativistic field theories in an important way: systems such as the KdV equation are integrable, possessing an infinite number of conserved quantities, whereas most relativistic field theories with soliton solutions are not integrable systems. As a result, during scattering, solitary waves in the KdV system exhibit exact elasticity, experiencing only a time delay or advance, while most soliton scatterings in relativistic field theories do not. 

Many of the solitons in relativistic field theories are static and endowed with nontrivial topology \cite{Manton:2004tk, Ryder:1985wq}. For instance, in 1+1D with a single scalar field, the Sine-Gordon model allows for the existence of kinks. A kink is topologically nontrivial because its field value asymptotically approaches different degenerate vacua at opposite spatial infinities, giving it a topological charge. This topological charge stabilises the kink, as an infinite amount of energy would be required to shift the field from one vacuum state to another in the asymptotic region. In higher dimensions, however, a simple argument due to Derrick \cite{Derrick:1964ww}, which is reviewed in Appendix \ref{sec:Derrick}, precludes the existence of localized, static and stable solutions composed solely of canonical scalar fields. 

One way to circumvent the constraints of Derrick's theorem is by introducing additional gauge fields. This approach has led to the discovery of various topological solitons in higher dimensions (see \cite{Manton:2004tk} for a pedagogical review), which find a variety of applications in particle physics and cosmology. Examples of topological solitons include Nielsen-Olesen vortices (or cosmic strings) \cite{Nielsen:1973cs} and the 't Hooft-Polyakov monopole \cite{tHooft:1974kcl, Polyakov:1974ek}. Similar to the Sine-Gordon kinks, these static solitons are topologically charged by a nontrivial mapping between spatial infinity and the internal field space.  On the other hand, since Derrick's theorem applies only to static solutions, another way to evade it, as already pointed by Derrick \cite{Derrick:1964ww}, is to search for time-periodic solutions, even without adding gauge fields, and this leads to non-topological solitons \cite{Friedberg:1976me, Friedberg:1977xf, Coleman:1985ki, Lee:1991ax, Rosen:1968mfz, Kusenko:1997zq, Kusenko:1997si, Enqvist:1997si, Kasuya:1999wu, Kasuya:2000wx}, which is the main focus of this review.

Non-topological solitons are localized solutions whose field values are stationary rather than strictly static. These solitons do not have to be topologically charged, and are stable due to the attractive nature of the potential and some conserved internal charges. For example, in the simple case of a complex scalar $\varphi$ charged under a global U(1) symmetry, there exist spherically symmetric, localized, stable lumps that go like
\be
\varphi(x)= f(r) e^{-i\oi t} ,
\ee
where the constant $\oi$ is the rotation frequency in the internal field space and $f(r)$ is a localized function that exponentially decays at large $r$. These objects are often referred to as Q-balls \cite{Coleman:1985ki}. In this review, we will mostly use the terms non-topological soliton and Q-ball interchangeably. 

It is worth pointing out that although the field exhibits time dependence, the energy density of a non-topological soliton remains static. In the absence of a stabilising conserved internal charge, an attractive potential still allows for the formation of localized lumps that are approximately periodic, which can occur with a real scalar field \cite{Bogolyubsky:1977tc,  oscillon2}. These ultimately unstable but long-lived lumps nowadays usually go under the name of oscillons \cite{Gleiser:1993pt, Copeland:1995fq}. Despite their obvious differences, many aspects of non-topological quasi-solitons/oscillons closely resemble those of non-topological solitons/Q-balls, making it appropriate for a combined review.

To understand these highly non-linear objects, a slew of analytical and numerical methods have been applied, dating back from their initial conception. Driven in part by advances in computational methods and resources, recent years have seen substantial progress in uncovering the diverse properties of non-topological solitons and quasi-solitons as well as their roles in particle physics and cosmology. We will review the well-established aspects of these nonperturbative structures, as well as the evolving landscape of their phenomena.

As stable or long-lived nonlinear solutions to relativistic field theories, Q-balls and oscillons, along with their complex dynamics, provide valuable insights into the nonperturbative aspects of these theories (see  \S\,\! \ref{sec:Q-balls} and \ref{sec:oscillon} for detailed references). They exist in many nonlinear field theories that exhibit attractive interactions. For a canonical scalar field theory, such attractiveness is characterized by the interacting part of the potential dipping below zero over a certain range of field values. The richness of non-topological (quasi-)solitons is further highlighted by their couplings to other fields, leading to a plethora of intriguing phenomena. The basic forms of Q-balls and oscillons are spherically symmetric and have been extensively studied. In spherical symmetry, a wide range of analytical and numerical methods can be effectively applied, leading to well-established conditions regarding their existence, stability, scattering and quantum properties. Conversely, the exploration of more complex structures of (quasi-)solitons and their interactions often relies heavily on numerical simulations, which have nevertheless revealed many fascinating phenomena, some of them very recently.

Q-balls and oscillons have many applications in modern particle physics and cosmology (see  \S\,\! \ref{sec:applications} for detailed references). Q-balls naturally arise in the supersymmetric extensions of the Standard Model and can form during the fragmentation of the Affleck-Dine condensate in the early universe, a process that may generate a stochastic gravitational wave background. Their decays may also induce additional gravitational wave backgrounds. They are also considered potential dark matter candidates and could even give rise to primordial black holes. Additionally, non-topological solitons serve as effective models for hadrons. Current observational constraints suggest that the inflaton potential is compatible with oscillon formation, indicating that oscillons may be abundantly generated during the preheating phase following inflation, which can also lead to the production of a stochastic gravitational wave background. Associated with oscillon formation are also the lower-frequency induced gravitational wave backgrounds from the oscillon decay and the oscillon isocurvature perturbations. Furthermore, oscillons may form at lower energies through preheating-like condensate fragmentation, and also contribute to the formation of primordial black holes. Many of the gravitational wave backgrounds from Q-balls and oscillons could be detectable in current or upcoming GW experiments.

Both topological and non-topological solitons are usually constructed and evolved in the classical limit. This is justified because solitons are field configurations where the occupation numbers for the constituent quantum modes are very large, in which case the classical approximation is a rather good one \cite{Khlebnikov:1996mc, Berges:2004yj}. Quantization of the classical solitons can be achieved via the functional method \cite{Dashen:1974ci, Gervais:1974dc, Goldstone:1974gf} or canonical quantization \cite{Christ:1975wt, Tomboulis:1975gf, Creutz:1975qt}. Nowadays, quantum solitons and their time evolution can also be simulated in real-time using lattice field theories \cite{Salle:2003ju, Borsanyi:2007wm, Tranberg:2013cka, Xie:2023psz}. Typically, as the nonlinear couplings in the theory approach zero, the classical solutions increasingly serve as accurate approximations of the solitons.

In the presence of strong gravitational back-reaction, non-topological solitons, in which the metric field plays a significant role, are referred to as boson stars or oscillatons. As gravity is famously attractive, these objects can form even without attractive matter interactions, so their properties can differ significantly from those of the non-topological solitons in flat space. In this review, we primarily focus on non-topological solitons without strong gravitational effects, except for briefly mentioning these objects in scenarios directly connected to Q-balls and oscillons. The literature on boson stars and oscillatons is extensive and has been thoroughly reviewed in other works \cite{Lee:1991ax,  Jetzer:1991jr,  Liebling:2012fv, Visinelli:2021uve}. 

The rest of this review is organized as follows. We shall first review the various perturbative and nonperturbative field-theoretical aspects of non-topological solitons and quasi-solitons in  \S\,\!  \ref{sec:Q-balls} and  \S\,\!  \ref{sec:oscillon} respectively. We then discuss their applications in models of particle physics and cosmology in  \S\,\! \ref{sec:applications}. The subsection titles are crafted to be self-explanatory regarding their contents. Without further ado, let us dive into the fascinating field of Q-balls and oscillons.

\section{Q-balls}
\label{sec:Q-balls}

In this section, we will review the properties of non-topological solitons as nonlinear extended structures in field theories, focusing on the various perturbative, nonperturbative and dynamical aspects of these localized objects, deferring the discussions of their roles in cosmology and particles physics to  \S\,\!  \ref{sec:applications}. As noted in the introduction, we shall use the term Q-balls in a broad sense, interchangeably with non-topological solitons.

\subsection{Basics of a Q-ball}
\label{sec:Qballbasics}

Non-topological solitons can easily arise in relativistic field theories  \cite{ Friedberg:1976me, Friedberg:1977xf, Coleman:1985ki, Lee:1991ax, Rosen:1968mfz}. One of the simplest models realizes them involves a U(1) symmetric complex scalar field $\varphi$ described by Lagrangian
\be
\label{originLag}
\mathcal{L}=-\partial_\mu \varphi^{\dagger} \partial^\mu \varphi-V(\varphi) ,
\ee
where 
\be
\label{potV0}
V (\varphi) =  m^2 |\varphi|^2 + \li |\varphi|^4 + g |\varphi|^6 + ...
\ee 
is an attractive potential, with $m^2$, $\li$, $g$, etc., being coupling constants of the theory. The specific conditions for a potential to be attractive, along with the requirements necessary for a stable soliton to exist, will be discussed momentarily. Without loss of generality (in the absence of gravity), the minimum of $V(\varphi)$ can be assumed to be at $\varphi=0$:
\be
\left.\f{\d V}{\d \varphi}\right|_{\varphi=0}=0 .
\ee
Note that the Minkowski metric is chosen to be $\eta_{\mu\nu}={\rm diag}(-1,+1,...,+1)$, and  natural units with $\hbar=c=1$ are adopted.

The Noether currents associated with the global U(1) symmetry $\varphi \to \varphi + i \ai \varphi$ and the translation symmetries $\varphi(x) \to \varphi(x) + \bi^\mu\pd_\mu\varphi(x)$, $\ai$ and $\bi^\mu$ being infinitesimal constants, are  given respectively by
\bal
j_\mu & =i\left(\varphi^{\dagger} \partial_\mu \varphi-\partial_\mu \varphi^{\dagger} \varphi\right), \\
T_{\mu\nu} & =\partial_\mu \varphi^{\dagger} \partial_\nu \varphi+\partial_\nu \varphi^{\dagger} \partial_\mu \varphi  + \eta_{\mu\nu}\mc{L}
\eal
The conserved charges associated with these currents include the U(1) charge, energy and angular momentum:
\bal
Q=\int \d^d x j_0,~~
E=\int \mathrm{d}^d x T_{t t},~~
L=\int \mathrm{d}^d x T^t{}_\phi .
\eal
Every particle either has particle number +1 or $-1$, so $Q$ is also the difference between the number of positively charged particles and negatively charged particles (in the quantized theory). The quantities above are expressed in $d+1$ dimensions, as both 3+1D and other spacetime dimensions are extensively studied in the literature.

As we will soon see, a basic Q-ball is an extended, periodically oscillating field configuration with a conserved total charge $Q$ and a core that is essentially a spherically shaped ball, hence the name Q-ball as coined by Coleman \cite{Coleman:1985ki}. Before the coining of Q-balls, Friedberg, Lee and Sirlin demonstrated that stable non-topological solitons can exist even quantum mechanically in a renormalisable model involving a complex scalar and a real scalar \cite{Friedberg:1976me} (see \S\,\! \ref{sec:FriedbergLeeS} for more details). Integrating out the heavy real scalar field in this renormalisable model, one gets a U(1) scalar theory more or less resembling that of \eref{originLag}. Ref \cite{Coleman:1985ki} rigorously established that the Q-ball solution minimizes the energy functional, thereby also confirming its absolute stability, even at the quantum level. The idea of the existence of solitonic solutions in a U(1) scalar theory dates back to as early as Rosen \cite{Rosen:1968mfz}, who focused on the more mathematical aspects of iteratively constructing localized spherical solutions and did not address the important issue of stability. We choose the theory \eqref{originLag} as the fiducial model for its simplicity and popularity (see \cite{Lee:1991ax, Nugaev:2019vru} for previous reviews). However, many of the following analyses can be easily extended to the other models, and many central results are quite similar.

In general, a collection of charges may tend to disperse throughout space, lacking a distinct localized structure, unless the potential $V(\varphi)$ is attractive, in which case charges can coalesce to form a Q-ball. This intuitive picture is supported by the fact that a Q-ball solution corresponds to the minimum of the energy functional. To see this, for an isolated system with a conserved total charge $Q$, we can write the total energy as
\be
E_Q= \int \mathrm{d}^d x T_{t t}  + \oi \( Q- \int \d^d x j_0 \) ,
\ee 
where the charge conservation is made explicit by introducing the Lagrange multiplier $\oi$. Note that $\oi$ is independent of the spacetime coordinates as it enforces a global constraint. Parametrizing the complex field as $\varphi=a(x) e^{i b(x)}$, $a(x)$ and $b(x)$ being real functions of spacetime coordinates, the total energy can be written as
\bal
E_Q&= \omega Q+\int \mathrm{d}^d x\Big[\dot{a}^2+(\nabla a)^2+a^2(\dot{b}+\omega)^2
+a^2(\nabla b)^2  \nn
&~~~~ +V(|a|)- \omega^2 a^2\Big],
\label{Efunctional}
\eal
where $\dot{}$ denotes a derivative with respect to time. At a given time, we can choose $a$, $b$, $\dot{a}$ and $\dot{b}$ independently. From the energy functional \eqref{Efunctional}, we see that the configuration with the minimal energy necessarily satisfies $\dot{a}=0$, $\dot{b}=-\omega$ and $\nd b=0$. (Note that choosing $\nd a=0$ does not necessarily lower the energy because of the presence of $V(|a|)- \omega^2 a^2$.) This implies that the minimal energy configuration must take the form: $\varphi=a(\bfx) e^{-i \oi t}$, where we have neglected an unimportant phase $e^{i\oi t_0}$, corresponding to choosing an appropriate initial time. With this reduced form, the energy functional becomes 
\be
\label{EQ001}
E_Q = \omega Q+\int \mathrm{d}^d x\Big[ (\nabla a)^2  +V(|a|)- \omega^2 a^2\Big] .
\ee
Since we are seeking a configuration $a(\bfx)$ that vanishes at infinity, the spherical rearrangement theorem tells us that the minimal energy configuration must be of a spherical form
\be
\label{phidef0}
\varphi(x)= f(r) e^{-i\oi t} ,
\ee
which minimizes the gradient terms while keeping the non-gradient terms unchanged \cite{Coleman:1985ki}. Thus, the Q-ball solution can be obtained by minimizing the energy functional of the form
\be
\label{EnergywithQ}
E_Q = \omega Q+ {\cal V}_{d-1} \!\int \mathrm{d} r  r^{d-1} \Big[(f')^2  +V(|f|)-  \omega^2 f^2\Big],
\ee
where $'$ is a derivative with respect to $r$ and ${\cal V}_{d-1}$ is the volume of the unit $(d-1)$-sphere. As a necessary condition, the minimal energy solution must extremize the energy functional with respect to the variations of $f(r)$, so the localized Q-ball profile must satisfy\,\footnote{In \eref{EnergywithQ}, the charge conservation is still enforced by $\oi$, which is now an auxiliary variable, rather than a Lagrangian multiplier. That is, varying \eref{EnergywithQ} with respect to $\oi$ gives the charge conservation, which can be substituted back into \eref{EnergywithQ} to eliminate $\oi$, if desired.
}
\be
\label{fEoM}
f'' + \frac{d-1}{r} f' + \omega^2 f -  \f12 \frac{\partial V}{\partial f} =0 .
\ee
Viewing $r$ as the role of ``time'', this is analogous to a mechanical system where a ``particle'' moves in the direction of $f$, subject to an effective potential 
\bal
V_{\rm eff} (|f|) &= \oi^2 f^2 - V(|f|)
\nn
&= (\oi^2-m^2) f^2 - \li f^4 - g f^6 - ... ,
\label{Veffdef}
\eal
along with ``time-dependent'' friction that dies down as $1/r$. As we will discuss later, for certain potentials, the extremal profile corresponds to the minimum of the energy functional. It is worth pointing out that, as we see from \eref{phidef0}, a Q-ball uniformly rotates in the internal field space, in contrast to a static soliton.

\subsubsection{Existence conditions}
\label{sec:QballExist}

\begin{figure}
	\centering
		\includegraphics[height=3.7cm]{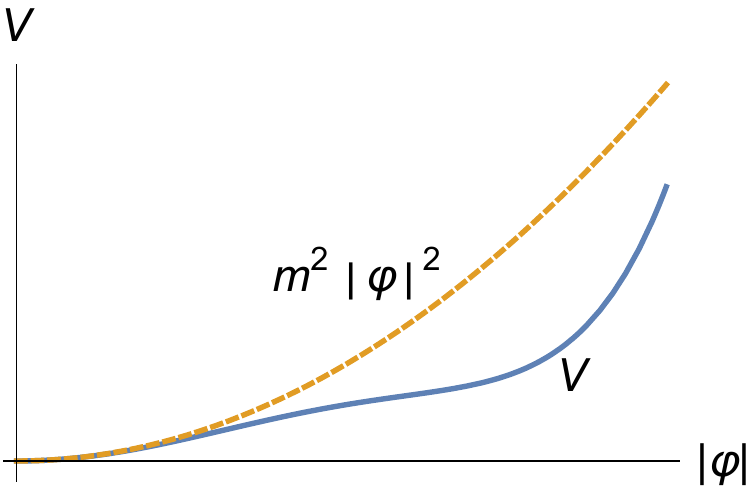}
		
		~\\
		
		\includegraphics[height=3.7cm]{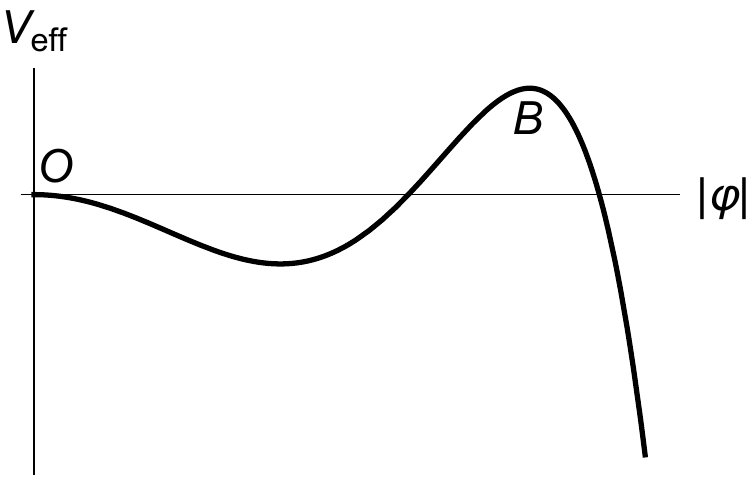}
	\caption{An example of a Q-ball-supporting potential $V(|\varphi|)$ and its associated effective potential $V_{\rm eff}(|\varphi|)$. To support a Q-ball solution, $V$ must dip below the quadratic mass term $m^2|\varphi|^2$ over some range (see \eref{Vint}), thereby creating an attractive potential. From $V_{\rm eff}$'s viewpoint, {\bfO} must be a local maximum at $\varphi=0$, and peak {\bfB} must be higher than peak {\bfO}. The Q-ball profile $f(r)$ corresponds to a ``particle'' rolling down from a point near peak {\bfB} at ``time'' $r=0$ to stop exactly at peak {\bfO} at ``time'' $r=\infty$. }
 \label{fig:VVeff}
\end{figure}

Now, we are ready to establish the conditions on the potential $V(\varphi)$ for a Q-ball to exist. By regularity of the solution at $r=0$, the ``particle'' must initially have a vanishing ``velocity''
\be
\left.\f{\d f}{\d r}\right|_{r=0}=0 ,
\ee
and the locality of the solution requires that the ``particle'' finally stop at the origin in the field space
\be
f(r\to \infty ) \to 0  .
\ee 
So a viable solution $f(r)$ can be visualized as follows: the ``particle'' begins at a nonzero $f$, descends down into the valley of the effective potential $V_{\rm eff}$ and then climbs up to stop at exactly the peak at the origin. For the origin to be a peak, from Eq.~\eqref{Veffdef}, we see that a viable $\oi$ must be smaller than $m$,
\be
|\oi|  < \oi_+ \equiv m ,
\ee 
{\it i.e.,}, the mass of the scalar sets an upper bound on the internal frequency of the Q-ball \cite{Coleman:1985ki}. 

On the other hand, for the effective potential to have a valley, the (nonlinear) interacting part of the potential $V_{\rm int}(|f|)\equiv V(|f|)-m^2 f^2  = \li f^4 + g f^6+...$ must dip below zero in some region \cite{Lee:1991ax}:
\be
\label{Vint}
V_{\rm int}(|f|\neq 0)<0 .
\ee
In other words, the total potential should ``open up'' wider than the quadratic potential away from the minimum; see Figure \ref{fig:VVeff} for an exemplary illustration. As a simple example, this can be achieved with $\li<0$. Essentially, negative interacting terms in potential exert attractive forces. This is because such terms allow the energy of a localized configuration to be lower than that of a homogeneous configuration, as a localized configuration can access large field values in its core and large field values make negative interacting terms sizable. Physically, large field values correspond to particles coalescing to localized structures. 

For a potential that is bounded from below, the condition of $V_{\rm int}(|f|\neq 0)<0$ implies that 
\be
\label{oiminusDef}
\oi_-^2  \equiv {\rm min}\(\f{V(|f|)}{f^2}\)  \text{is reached at } f\neq 0, 
\ee
which is another commonly used criterion to inspect whether a potential support a Q-ball solution \cite{Coleman:1985ki}. $\oi_-$ actually sets the lower bound on a viable internal frequency $\oi$ for the Q-ball:
\be
\oi_-  <  |\oi|  .
\ee
This is because if $\oi<\oi_-$, then the effective potential $V_{\rm eff}$ will be always negative, meaning that $V_{\rm eff}$ will not have a $f\neq 0$ peak that is higher than the peak at $f=0$. In such a case, the ``particle'' will not be able to climb up to the $f=0$ peak, and thus cannot support a Q-ball solution. 

In summary, a necessary condition for a Q-ball to exist is that its internal frequency must lie within the range
\be
\oi_-  <  |\oi| < \oi_+ .
\ee
In 2+1D, this entire range turns out to support a stable Q-ball, but in 3+1D the higher end of this range (near $\oi_+$) does not support stable Q-balls (cf.~Figure \ref{fig:EvsQ}). A typical Q-ball supporting potential and its corresponding effective potential are shown in Figure \ref{fig:VVeff}.

\subsubsection{Radial profile}

\begin{figure}
	\centering
		\includegraphics[height=4.3cm]{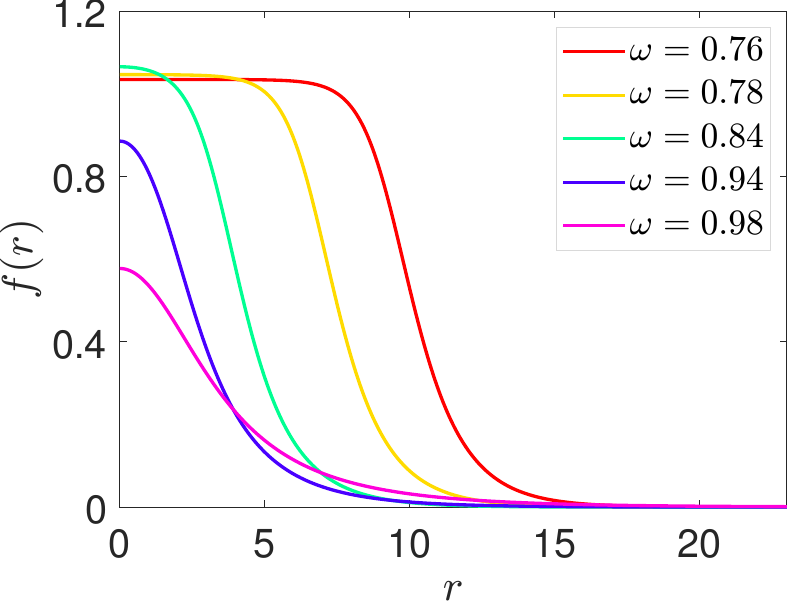}~~~~~

	        ~~\includegraphics[height=4.55cm]{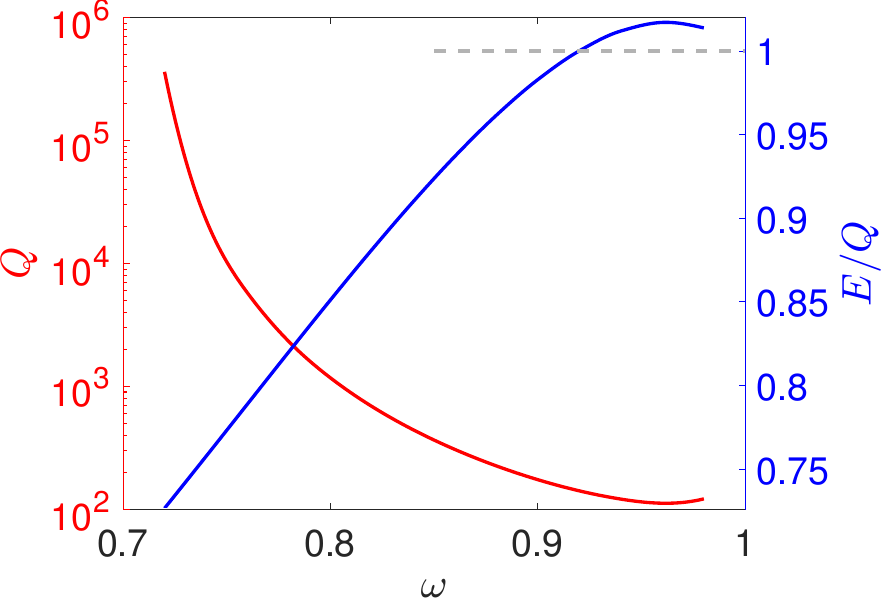}
	\caption{({\it top}) A few example profile functions $f(r)$ for the potential $V(|\varphi|)=|\varphi|^2-|\varphi|^4+g|\varphi|^6$ in 3+1D with $g=\frac{1}{2}$.  A Q-ball with $\oi$ close to the mass $m=1$ has a thick wall, while an $\oi$ close to $\oi_-$ leads to a thin-wall Q-ball. ({\it bottom}) Dependence of $Q$ and $E/Q$ on $\oi$ in this 3+1D model. The minimum of $Q$ is reached slightly below $\oi=m=1$.}
 \label{fig:profiles}
\end{figure}

If $\oi$ is within $\oi_- < |\oi| < \oi_+$, it is always possible to find a solution to \eref{fEoM} satisfying the initial and final conditions mentioned above \cite{Coleman:1985ki}. Note that we seek a solution where the ``particle'' starts at $r=0$ with zero velocity near peak {\bfB} in Figure \ref{fig:VVeff}, rolls down the valley and stops exactly at peak {\bfO} when $r\to \infty$. If we release the ``particle'' too low, it cannot reach peak  ${\bfO}$, which corresponds to the case of undershooting. On the other hand, if we release the ``particle'' very close to peak {\bfB}, it will stay near peak {\bfB} for a long time, and when it starts to roll down the valley, the friction term in \eref{fEoM} already becomes negligible, so the ``particle'' can overshoot peak {\bfO}. By continuity, there must be a place to release the ``particle'' such that it exactly stops at peak ${\bfO}$. Generally, such a solution $f(r)$ can only be obtained numerically, for instance, by a shooting method, which is essentially a refinement of the above argument. A few typical profile functions are shown in Figure \ref{fig:profiles}.

Note that, classically, the potential 
\be
\label{phi6000}
V(|\varphi|)=|\varphi|^2-|\varphi|^4+ g |\varphi|^6
\ee
used in Figure \ref{fig:profiles} (but with a generic $g$) actually covers all the potentials \eqref{potV0} when truncated to the sextic order:
\be
\label{phi6mlig}
V (|\varphi|) =  m^2 |\varphi|^2 + \li |\varphi|^4 + g |\varphi|^6 .
\ee
This is because we can perform the following of rescalings 
\be
\label{variableScaling0}
x^\mu \to \f{x^\mu}{m}, ~~~ \varphi \to \frac{m \varphi}{ (-\lambda)^{1 / 2} },~~~ g \to \f{g \lambda^2}{m^2} 
\ee
to reduce the action with (\ref{phi6mlig}) to the action with (\ref{phi6000}) multiplied by constant $m^{4-D}/|\li|)$, $D$ being the dimension of the spacetime. In other words, the classical solutions obtained with potential \eqref{phi6000} can be used to generate all the solutions for the models with different values of $m$, $\li$ and $g$, by inverting the scalings of \eqref{variableScaling0}. Quantum mechanically, however, rescaling the action $S\to S m^{4-D}/|\li|)$ is equivalent to rescaling the reduced Planck constant by $\hbar\to \hbar |\li|/m^{4-D}$, say, in the path integral formulation. Thus, quantum mechanically, the potentials with different $m$, $\li$ and $g$ are generally inequivalent and should be treated separately. Indeed, quantum corrections are generally non-negligible if the nonlinear couplings $\li$ and $g$ are not small \cite{Lee:1991ax, Xie:2023psz}.

There are also solutions where the ``particle'' goes over the peak into negative values of $f$ and then returns to stop at the peak at the origin, or even oscillating multiple times in the valleys of the potential before that. However, these multi-nodal solutions, or excited Q-balls, have higher energies and are unstable. To see this, suppose there is a node $f(r_n)=0$ at $r_n$, across which $f(r)$ changes its sign. Consider a modified profile $\tilde f(r)$ by flipping one side of the node from $f$ to $-f$. According to \eref{EnergywithQ}, the energy of the new profile $\tilde f(r)$ must be the same as that of $f(r)$. However, the function $\tilde f(r)$ has a cusp, so a lower energy profile can be obtained by smoothing out the cusp \cite{Lee:1991ax}. These radially excited Q-balls have been explored numerically in \cite{Mai:2012cx}, and analytical approximations to them have also been developed \cite{Almumin:2021gax}.

Substituting the Q-ball solution \eqref{phidef0} into the definition of total charge $Q$ (or equivalently varying \eref{EnergywithQ} with respect to $\oi$),  we get
\be
\label{Qdef1}
Q=2 \oi  {\cal V}_{d-1} \int  \d r r^{d-1}   f^2 .
\ee
We see that the sign of $Q$ is determined by $\oi$. Sometimes, we shall refer to a Q-ball with a negative charge as an anti-Q-ball, especially when it is necessary to distinguish it from one with a positive charge. An anti-Q-ball rotates in the opposite direction in the internal field space. In the following of this paper, we shall often focus on the case of 
\be
\oi>0 ~~{\rm and}~~ Q>0 
\ee
when discussing generic properties of one Q-ball. This choice is made without loss of generality, owing to the U(1) symmetry.

Asymptotically at large $r$, $\varphi$ is small, so the linearized equation of motion for $\varphi$ provides a good approximation to compute the tail of the Q-ball profile. Since $\oi<m$, the asymptotic Q-ball solution is given by
\be
\varphi (r\to \infty,t) \propto  \frac{1}{r^{(d-1)/2}} e^{-\sqrt{m^2-\oi^2} r -i\oi t}  .
\ee
So the Q-ball profile has a tail that falls off exponentially with $r$. In comparison, if $\oi>m$, we would have a spherical wave solution
\be 
\varphi (r\to \infty,t) \propto  \frac{1}{r^{(d-1)/2}} e^{-i\sqrt{\oi^2-m^2} r -i\oi t}  .
\ee
In some sense, it may be argued that the Q-ball solution is the analytic continuation of the wave solution from $\oi>m$ to $\oi<m$ \cite{Lee:1991ax}. 

For a non-topological soliton solution, the Virial theorem can be a useful tool  \cite{Friedberg:1976me, Kusenko:1997ad}. To obtain the Virial expression for a Q-ball, we write its energy as
\bal
E &= \langle \oi^2 f^2 + (\nd f)^2 +V(f) \rangle
\\
&= \f{Q^2}{ 4\langle f^2\rangle} + \langle (\nd f)^2 \rangle + \langle V(f) \rangle ,
\eal
where the averaged is defined as $\langle \cdots \rangle\equiv \int \d^d x (\cdots)$ and we have used \eref{Qdef1} in the second equality. Now, since the Q-ball solution is the stationary point of the energy functional $E[f(\bfx)]$, we must have $\d E[f(\ai\bfx)] /\d \ai|_{\ai=1}=0$, which leads to
\be
 d\, \langle V(f) \rangle= (2-d) \langle (\nd f)^2 \rangle + \f{d\,Q^2}{ 4\langle f^2\rangle}  .
\ee
This is essentially the same argument that leads to Derrick's theorem for a static solution. From this Virial theorem, we can also see that it is the presence of the $Q^2$ term that allows the Q-ball solution to evade the Derrick theorem.

\subsubsection{Stability}
\label{sec:QballStab}

The stability of a Q-ball can be characterized by the $E$-$Q$ (energy-charge) curve, which can be obtained by numerically computing a series of Q-balls with different charges. An important feature of this curve is that its gradient is given by the internal frequency $\oi$:
\be
\label{dEdQoi}
\f{\d E}{\d Q} = \oi .
\ee
This relation can be derived by differentiating \eref{EnergywithQ} with respect to $Q$. In doing this, one should note the dependence of $\oi$ and $f$ on $Q$ and additionally make use of \eref{Qdef1} along with the equation of motion for $f(r)$.

\begin{figure}
	\centering
		\includegraphics[height=4.5cm]{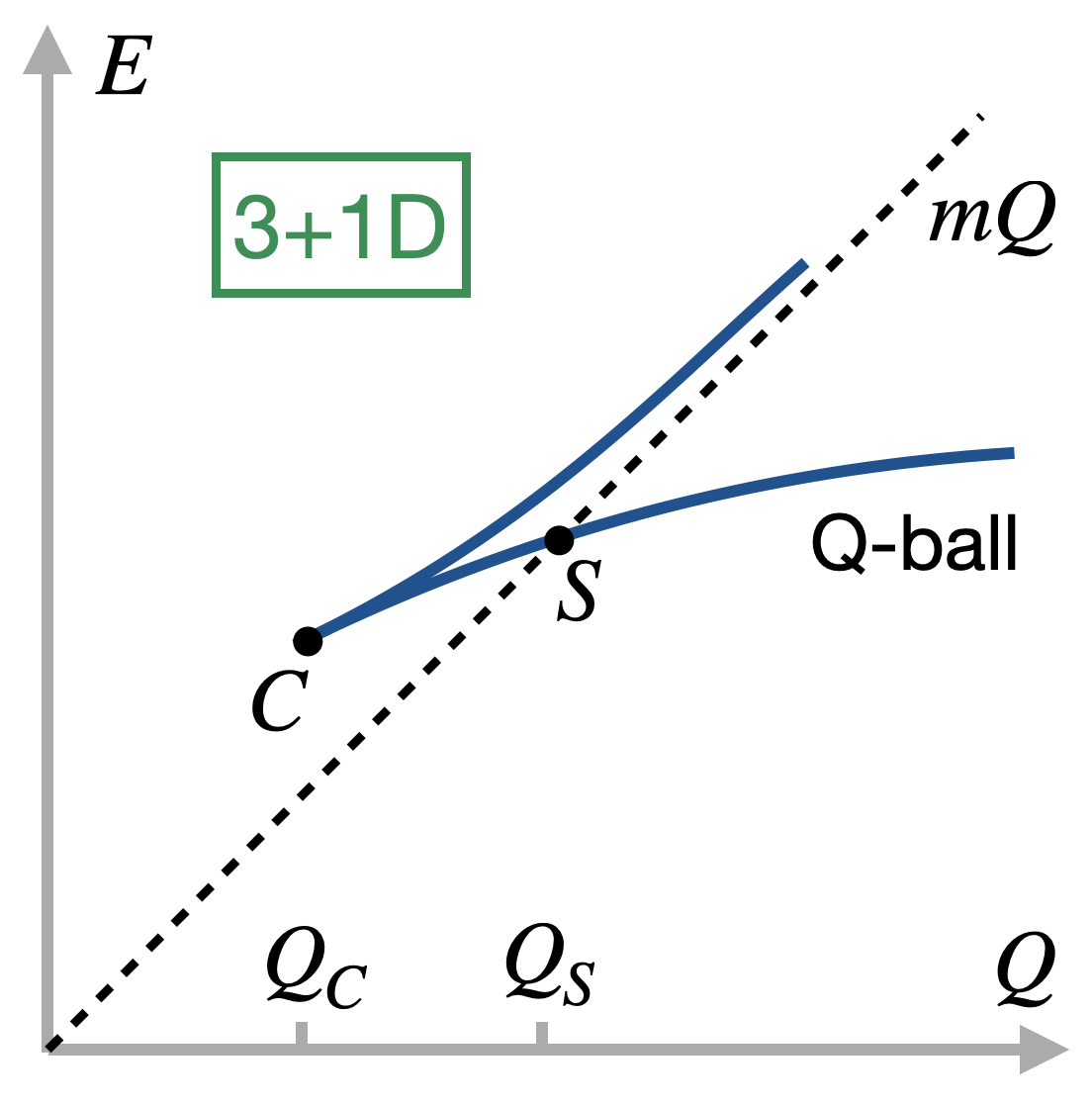}
	\caption{Schematic plot of the $E$-$Q$ curve of the Q-ball solutions in 3+1D for the case of a negative quartic coupling $\li<0$, assuming, without loss of generality, that $Q>0$. For the lower energy branch, when $Q_C<Q<Q_S$, the Q-ball is only classically stable. When $Q>Q_S$, the Q-ball is also stable against perturbatively emitting quanta.}
 \label{fig:EvsQ}
\end{figure}
 
Let us now focus on the case where the quartic coupling is negative
\be
\li<0 .
\ee
A schematic plot of the $E$-$Q$ curve in 3+1D can be found in Figure \ref{fig:EvsQ} for a Q-ball with $Q>0$ \cite{Lee:1991ax, Friedberg:1976me}. When $\oi$ is close to the mass $m$, the energy of the Q-ball is greater than the energy of $Q$ free particles: $E>Qm$, which means that the Q-ball is unstable. The point $C$ marks the critical point where the difference $E-Qm$ begins to decrease as $\oi$ is decreased from $m$. So there are two branches of solutions when $Q>Q_C$. Only the lower branch is potentially stable. For a given $Q$, the energy of a solution on the upper branch is higher than both the corresponding solution in the lower branch and the plane waves/free particle solution (the dashed line), so the upper branch represents the peak of an energy barrier between two locally stable solutions. From Figure \ref{fig:EvsQ}, we see that $Q(\oi)$ also reaches its minimum at the critical point $C$, and from Figure \ref{fig:profiles}, we see that $Q$ decreases to its minimum as $\oi$ increases from small values. So for the Q-ball to be on the potentially stable lower branch, we need 
\be
\label{dQdoiQball}
\f{\d Q}{\d \oi}\le 0 .
\ee
When the Q-ball on the lower branch is sufficiently large with $Q>Q_S$, the energy of the Q-ball is lower than that of $Q$ free particles. In this case, the Q-ball is stable even against perturbatively emitting particles at the quantum level. Note that, barring unusual behaviour of the potential at large field values, this holds for any $Q$ greater $Q_S$. To see this, suppose we have a $Q$ satisfying $Q>Q_S$, meaning the corresponding energy is below the $mQ$ line ($E<mQ$) in Figure \ref{fig:EvsQ}. We can always add free particles at infinity, which adds $\Delta Q$ to $Q$ and add $m\Delta Q$ to $E$. This leads to a configuration (with a larger $Q$) that is below the $mQ$ line  $E+m\Delta Q < m(Q+\Delta Q)$, and the corresponding Q-ball solution can only have less energy. (In the presence of gravity, however, the charge of a non-topological soliton/boson star is limited by an upper bound beyond which it becomes unstable \cite{Liebling:2012fv}.) See \cite{Sakai:2007ft} for a connection between the stability of Q-balls and the catastrophe theory.

In fact, if the potential $V$ does not dip too much below the mass term $m|\varphi|^2$, {\it i.e.}, $V_{\rm eff}(|\varphi|) \le \min \left(a, b |\varphi|^c\right)$ for some positive numbers $a,b>0,\,c>2$, Coleman was able to rigorously prove that when $Q>Q_S$ the Q-ball solution is the absolute minimum of the energy functional \cite{Coleman:1985ki}. 

On the other hand, at any point along the curve between points $C$ and $S$, the mass of the Q-ball exceeds that of $Q$ free particles, rendering the soliton quantum mechanically unstable. Nevertheless, the Q-ball turns out to be classically stable \cite{Lee:1991ax, Friedberg:1976me, Smolyakov:2019cld}. In particular, it is stable against emitting infinitesimal perturbations. Since the classical solution provides an accurate approximation in the weak coupling limit, these Q-balls are long-lived even when quantum corrections are considered. The existence of a spike (point $C$) can be inferred from  \eref{dEdQoi} and the fact that $\d E/\d Q>0$ as $\oi$ approaches both $0^+$ and $m^-$. Complementing these earlier analyses, the classical \cite{Battye:2000qj, Campanelli:2007um, Tsumagari:2008bv, Copeland:2009as} and quantum \cite{Tranberg:2013cka, Xie:2023psz} stability of Q-balls have been demonstrated through finite-difference simulations in various dimensions. 

The $E$-$Q$ curve depends sensitively on the spatial dimensions \cite{Lee:1991ax, Friedberg:1976me, Gleiser:2005iq}. In 2+1D, the $E$-$Q$ curve lacks the section above the dashed $E=mQ$ line, and the Q-ball curve intersects the $E=mQ$ line at some finite $Q_S$. In 1+1D, the section above the $E=mQ$ line is also absent, but now the Q-ball curve intersects the $E=mQ$ line at $Q=0$. These differences can be understood analytically \cite{Lee:1991ax}. To this end, note that an improved approximation of the profile function at large $r$ as $\oi$ approaches $m$ from blow is
\be
f(r)\sim \f{(m^2-\oi^2)^\f12}{r^{(d-1)/2}} e^{-\sqrt{m^2-\oi^2} r},~~~~\oi \to m^ - ,
\ee
where the pre-factor $(m^2-\oi^2)^{1/2}$ is obtained by seeking the zero of the effective potential $V_{\rm eff}(f)=0$, truncated at quartic order. Plugging this improved expression for $f$ into \eref{Qdef1}, we find that
\bal
Q  \sim 2m (m^2-\oi^2)  (m^2-\oi^2)^{-\f{d}2}\mc{O}(1), ~~\oi \to m^- .
\eal
So when $d\geq 3$, $Q$ goes infinity as $\oi \to m^-$. On the other hand, when $d=2$, $Q\to 2m \mc{O}(1)$ as $\oi \to m^-$; when $d=1$, $Q\to  0$ as $\oi \to m^-$.

As Q-balls are dynamically stable, they are attractor solutions and can naturally form from a wide range of initial configurations. This allows them to have many applications in the early universe, where the initial conditions often consist of stochastic fluctuations, as will be discussed in  \S\,\!  \ref{sec:applications}.

\subsubsection{Analytical solutions}
\label{sec:analySol}

It is instructive to illustrate some of the salient properties of non-topological solitons with analytical solutions. Let us first consider a 1+1D model with a generic $|\varphi|^6$ potential
\bal
V & =  m^2 |\varphi|^2 + \li |\varphi|^4 + g |\varphi|^6
\\
&=\f{m^2  |\varphi|^2}{1+\varepsilon^2} \left[ \left(1-\eta^2  |\varphi|^2 \right)^2+\varepsilon^2\right] ,
\eal
where we have traded the couplings $\li$ and $g$ for $\eta^2=2g/(-\li)$ and $\varepsilon^2=(4gm^2-\li^2)/\li^2$. In 1+1D, the friction term in \eref{fEoM} is absent, so we can integrate to get an analytical form for the profile function $f(x)$ \cite{Lee:1991ax}
\be
\varphi(t,x)=\f1{\eta} \sqrt{\f{a}{1+\sqrt{1-a} \cosh y}} e^{-{i} \omega t},
\ee
where we have defined $a= {(1+\varepsilon^2 ) (m^2-\omega^2 )}/{ m^2 }$ and $y=2 \sqrt{m^2-\omega^2}(x-\xi)$, $\xi$ being the central location of the soliton. We see that the soliton solution has a simple pole in the coupling $\eta$. This singularity means that the soliton solution cannot be captured by a perturbative expansion in $\eta$. 

In fact, this is a generic feature for a potential $V(\varphi)$ in which all the nonlinear couplings scale universally. In such cases, we can introduce a rescaled field $\varphi=\eta^{-1}\bar\varphi$ and a rescaled potential $V(\varphi) = \eta^{-2} \bar{V}(\bar{\varphi})$, where $V$ is assumed to have a minimum at $\varphi=0$. With these rescalings, the universal coupling can be factored out of the Lagrangian (\ref{originLag}):
\be
\mc{L} =  \eta^{-2}\bar{\mc{L}} = \eta^{-2} (-\partial_\mu \bar\varphi^{\dagger} \partial^\mu \bar\varphi- \bar{V}(\bar\varphi)) .
\ee
Thus, the equation of motion for $\bar\varphi$ is independent of $\eta$, which means that the solution $\varphi$ has a simple pole at $\eta=0$. This also implies that a weaker coupling typically results in a larger field amplitude for the soliton. Additionally, since $\eta$ contributes an overall factor $\eta^{-2}$ in front of the rescaled action, in the path integral quantization, this overall factor is always accompanied by $\hbar$ in the phase: $e^{i\int \d^{d+1} x \bar{\mc{L}}/(\hbar \eta^2)}$. So, loosely speaking, a smaller $\eta$ results in smaller quantum effects, allowing for a better approximation by the classical solutions. This can be demonstrated with nonperturbative lattice simulations of Q-balls in higher dimensional spacetimes \cite{Xie:2023psz}.

For some potentials in 1+1D, where analytical solutions can also be obtained, Q-balls may exhibit a split profile \cite{Bazeia:2016xrf}, compact support \cite{Bazeia:2016wco} or a tail that decays faster than exponentially \cite{Bazeia:2019ymk}. In 3+1D, analytic solutions can be obtained for (piecewise) parabolic potentials \cite{Theodorakis:2000bz, Gulamov:2013ema}.

Remarkably, an analytic solution can also be found in 3+1D for a potential with a logarithmically modified mass term \cite{Enqvist:1997si, Enqvist:1998en}
\be
\label{logVnon}
V(\varphi)= m^2 |\varphi|^2\left[1+K \ln \left(\frac{|\varphi|^2}{ M^2}\right)\right]  ,
\ee
where $M$ is a mass scale and $K$ must be negative for the existence of Q-balls. This is essentially a quantum effective potential taking into account the running of the mass, typically arising along some flat directions lifted by soft masses in supersymmetric extensions of the Standard Model (see \S\,\! \ref{sec:gravitySUSY}). This model also has a direct bearing in the Affleck-Dine baryogenesis \cite{Dine:2003ax} in the early universe, as will be discussed in more detail later. The potential \eqref{logVnon} is not bounded from below, so a higher order term ${\lambda_n|\varphi|^{2n-2}}/{\Lambda^{2 n-6}}$ should be added in principle for vacuum stability. However, as $\Lambda$ is a large scale, the $\li_n$ term is usually negligible in the Q-ball dynamics. Without the $\li_n$ term, the basic Q-ball solution of this logarithmic potential can be obtained analytically \cite{Bialynicki-Birula:1975nws}
\bal
\label{gaussianQball}
 &~~~~~~~~~~~~~~~~\varphi(r,t)  =  A e^{-\f{r^2}{2\si^2}} e^{i \oi t}
\\
A&=M e^{(m^2-\omega^2-2 m^2 K) / (-2 m^2 K)}, ~~ \si=(-Km^2)^{-1/2} ,
\nonumber
\eal
which has a Gaussian profile. The variance of the Gaussian profile can be used to define the radius of the Q-ball: $R\sim (-Km^2)^{-1/2}$. The energy to the charge ratio of such a Q-ball is given by
\be
\label{EQgaussian}
\f{E_Q}{Q} = \oi - \frac{K m^2}{2 \oi} .
\ee
We see again that in the limit of small coupling $|Km^2|\to 0$, the amplitude, as well as the width, of the solution becomes very large.

\subsubsection{Thin-wall Q-ball}
\label{sec:thinwall}

A particularly interesting type of Q-ball appears when the profile function $f(r)$ of the Q-ball closely resembles a step function, remaining nearly constant within a radius and dropping to zero outside:
\be
\label{thinwallansatz}
f(r)= 
\begin{cases}
f, & r<R ,
\\ 
0, & r \geq R.
\end{cases}
\ee
This type of Q-ball, known as a thin-wall Q-ball \cite{Coleman:1985ki}, occurs when the internal frequency of the Q-ball, $\oi$, is close to the lower limit $\oi_-^2= {\rm min}\({V(|f|)}/{f^2}\)$, the trend of which can be seen in Figure \ref{fig:profiles}. As $\oi$ approaches $\oi_-$, the total charge of the Q-ball goes to infinity. Inside such a Q-ball, the energy density is almost uniform, and this state is sometimes referred to as Q-matter. 

A thin-wall Q-ball can be readily analyzed analytically, which makes one of the main existence conditions (\ref{oiminusDef}) more explicit. Since a thin-wall Q-ball is large, we can neglect the surface tension of the ball. Additionally, the interior is almost homogeneous, so we can neglect the spatial gradient terms in the energy functional. Again assuming the Q-ball is of the form (\ref{phidef0}), we can write the energy of the ball as
\bal
\label{Eapprox0}
E  = \int \d^3 x \[ \dot{\varphi}^{\dagger} \dot{\varphi} +V(|\varphi|) \] 
= \oi^2f^2 \mc{V} + V \mc{V} .
\eal
where $\mc{V}$ is the volume of the Q-ball within radius $R$. On the other hand, from \eref{Qdef1}, we get $Q =  2\oi f^2 \mc{V}$. As we are looking for the minimum of the energy for a fixed $Q$, we can write the energy as
\be
E  =  \f{Q^2}{4f^2 \mc{V}} + V \mc{V} .
\ee 
In the above equation, $\mc{V}$ is an independent parameter that can be varied to minimize the energy, which leads to
\be
E|_{\mc{V}= Q/(2f\sqrt{V})}=  Q \sqrt{V/f^2} .
\ee
Then, for fixed $Q$, the minimum energy is obtained if $V/f^2$ has a minimum, which is just the existence condition (\ref{oiminusDef}). So for a large thin-wall Q-ball, its energy is given by \cite{Coleman:1985ki}
\be
\label{EtwQ}
E_\text{thw}=  Q \sqrt{{\rm min}(V/f^2)} \equiv Q \sqrt{V(|f_0|)/f_0^2} ,
\ee
where $f_0$ is fixed by the form of the potential, independent of any initial conditions.
Also, for a thin-wall Q-ball, the internal frequency $\oi$ approaches $\oi_-=\sqrt{V(|f_0|)/f_0^2}$, independent of $Q$.

The charge of a thin-wall Q-ball is greater than the critical charge $Q_S$ (cf.~Figure \ref{fig:EvsQ}), so it is stable against dissipating into free particles. This is readily seen from \eref{EtwQ}, which gives the energy per unit charge within a thin-wall Q-ball
\be
\label{twEQ}
E_{\rm thw}/Q = \sqrt{V(f_0)/f_0^2} .
\ee
This obviously is less than the mass of a free $\varphi$ particle, $m$, since $\oi_- <\oi_+=m$ is required for the Q-ball solution to exist.

It is instructive to consider a very small energy-charge ratio $E_{\rm thw}/Q$, which corresponds to a small $\oi_-$. As $\oi_-$ goes to zero, we have $V(f_0)\to 0$, and the thin-wall Q-ball becomes a degenerate vacuum at $f_0$. From this perspective, for a finite $\oi_-$, the interior of a Q-ball is much like a false vacuum, so one may wonder whether quantum tunneling can occur within a large Q-ball and destablize it. This is impossible because quantum tunneling must conserve $Q$ and $E$ when transiting from one configuration to another, but the Q-ball solution is the only configuration for such a setup, since $E^\text{thw}_{Q}$ is the minimum of $E$ for a given $Q$ \cite{Coleman:1985ki}.

Now, let us compute the energy from the surface tension of a thin-wall Q-ball. After all, it is the surface tension that makes the Q-ball spherically symmetric. To this end, we can approximate $\oi$ with $\oi_-$ and neglect the friction term in \eref{fEoM} (due to the large $r\simeq R$ suppression), and then \eref{fEoM} can be written as
\be
\frac{\d^2 f}{\d r^2}   =  \f12 \frac{\partial  (V -\omega_-^2 f^2)}{\partial f}  ,
\ee
In the mechanical analogy, this describes a ``particle" subject to a conserved potential $\f12(\omega_-^2 f^2-V)$, so it has the following ``energy conservation''
\be
\label{fVsurface0}
 \(\frac{\d f}{\d r}\)^2   =   V -\omega_-^2 f^2  ,
\ee
where we have taken into account that the ``particle's energy'' vanishes at $r=\infty$. The above equation allows us to integrate to define the radius of the Q-ball
\be
R-r=\int_{F}^f { \mathrm{d} f}/{ \sqrt{ V -\omega_-^2 f^2 }}  ,
\ee
where $f$'s boundary value $F$ is defined to satisfy the relation $\int \mathrm{d}^3 x f^2=\frac{4}{3} \pi R^3 f_0^2$. Therefore, in addition to  the volume energy (\ref{Eapprox0}), a thin-wall Q-ball has the surface energy
\bal
E_{\rm surf} &=  4 \pi R^2 \int_0^R \mathrm{d} r\left[\left(f'\right)^2+V -\omega_-^2 f^2\right]
\\
&=8 \pi R^2 \int_0^{f_0} \! \mathrm{d} f   \sqrt{V -\omega_-^2 f^2} ,
\eal
where \eref{fVsurface0} has been used to obtain the second equality above.

In the thin-wall limit, it is also possible to analytically calculate the low-lying vibration modes of the Q-ball, which correspond to the lowest excited levels of a large Q-ball in the quantum theory  \cite{Coleman:1985ki} (see \cite{Kovtun:2018jae} for the vibration modes of Q-balls in a piecewise parabolic potential). These modes are associated with the internal rotation and the translation normal to the surface of the ball, and can be obtained from the small wave-vector limit of the bulk sound wave and the surface wave respectively. They are zero modes in the sense that their frequencies go to zero as the wave-vector approaches zero.

An improved thin-wall ansatz has recently been developed, which deals more carefully with the transition region around $R$ and leads to excellent analytical approximations to the Q-ball solutions beyond the strict thin-wall limit \cite{Heeck:2020bau, Heeck:2022iky}. For the sextic potential, the accuracy is at least about 10\% and much smaller for large Q-balls.

\subsubsection{Non-analytic potentials and other models}

\label{sec:nonAnalyPot}

Within a U(1) symmetric scalar theory, if non-analytic potentials are allowed, stable Q-balls can exist in a potential without higher dimensional operators. In fact, in those theories, stable Q-balls with arbitrarily small charges, sometimes called {\it thick-wall} Q-balls\footnote{Thick-wall Q-balls may generally refer to Q-balls with thick-wall profiles, {\it i.e.,} Q-balls that are not of the thin-wall type.}, can be found \cite{Kusenko:1997ad, PaccettiCorreia:2001wtt}. On the other hand, in theories with multiple scalars, it is straightforward to construct analytical, cubic Lagrangian terms. Indeed, cubic terms are generically present in the supersymmetric extensions of the Standard model, which may lead to the production of small Q-balls with baryon or lepton numbers in the early universe \cite{Kusenko:1997zq}. The thick-wall limit has been analyzed for theories involving a U(1) scalar plus multiple real scalars \cite{Lennon:2021zzx}, multiple U(1) scalars \cite{Lennon:2021fde} and scalars with non-canonical kinetic terms \cite{Lennon:2021uqu, Faundez:2023svu, Kuniyasu:2016tse}. 

Suppose that the potential in Lagrangian (\ref{originLag}) in 3+1D is given by 
\be
V(|\varphi|) = m^2 |\varphi|^2 + \li |\varphi|^p + g |\varphi|^4 .
\ee
For stable Q-balls to exist, we require $\li<0$ and $g>0$, with $p$ being a positive number within the range $2<p<4$. For example, a negative cubic term may arise from finite temperature corrections \cite{Pearce:2022ovj}. Generic Q-ball solutions in this model can again be obtained by solving \eref{fEoM} numerically. However, here let us focus on the small $Q$ limit where the wall of the Q-ball is very thick, which can be analyzed semi-analytically \cite{Kusenko:1997ad}, inspired by a similar scenario in false vacuum decay \cite{Linde:1981zj} (see \cite{Espinosa:2023osv} for a novel approach to computing the Q-ball solutions derived from a new technique to compute the tunneling bounces.). The main observation is that in the thick-wall limit the influence of the $|\varphi|^4$ term in the potential is negligible for computing the Q-ball solution. This can be easily understood by considering the picture of a particle rolling in the effective potential (see  \S\,\! \ref{sec:QballExist}). In the thick-wall limit, $\oi$ approaches $m$ from below, so the negative (valley) part of $V_{\rm eff}$ becomes very small and thus even the maximum value of $f$ within the Q-ball is close to zero. (Because of the small field values, quantum effects could be significant in these thick-wall Q-balls \cite{Graham:2001hr}. The semi-classical decay rate of a small Q-ball has been computed using a method similar to that of the Euclidean bounce in the false vacuum decay \cite{Levkov:2017paj}; see \S\,\! \ref{sec:QballQuantTun}.)

Within the classical approximation and neglecting the $|\varphi|^4$ term, by defining dimensionless variables $\hat f$ and $\hat r$ with appropriate powers of $|\li|$ and $m^2-\omega^2$, we can express the energy (\ref{EnergywithQ}) as
\be
{E}=\omega Q+\left(m^2-\omega^2\right)^{\f{6-p}{2 p-4}}|\li|^{-\f{2}{p-2}} E_p ,
\ee
where the minimum of $E_p= 4\pi\! \int\! \mathrm{d} \hat{r}  \hat{r}^2  \big[ (\pd_{\hat{r}} \hat{f})^2  + \hat{f}^2-\hat{f}^p \big]$, if needed, can be computed numerically for any given $p$. Then, to obtain the energy of the thick-wall Q-ball, we only need to minimize $E$ with respect to $\oi$ (and use this minimization condition to eliminate $|\li|^{-\f{2}{p-2}} E_p$ in favour of $\oi$) to get
\be
E_Q=m Q\left[\f{\oi}{m}+ \frac{p-2}{6-p} \frac{\left(m^2-\oi^2\right)}{m\oi}\right] .
\ee
(If we eliminated $\oi$ in favour of $|\li|^{-\f{2}{p-2}} E_p$, we would, on the other hand, get a perturbative series in the smallness of $Q$.) Now, for the Q-ball to be stable (perturbatively at the quantum level), the quantity in the above square bracket must be less than 1. Combining with the existence condition $w<m$, a simple algebraic analysis allows us to infer that $p$ must fall in the range  \cite{PaccettiCorreia:2001wtt,Lennon:2021uqu}
\be
2 < p < \f{10}3 
\ee
for the thick-wall Q-ball (a Q-ball with a very small charge) to exist. This is consistent with the standard Q-ball with $p=4$ discussed in \S\,\! \ref{sec:Qballbasics}, in which case there is a minimum charge, $Q_S$, for the Q-ball to be stable (see Figure \ref{fig:EvsQ}).

Another widely studied non-analytic potential is one with a logarithmically modified mass term, arising from an effective potential generated by loop corrections, as already mentioned in \S\,\! \ref{sec:analySol} around \eref{logVnon}, with further discussions on its applications in the early universe to follow.

In a U(1) symmetric theory, Q-balls with a parametrically small range of possible charges and energies, thus more particle-like, can be realized by designing the potential to decrease at large field values. \cite{Nugaev:2013poa}. On the other hand, nonlinear sigma(-like) models such as those with the $CP^N$ (the coset SU(N+1)/(SU(N)$\otimes$U(1))) target space are interesting for the existence of topological solitons in lower spacetime dimensions. Refs.~\cite{Klimas:2017eft, Klimas:2021eue, Klimas:2022ghu} studied Q-balls, Q-shells and their excited states in the 3+1D $CP^N$ model with an odd number of scalars, extended by a potential, and found that the energy and charge relations of these objects behave like $E\sim Q^{5/6}$. Q-balls and Q-shells from gauged $CP^N$-like models have also been investigated \cite{Klimas:2023zxm}. Moreover, Q-balls are found to exist in a truncated chiral Lagrangian if the pseudo-Nambu-Golstone bosons are coupled to a Higgs field \cite{Bishara:2021fag}.

\subsection{Complex Q-balls}
\label{sec:complexQballs}

As extended objects without topological restrictions, the spatial configurations of Q-balls can deviate from spherical symmetry, exhibiting a range of complex structures.

\subsubsection{Spinning Q-balls}
\label{sec:spinningQball}

There are, surprisingly, no known spinning topological solitons in 3+1D Minkowski space (see \cite{Sutcliffe:2023xau} for 2+1D Q-lumps, which are spinning topological solitons that, like Q-balls, are not static but stationary). For instance, it has been demonstrated that the 't Hooft-Polyakov monopole cannot be endowed with a small angular momentum \cite{Heusler:1998ec}, although the possibility of fast rotation remains open. In contrast, spinning non-topological solitons have been found  \cite{Volkov:2002aj, Kleihaus:2005me, Kleihaus:2007vk, Campanelli:2009su}. Interestingly, for a given total charge $Q$, the angular momentum of the spinning Q-ball is quantized.

Consider the simplest U(1) symmetric model (\ref{originLag}). The 3+1D spinning Q-ball solution takes the form
\begin{equation}
\label{spinQball0}
\varphi = f(r, \theta) e^{-i\left(\omega t-\tilde m \phi\right)},
\end{equation}
where $\thi$ and $\phi$ are the spherical angles and $\tilde m$ is an integer. When $\tilde m=0$ and $f$ depends only on $r$, the solution reduces to the standard spherical Q-ball, which only rotates in the internal field space spanned by the real and imaginary components of $\varphi$. For $\tilde m\neq 0$, the phase depends on the azimuthal angle $\phi$, so the Q-ball field also, in a sense, rotates in real space with angular phase velocity $\Omega_Q=\oi/\tilde m$. In fact, as mentioned above, the angular momentum of the spinning Q-ball has to be {\it integer multiples} of the total charge $Q$:
\be
J=2 \pi \int \mathrm{d} r \mathrm{d} \theta r^2 \sin \theta T_{0 \phi}=\tilde m Q .
\ee
(A Q-ball can, in a sense, rotate slowly with a small angular momentum for a given total charge if, instead of \eref{spinQball0}, one considers a generic harmonic expansion on top of the spherical Q-ball solution \cite{Almumin:2023wwi}. However, these slowly spinning Q-balls are unstable, constantly emitting energy and lacking a well-defined angular velocity, and are thus more akin to oscillons; see \S\,\! \ref{sec:complexOsc}.)

To obtain the spinning Q-ball solution, we can expand $f(r, \theta)$ in terms of partial waves
\be
\label{fellExp}
f(r, \theta)=\sum_{\ell=|\tilde m|}^{\infty} f_{\ell}(r) P_{\ell}^{\tilde m}(\cos \theta) ,
\ee
where $P_{\ell}^{\tilde m}(\cos \theta)$ is the associated Legendre function. Substituting the above expansion into the equations of motion and expanding the nonlinear terms in the basis of $P_{\ell}^{\tilde m}(\cos \theta)$, we get a system of coupled ordinary differential equations (ODEs) for $f_\ell$. To compute the soliton solution, the boundary conditions are also needed, which can be obtained by taking the $r\to 0$ and $r\to \infty$ limits of the $f_\ell$ equations of motion:
\be
f_{\ell}(r \rightarrow \infty) \rightarrow 0, \quad f_{\ell}(r \rightarrow 0) \rightarrow\left(\kappa_{\ell} r\right)^{\ell} ,
\ee
with $\kappa_\ell$ being constant. From these boundary conditions, we see that a spinning Q-ball is hollow in the center, unlike the basic Q-ball, whose energy density peaks in the center. Then, truncating $\ell$ at a finite order $\ell_{\rm max}$, the solution can be obtained by, for example, a high dimensional shooting method. In practice, the energy of the spinning Q-ball converges with only a few partial waves.

\begin{figure}[tbp]
\centering
\includegraphics[width=0.23\textwidth]{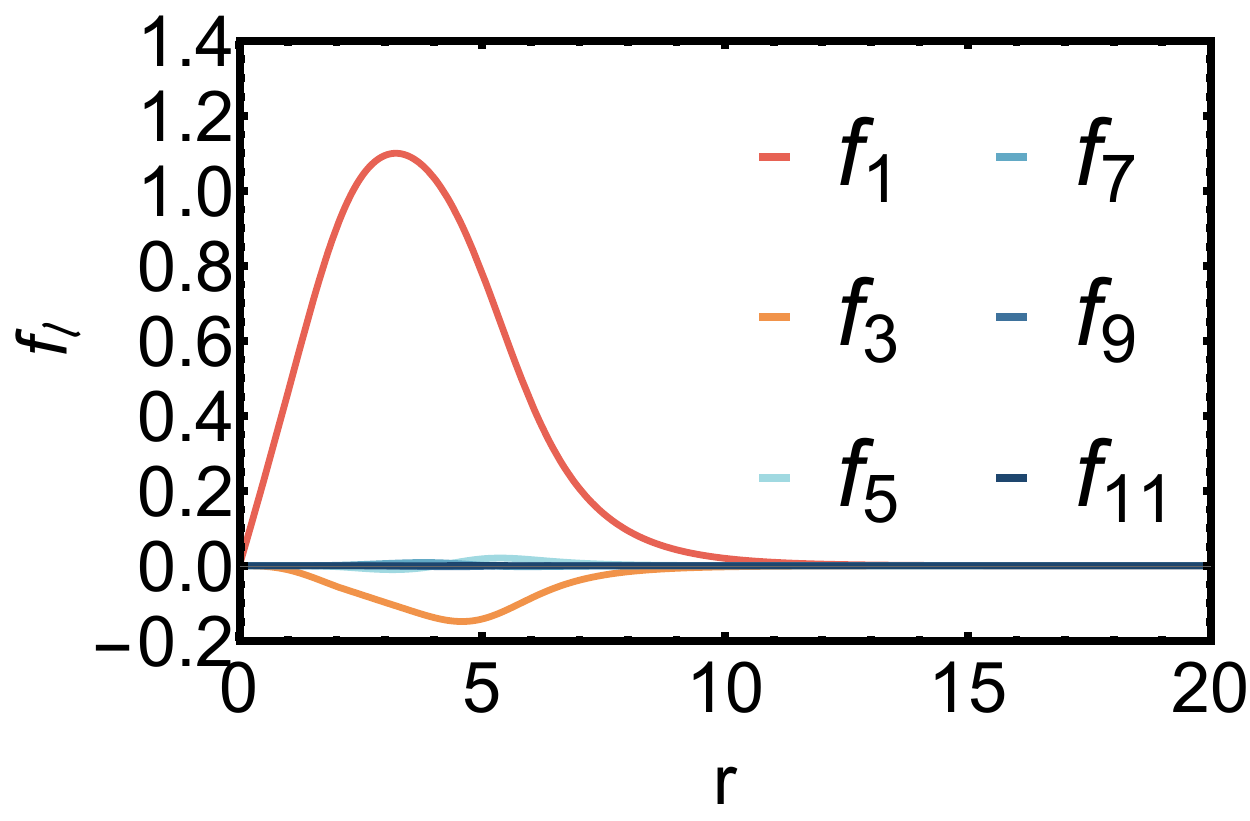}
\includegraphics[width=0.23\textwidth]{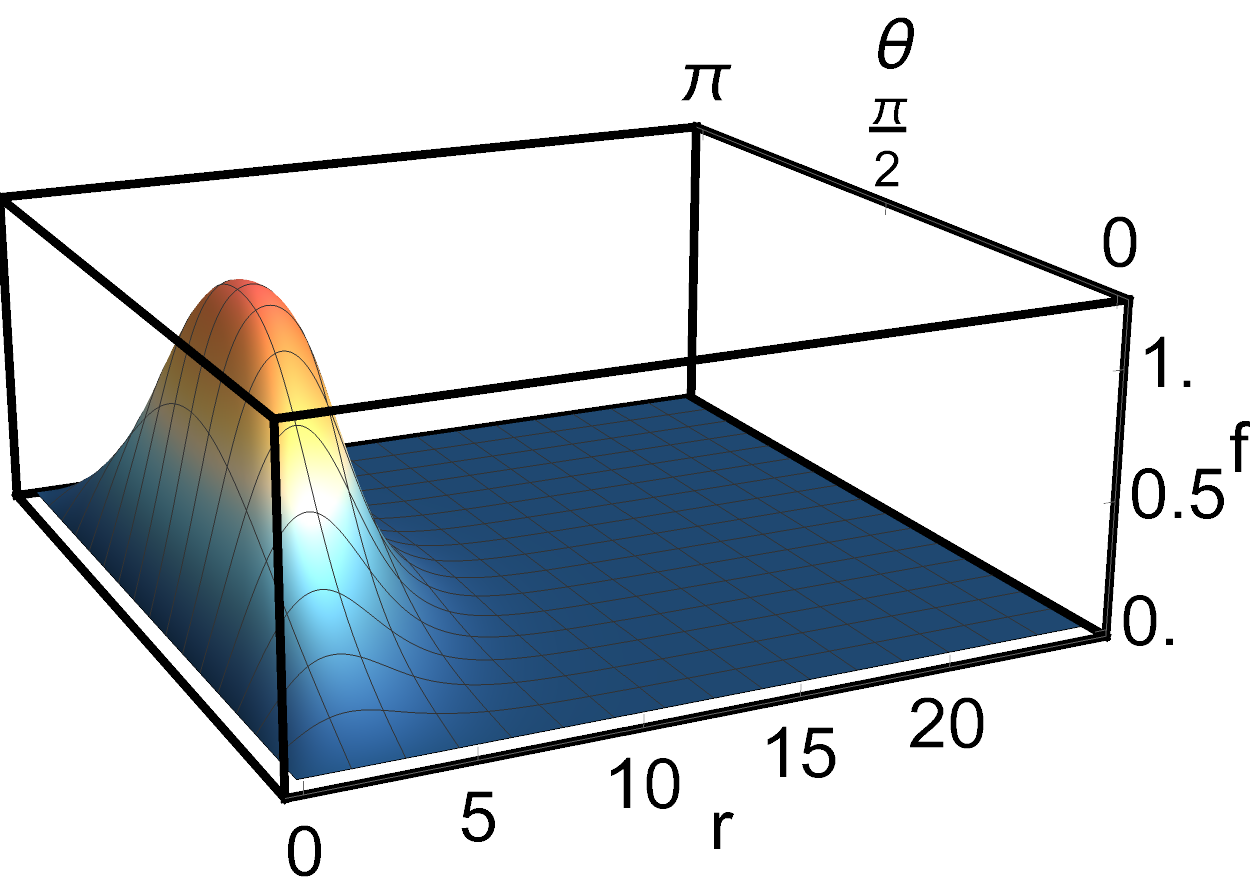}

\includegraphics[width=0.23\textwidth]{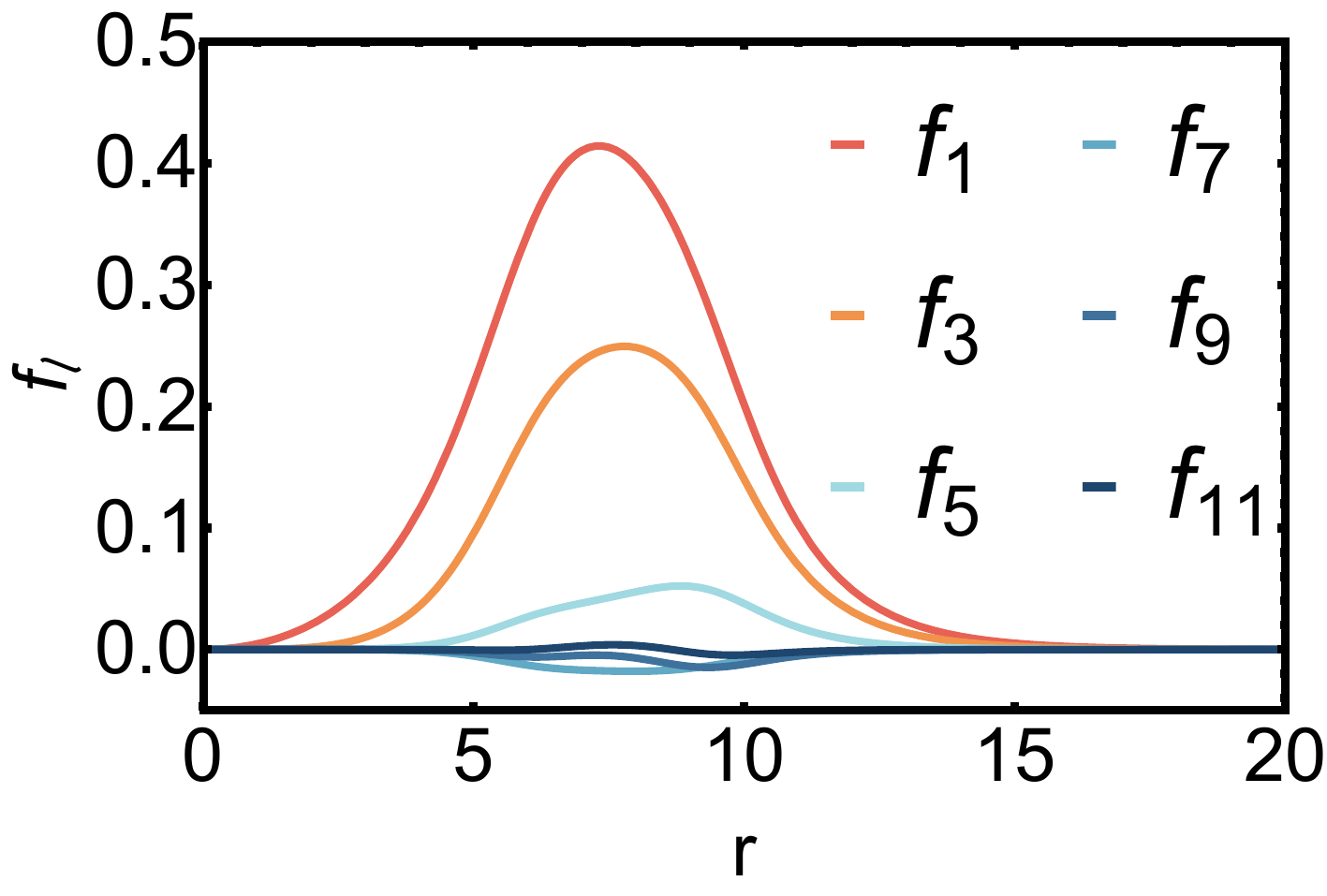}
\includegraphics[width=0.23\textwidth]{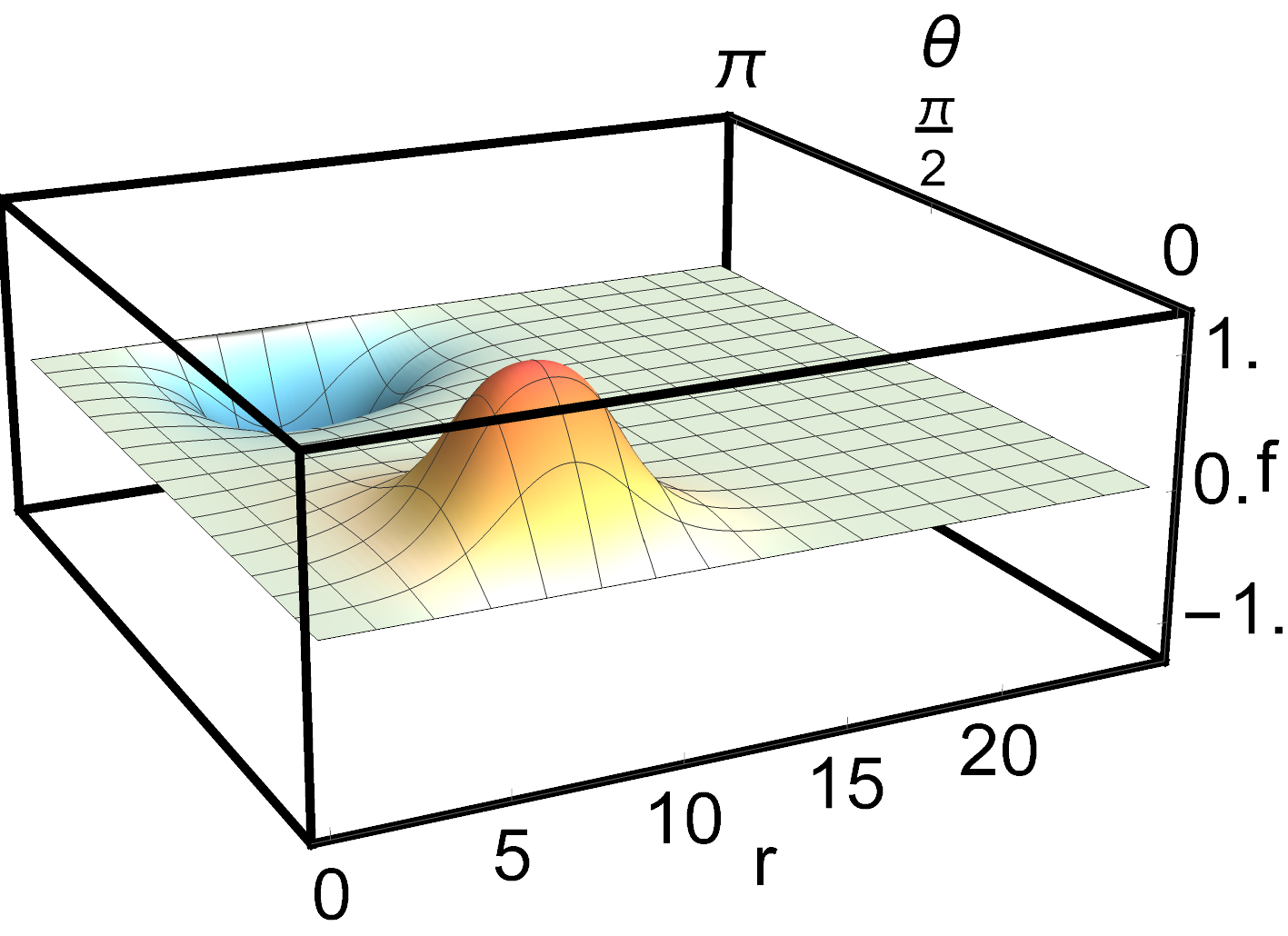}
\caption{Radial profiles for the leading partial waves $f_\ell$ ({\it left}) and total angular dependent profiles $f(r,\thi)$ ({\it right}) of an even-parity ({\it top}) and odd-parity ({\it bottom}) spinning $Q$-ball for potential $V(|\varphi|)=|\varphi|^2-|\varphi|^4+\f13 |\varphi|^6$. We also choose $\tilde m = 1, \omega = 0.76$. }
\label{fig:spinQball}
\end{figure}

Depending on whether $\varphi$ is a scalar field ($p=0$) or a pseudo-scalar field ($p=1$), the $f(r, \theta)$ function has different parity: $f(r, \pi-\theta)=(-1)^p f(r, \theta)$. Since the theory is invariant under parity and $P_{\ell}^{\tilde m}(-\cos \theta)=(-1)^{\ell+\tilde m} P_{\ell}^{\tilde m}(\cos \theta)$, the $f_{\ell=|\tilde m|+p+2k\neq 0}$ modes with $k=0,1,2,...$ vanish identically. So there are two kinds of spinning Q-balls with different parities. While the energy density of an even-parity Q-ball peaks at the equator, that of an odd-parity Q-ball vanishes at the equator. Examples of the even-parity and odd-parity Q-balls are shown in Figure \ref{fig:spinQball}. In these examples, the energy and charge of the even-parity Q-ball mostly converge already with the leading 3 partial waves, while for the odd-parity Q-ball, the first and second partial wave are of a similar magnitude and the energy and charge mostly converge with the leading 4 partial waves.

In 2+1D, the profile function of a spinning Q-ball only depends on the radius:
\begin{equation}
\varphi = f(r) e^{-i\left(\omega t-\tilde m \phi\right)}  .
\end{equation} 
So, in this case, one only needs to solve an ODE to obtain the spinning Q-ball solution, similar to the basic Q-ball scenario. Morphologically, the spinning Q-ball in 2+1D resembles a ring. 

When gravity is included, there exist rotating boson stars \cite{Yoshida:1997qf}, which were actually discovered before the spinning Q-balls. However, the spinning Q-balls discussed above are not the weak gravity limit of the rotating boson stars of Ref \cite{Yoshida:1997qf}. One major difference is that the scalar part of the model of \cite{Yoshida:1997qf} is a free field, so the nonlinearities that allow the boson star to rotate come from the metric field and its interaction with the scalar.

\subsubsection{Composite Q-balls}
\label{sec:compositeQballs}

Generally, for a theory where basic spherical Q-balls exist, there also exist a tower of composite Q-balls within which positive and negative charges co-exist and swap with each other quasi-periodically \cite{Copeland:2014qra, Xie:2021glp, Hou:2022jcd} (see \cite{Xie:2023psz} for quantum corrections). These charge-swapping Q-balls (CSQs) are not absolutely stable but are long-lived, and may be thought of as the ``molecular'' states of Q-balls. 

Let us take the polynomial potential \eqref{potV0} as an example, truncated up to the $|\varphi|^6$ order for simplicity. A CSQ can be prepared by initially superposing a number of basic Q-balls \cite{Copeland:2014qra}
\be
\label{CSQinitial}
\varphi(t_0,\bfx)=\sum_n  f(|\mathbf{x}-\mathbf{d}_n|)\, e^{-i (\omega_n t_0+\alpha_n)}
\ee
such that the constituent Q-balls have sufficient overlap, where $\mathbf{d}_n$ denotes the central location of the $n$-th Q-ball and $\oi_n$ and $\alpha_n$ are the corresponding frequencies and phases. A simple choice for $\oi_n$ is to set them all equal in magnitude while allowing them to differ by a sign depending on their spatial locations. Of course, the configuration \eqref{CSQinitial} is not stationary, and it will radiate away some energy in the initial relaxation stage and settle down to the quasi-stable CSQ, which has a distinct multipolar structure. While use of the basic Q-balls via superposition \eqref{CSQinitial} is a reliable way to construct these ``molecular'' Q-balls, other non-Q-ball configurations may also be used in the initial setup, which reflects the fact that CSQs are local attractor solutions in the phase space \cite{Xie:2021glp}.

\begin{figure}[tbp]
\centering
\includegraphics[width=0.47\textwidth]{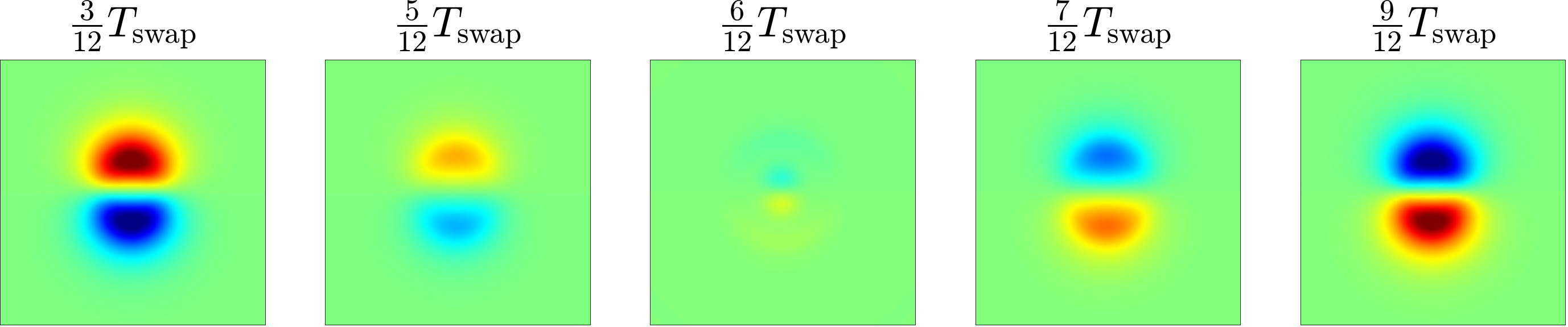}

\includegraphics[width=0.47\textwidth]{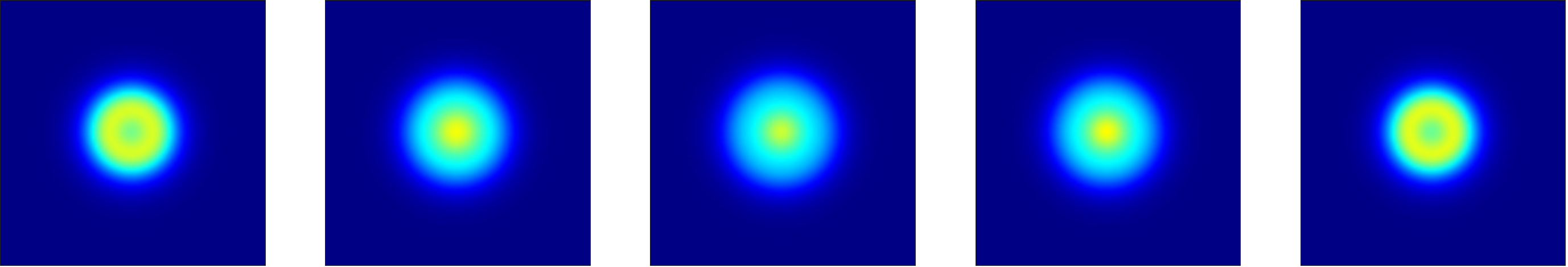}

\includegraphics[width=0.41\textwidth]{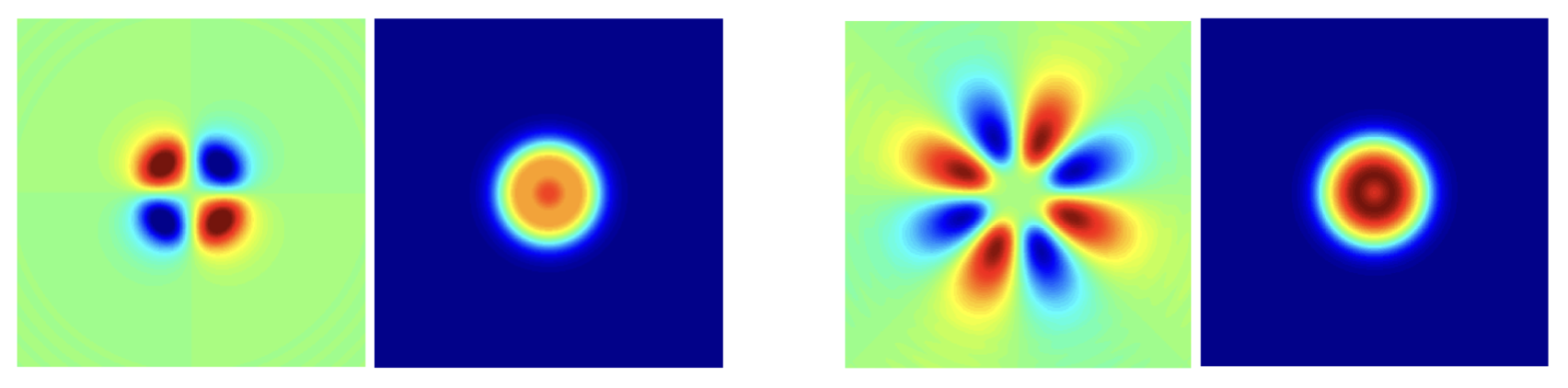}
\caption{Evolution of the charge density ({\it top}) and the energy density ({\it middle}) for a properly formed dipolar charge swapping Q-ball (CSQ) in 2+1D. $T_{\rm swap}$ is the period for a charge-swapping process. In the top row, red is for positive charges, while blue is for negative charges. ({\it bottom}) Examples of higher multipolar CSQs in 2+1D. The (mostly) spherical configurations are the energy densities for the corresponding multipolar charge densities to the left.}
\label{fig:CSQ_CE}
\end{figure}

For the simplest dipolar CSQ, once properly formed, a typical evolution of the charge distribution in a half swapping period and the corresponding evolution of the energy density distribution can be found in Figure \ref{fig:CSQ_CE}. We see that while the energy density is mostly spherically symmetric, the positive and negative charges swap their spatial locations violently as time goes by, and in between there are periods where the charge density is negligible across the space.

\begin{figure}[tbp]
\centering
\includegraphics[width=.4\textwidth,trim=90 260 110 280,clip]{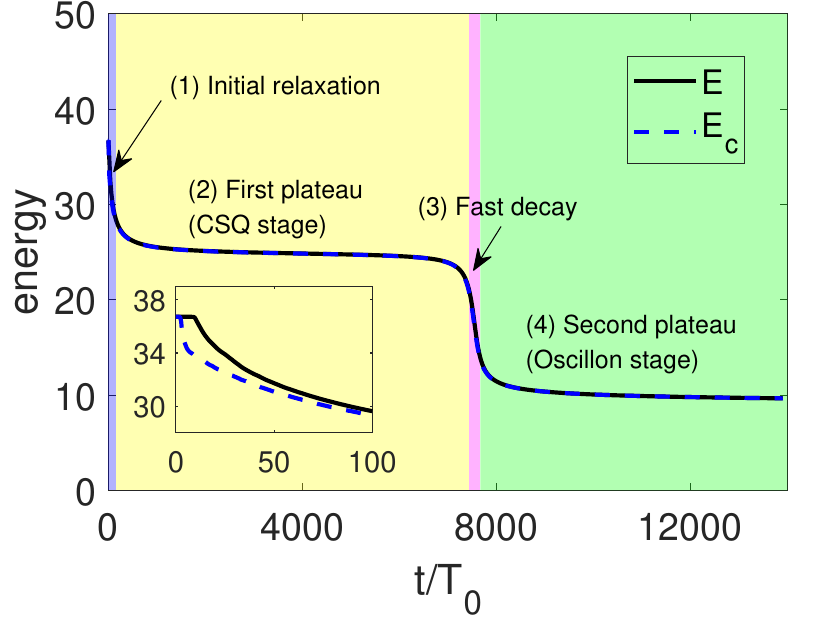}
\caption{Evolution of the total energy for a dipolar CSQ. $E$ is the energy integrated over the whole simulation box and $E_c$ is the energy within a large circular disk.  $t$ is in units of the oscillation period of the field $T_0=2\pi/m$. The inset of the top plot shows the differences between $E$ and $E_c$ in the initial relaxation stage. This figure is from Ref \cite{Xie:2021glp}.}
\label{fig:CSQ_TETC}
\end{figure}

Figure \ref{fig:CSQ_TETC} shows a typical evolution of the total energy for a dipolar CSQ in 2+1D. We see that, after the initial preparation via \eref{CSQinitial}, the configuration goes through 4 stages: 
\begin{itemize}
\item {\it Initial relaxation}, when the configuration quickly relaxes to a composite Q-ball, radiating away some excess energy;  
\item {\it CSQ stage}, when a composite Q-ball is properly formed with a very stable total energy and the charges within one ball swap with an almost constant frequency (see Figure 10 of \cite{Xie:2021glp}), the latter providing a convenient way to define the lifetime of a CSQ; 
\item {\it Fast decay}, when (the envelope of) the total positive charge of a CSQ decays exponentially  (see Figure 9 of \cite{Xie:2021glp}) and a big portion of its energy dissipates within a short period of time; 
\item {\it Oscillon stage}, when the leftover configuration has negligible charge and essentially becomes an oscillon (see \S\,\! \ref{sec:oscillon}).
\end{itemize}
The 3+1D case is very similar but generally the CSQ stage is shorter compared to the 2+1D case, as there are more decay modes in higher dimensions. By the rescalings \eqref{variableScaling0}, for the sextic potential, there is only the $|\varphi|^6$ coupling to choose, and under these rescalings, the lifetime of a CSQ increases with the $|\varphi|^6$ coupling.

These simulations are carried out in a lattice with a finite size. Since such a composite Q-ball is prepared with an imprecise method, which shreds some energy initially, and it is very long-lived, to accurately determine the lifetime of a CSQ, it is important to implement boundary conditions to efficiently absorb as much perturbations as possible in order to accurately determine the lifetime of a CSQ. To this end, Ref \cite{Xie:2021glp} compared a few absorbing boundary conditions, and found that an appropriately chosen second order Hidgon's condition, which have two selectable phase velocities, is excellent at absorbing unwanted perturbations on the box/Cartesian boundaries. The second order Hidgon's conditions are then used to chart the lifetimes of CSQs.

There are also higher multipolar Q-balls; see the bottom plot of Figure \ref{fig:CSQ_CE} for the explicit examples of a quadrupole and an octupole. For the polynomial potential in 2+1D, the lifetimes of the quadrupole and the octupole turn out to  be similar to that of the dipole. For certain choices of the sextic coupling $g$, the higher multipoles can even have longer lifetimes, the reason of which remains unclear. However, in 3+1D, the lifetimes of the higher multipoles are significantly shorter, at least for cases where the initial setups are not fine-tuned.
  
Another carefully examined potential for CSQs is the logarithmic potential of \eqref{logVnon} \cite{Copeland:2014qra, Hou:2022jcd}. Indeed, (messy forms of) CSQs can naturally arise during the Affleck-Dine fragmentation process (see  \S\,\!  \ref{sec:QballFormedAD}) under this potential \cite{Hou:2022jcd}. In this case, due to the softness of the logarithmic interaction, all the multipolar structures are extremely stable in either 3D or 4D. For example, it has been established that the lifetime of the dipole in this model (with $K=-0.1$) exceeds the following lower bounds
\bal
(\text{lifetime})_{\log \mathrm{CSQ}}^{3+1 \mathrm{D}}& \gtrsim 4.6 \times 10^5 / {m} ,
\\
(\text{lifetime} )_{\operatorname{logCSQ}}^{2+1 \mathrm{D}} & \gtrsim 2.5 \times 10^7 / {m} .
\eal
Also, in the initial relaxation process of this model, little energy is emitted, and the charge-swapping period $T_{\rm swap}$ only slowly decreases by a small amount in the early CSQ stage. Ref \cite{Hou:2022jcd} also comprehensively charts the parameter space for forming these composite Q-balls from scatterings for different frequencies of the constituent Q-balls, initial separations, initial velocities and impact parameters.

The quantum corrections of CSQs have been computed with the inhomogeneous Hartree approximation \cite{Xie:2023psz}, as will be discussed in more details in \S\,\! \ref{sec:quantumQball}. We will see that the classical charge-swapping behavior survives at the quantum level in the weak coupling limit, as one may expect. The dynamics of CSQs under strong gravitational effects have also been investigated recently with full numerical relativity \cite{Jaramillo:2024smx}, where a structure of this type on the Galactic scale has been proposed as a potential explanation for the observed anisotropic distribution of satellite galaxies.

For the logarithmic potential \eqref{logVnon}, Q-balls with approximate elliptical orbits in field space are found to be long-lived \cite{Hasegawa:2019bbo}. Axi-symmetric Q-chain solutions have been constructed in the Fridberg-Lee-Sirlin model \cite{Jaramillo:2024cus} as well as in the gauged version \cite{Loiko:2020htk}. However, these Q-chains beyond the dipoles are unstable, at least in the ungauged version. A soliton can carry charges associated with both topology and Noether symmetries, exhibiting various interesting properties \cite{Ward:2003un, Shnir:2011gr, Loginov:2019rwz, Bai:2021mzu, Alonso-Izquierdo:2023xni, Garcia:2023hrn}. Ref \cite{Nugaev:2016wyt} examined Q-hole solutions, which are dips in charge density within a homogeneous background in a U(1) symmetric theory, sometimes referred to as ``dark solitons'', and investigated their classical stability.

\subsection{Q-ball interactions}

In the previous subsections, we have focused on Q-balls of various kinds that are isolated in space. Q-balls of course can interact with each other and with ambient waves, which is the subject of this subsection.

\subsubsection{Interactions between Q-balls}

A natural line of inquiry is to study the interactions between two Q-balls. For this, except for a few special models in 1+1D, one usually resorts to lattice simulations that solve the partial differential equations with a finite difference method in a lattice \cite{Battye:2000qj, Axenides:1999hs, Hou:2022jcd}. For a single Q-ball, one can always boost to its center of mass frame where the ball is spherical. For two Q-balls, there can be a relative velocity between them, and they may scatter with a non-zero impact parameter. Furthermore, there can be differences in internal field space: they can have different types of charges, and even for the same type of charge, they can have a different internal frequency $\oi$ or a relative phase $\ai$. For example, the two Q-balls can be $\varphi_1=f_1(\bfx-\bfx_1) e^{i\oi_1 t +i\ai_1}$ and $\varphi_2=f_2(\bfx-\bfx_2) e^{i\oi_2 t +i\ai_2}$. All of these differences strongly affect the outcome of the scattering between two Q-balls. 

\begin{figure}[tbp]
\centering
\includegraphics[width=0.27\textwidth]{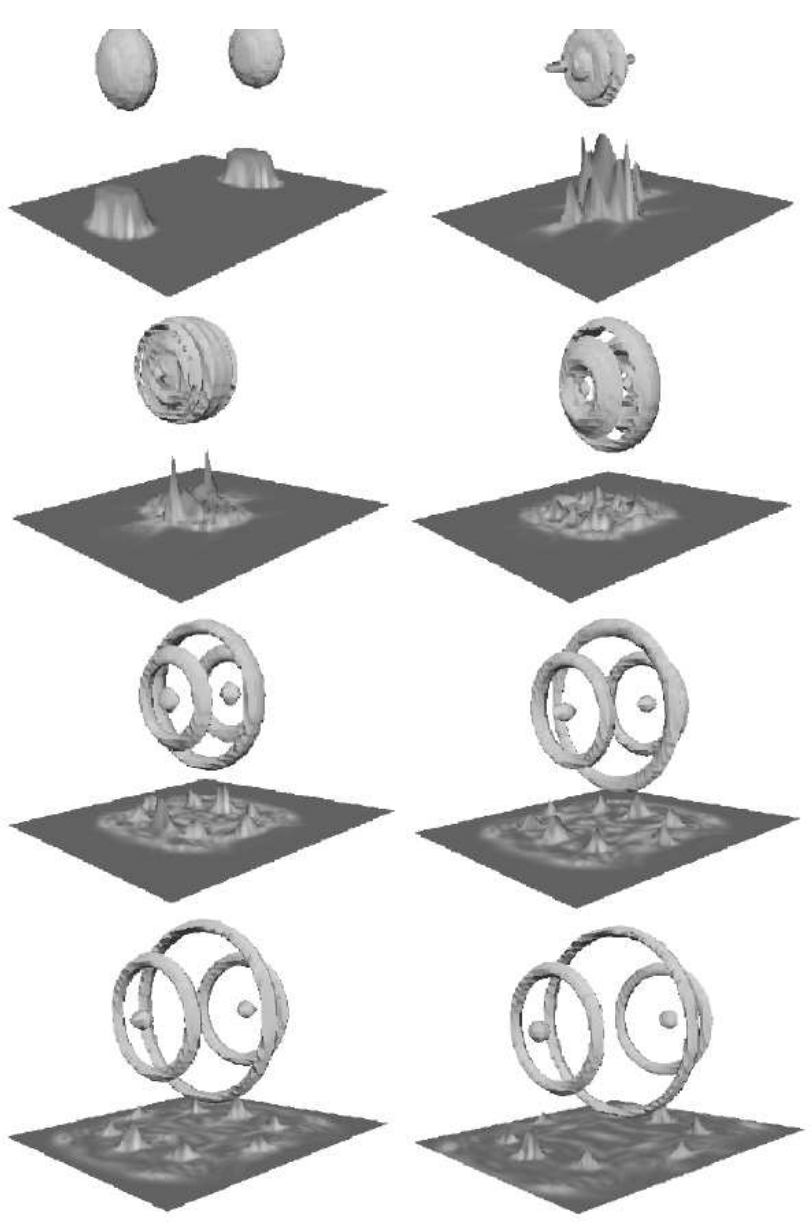}
\caption{Snapshots of the charge density at time $2mt= 0, 27, 33, 39, 45, 60, 75, 90$ for two equal-charge Q-balls colliding with high initial velocities ($v=0.8$). Plotted are 3D isosurfaces and 2D slices along the central line of the collision. This figure is from Ref  \cite{Battye:2000qj}.}
\label{fig:QballCollide}
\end{figure}

Generally, the scattering process is highly nonlinear and exhibits a wide range of patterns and behaviours, including merger, fission, repulsion, annihilation, charge-transfer and so on, or a combination of them \cite{Battye:2000qj}. However, it is possible to identify a few general trends in these scattering processes. First of all, if the relative phase is close to $\ai_1-\ai_2\simeq 0$ (in-phase), the two Q-balls tend to attract each other, and a repulsive force is generally expected if $\ai_1-\ai_2\simeq\pi$ (out-of-phase). If the relative velocity is extremely large, the two Q-balls tend to pass through each other without much changes to them. For small Q-balls, the number of spatial dimensions does not strongly affect the scattering. However, for large Q-balls, fission is more likely to take place during the scattering, and the dimensionality affects how this happens. In 1+1D,  this is simply via emission of small Q-balls or charged lumps along the one spatial dimension. In higher dimensions, fission can often happen in the dimensions perpendicular to the direction of incidence, similar to the behaviour of topological solitons \cite{Ruback:1988ba}. In 3+1D, there are two directions perpendicular to the direction of incidence, so one or multiple rings can be generated during the scattering; see Figure \ref{fig:QballCollide}. 

For two well-separated Q-balls in 1+1D, the forces between them can be analytically computed with the tail interaction method \cite{Bowcock:2008dn, Gorshkov:1981ke}. Generally, the force on a soliton can be computed by evaluating the time derivative of the soliton momentum within a region encompassing the soliton. In 1+1D, one can use the Lorentz-invariant momentum conservation to convert the time derivative of the momentum density, $\pd_t p$, into a spatial derivative, $\pd_x j^p = \pd_t p$. With this, the integration over the soliton encompassing interval $[x_1,x_2]$ can be trivially carried out. Thus, the force on a soliton is only determined by the end points of the interval, $x_1$ and $x_2$. Choosing the end points to be far from the center of the soliton, the analytical asymptotic solution can be used. For this method to work, there should be no net flow of momentum from the interval. This can be guaranteed if the two Q-balls are of the same internal frequency. Such a computation can explain the above numerical observation that in-phase Q-balls attract and out-of-phase Q-balls repel. 

For Q-balls with opposite charges, the scattering can generate charge-swapping Q-balls, and the parameter space of the relative velocity and the impact parameter for this to happen has been charted in \cite{Hou:2022jcd}. The quantum corrections to the head-on collisions of a Q-ball and an anti-Q-ball have been computed with the inhomogeneous Hartree approximation \cite{Xie:2023psz}. The relativistic head-on collisions of classical U(1) gauged Q-balls have also been investigated assuming axi-symmetry \cite{Kinach:2024hfa} and in 3+1D \cite{Kinach:2024qzc}. It is found that the results also strongly depend on the relative velocity, phase and charge, but differ significantly from the un-gauged Q-balls when the gauge coupling is large. Ref \cite{Hong:2024uxl} investigated the gravitational wave production from Q-ball collisions with various relative velocities.

\subsubsection{Q-ball superradiance}

Superradiance refers to a range of physical phenomena in which radiation becomes amplified through its interaction with a physical system. The term was originally coined by Dicke to describe amplification of radiation due to coherence in a gas of emitters \cite{Dicke:1954zz}, and to date, coherent superradiance is still an active area of research \cite{MassonSuperrad}. Phenomena such as Cherenkov radiation, Mach shocks and critical speed for superfluidity can be viewed as inertial motion superradiance \cite{Bekenstein:1998nt}. Rotational superradiance refers to a phenomenon where a wave incident on a rotating object gets amplified. Thanks to the existence of the ergoregion, rotating/Kerr black holes provide an important example of rotational superradiance, which finds various applications in gravitational and particle physics such as gravitational waves and ultra-light particle detections (see \cite{Brito:2015oca} for a comprehensive review). Many of these superradiant phenomena are implied by the second law of thermodynamics \cite{Bekenstein:1998nt}.

\begin{figure}[tbp]
\centering
\includegraphics[width=0.45\textwidth]{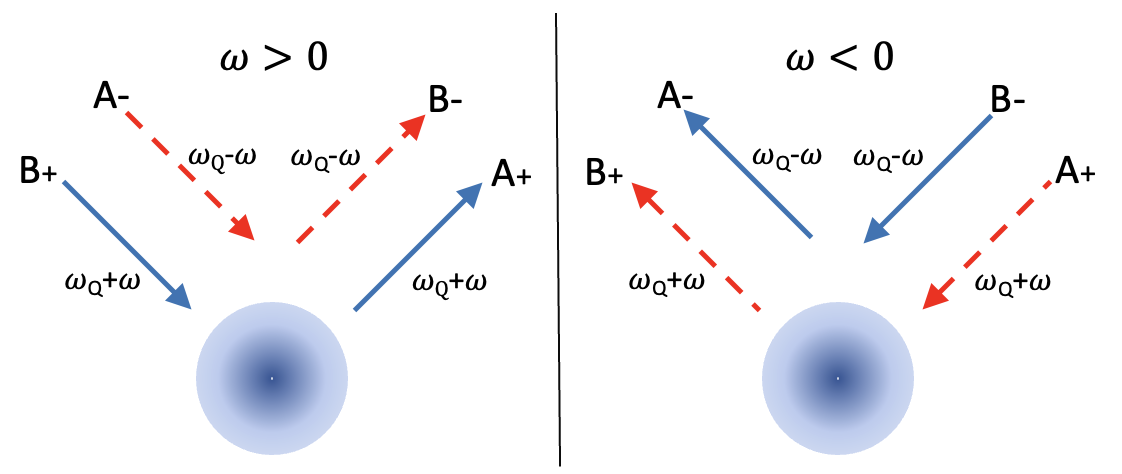}
\caption{Whether $A_\pm$ and $B_\pm$ represent ingoing or outgoing waves scattering on or off the Q-ball depends on the sign of $\oi$ (see \eref{etaAB0}). }
\label{fig:qballsuper0}
\end{figure}

A spherical Q-ball does not rotate in real space, but it rotates in internal field space. It is found that the internal rotation of a spherical Q-ball can also induce superradiance for incident waves \cite{Saffin:2022tub}. To see this, we consider a minimal case of two perturbative spherical waves (with frequencies $\oi_\pm = \oi_Q\pm \oi$) scattering off the Q-ball
\bal
\varphi(x)&=f(r)e^{-i\oi_Q t}+ \eta_{+} e^{-i \omega_{+} t}+\eta_{-} e^{-i \omega_{-} t} ,
\eal
where $\eta_\pm = \eta_\pm(\omega, r)$ satisfy 
\be
\eta_{ \pm}^{\prime \prime}+\frac{d-1}{r} \eta_{ \pm}^{\prime}+\left(\omega_{ \pm}^2-U_+ \right) \eta_{ \pm}-U_- \eta_{\mp}^*=0 ,
\ee
along with the boundary conditions 
\be
\label{etaAB0}
\eta_{ \pm}(\omega, r \rightarrow \infty) \rightarrow \frac{A_{ \pm}e^{i k_{ \pm} r}}{(k_{ \pm} r)^{({d-1})/2}} +\frac{B_{ \pm}e^{-i k_{ \pm} r}}{(k_{ \pm} r)^{({d-1})/2}}  .
\ee
Here $k_{ \pm}=(\omega_{ \pm}^2-m^2)^{1/2}$ and $U_\pm=\frac{\mathrm{~d}^2 V}{\mathrm{~d} f^2}\pm\frac{1}{f} \frac{\mathrm{~d} V}{\mathrm{~d} f}$. In other words, we are treating the ingoing and outgoing spherical waves along with their changes to the Q-ball perturbatively. Further taking into account the regularity conditions for $\eta_\pm$ near the center $r=0$, this set of ODEs can be solved as an initial value problem. The minimal case above involves two fixed-frequency waves, because the complex scalar system has two coupled degrees of freedom that can not be diagonalized at the level of equations of motion around the Q-ball background. Therefore, in general, in this minimal setup, there are two ingoing waves and two outgoing waves, whose amplitudes at infinity are given by $A_\pm$ and $B_\pm$. Depending on the sign of $\oi$, the ingoing and outgoing amplitudes are illustrated in Figure \ref{fig:qballsuper0}.

\begin{figure}[tbp]
\centering
\includegraphics[width=0.4\textwidth]{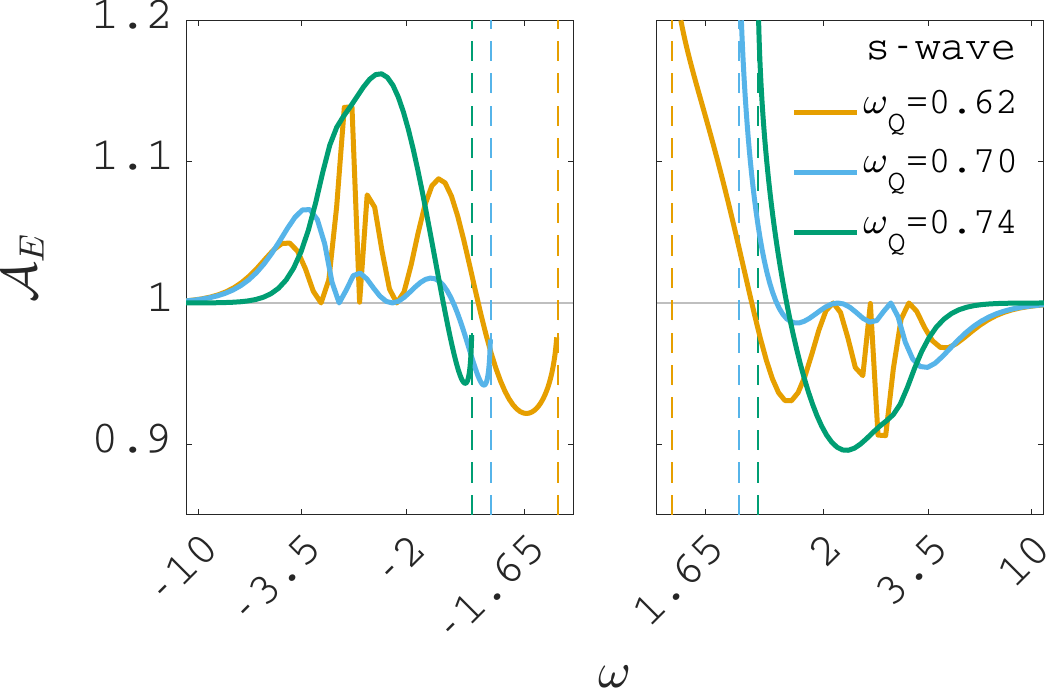}
\caption{Spectra of the energy amplification factor $\mc{A}_E$ (see \eref{AEdef}) for a $\eta_+$ s-wave scattering off a 3D spherical Q-ball. The $\eta_+$ has frequency $\oi_+=\oi_Q+\oi$, where $\oi_Q$ is the frequency of the internal rotation of the Q-ball. This figure is from Ref \cite{Saffin:2022tub}. }
\label{fig:qballsuperAE3D}
\end{figure}

Let us focus on the case of 3+1D. To assess how the interaction with the Q-ball changes the incident waves, we can compute the average energy contained in a sufficiently wide shell region at large $r$:
\be
E_{\circledcirc} \!\propto  \frac{\omega_{+}^2}{k_{+}^2} \left|A_{+}\right|^2\!+ \frac{\omega_{+}^2}{k_{+}^2} \left|B_{+}\right|^2\! + \frac{\omega_{-}^2}{k_{-}^2}\left|A_{-}\right|^2+ \frac{\omega_{-}^2}{k_{-}^2} \left|B_{-}\right|^2 ,
\ee
with each term corresponding to the energy for each wave. So one may define an energy amplification factor for the outgoing waves against the ingoing waves: 
\be
\label{AEdef}
\mathcal{A}_E=\left(\frac{\frac{\omega_{+}^2}{k_{+}^2}\left|A_{+}\right|^2+\frac{\omega_{-}^2}{k_{-}^2}\left|B_{-}\right|^2}{\frac{\omega_{+}^2}{k_{+}^2}\left|B_{+}\right|^2+\frac{\omega_{-}^2}{k_{-}^2}\left|A_{-}\right|^2}\right)^{\operatorname{sign}(\omega)} .
\ee
For example, if we prepare the boundary conditions such that there is only one ingoing wave $\eta_+$, a few example spectra of the energy amplification factor are shown in Figure \ref{fig:qballsuperAE3D}. We see that there are regions of the spectra where $\mathcal{A}_E>1$, meaning that superradiant amplification is achieved. Note that although there is only one ingoing wave, after interaction with the Q-ball, there are both the $+$ and $-$ types of outgoing waves.

Because of the U(1) symmetry of the complex scalar, the particle number of the outgoing waves is equal to that of the ingoing waves \cite{Saffin:2022tub}
\be
\label{superparticleCon}
\f{\left|A_{+}\right|^2}{k_+}+\f{\left|B_{-}\right|^2}{k_-}=\f{\left|B_{+}\right|^2}{k_+}+\f{\left|A_{-}\right|^2}{k_-} .
\ee
Therefore, the reason why the energy of the incident waves gets amplified in the scattering is because some of the low energy modes are turned into higher energy modes after interaction with the Q-ball, and as a result, the total energy of the outgoing modes becomes greater than that of the ingoing modes, despite that the number of (perturbative) modes remain unchanged during the scattering. This differs from the standard rotational superradiance in which there is only a single mode involved and the energy of this mode gets enhanced after the scattering. In contrast, in the Q-ball case, there are at least two coupled modes, making it natural to tally the energies of the two modes to determine wether amplification occurs. 

The particle number conservation also allows us to find an analytical criterion for when superradiance occurs. By simply comparing \eref{AEdef} and \eref{superparticleCon}, we see that $\mc{A}_E=1$ if 
\be
\f{\omega_{+}^2}{k^2_{+}}=\f{\omega_{-}^2}{ k^2_{-}} .
\ee
This equation gives rise to values of $\oi$ that delineate the regions of superradiance and non-superradiance, aligning perfectly with plots such as Figure \ref{fig:qballsuperAE3D}. 
One can also evaluate the amplification of the U(1) charge and find that amplification occurs only if both of the two ingoing modes are involved. 

As the scattering involves two modes with different propagation speeds, the energy fluxes of the waves, which describe the energy exchange with the exterior, generally differ from the $E_{\circledcirc}$ energy, which is the observable energy localized in a far away region. Essentially, $E_{\circledcirc}$ is calculated with the $tt$ component of the energy-momentum tensor on a large-radius annular region, while the energy flux is calculated with the $tr$ component in the annular region. This allows one to define another energy amplification factor \cite{Cardoso:2023dtm} 
\be
\mathcal{A}_{EF}=\left(\left|\frac{\frac{\omega_{+}}{k^2_{+}} A_{+}^2-\frac{\omega_{-}}{k_{-}^2} B_{-}^2}{\frac{\omega_{-}}{k_{-}^2} A_{-}^2-\frac{\omega_{+}}{k^2_{+}} B_{+}^2}\right|\right)^{\operatorname{sign}(\omega)},
\ee
$\mathcal{A}_{EF}$ can differ significantly from $\mathcal{A}_{E}$ when $\oi$ is near the mass gap $m$, but away from the mass gap, $\mathcal{A}_{EF}\simeq\mathcal{A}_{E}$. 
For the energy flux amplification factor, the scattering satisfies the celebrated Zel'dovich rotational superradiance criterion: 
\be
\oi_{\cal I} <\oi_Q,
\ee
if there is only one ingoing mode with ${\cal I}=+$ or $-$, which can also be easily understood by observing the particle number conservation \eqref{superparticleCon}. 

A Q-ball can also spin in real space \cite{Volkov:2002aj}, in addition to the internal rotation. One can also compute the amplification factors for the angular momentum density and angular momentum flux \cite{Saffin:2022tub, Cardoso:2023dtm, Zhang:2024ufh}. In 3+1D, both the spinning Q-ball itself and its perturbative waves need to be dealt with partial wave expansion, which is numerically much heavier to compute \cite{Zhang:2024ufh}. Some analytical bounds on the amplification factors along with their asymptotic behaviors have also been derived in  \cite{Zhang:2024ufh}. The amplification criteria for the angular momentum density and the one-ingoing-mode angular momentum flux are also easy to establish using the particle number conservation \eqref{superparticleCon}, but the interplay between superradiant effects from real space and internal space rotations remain largely unclear. The Q-ball type of superradiance has also been generalised to the case of Newtonian boson stars \cite{Cardoso:2023dtm} and (relativistic) boson stars \cite{Gao:2023gof}. Fully nonlinear time-domain simulations of the Q-ball \cite{Cardoso:2023dtm} and boson star \cite{BSStime} superradiant amplification processes have also been carried out, complementing the results of the perturbative analyses.

\subsection{Quantum corrections}
\label{sec:quantumQball}

Q-balls are highly nonperturbative configurations that have large particle numbers for the relevant modes. Generally, as mentioned, when a bosonic field has high occupation numbers for the pertinent modes, it can be usually approximated by a classical field \cite{Khlebnikov:1996mc, Berges:2004yj}. Indeed, most of the studies on Q-balls are carried out in the classical limit, similar to most studies on topological solitons \cite{Manton:2004tk}.  Large occupation numbers lead to a large field amplitude for the soliton, which however replies on the theory having small nonlinear couplings \cite{Lee:1991ax, Xie:2023psz}. (This is similar to the case of the gravitational field, where the classically approximation is almost always valid because the couplings in Einstein gravity are suppressed by the large Planck mass.) When the couplings increase, it becomes important to include the quantum corrections. For nonperturbative configurations, such as Q-balls and their composite structures and interactions, similar to classical studies, it often requires real-time lattice simulations, but now within the framework of quantum theory. 

\subsubsection{Inhomogeneous Hartree approximation}
\label{sec:QballInhomo}

Instead of solving the classical, c-number equations of motion, in the quantum theory, we can solve the corresponding Heisenberg equations of motion. The inhomogeneous Hartree approximation has been used to solve these quantum equations of motion \cite{Tranberg:2013cka, Xie:2023psz}. In this approach, we initially identify the one-point functions/mean fields with the classical Q-ball fields, and approximate the quantum fluctuations on top of the classical solution with Gaussian fields. Specifically, denoting $\Phi_1$ and $\Phi_2$  as the real and imaginary components of the U(1) quantum field  respectively, we can derive equations of motion for the mean fields 
\be
\phi_i(x)=\< \Phi_i(x)\>
\ee
and the connected two-point functions 
\be
G_{ij}(x,y)=  \<[\Phi_i(x)- \phi_i(x) ] [\Phi_j(y)- \phi_j(y) ] \> ,
\ee 
which are c-numbers, by simply setting all the connected higher-point correlation functions to zero. By this truncation, which amounts to keep the leading order of the 2-particle irreducible (2PI) expansion \cite{Berges:2004yj}, these equations of motion are self-contained and schematically given by
\bal
\label{quantumEoM1}
 \left[-\partial_x^2+M_{11}^2(x)\right] \phi_1(x)+M_{12}^2(x) \phi_2(x)&=0 ,
 \\ 
 \left[-\partial_x^2+M_{22}^2(x)\right] \phi_2(x)+M_{21}^2(x) \phi_1(x)&=0 ,
 \\
 \left[-\partial_x^2+\bar{M}_{11}^2(x)\right] \!G_{11}(x, y)+\bar{M}_{12}^2(x) G_{21}(x, y)&=0 ,
  \\ 
  \left[-\partial_x^2+\bar{M}_{11}^2(x)\right] \! G_{12}(x, y)+\bar{M}_{12}^2(x) G_{22}(x, y)&=0  ,
  \\ 
  \left[-\partial_x^2+\bar{M}_{22}^2(x)\right]\! G_{22}(x, y)+\bar{M}_{21}^2(x) G_{12}(x, y)&=0 ,
   \\ 
  \label{quantumEoM6}
   \left[-\partial_x^2+\bar{M}_{22}^2(x)\right]\! G_{21}(x, y)+\bar{M}_{21}^2(x) G_{11}(x, y)&=0 .
\eal
Here ${M}_{ij}^2$ and $\bar{M}_{ij}^2$, apart from containing the couplings of the theory, are only functions of $\phi_i(x)$ and $G_{ij}(x,x)$, since Gaussianity means that the higher-point functions can be written in terms of the one-point and two-point functions. The two-point functions $G_{ij}(x,x)$ are evaluated at the same spacetime point, so they contain ``divergences'', which, on a lattice, manifest as large numbers due to the inherent regularisation from the lattice spacing. To compare with the classical simulations, it is important to renormalise the couplings of the theory.  In practice,  Eqs.~(\ref{quantumEoM1}-\ref{quantumEoM6}) are numerically expansive to solve on a lattice, as $G_{ij}$'s are a set of functions that depend on two points, $x$ and $y$, for an inhomogeneous background, which is the classical Q-ball at the initial time.

In fact, for many purposes, we do not need to know the generic two-point $G_{ij}(x,y)$, since ${M}_{ij}^2$ and $\bar{M}_{ij}^2$ only depends on $\phi_i(x)$ and $G_{ij}(x,x)$. To improve the numerical efficiency, one can adopt the stochastic ensemble method to approximate the quantum correlation functions \cite{Borsanyi:2007wm, Borsanyi:2008eu}. The key point of this method is that, since the quantum fields on top of the mean field are assumed to be Gaussian
\be
\Phi_i(x)- \phi_i(x)  \sim \sum_\bfk \left[a_{\mathbf{k}}^i f_{\mathbf{k}}^i(x)+a_{\mathbf{k}}^{i \dagger} f_{\mathbf{k}}^{i *}(x)\right] ,
\ee
the classical stochastic fields 
\be
\varphi^c_i\sim \sum_\bfk \left[c_{\mathbf{k}}^i f_{\mathbf{k}}^i(x)+c_{\mathbf{k}}^{i *} f_{\mathbf{k}}^{i *}(x)\right]
\ee
have the same ensemble two-point functions, 
\be
G^c_{ij}(x,x) =  \< \varphi^c_i(x)\varphi^c_j(x) \>_e - \< \varphi^c_i(x)\> \<\varphi^c_j(x) \>_e  ,
\ee
as the quantum two-point functions $G_{ij}(x,x)$, where quantum operators $a_{\mathbf{k}}^i$ and $a_{\mathbf{k}}^{i\dagger}$ satisfy the canonical commutation relations and c-numbers $c_{\mathbf{k}}^i$ and $c_{\mathbf{k}}^{i*}$ follow the corresponding normal distributions. Here $\< ~\>_e$ is the classical ensemble average, which can be computed by evolving many copies of the following classical equations of motion
\bal
 \left[-\partial_x^2+M_{11}^2(x)\right] \varphi^c_1(x)+M_{12}^2(x) \varphi^c_2(x)&=0 ,
 \\ 
 \left[-\partial_x^2+M_{22}^2(x)\right] \varphi^c_2(x)+M_{21}^2(x) \varphi^c_1(x)&=0 ,
\eal
from different stochastic realisations. If the number of copies needed to achieve the required accuracy is smaller than the number of the lattice sites, we will have speeded up numerically. Indeed, for the Q-ball case, the speed-up factor is quite sizable.

As mentioned previously, the scalings \eqref{variableScaling0} will re-scale the action by an overall factor. Therefore, while these scalings link classical equivalent theories, quantum mechanically, these linked theories are physically distinct, as can be confirmed explicitly with the simulations \cite{Xie:2023psz}. 

For the basic, spherical Q-balls, quantum stability conditions have been established through perturbative analyses around the classical background. The inhomogeneous Hartree simulations have confirmed these results in both 3D and 2D \cite{Tranberg:2013cka, Xie:2023psz}, especially for the large thin-wall cases. For the smaller Q-balls, quantum effects are significant. Intriguingly, while the total charge conservation is still observed, there is a large oscillation between the contribution from the mean field and that from the two-point functions. However, it is unclear whether this is an artifact of the inhomogeneous Hartree approximation or a genuine quantum effect.  

For well-separated multiple Q-balls as well as charge-swapping Q-balls, the conclusion is similar: the classical results are a good approximation when the amplitude of the Q-ball is sufficiently large or equivalently when the nonlinear couplings---particularly the quartic coupling---are sufficiently weak \cite{Xie:2023psz}. It is also found that the oscillation between the mean fields and the quantum modes tends to be quenched by collisions between Q-balls.

\subsubsection{Tunnelling of classically stable Q-balls}
\label{sec:QballQuantTun}

When the charge $Q$ of a Q-ball is slightly below the stability threshold $Q_S$ but above the the critical charge $Q_C$ (see Figure \ref{fig:EvsQ}), the Q-ball is classically stable but quantum mechanically unstable \cite{Friedberg:1976me, Lee:1991ax, Smolyakov:2019cld}. This means that in the space of the energy functional, there is an energy barrier between the lower branch Q-ball solution and the solution of dissipated $Q$ free particles. This barrier is peaked at the saddle-point Q-ball solution ({\it i.e.}, the upper branch Q-ball in Figure \ref{fig:EvsQ}), and the solution of free particles has a lower energy than the lower branch Q-ball. If the nonlinear couplings are small, the lower branch metastable Q-ball can be very long-lived, but the question is how long can it live? As usual, its lifetime can be determined by the quantum tunnelling rate between the two solutions \cite{Levkov:2017paj}.

Similar to false vacuum decay \cite{Coleman:1977py}, the tunnelling rate at leading semi-classical order
\be
\Gamma_Q =A_Q e^{-F_Q} ,
\ee
particularly the decay exponent $F_Q$, can be computed in the Euclidean formalism by the bounce solution. Of course, there are major differences between the Q-ball decay and false vacuum decay. For example, for the Q-ball solution, $\varphi$ and $\varphi^\dagger$ contain $e^{i\oi t}$ and  $e^{-i\oi t}$, which in Euclidean time $\ti=it$ can grow unbounded from above as $\ti\to -\infty$. Because the Q-ball is spherically symmetric in the spatial dimensions, we should look for a tunnelling solution that is spatially spherically symmetric. Also, since the total charge is conserved, one should select quantum states that conserve the charge in the transition amplitude. In the Euclidean signature, $\varphi$ and $\varphi^\dagger$ must be treated as independent real functions, but when continued to the Minkowski signature, they become complex conjugates of each other.

Ref \cite{Levkov:2017paj} first constructs quantum Q-ball states, and derives a path integral representation of the decay probability that conserves the charge. This path integral is evaluated in Euclidean time essentially by construction, and involves the Q-ball charge and energy in addition to the standard action in the phase, which enforces some specific boundary conditions on the fields. The decay exponent is then found to be
\be
F_Q=S_E\big[\varphi_{B}, {\varphi}^\dagger_{B}\big]-S_E\big[\varphi_Q, {\varphi}^\dagger_Q\big]+\eta_0 Q ,
\ee 
where the first and second terms are the Euclidean actions for the bounce solution and the Q-ball solution respectively, and $\eta_0$ is a parameter that enforces the charge conservation and must be minimized in the path integral. Computing the bounce solutions numerically, one finds that $F_Q=0$ when $Q=Q_C$. Typically, when $Q\sim Q_C$ is very large, we have
\be
F_Q\sim Q_C .
\ee
On the other hand, a rough estimate of the decay prefactor is given by$A_Q\sim m Q_C^{1/2}$. Therefore, these metastable Q-balls can be extremely long-lived, even longer than the age of the universe.

Ref \cite{Levkov:2017paj} also computed the tunnelling rate of the metastable Q-ball at a finite temperature. In the Euclidean signature, this can be achieved by restricting the Euclidean time $\tau$ to a finite range, $-\beta / 2<\tau<\beta / 2$, and identify $T=\beta^{-1}$ as the temperature. As expected, thermal fluctuations decrease the lifetime, as they make it easier to overcome the energy barrier. At high temperatures, the bounce solution ceases to exist, and the thermal activation (``jumping over'' the energy barrier) dominates, for which the decay exponent can be computed using the saddle-point energy configuration.

\subsection{Coupling to other fields}

The complex scalar in (\ref{originLag}) can naturally couple to a fermionic field through a Yukawa interaction, for instance, if the global symmetry corresponds to lepton or baryon number conservation (see \S\,\! \ref{sec:existMSSM}). This introduces a decay channel for the Q-ball, and, interestingly, this decay process only occurs via evaporation on the surface of the Q-ball \cite{Cohen:1986ct}. On the other hand, when a fermionic degree of freedom is present, non-topological solitons can actually exist with one extra real scalar \cite{Bardeen:1974wr, Friedberg:1976eg}.

The complex scalar can also couple to other bosonic fields. The Friedberg-Lee-Sirlin model (see \S\,\! \ref{sec:FriedbergLeeS}) is a renormalisable theory where the complex scalar is coupled to a real scalar. The natural way to couple the complex scalar to a U(1) vector field is to promote the global symmetry to a gauged one. Also, the complex scalar can be (minimally) coupled to gravity, in which context the soliton solution is usually called a boson star, if the gravitational effects are significant. We will discuss all the above cases in this subsection (see \cite{Nugaev:2019vru} for a recent review), except for boson stars, which consist of a vast literature and have been comprehensively reviewed elsewhere \cite{Lee:1991ax,  Jetzer:1991jr,  Liebling:2012fv, Visinelli:2021uve}. Additionally, Q-balls can form with extra dimensions \cite{Demir:2000gj}, on the world-volume of D$p$ branes \cite{Abel:2015tca} or around the Ellis-Bronnikov wormhole geometry \cite{Blazquez-Salcedo:2022kaw}. In an anti-de Sitter space, a stable Q-ball has a maximum radius, beyond which a Q-ball can decay via a third-order phase transition \cite{Rajaraman:2023ygy}.

\subsubsection{Surface evaporation via fermions}
\label{sec:fermionEvap}

Consider the extension of Lagrangian (\ref{originLag}) by the following terms \cite{Cohen:1986ct}
\be
\mc{L}_{\rm ferm} = i \psi^\dagger \si^\mu \pd_\mu \psi - i g(\varphi \psi^\dagger \si_2\psi^*  -  \varphi^\dagger \psi^T \si_2\psi) ,
\ee
where $\psi$ is a Weyl spinor and $\si^\mu$'s denote the extended Pauli matrices. In the context of lepton Q-balls (L-balls), the fermion might be a neutrino field. The Yukawa coupling destabilises the Q-ball, as a scalar particle can now decay into a pair of fermions. Ref \cite{Cohen:1986ct} computed the decay rate of this process for a Q-ball with a large radius, and found that the decay only takes place on the surface of the Q-ball, characterizing it as an evaporation process. The effects of the Q-ball's size and the fermion mass on the evaporation have also been analysed \cite{Clark:2005zc, Multamaki:1999an}.

The reason that the decay of a large Q-ball occurs via surface evaporation becomes clear upon noting that Q-matter ({\it i.e.,} a homogeneous Q-ball with an infinite radius) is stable against fermionic decay. To see this, one can evaluate the amplitude from the Q-matter (vacuum) to itself \cite{Cohen:1986ct}
\be
\< Q|S|Q\> = e^{W} ,
\ee
where $W$ is the sum of all connected vacuum diagrams. Treating the Q-matter as a classical background, we can perform the path integral over the fermionic fields via a functional determinant, as the integral is Gaussian. All the vacuum diagrams are one-loop, and $W$ can be written as a trace of a sum of powers of the fermionic propagator. If all the singularities of the propagator lies on the real axis of $k^0$ (for fixed wave vector $k^i$), $W$ is purely imaginary, as in this case the $k^0$ integration contour, which is part of the trace, can be deformed to the imaginary axis.  The Q-matter can decay only if $W$ has a real part. 

The singularities of the fermionic propagator occur at the normal frequencies of the fermionic fields. Let us assume that the Q-matter takes the form $\varphi=f e^{-i \oi t}$, $f$ and $\oi$ being constant. To find the normal modes, we can substitute the ansatz $\psi={e}^{-i \omega t / 2} {e}^{-i k \cdot x} u$ and $i\si_2\psi^*={e}^{i \omega t / 2} {e}^{-i k \cdot x} v$ into the equations of motion for the fermions, which reduce to
\be
\label{fermiEoM0}
\left[\begin{array}{cc}k^0+\frac{1}{2} \omega- \si_i k^i & -g f \\ -g f & k^0-\frac{1}{2} \omega+\si_i{k}^i \end{array}\right]\binom{u}{v}=0 .
\ee
The normal frequencies $k^0$ can be obtained by setting the determinant of the coefficient matrix in \eref{fermiEoM0} to zero. It is easy to see that all the normal frequencies for any $k^i$ are real, so the Q-matter, {\it i.e.,} the interior of the Q-ball, can not decay into fermionic pairs.

On the other hand, if the Q-ball is coupled to a complex scalar, decay does occur at the interior of the Q-ball. This can also be seen by evaluating the normal modes of that system. Essentially, it is the Pauli exclusion principle that prevents the pair production from happening in the interior: fermion production inside the Q-ball fills the dynamically relevant states (``filling the overflowed Dirac sea''), and this then prevents further fermion production inside the ball because of the exclusion principle. (Also, due to the exclusion principle, it is energetically disfavoured to form Fermi gas cavities via quantum tunnelling within the Q-matter, at least in the semi-classical limit.) Considerations along this line can be used to put an upper bound on the rate of pair production per unit area
\be
\label{upperBEvap}
\frac{\text{evaporation rate}}{\text{surface area}} \leqslant \frac{\omega^3}{192 \pi^2} .
\ee

\begin{figure}
	\centering
		\includegraphics[height=5.5cm]{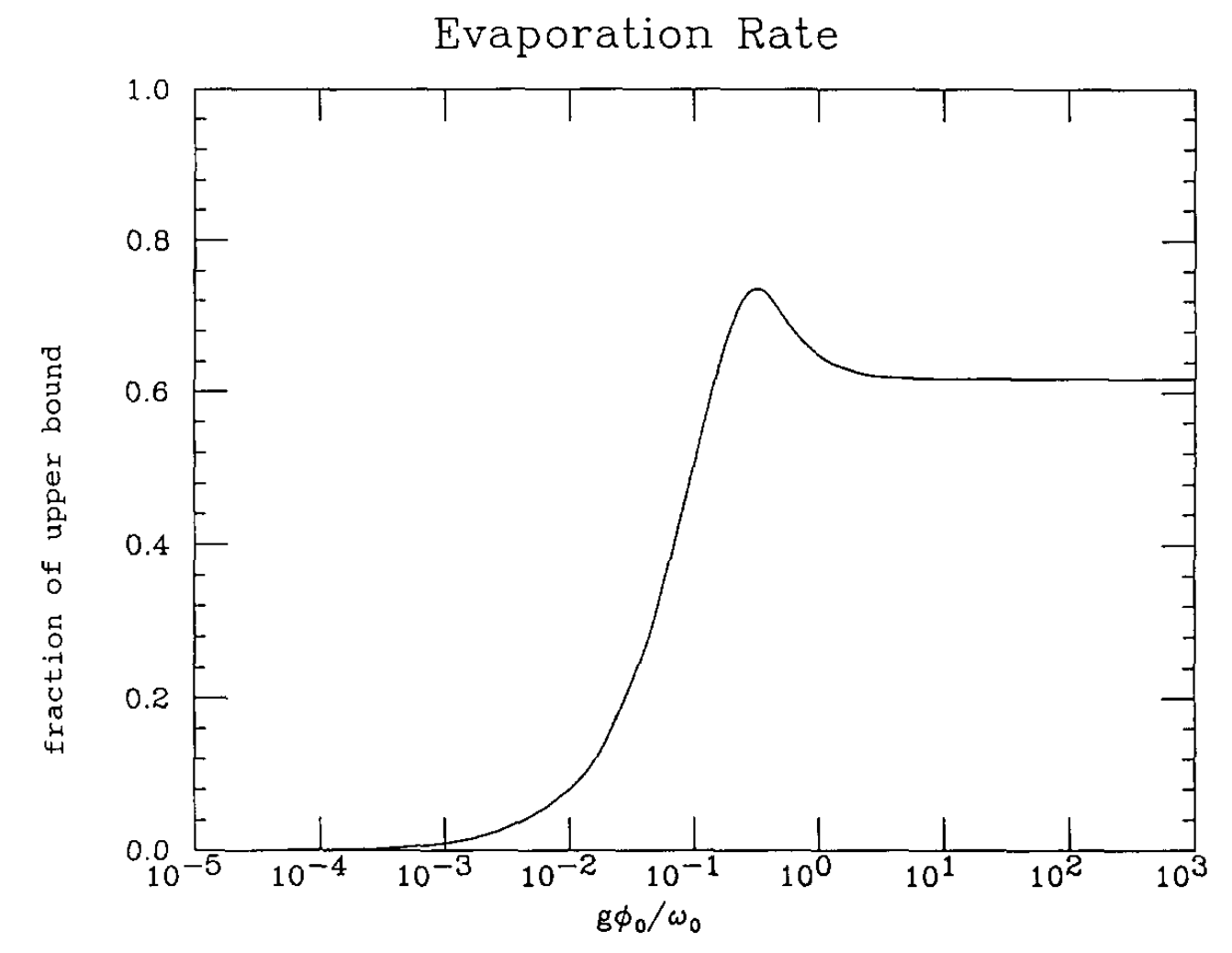}
	\caption{Quantum evaporation rate per unit area of a Q-ball. This figure is from Ref \cite{Cohen:1986ct}.}
 \label{fig:colemanEvap}
\end{figure}

To compute the decay rate in the large Q-ball limit, one should begin with a finite Q-ball and take the large radius limit towards the end, so as to avoid ambiguities in matching the fermionic solutions at the Q-ball boundary. For simplicity, a thin-wall Q-ball is considered: $\varphi(t,r)=f e^{-i \oi t}\thi(R-r)$, $\thi(z)$ being the Heaviside step function. Again, treating the Q-ball as a classical background, the quantum fermion fields can be expanded in terms of scattering solutions to the fermionic field equations, with the expansion coefficients being the creation and annihilation operators. For example, in the exterior of the Q-ball ($r>R$), one can use the following set of scattering solutions  \cite{Cohen:1986ct}
\bal
 &\psi_{k_{+}, j, m}= {e}^{-i k_{+} t}\left[u_{k_+,j,m}^{(2)}\left({\bsb r}\right)+R_{k_{+}, j} u_{k_{+}, j, m}^{(1)}\left({\bsb r}\right)\right], 
 \\ 
&i\si_2\psi^*_{k_{+}, j, m} = {e}^{i k_{-} t}  T_{k_+, j} u^{(2)}_{k_{-}, j,m} (\bsb r) 
\eal
and their symmetric counterparts as the basis functions to expand the quantum fields, where $k_\pm = \oi/2 \pm k^0$ are the normal frequencies of the fermionic modes and $u_{k_\pm,j,m}^{(1,2)}\left({\bsb r}\right)$ are partial wave solutions to the fermionic field equations. In this scattering setup, each pair of the solutions describes an ingoing wave with frequency $k_+$ scattering off the Q-ball and two outgoing waves coming out with frequencies $k_+$ and $k_-$, where $R_{k_{+}, j}$ and $T_{k_{+}, j}$ are the amplitudes for the wave reflection and transmutation respectively. The quantum field can be written as
\be
\psi = \sum_{j,m}\int \d k_+  \,a^{\rm in}_{k_+,j,m}  \psi_{k_{+}, j, m} + \cdots .
\ee
Since only the ingoing wave of $\psi_{k_{+}, j, m}$ survives in the far past, $a^{\rm in}_{k_+,j,m}$ is the annihilation operator for an ingoing fermion and anti-commutes with its hermitian conjugate according to the usual free-field quantisation rule. Similarly, one can define the annihilation operators for the outgoing waves $a^{\text{out }}_{k_{+}, j, m}$, and the Bogoliubov transformation between the two sets of operators are given by
\be
a^{\text{out }}_{k_{+}, j, m}=R_{k_{+}, j} a^{\text {in }}_{k_{+}, j, m}+(-1)^{m+\f12} T^*_{k_{-}, j} a^{\text {in}\dagger}_{k_{-}, j,-m} .
\ee
This allows one to compute the average number of outgoing fermions $N_{\rm out} = \sum_{j,m} \int_0^{\omega} \!\!\mathrm{d} k\,  a^{\text {out}\dagger}_{k, j, m} a^{\text {out }}_{k, j, m}$ emitted in unit time in the state of the ingoing wave vacuum $|0\rangle_{\rm in}$, which is given by
\be
\f1{\mc T} ~~\big\langle_{\rm\!\!\!\!\!\!\!\! in} ~0\big| N_{\rm out}\big| 0\big\rangle_{\text {in }}=
\sum_j \int_0^{\omega} \!\!\f{\mathrm{d} k}{2\pi} (2j+1) |T_{k, j}|^2  .
\ee
To really compute this decay rate, one also needs to solve the field equations within the Q-ball, and match with the exterior solutions at the boundary of the Q-ball. In the limit where the Q-ball radius goes to infinity, the evaporation rate per unit area is plotted against $gf/\oi$ in Figure \ref{fig:colemanEvap}. It is found that the rate plateaus around 60\% of the upper bound (\ref{upperBEvap}) when $gf/\oi \gtrsim \mc{O}(1)$. In the weak coupling limit, an analytical result has also been found \cite{Cohen:1986ct}:
\be
\frac{\text{evaporation rate}}{\text{surface area}} =3 \pi \frac{g f}{\omega} \frac{\omega^3}{192 \pi^2} .
\ee

\subsubsection{Fermionic non-topological solitons}
\label{sec:fermiSoliton}

In the presence of fermions, a non-topological soliton can form with an additional real scalar field. In this case, the configuration of the scalar can be static, while the fermionic configuration is time dependent. A specific model of such kind was discussed when the SLAC bag model was introduced \cite{Bardeen:1974wr}, and a general formalism was set up in \cite{Friedberg:1976eg} and then applied to hadronic models \cite{Friedberg:1977xf, Friedberg:1978sc, Goldflam:1981tg, Cahill:1985mh} (see \S\,\! \ref{sec:FriedLee}).  See \cite{Xie:2024mxr} for a recent discussion for general potentials and new analytical limits, utilizing the mean-field approximation.

Suppose that there are $N$ Dirac fermions of the same mass $\psi_n$, $n=1,2,...,N$, coupled to a real scalar $\si$ via a Yukawa coupling \cite{Friedberg:1976eg}:
\bal
\label{fermionNTSLag}
\mc{L} &= -\f12 \pd_\mu \si \pd^\mu \si -V(\si) + i \bar\psi \slashed{\pd} \psi - (m+ g \si)\bar\psi \psi .
\eal
where $\psi = (\psi_1,\psi_2,...,\psi_N)^T$ and the scalar potential can be chosen to be renormalisable, $V(\si)=\frac{1}{2} a \si^2+\frac{1}{3!} b \si^3+\frac{1}{4!} c \si^4$, with a minimum at $\si=0$. Similar to the case of scalar Q-balls, this model also has a global symmetry, which is now U(N) and in the fermionic sector. For example, the overall U(1) part of the symmetry $\psi_n\to e^{i\ai} \psi_n$, $\ai$ being a constant, gives rise to a conserved integer charge
\be
Q_f = \int  \d^3 x  \sum_n \psi_n^\dagger \psi_n .
\ee
This is just the conservation of the total fermion number, or the difference between the numbers of fermions and anti-fermions. Because of the Pauli exclusion principle, we have $|Q_f|\leq N$. 

For the bosonic Q-ball, use of the classical solution is justified if the characteristic modes of the soliton have high occupation numbers. However, for a fermionic field, each mode can only have occupation number 0 or 1, so the same justification does not apply, and the problem is no longer purely classical. To proceed, one can expand the scalar around the classical solution $\si_c(\bfx)$ with the Fourier series
\bal
 \delta\si(t,\bfx)&=  \si (t,\bfx) - \si_c(\bfx)
 \\
 &= \int \f{\d^3k}{(2\pi)^3 2\oi_\bfk} (c_\bfk(t) e^{i\bfk\cdot \bfx} + c^\dagger_\bfk(t) e^{-i\bfk\cdot \bfx}) ,
\eal
and expand the fermionic fields with a complete set of static energy eigenstates $u_\ell(\bfx)$ and $v_\ell(\bfx)$
\be
\psi_n (t,\bfx)=\sum_{\ell=1}^{\infty}\left[a_{n,\ell}(t) u_\ell(\bfx)+b_{n,\ell}^{\dagger}(t) v_\ell(\bfx)\right] ,
\ee
where $u_\ell$ and $v_\ell$ satisfy
\bal
\label{uellEq0}
\gi^0(-i\gi^j \pd_j +  m+g\si_c) u_\ell &=\epi_\ell u_\ell ,
\\
\gi^0(-i\gi^j \pd_j +  m+g\si_c) v_\ell &= -\epi_\ell v_\ell ,
\\
{\rm with~~~}0<\epi_1 <\epi_2<  \epi_3 &< \cdots , \nonumber
\eal
and the annihilation operators $c_\bfk$, $a_{n,\ell}$ and $b_{n,\ell}$ and their hermitian conjugates satisfy the standard canonical commutation and anti-commutation relations respectively \cite{Friedberg:1976eg}.

To compute the leading semi-classical approximation, we can assume that all the excited $Q_f$ fermionic fields are in the lowest positive energy state $u_1$ with energy $\epi_1$, assuming that all the negative energy states $v_\ell$ have been ``filled''. Then, the c-number field equations to solve are
\bal
-\nd^2\si_c + \f{\d V(\si_c)}{\d \si} + g Q_f \bar\psi_c \psi_c &=0 ,
\\
(-i\gi^\mu \pd_\mu +  m+g\si_c ) \psi_c &= 0 ,
\eal
where we have defined $\psi_c\equiv u_1(\bfx)e^{-i\epi_1 t}$ and the second equation above is simply a rewriting of \eref{uellEq0}. Thus, although the fermionic fields do not form coherent classical configuration in the same way as the boson fields, the leading semi-classical approximation for this system still amounts to solving the classical field equations. Due to the particle-antiparticle symmetry, one may focus on the case of $Q_f>0$.

Generally, for a sufficiently large particle number $Q_f$ ({\it i.e.,} when $Q_f>Q_S$), the lowest energy state for such an isolated system is a non-topological soliton, rather than a plane wave \cite{Friedberg:1976eg}. Ref \cite{Friedberg:1976eg} also proved that, as $Q_f \to \infty$, the energy of the soliton is bounded by
\be
E < \frac{4 \sqrt{2}}{3} \pi  Q_f^{\f34}V({-m}/{g})^{\f14}   .
\ee
On the other hand, if the mass of the fermion, $m$, is greater than a critical value $m_c$, the critical particle number $Q_S$ above can be smaller than 1. That is, in this case, the soliton has even a lower energy than the one fermion state above the vacuum in the semi-classical approximation. For a weakly coupled theory satisfying $g^2/4\pi\ll 1$ and $c\sim \mc{O}(g^2)$, the critical fermion mass is bounded by
\be
m_c < \f{729 \pi a^{\f12}}{ 32g^2} .
\ee
There is generally also a critical value $Q_C<Q_S$ such that when $Q_f>Q_C$ the soliton is stable against infinitesimal perturbations. In fact, the $E$-$Q_f$ plot of the fermionic soliton is very similar to that of the scalar Q-ball (see Figure \ref{fig:EvsQ}).

Ref \cite{Friedberg:1976eg} also studied the quantum corrections on the fermionic soliton for the case where the scalar self-interactions are absent, $b=c=0$ in the canonical formalism. By comparing the quantum corrected soliton energy to the semi-classical one, it is found that, in a weakly coupled theory, as expected, the semi-classical approximation is accurate when $Q_f\gg 1$. However, even when the fermion particle number is as low as $Q_f=2$, the quantum correction to the soliton energy is only about 20\%. Furthermore, in the nonrelativistic limit of a strongly coupled theory, the quantum corrections are negligible for any $Q_f$. 

Non-topological solitons also exist when the fermions in Lagrangian \eqref{fermionNTSLag} are coupled to a U(1) gauge field if the gauge coupling is sufficiently small and the ball is not too large \cite{Lee:1988ag, Levi:2001aw}.

\subsubsection{Gauged Q-balls}
\label{sec:gauge}

Another natural way to couple to other fields is by promoting the global symmetry to a gauge symmetry:
\be
\label{gaugedQball00}
\mathcal{L}=-(D_\mu \varphi)^{\dagger} D^\mu \varphi-V(|\varphi|) -\frac14 F^\mn F_\mn ,
\ee
where $D_\mu=\pd_\mu -i eA_\mu$ is the gauge covariant derivative and $F_\mn$ is the field strength of the U(1) gauge field. Just as in electrodynamics where like charges repel, coupling the scalar field to a gauge field introduces repulsion within the Q-ball (which is essentially a bag of charges). This means that very large, gauged Q-balls are generally unattainable. Nevertheless, smaller gauged Q-balls, including some thin-wall Q-balls, are stable \cite{Lee:1988ag, Kusenko:1997vi, Benci:2010cs}. In fact, inside the Q-ball, the U(1) gauge symmetry is spontaneously broken, so the vector field becomes massive within the ball. As a result, for certain scalar self-interactions, the charges are pushed to the surface shell of a Q-ball when its size is larger than the Compton wavelength of the massive vector ($m_V^{-1}$). Beyond this surface shell (with width $\sim m_V^{-1}$), the Q-ball behaves as a superconductor. 

Gauged Q-ball solutions had been perturbatively constructed from the ungauged Q-balls as early as \cite{Rosen:1968zwl}, which however did not discuss the stability of the solutions. More recently, Ref \cite{Gulamov:2013cra, Gulamov:2015fya} revisited the theory of gauged Q-balls and showed that the relation \eqref{dEdQoi} also holds for a gauged Q-ball, with which it is argued that the maximal charge for a stable gauged Q-ball derived in \cite{Lee:1988ag} is inconsistent. On the other hand, the stability criterion \eqref{dQdoiQball} for a global Q-ball is found to be inapplicable in the case of a gauged Q-ball \cite{Panin:2016ooo}. If a potential supports global Q-balls, generally, it also supports gauged Q-balls \cite{Heeck:2021gam}. Interestingly, some of the properties of gauged Q-balls such as the internal frequency/charge/energy and radius relation can be analytically approximated by mapping to the corresponding global Q-balls \cite{Heeck:2021zvk}.  Generalisation to the case with an additional Proca mass term for the vector filed has also been considered \cite{Heeck:2021bce}.  Some stress stability criteria have been derived using the energy-momentum tensor for the single scalar gauged theory as well as the gauged Friedberg-Lee-Sirlin model \cite{Loiko:2022noq}. Spinning Q-balls in the gauged Friedberg-Lee-Sirlin model have been found in \cite{Loiko:2019gwk}. Fully nonlinear simulations of (single scalar) gauged Q-balls have been performed in axi-symmetry \cite{Kinach:2022jdx} and in 3+1D \cite{Kinach:2024qzc}, which identified some non-spherical unstable modes. These instabilities have also been analyzed perturbatively \cite{Rajaraman:2024erp}, based on the beyond the thin-wall approach of \cite{Heeck:2020bau, Heeck:2021zvk}. Excited gauged Q-balls are explored in \cite{Loginov:2020xoj}, and the gauge-global Q-ball mapping relations have been generalised to the case of these excited Q-balls \cite{Almumin:2024dem}. Properties of gauged Q-balls have also been examined in details for some specific potentials such as the signum-Gordon potential \cite{Arodz:2008nm} and the sextic potential \cite{Han:2023uyo}, and in models with more fields \cite{Loginov:2015rya, Ishihara:2018rxg, Forgacs:2020vcy} and supersymmetry \cite{Loginov:2023wmh}. 

Intuitively, a gauged Q-ball is stable if the ``electrostatic" energy from the Coulomb repulsion, which may be estimated by $\sim e^2 Q^2/R$ ($R$ being the radius of the Q-ball), is subordinate to other forms of energy \cite{Lee:1988ag}. Thus, generally, if $e$ is very large, a gauged Q-ball ceases to exist.  Also, for a gauged Q-ball, the decay rate into fermions is enhanced compared to that of a global Q-ball, at least in the leading semiclassical approximation \cite{Hong:2017uhi}. This is due to the charges being concentrated near the surface for a gauged Q-ball, because of the ``electric'' repulsion.

For a favorable gauge coupling, utilizing the spherical, stationary ansatz $\varphi(r,t)=f(r)e^{-i\oi t}$ and $A_\mu=A_\mu(r)$, the gauged Q-ball in 4D can be obtained by extremizing the following Lagrangian  
\be
L = 4\pi\!\! \int\!\! \d r \,r^2 \left[- (f')^{2}+\frac{(g')^{2}}{ 2e^2} + f^2 g^2-V(f)\right],
\ee
where $g(r)=\omega+e A_0(r)$ and a prime denotes a derivative with respect to $r$. We see that a gauged Q-ball solution only depends on the scalar potential of the gauge field. In the thin-wall limit where $f$ is constant (denoted as $f$ with slight abuse of notation) within radius $R$ and zero for $r>R$, the equations of motion can be solved analytically when $efR\ll 1$. In this case, the energy of the gauged Q-ball is given by \cite{Lee:1988ag}
\be
\label{Qmaxsi}
E=Q\left[\frac{V(f)}{f^2}\right]^{1 / 2}+\frac{3 e^2 Q^2}{20 \pi R}  .
\ee
The first term here is that of the global thin-wall Q-ball discussed in \S\,\! \ref{sec:thinwall}, and the second term confirms our intuition above about the additional energy from the Coulomb repulsion. For fixed $Q$, $f$ and $R$ are interdependent, so the energy of a gauged Q-ball is obtained by minimizing  $E$ with respect to, say, $f$.

For a sextic potential $V(f)=m^2 f^2 -f^4 + \f{4\eta^2}{3m^2}  f^6$, 
the Compton wavelength of the U(1) vector inside the thin-wall Q-ball is $m_V^{-1} =  {2\eta}/{(\sqrt{3}m e)}$. For a large $\eta$, the Compton wavelength is much greater than the radius of the gauged Q-ball, in which case the inside of the gauged Q-ball is not superconducting. Essentially, in this case, the scalar self-interaction dominates over the Coulomb repulsion, and the charges are mostly uniformly distributed within the ball.

For a smaller $\eta$, however,  the Coulomb repulsion can be quite important, and charges are pushed to the surface as a result, which can be seen by solving the equations of motion numerically. For some extreme cases, the inside of the gauged Q-ball becomes superconducting. One of such cases is the choice of $\eta^2=3/16$, where the corresponding global Q-ball would have zero energy inside the ball. For this case, one can adopt a thin-shell approximation, and find that the radius and the energy of the gauged Q-ball are respectively \cite{Lee:1988ag}
\be
R=\frac{1}{2^{1 / 6}\mu}\left(\frac{e Q}{8 \pi}\right)^{2 / 3},~~~ {E}\simeq \frac{3m(e Q)^{4 / 3}}{2^{11 / 6} \pi^{1 / 3}},
\ee 
in the limit of $efR\gg 1$, and the corresponding width of the shell is $2\sqrt{2}m^{-1}$. So for a weak coupling, the radius $R$ of the gauged Q-ball is much larger than the shell width, which is of the order of $m_V^{-1}$. The bulk of the gauged Q-ball now corresponds to a lump of superconductor.

To increase the total charge of a gauged Q-ball, one can introduce two U(1) scalars (or a U(1) scalar and $N$ fermions, modeled as a relativistic gas) with opposite charges so that the Coulomb attraction and repulsion can balance out, leading to a stable large gauged Q-ball \cite{Anagnostopoulos:2001dh}. For the case of a U(1) scalar and $N$ fermions, the Q-ball energy has also been estimated by directly solving the Dirac equation as well as via a variational approach \cite{Levi:2001aw}. In 2+1D, gauged Q-balls do not exist as the energy of the ball would diverge for the Lagrangian \eqref{gaugedQball00}. This changes if one adds a Chern-Simons term, $\epi^{\ai\bi\gi} A_\ai \pd_\bi A_\gi$, to the Lagrangian, as this term introduces an effective mass for the gauge field away from the ball \cite{Deshaies-Jacques:2006clf}.

\subsection{Renormalisable models}

Due to the U(1) symmetry and the ``open-up'' requirement of the potential, the simplest Q-ball supporting Lagrangian requires a $|\varphi|^6$ term, if we require the potential to be bounded from below for stability reasons. In 2+1D, the $|\varphi|^6$ operator is marginal, so the $|\varphi|^6$ model in 2+1D is renormalisable. However, in 3+1D, the $|\varphi|^6$ operator is an irrelevant higher dimensional operator, and comes with a $1/\Lambda^2$ of the EFT cutoff $\Lambda$. From the modern perspective, there is nothing wrong with the presence of higher dimensional operators \cite{Wilson:1971bg, Weinberg:1978kz} (see \cite{Burgess:2020tbq} for an introduction of EFTs), as long as the Wilson coefficients are within a reasonable range \cite{Adams:2006sv, Tolley:2020gtv, deRham:2022hpx} and the EFT validity is not violated when using the EFT. For the Q-ball solutions, the EFT validity requires that, for example, the energy density of the Q-ball not exceed that of the cutoff scale. Generally, absent of fine-tunings, it also requires that the contributions from the higher dimensional terms should not exceed those of the lower dimensional terms. Nevertheless, it is easy to construct non-topological solitons with only renormalisable terms by including more fields. In fact, the Friedberg-Lee-Sirlin model discussed below and  consisting of a complex scalar plus a real scalar, was introduced earlier than the single complex scalar model.

\subsubsection{Friedberg-Lee-Sirlin model}
\label{sec:FriedbergLeeS}

A renormalisable 3+1D model that supports non-topological solitons can be constructed by including a real scalar $\chi$ in addition to the U(1) symmetric complex scalar $\varphi$ \cite{Friedberg:1976me} 
\bal
\label{LFriedbergLee}
\mc{L} &= - \pd_\mu \varphi^\dagger \pd^\mu \varphi - \f12 \pd_\mu\chi \pd^\mu \chi 
\nn
&~~~~ -\li^2 \chi^2  |\varphi|^2 -\frac{g^2}{8} \left(\chi^2-\chi_0^2\right)^2 ,
\eal
where $\li$, $g$ and $\chi_0$ are constants. The mass of $\chi$ is $\mu=g\chi_0/\sqrt{2}$ and the mass of $\varphi$ is $m=\li\chi_0$. The conserved charge is given by $Q=i\int\d^3 x \left(\varphi^{\dagger} \dot{\varphi} -  \dot{\varphi}^{\dagger} \varphi\right)$.

In this model, one seeks a soliton solution of the following form
\be
\chi = \f{\mu}{g} A(r),~~~ \varphi = \f{\mu}{\sqrt{2}g}B(r)e^{-i\oi t}  ,
\ee
where we have introduced dimensionless profile functions $A(r)$ and $B(r)$. For a given total charge $Q$, it is easy to find a configuration that has a lower energy than that of free $\varphi$ particles, $E<Qm$, if $Q$ is greater than some minimum charge, $Q>Q_S$. For example, one can choose $A=0,\,B=B_0 \sin \oi r/r$ for $r\le R$ and $A=1-e^{-(r-R)/l},\,B=0$ for $r\ge R$, where $\oi$, $R$ and $l$ are constant. This configuration can actually lead to an accurate estimate $Q_S$ in the limit $\kappa\equiv m/\mu\to 0$:
\be
Q_S \sim\left(\frac{4 \pi}{3 \kappa}\right)^4 \frac{1}{2 g^2}  .
\ee
In this limit, $Q$ also tends to infinity, as $Q>Q_S\to \infty$. Similar to the case in Figure \ref{fig:EvsQ}, a non-topological soliton with a $Q$ greater than $Q_S$ is absolutely stable \cite{Friedberg:1976me}. (However, in the presence of a $\chi|\varphi|^2$ term, which is allowed if $\chi$ is not $\mathbb{Z}_2$ symmetric $\chi\to-\chi$, Q-balls with very large charges become unstable \cite{Hamada:2024pbs}.) In the limit $\kappa\to \infty$, $Q_S$ is found to be bounded by 
\be
Q_S<\f{111.8}{(g^2\kappa^3)} ,
\ee
which approaches zero as $\kappa\to \infty$. So, if $g$ is not too small, we can have $Q_S<1$. In the quantum theory, $Q$ is discrete, this means that the soliton is stable for any $Q$.

On the other hand, when $\oi$ approaches $m$ from below, the energy of the soliton solution becomes greater than that of $Q$ free $\varphi$ particles
\be
E=Qm \(1+\f{\xi^2}{2\kappa^2} + \mc{O}(\xi^4)\) ,
\ee
where $\xi\equiv [\kappa^2-(\oi/\mu)^2]^{1/2}$. So the soliton in this case is unstable. Since $Q$ goes to infinity both as $\oi\to 0^+$ and $\oi\to m^-$ and the soliton solution satisfies the relation
\be
\f{\d E}{\d Q} =\oi  ,
\ee
there must be a spike at some critical value $Q_C$ on the soliton curve in the $E$-$Q$ plot. Therefore, the $E$-$Q$ plot of model (\ref{LFriedbergLee}) has a similar structure as Figure \ref{fig:EvsQ} of the single-field, EFT model. 

Similarly, for the lower energy branch, the solitons between $Q_C$ and $Q_S$ are classically stable. To show this, one can expand around the soliton solution $A\to A+\delta A,\,B\to B+\delta B$ and
evaluate the second-order perturbative energy $\delta^2 E$ for a fixed $Q$ \cite{Friedberg:1976me}:
\be
\label{E2ndclassical}
(\delta^2 E)_Q  =\frac{\mu}{2g^2} \!\! \int \!\!{\psi}^T H \psi \d^3 \rho  +\f{2 \mu \nu^3}{Q g^4}\left[\int \!\! B\delta B \d^3 \rho\right]^2 \!\!,
\ee
where $\bsb\rho\equiv \mu \bfx$, $\psi=(\delta A, \delta B)^T$ and $H$ is defined as
\be
H=-{\nabla}^2+\left(\begin{array}{cc}
\kappa^2 B^2+\frac{3 A^2-1}{2} & 2 \kappa^2 A B \\
2 \kappa^2 A B & \kappa^2 A^2-\nu^2 
\end{array}\right)  .
\ee
It has been rigorously proven that, although $H$ has one negative eigenvalue, this negativity does not outweigh the second term in \eref{E2ndclassical}, As a result, $(\delta^2 E)_Q\geq 0$, largely due to the fact that $Q^{-1}\d Q/\d \oi <0$ for the lower energy branch.

Ref \cite{Friedberg:1976me} further studied the quantum stability of the non-topological soliton with canonical quantization, and found that the lower energy branch of the $E$-$Q$ plot is only stable if $Q>Q_S$. The calculation is similar to the above classical one in spirit, both evaluating the quadratic Hamiltonian, but the quantum calculation is technically much more challenging. First of all, now, one needs to expand around the classical soliton with the concrete vibrational modes
\bal
\label{chiExpansionQ}
 \chi &=\frac{\mu}{g} A(\bsb{\rho})+\sum_{n=5}^{\infty} q_n(t) \alpha_n(\bsb{\rho}), 
 \\
 \label{phiExpansionQ} 
  \varphi&=\frac{1}{\sqrt{2}}\left[\frac{\mu}{g} B(\bsb{\rho})+\sum_{n=5}^{\infty} q_n(t) \beta_n(\bsb{\rho})\right] e^{-i \zeta(t)} ,
\eal
where we have defined $\bsb\rho=\mu(\bfx-\bfR)$, $q_n$'s are the coordinates of the vibrational modes and the mode functions are chosen to be orthonormal
\be
\int\left[\alpha_N \alpha_{N^{\prime}}+\frac{1}{2}\left(\beta_N^* \beta_{N^{\prime}}+\beta_N 
\beta_{N^{\prime}}^{\prime}\right)\right] d^3 x=\delta_{N N^{\prime}} .
\ee
The vibrational modes must then be separated from the modes associated with the collective motions, $R_1$, $R_2$, $R_3$ and $\zeta$, which correspond to translations in 3D space and in internal field space, by imposing the constraints
\bal
\int\left[\alpha_n {\nabla} A+\frac{1}{2}{\nabla} B\left(\beta_n+\beta_n^*\right)\right] \d^3 \rho&=0 ,
\\
\int\left(\beta_n-\beta_n^*\right) B \d^3 \rho&=0 .
\eal
To canonically quantize the system, the conjugate momenta operators can be introduced as follows
\be
P_j=-i {\nabla}_j, ~~~ Q=-i \frac{\partial}{\partial \zeta},~~~
p_n=-i \frac{\partial}{\partial q_n}  .
\ee
The Hamiltonian $\mc{H}= \mc{H}_0 + \mc{H}_1 + \mc{H}_2 + \cdots$ can be obtained perturbatively in terms of the small vibrational coordinates $q_n$ and momenta $p_n$, and then used to solve the Schroedinger equation $\mc{H}|~\rangle = E|~\rangle$. In the expansion (\ref{chiExpansionQ}) and (\ref{phiExpansionQ}), we have identified $\mc{H}_0$ with the classical soliton energy, so one should impose $\mc{H}_1|~\rangle=0$ to $\mc{O}(g^{-2})$ and $Q|~\rangle = Q_c|~\rangle$ to $\mc{O}(g^{-1})$, where the $Q_c$ is the charge of the classical soliton and now must be an integer. A static quantum soliton also satisfies the condition $P_i |~\rangle=0$. To determine the quantum stability of the soliton, one must evaluate all the normal modes of $\mc{H}_2$. After a lengthy calculation, which involves evaluating the normal modes for the various components of $\mc{H}_2$ and partly recycles the results in the classical stability analysis mentioned above, it is found that all the normal frequencies are real if $Q>Q_S$ in the lower energy branch, meaning that the soliton is quantum mechanically stable. For $Q_C<Q<Q_S$, the soliton is quasi-stable, but it is long-lived if $g$ is small, as its decay rate is suppressed by $e^{-1/g^2}$.

In the regime where the mass of $\chi$ is much greater than that of $\varphi$ ($\mu\gg m$), the $\chi$ field remains mostly inactive at low energies, in which case, classically, the Friedberg-Lee-Sirlin model (\ref{LFriedbergLee}) can be approximated, to a reasonably good extent, by an effective theory involving only the field $\varphi$ with the following simple potential \cite{Kim:2023zvf}
\be
\label{FLSpot77}
V_{\rm eff}(\varphi)=
\begin{cases}
m^2|\varphi|^2-\frac{2\li^4}{g^2}|\varphi|^4 , & |\varphi|<\f{g\chi_0}{2\li} ,
 \\ 
\frac{g^2 \chi_0^4}{8}, & |\varphi|>\f{g\chi_0}{2\li} .
\end{cases}
\ee 
This effective potential is derived by substituting the $\chi$ equation of motion into the Lagrangian, with the $\chi$ kinetic term neglected both in the $\chi$ equation of motion and the Lagrangian. More specifically, this is obtained by setting $\chi^2=\chi_0^2-\frac{4\li^2}{g^2}|\varphi|^2$ in the Lagrangian for $|\varphi|<\f{g\chi_0}{2\li}$ and $\chi=0$ for $|\varphi|>\f{g\chi_0}{2\li}$. As a renormalisable theory, the Friedberg-Lee-Sirlin model can be extended by adding the $m_\varphi^2|\varphi|^2$ and $h |\varphi|^4$ terms, assuming the $\mathbb{Z}_2$ symmetry for $\chi$. The one-loop Coleman-Weinberg potential of this extended Friedberg-Lee-Sirlin model has been computed in the limit of a large mass hierarchy between the $\chi$ and $\varphi$ fields, $\mu\gg m$ \cite{Kim:2024vam}, and the effective potential for $\varphi$ similar to \eref{FLSpot77} has also been obtained for this quantum corrected model. It is found that the quantum corrections in this limit can induce a lower bound on the Q-ball frequency, and for large Q-balls, they can flatten the $E$-$Q$ curve from $E\sim Q^{3/4}$ to $E\sim Q$. Some analytic approximations to the solitons in the Friedberg-Lee-Sirlin model are obtained in \cite{Heeck:2023idx, Zhong:2018hwm}, and spinning Q-balls are found in \cite{Loiko:2018mhb}. Q-balls arising from a bi-scalar theory with a Henon-Heiles-like potential, as well as from its gauged counterpart, have been investigated \cite{Brihaye:2024oji}.

\subsubsection{Non-Abelian Q-balls}

Another class of renormalisable theories that support Q-balls is to generalise the global symmetry from U(1) to a non-Abelian group, as a non-Abelian symmetry allows cubic potential terms, with which the potential can dip below the quadratic term \cite{Safian:1987pr}. 

Let us take the SU(3) symmetry for an example. Consider a $3\times 3$ matrix scalar field $\Phi$ that is traceless and Hermitian ({\it i.e.,} in the adjoint representation), whose Lagrangian 
\be
\mc{L}= {\rm Tr}\left[ - \frac{1}{2} \pd_\mu \Phi \pd^\mu \Phi - V(\Phi) \right],
\ee
is invariant under the global symmetry
\be
\Phi \to U\Phi U^\dagger,~~~U\in \text{SU(3)} ,
\ee
where the potential is given by
\be
V(\Phi)=\frac{1}{2} \mu^2 \Phi^2+\frac{1}{3!} g \Phi^3+\frac{1}{4!} \lambda \Phi^4   .
\ee
It is assumed that $\mu^2>0$, $\li>0$, and $9 \lambda \mu^2>g^2$ so that the minimum of the energy is given by $\Phi=0$. Since SU(3) contains 8 independent symmetries, there are 8 conserved charges $Q_a$. Dotting $Q_a$ into the SU(3) generators, $T_a$, we can obtain a $3\times 3$ traceless and Hermitian charge matrix
\be
\mc{Q} = i \int \mathrm{d}^3 x[\dot{\Phi}, \Phi] .
\ee
Since the $\mc{Q}$ matrix is conserved, it can be, without loss of generality, taken to be diagonal and parametrized as follows
\be
\mc{Q}={\rm diag}(-Q_1,-Q_2, Q_1+Q_2) .
\ee
Due to the symmetry between positive and negative charges, we can focus on the case where the largest charge is positive, and also choose $Q_1,Q_2\ge 0$. For a given $\mc{Q}$, the non-Abelian Q-ball takes the form
\be
\Phi = e^{i\Omega t}F(r)  e^{-i\Omega t} ,
\ee
where now $\Omega$ and $F$ are $3\times 3$ matrices. 

In the thin-wall limit, it is straightforward to demonstrate that for a Q-ball solution with $Q_1,Q_2$ strictly positive, a multi-Q-ball solution will have a smaller energy \cite{Safian:1987pr}. Consequently, the Q-ball is quasi-stable and can fission into smaller Q-balls, as can be confirmed numerically \cite{Safian:1988cz}. Thus, stable Q-balls are those for which $Q_1=0$ or $Q_2=0$, meaning the charge matrix is degenerate, ${\rm det}(\mc{Q})=0$.
These stable Q-balls are actually unitarily equivalent to SO(3) Q-balls. (Since SO(3) is a subgroup of SU(3), SO(3) Q-balls are automatically solutions in the SU(3) model.) For these Q-balls to be stable, the couplings must satisfy
\be
g^2>\mu^2 \lambda .
\ee
Ref \cite{Safian:1987pr} also argued that the surface waves in the non-Abelian model is the same as those of the Abelian case, and calculated the spectrum of the sound waves.

\section{Oscillons}
\label{sec:oscillon}

Q-balls form due to the attractive nature of the potential in the theory, with their stability guaranteed by the conservation of the U(1) Noether charge. For a real scalar theory with a similar attractive potential, localized structures can still form, and, absent of a global internal symmetry, they are not stable but are long-lived \cite{Bogolyubsky:1977tc, oscillon2}, balanced by the dissipative quadratic terms and the attractive interactions. These localized structures, parallel to Q-balls in many aspects, are commonly known as oscillons \cite{Gleiser:1993pt, Copeland:1995fq} nowadays. In this section, we shall review the nonlinear properties of these non-topological quasi-solitons {\it per se}, deferring the discussions of their applications in cosmology and particle physics to  \S\,\!  \ref{sec:applications}.

\subsection{Basics of an oscillon}

\begin{figure}
	\centering
		\includegraphics[height=3.7cm]{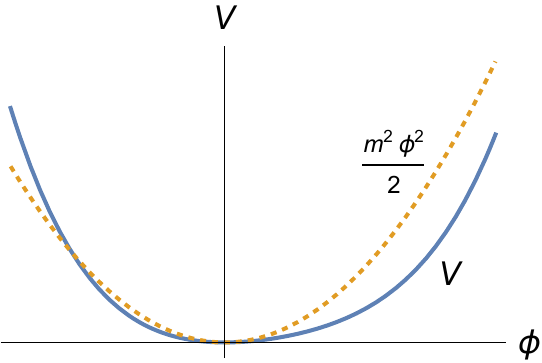}
	\caption{Example of an oscillon-supporting potential $V(\phi)$. Similar to the Q-ball case, to support an oscillon solution, $V$ must dip below quadratic mass term $m^2\phi^2/2$ in some range (see \eref{Vintosc}), leading to an attractive interacting potential. Different from the U(1) Q-ball case, now, for a real scalar, the potential can have odd powers of $\phi$, so the potential can be asymmetric with respect to the minimum.
	}
 \label{fig:VVeffosc}
\end{figure}

An oscillon is a long-lived, non-topological quasi-soliton that is approximately time-periodic. In the simplest case, oscillons arise in a real scalar field theory with an attractive potential:
\be
\label{originLagosc}
\mathcal{L}=-\f12\partial_\mu \phi \partial^\mu \phi-V(\phi) .
\ee
Similar to the case of Q-balls, for the potential $V(\phi)$ to be attractive, the interacting part of the potential $V_{\rm int}(\phi)\equiv V(\phi)-\f12 m^2 \phi^2 $ must dip below zero in some region:
\be
\label{Vintosc}
V_{\rm int}(\phi\neq 0)<0 ,
\ee
where $\phi=0$ is assumed to be the minimum or one of the minima of the potential. As the $\phi^3$ term is allowed in a real scalar theory, this condition can be satisfied by the $\phi^4$ theory $V(\phi)=\f12 m^2 \phi^2 + \li \phi^3 + g\phi^4$ with $\li<0$,  which is renormalisable even in 3+1D. See Figure \ref{fig:VVeffosc} for an example of an oscillon-supporting potential. 

A popular model that is widely studied is the symmetric double well potential \cite{Gleiser:1993pt, Copeland:1995fq}
\bal
\label{DWpotential}
V(\phi)&=\frac{m^2}{8 v^2}\left((\phi-v)^2-v^2\right)^2
\\
&=\frac{1}{2} m^2\phi^2-\frac{m^2\phi^3}{2 v}+\frac{m^2\phi^4}{8 v^2} ,  \nonumber
\eal
which has two minima/vacua at $\phi=0$ and $\phi=2v$ respectively. The existence of the central peak at $\phi=v$ makes it obvious that the condition \eqref{Vintosc} is satisfied. Without loss of generality, one may consider oscillons whose the large-$r$ field value approaches the left vacuum, and in \eref{DWpotential}, the field has been expanded around this left vacuum.  Classically, a scaling similar to that of \eref{variableScaling0} can eliminate the two parameters completely in the symmetric double well potential, reducing it to simply $V(\phi)=((\phi-1)^2-1)^2/8$. A general $\phi^4$ theory will have one free parameter in the Lagrangian after such a scaling. For a $\phi\to-\phi$ symmetric model whose vacuum is at $\phi=0$, similar to the simplest Q-ball model, a positive $\phi^6$ term must be included for vacuum stability
\be
\label{oscphi6pot}
V(\phi)=\frac{1}{2} m^2 \phi^2-\frac{\lambda}{4} \phi^4+\frac{g^2}{6 m^2} \phi^6 ,
\ee 
as the $\phi^4$ term must now be negative to satisfy \eqref{Vintosc}. This potential can give rise to flat-top oscillons \cite{Amin:2010jq, Amin:2010xe}. Another popular oscillon-supporting potential is the so-called generalised monodromy model
 \be
 \label{oscGmono}
V(\phi)=\frac{m^2 M^2}{2 \alpha}\left[\left(1+\frac{\phi^2}{M^2}\right)^\alpha-1\right] ,
 \ee
where $m$, $M$ and $\ai$ are constants. This is motivated by model building in string/M theory \cite{Silverstein:2008sg, McAllister:2008hb} and finds interesting applications in the inflationary scenario in the early universe \cite{Amin:2011hj} (see \S\,\!  \ref{sec:oscUniverse} for more details). Indeed, current observational constraints require an inflaton potential that is flatter than a quadratic one \cite{Planck:2018vyg}, which corresponds to $\ai<1$. 

Oscillons can also exist in scalar theories with derivative interactions $\mc{L}\sim X + c_2 X^2+...$ with $X=-\f12(\pd\phi)^2$ and $c_2$ being a constant, even without a potential, which are dubbed K-oscillons \cite{Amin:2013ika}. Interestingly, for the model $\mc{L}\sim X + c_2 X^2$, the condition for small amplitude K-oscillons to exist coincides with the causality/positivity bound $c_2>0$ \cite{Adams:2006sv}. The galileon terms involve even higher-order derivatives, and the existence of oscillons arising from these terms has been explored in \cite{Sakstein:2018pfd}. Oscillons also exist in multi-scalar theories \cite{VanDissel:2020umg} or vector theories \cite{Jain:2021pnk, Zhang:2021xxa, Jain:2022kwq, Zhang:2023ktk, Wang:2023tly}---Massive vector theories in the Stuckelberg formalism also contain derivative interactions. Collisions between ultrarelativistic oscillons in 1+1D have been investigated in \cite{Amin:2014fua}. For model \eqref{oscphi6pot} with $\ai$ close to 1, the potential is nearly quadratic, for which one can perform a perturbative expansion in terms of $\ai-1$ to obtain a fairly accurate analytical approximation of the oscillon, provided that the (renormalized) mass of the field is carefully chosen \cite{Levkov:2023ncb}.

\subsubsection{Radial profile}

\begin{figure}
	\centering
		\includegraphics[height=3.7cm]{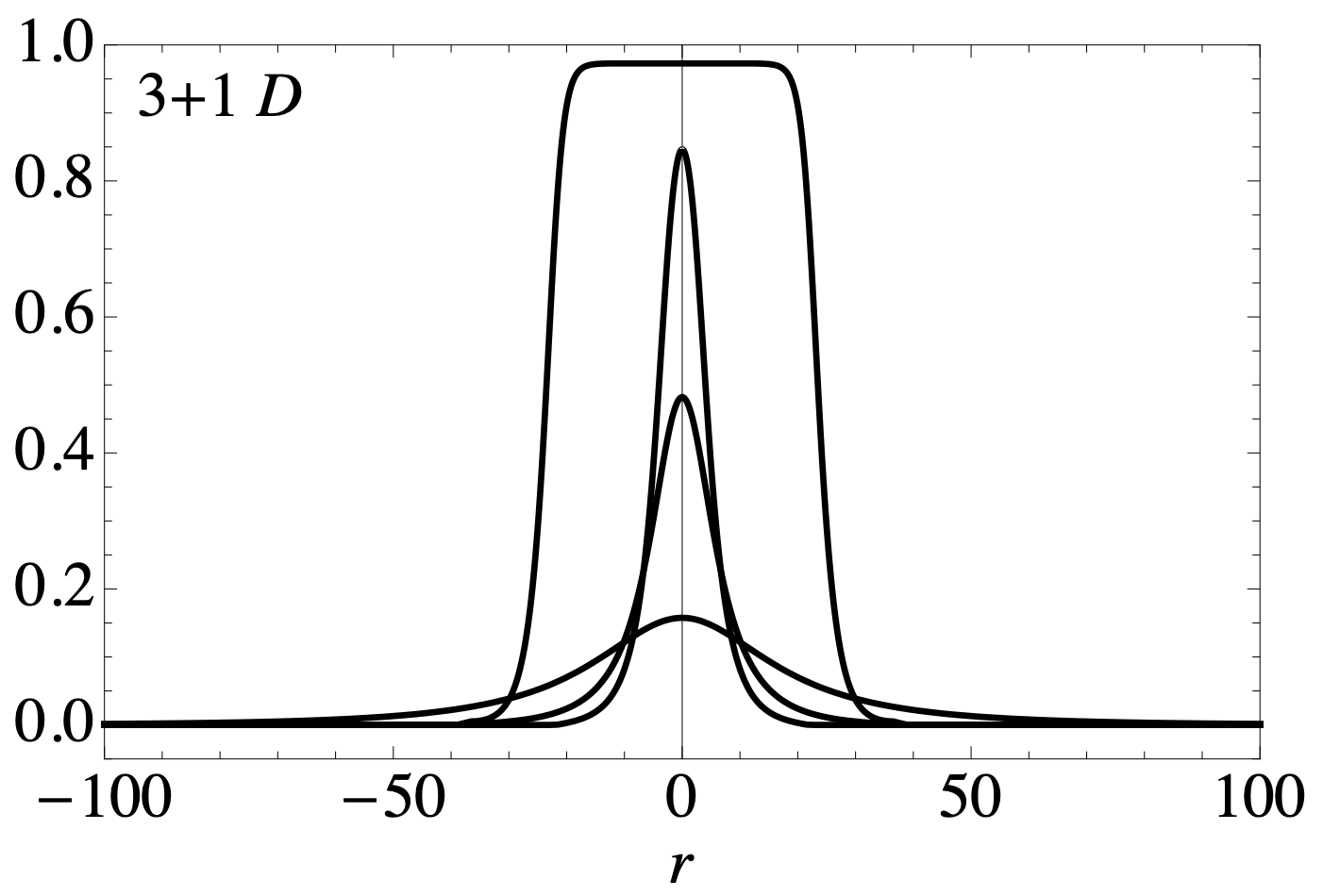}
	\caption{A few oscillon radial profiles $f_t(r)$ (in suitable units). The profile can be Gaussian-like or top-hat-like. The figure is from Ref \cite{Amin:2010jq}.}
 \label{fig:oscProfiles}
\end{figure}

Like a Q-ball, an oscillon is a localized lump of a scalar field that, as its name suggests, oscillates. However, unlike a Q-ball, an oscillon is not strictly periodic and generally cannot be factorized into spatial and temporal parts. A basic oscillon is spherically symmetric, with $\phi=\phi(t,r)$, as spherical symmetry reduces the gradient energy, according to the spherical rearrangement theorem. The equation of motion in ($d+1$)D is given by
\be
 \ddot{\phi} - \phi'' -\frac{d-1}{r} \phi'  +  \frac{\partial V}{\partial \phi}=0 .
\ee 
Schematically, the dominant oscillatory behavior of an oscillon goes like
\be
\phi(t,r)\sim f_t(r) \cos (\omega_t t) + \text{(higher frequencies)} ,
\ee
where $f_t(r)$ and $\omega_t$ weakly depend on time for a properly formed oscillon. The regularity condition at the center of the oscillon requires 
\be
\pd_r\phi(t,r=0)=0 .
\ee
With this regularity condition and a decaying asymptotic condition at large $r$, the dominant radial profile of an oscillon can be captured by truncating the equation of motion at leading order, reducing it to a case similar to that of a Q-ball. At a given instance, the oscillon profile appears Gaussian-like visually, though it is not exactly Gaussian. In fact, for the $\phi^6$ model (\ref{oscphi6pot}) with small $(\li/g)^2\ll 1$ for example, it is possible to have flat-top oscillons, which are top-hat shaped \cite{Amin:2010jq, Amin:2010xe}, much like thin-wall Q-balls;  see Figure \ref{fig:oscProfiles}.

Besides the dominant frequency, an oscillon necessarily contains many higher frequency modes.  While the dominant frequency is less than the mass of the field $m$, the higher frequencies are greater than $m$. By solving the linear equation of motion at large $r$, one can infer that the tail of an oscillon contains many of the following outgoing waves,   
\be
\phi_\Omega(t,r\to \infty) \propto  \f{1}{r^{(d-1)/2}} \cos (k r -\Omega t),
\ee
with different $\Omega^2=k^2+m^2$. If one integrates the energy of each of these waves to infinity, it diverges, so strictly speaking an oscillon is only quasi-local. This does not contradict the existence of oscillons, as an oscillon, starting from a lump with finite energy, is only meta-stable and does not exist for an infinite amount of time for the outgoing waves to propagate to infinity. Also, this is different from the case of a stable, strictly-local Q-ball whose field value decreases exponentially at large $r$. This is possible because for a complex scalar field, there exists an exact single-frequency solution whose frequency $\oi$ is less than the mass of the field $m$, thus giving rise to an imaginary wavenumber $k=\sqrt{\oi^2-m^2}$. Indeed, the profile of the oscillon's dominant-frequency mode also decreases exponentially at large $r$.

\begin{figure}
	\centering
		\includegraphics[height=4cm]{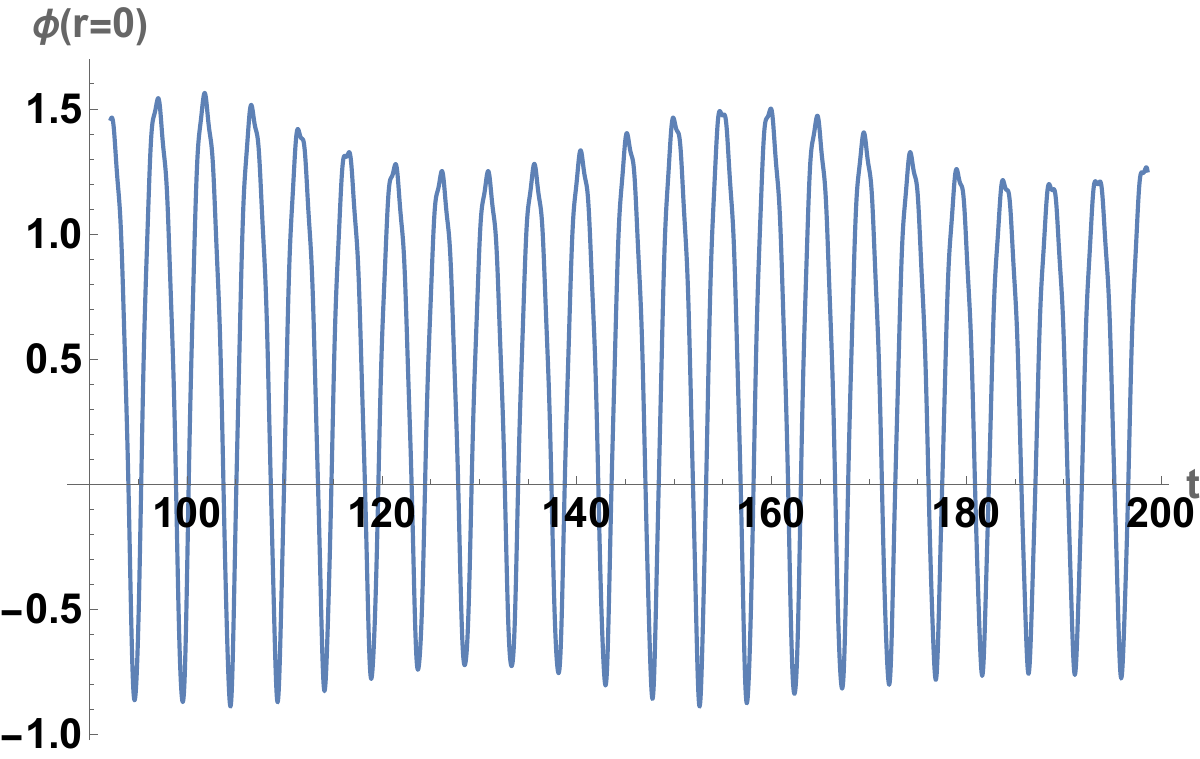}
	\caption{A typical evolution of the center of an oscillon starting from a Gaussian initial configuration for the double well $\phi^4$ potential.}
 \label{fig:20to40osc}
\end{figure}

\subsubsection{Evolution stages}
\label{sec:oscevolution}

As oscillons are quasi-stable attractors, they can form from quite generic initial conditions \cite{Gleiser:1993pt, Copeland:1995fq, Gleiser:1999tj, Fodor:2006zs, Hindmarsh:2006ur, Gleiser:2009ys, Amin:2010dc, Andersen:2012wg} (see \cite{Gleiser:2012tu} for a proposal to use the relative configurational entropy to quantify the emergence of oscillons). Depending on the initial conditions, the evolution of a quasi-stable oscillon may go through the following 3 stages:
\begin{itemize}
\item {\it Modulated amplitude}: If the initial configuration deviates significantly from the innate oscillon configuration, the oscillon's frequency and amplitude can change substantially with time in the initial relaxation process \cite{Gleiser:1993pt, Salmi:2012ta}. This manifests as a potentially large decrease in the energy of the initial lump, as well as in the slow and sizable modulations in the frequency and amplitude \cite{oscillon2}; see Figure \ref{fig:20to40osc} (and also Figures \ref{fig:oscvsqbPaul} and \ref{fig:HCphenv_energy}). The initial relaxation can be a quite  long process. It is justified to call an oscillating lump in this stage an oscillon because it is essentially an oscillon topped with some extra mildly unstable fluctuating modes \cite{Honda:2001xg, Wang:2022rhk}. In fact, an oscillon can essentially enters the fast decay stage without passing through the (strict) oscillon track stage. 

\item {\it Oscillon track}: Once properly formed, {\it i.e.,} on the oscillon's evolution track, the oscillation frequency $\oi_t$ of an oscillon slowly increases, continuously radiating away energy. The oscillating lump in this stage may be considered an oscillon in the strict/narrow sense. As a quasi-attractor, configurations close to an oscillon are naturally drawn towards this evolution track.

\item {\it Fast decay}: When the dominant oscillation frequency reaches some critical frequency that is close to the mass of the scalar field, $m$, the oscillon (or quasi-oscillon) quickly decays. Essentially, this is because at this point the dominant frequency is turned from a localized oscillation mode into a radiation mode.  
\end{itemize}

The lifetime of an oscillon depends on the number of the spatial dimensions, with the 2+1D oscillons typically living for much longer than the higher dimension ones. For example, for the symmetric double well potential $V=((\phi-1)^2-1)^2/8$, where $m$ has been chosen to be 1, starting from a Gaussian profile, a 2+1D oscillon can exist for more than $\mc{O}(10^6)/m$ and a 3+1D oscillon can exist for about $\mc{O}(10^3)/m$. Assuming spherical symmetry, results for tuning only the width of the Gaussian profile in various spatial dimensions can be found in \cite{Copeland:1995fq, Saffin:2006yk, Hindmarsh:2006ur, Honda:2001xg}, and those for turning both the width and amplitude in \cite{Andersen:2012wg}. In the latter 2D parameter-space survey, long-lived oscillons are found in up to 7D spacetime with a lifetime of about $\mc{O}(10^3)/m$. 

For some other potentials, especially those slightly shallower than the quadratic one in the range of field values relevant for the oscillon solutions, the lifetimes of 3+1D oscillons can be several orders of magnitude longer \cite{Zhang:2020bec}. Physically, these potentials exhibit weaker nonlinear interactions, leading to longer time scales. However, the potential does not need to be extremely close to the quadratic one. For example, in the generalized axion monodromy model, even with the choice of $\ai=1/2$ in \eref{oscGmono}, oscillons can live for $\mc{O}(10^8)/m$ in 3+1D. Oscillons with lifetimes exceeding the age of the universe can be constructed by judiciously choosing the potential and the initial and boundary conditions \cite{Cyncynates:2021rtf}.

\subsubsection{Quasi-breather approximation}

During the oscillon track stage, the dominant frequency of the oscillon can be treated as a constant over short-term evolution. Then, for the core of an oscillon, where the field values are substantial, the field at a given radius can be well approximated by a periodic function of time \cite{Honda:2001xg, Fodor:2006zs, Saffin:2006yk} (see also \cite{Amin:2010jq, Zhang:2020bec, Nagy:2021plv} and \cite{Fodor:2019ftc} for a review). This is often referred to as the quasi-breather \cite{Fodor:2006zs} approximation. A periodic (temporal) function can be expanded with a Fourier series. In the oscillon case the phase differences between the different Fourier modes are negligible, so we can adopt the following approximation for the core of the oscillon: 
\be
\phi(t, r)=\sum_{n=0}^{\infty} \phi_n(r) \cos (n \omega t) ,
\ee
where $\omega$ is smaller than $m$. Thus, different from the higher frequencies, the dominant frequency $\oi$ does not behave like a wave at large $r$. Substituting this ansatz into the equation of motion, we get a set of coupled equations for the different Fourier modes
\be
\nd^2 \phi_n+\left(n^2 \omega^2-m^2\right) \phi_n+N_n(\phi_m)=0 ,
\ee
where $N_n$'s are nonlinear terms in $\phi_m$, whose specific forms depend on the potential. The regularity condition at the center requires $\pd_r\phi_n(r=0)=0$. At large $r$, the tail of the quasi-breather is given by
\be
\phi_n(r\to\infty)\to
\begin{cases}
\dfrac{\alpha_n}{r^\f{d-1}2} e^{-k_n r}, & n\oi<m   ,
\\
{\tiny~~}  &
\\
\dfrac{\alpha_n}{r^\f{d-1}2} \sin (k_n r+\bi_n), & n\oi>m   ,
\end{cases} 
\ee 
where $k_n= |(n\oi)^2-m^2|^{1/2}$ and $\ai_n$ and $\bi_n$ are constants. We see that $\oi$ must be smaller than $m$ so that the dominant mode $n=1$ has an exponential tail. Typically, $\oi$ is also greater than $m/2$, in which case only the first two modes $n=0,1$ have exponential tails. With these boundary conditions, for a sufficient truncation of $n$, one can use a multi-dimensional shooting method to obtain a quasi-breather solution by minimizing either the amplitudes \cite{Fodor:2006zs} or the radiation \cite{Saffin:2006yk} at large $r$.

For a true breather (not quasi-breather) solution, we need $\ai_n=\bi_n=0$ for all modes with $n\oi>m$. These relations however are generally impossible to hold when coupled with the regularity conditions at the origin, except for some special cases such as the (integrable) sine-Gordon potential $V=1-\cos\phi$ in 1+1D. (True breathers have also been found in many nonlinear wave equations when the spatial dimensions are discretized \cite{FLACH1998181}.)

\begin{figure}
	\centering
		\includegraphics[height=6cm]{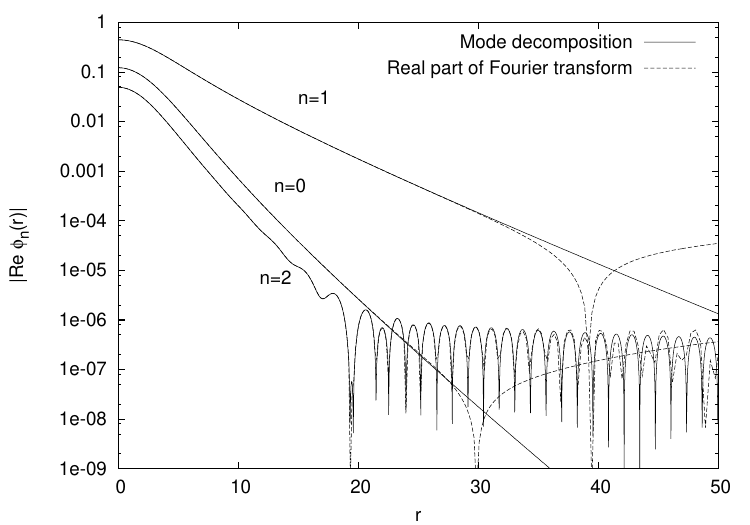}
	\caption{Comparison between the $\phi_n(r)$ Fourier modes of a quasi-breather (``Mode decomposition'') and those of the corresponding oscillon in the oscillon track stage (``Real part of Fourier transform'').  The figure is from Ref \cite{Fodor:2006zs}.}
 \label{fig:oscvsqb}
\end{figure}

Notice that the periodicity of the quasi-breather solution requires that the modes with $n\oi>m$ be standing waves. Thus, the energy within a sphere centered on the quasi-breather diverges as the sphere's radius goes to infinity. That is, the quasi-breather is also not strictly localized. This is not surprising, as a standing wave can be decomposed into two waves travelling in opposite directions. A quasi-breather is a solution in which some external source continuously pumps energy into the system from infinity such that the emitted outgoing waves are precisely combined with the incoming waves to form standing waves in the tail region. This marks the fundamental difference between an oscillon and a quasi-breather, as an oscillon is an object that only emits outgoing waves. This is also the reason why an oscillon cannot be strictly periodic, and the quasi-breather is only an approximation to the oscillon in the core region. In other words, when on the oscillon evolution track, the core of an oscillon can be viewed as adiabatically transitioning through a sequence of quasi-breathers, each with a slightly different dominant frequency $\oi_t$. See Figure \ref{fig:oscvsqb} for a comparison between the leading Fourier modes of a quasi-breather and those of the corresponding oscillon. We see that in the core region the quasi-breather is indeed an accurate approximation of the oscillon.

\begin{figure}
	\centering
		\includegraphics[height=4.2cm]{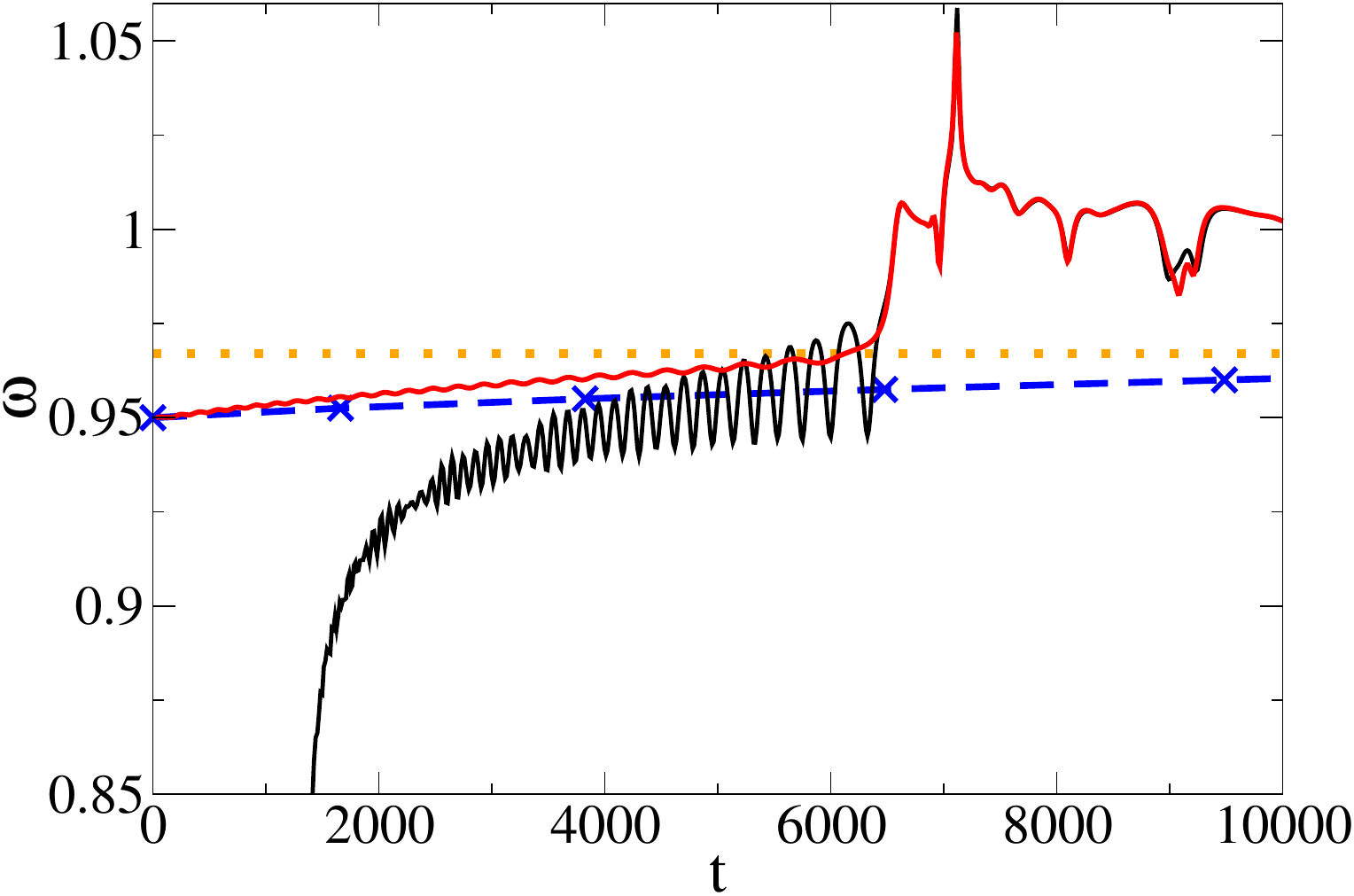}
		\includegraphics[height=4.2cm]{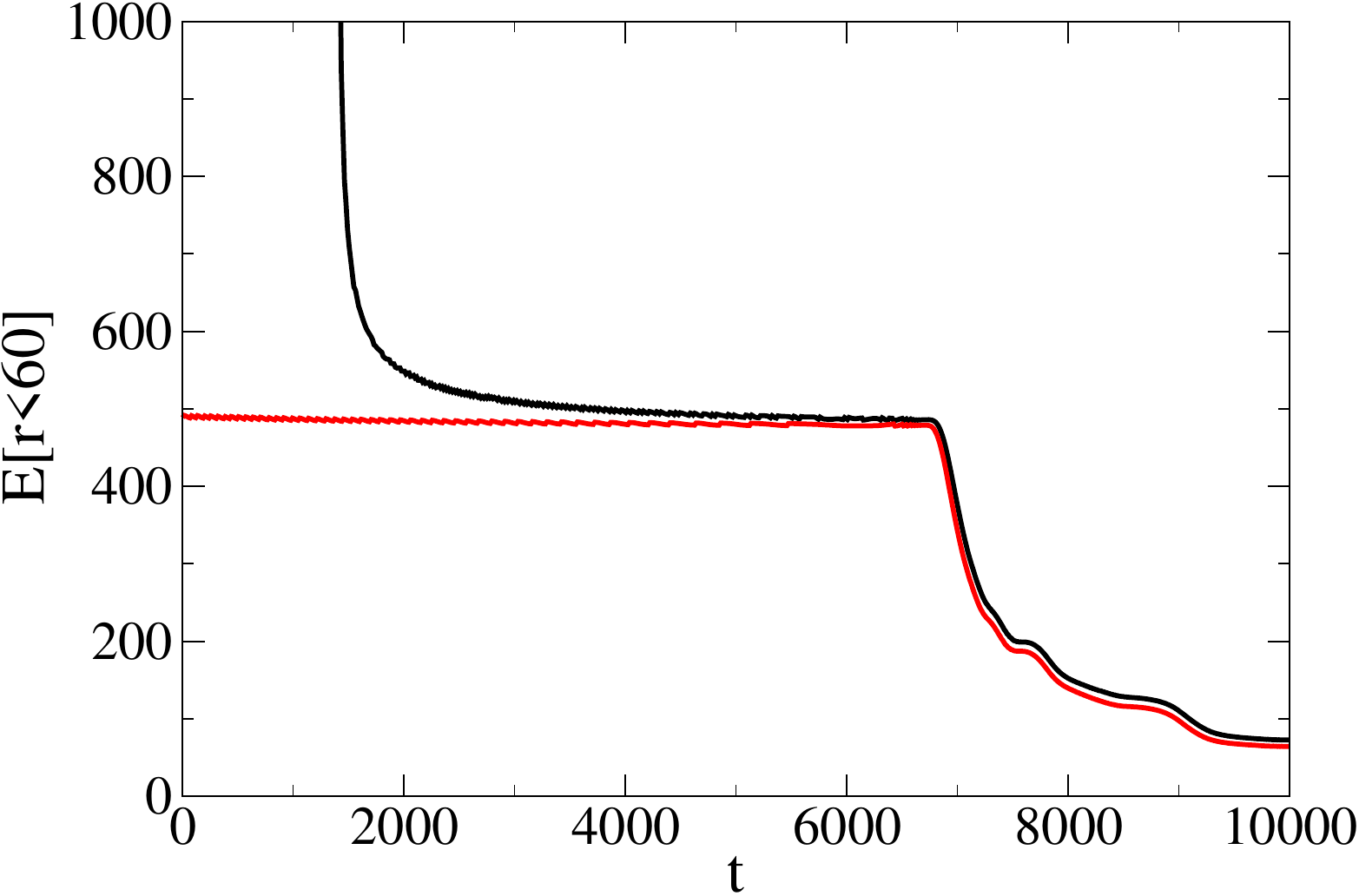}
	\caption{({\it top}) Comparison between the evolution of the oscillon's oscillation frequency from a Gaussian initial configuration (black) and that from a truncated quasi-breather initial configuration (red). The orange dotted line is the critical frequency corresponding to the lowest energy quasi-breather. The dashed blue line is the estimate obtained by computing the energy emission from the $n\oi>m$ modes with the quasi-breather approximation. ({\it bottom}) Comparison between the evolution of the energy within a shell of $r=60$ from a Gaussian initial configuration (black) and that from a truncated quasi-breather initial configuration (red). The figure is from Ref \cite{Saffin:2006yk}.}
 \label{fig:oscvsqbPaul}
\end{figure}

As mentioned above, the detailed evolution of an oscillon depends on the initial conditions, especially at the initial modulated amplitude stage. The quasi-breather approximation provides a set of initial conditions that closely resemble the inherent state of an oscillon. Thus, if one starts a simulation with the quasi-breather approximation, the energy emitted in the relaxation state is quite small, as compared to, say, that with an initial Gaussian configuration \cite{Fodor:2006zs, Saffin:2006yk}; see Figure \ref{fig:oscvsqbPaul} for  a comparison between the two. In other words, a quasi-breather essentially lies on the oscillon evolution track, whereas the Gaussian configuration is relatively distant from it by comparison. A quasi-breather has a minimal energy as $\oi$ varies, and the frequency of the minimal energy quasi-breather essentially sets the upper limit of the oscillon's frequency.

\begin{figure}
	\centering
		\includegraphics[height=6.5cm]{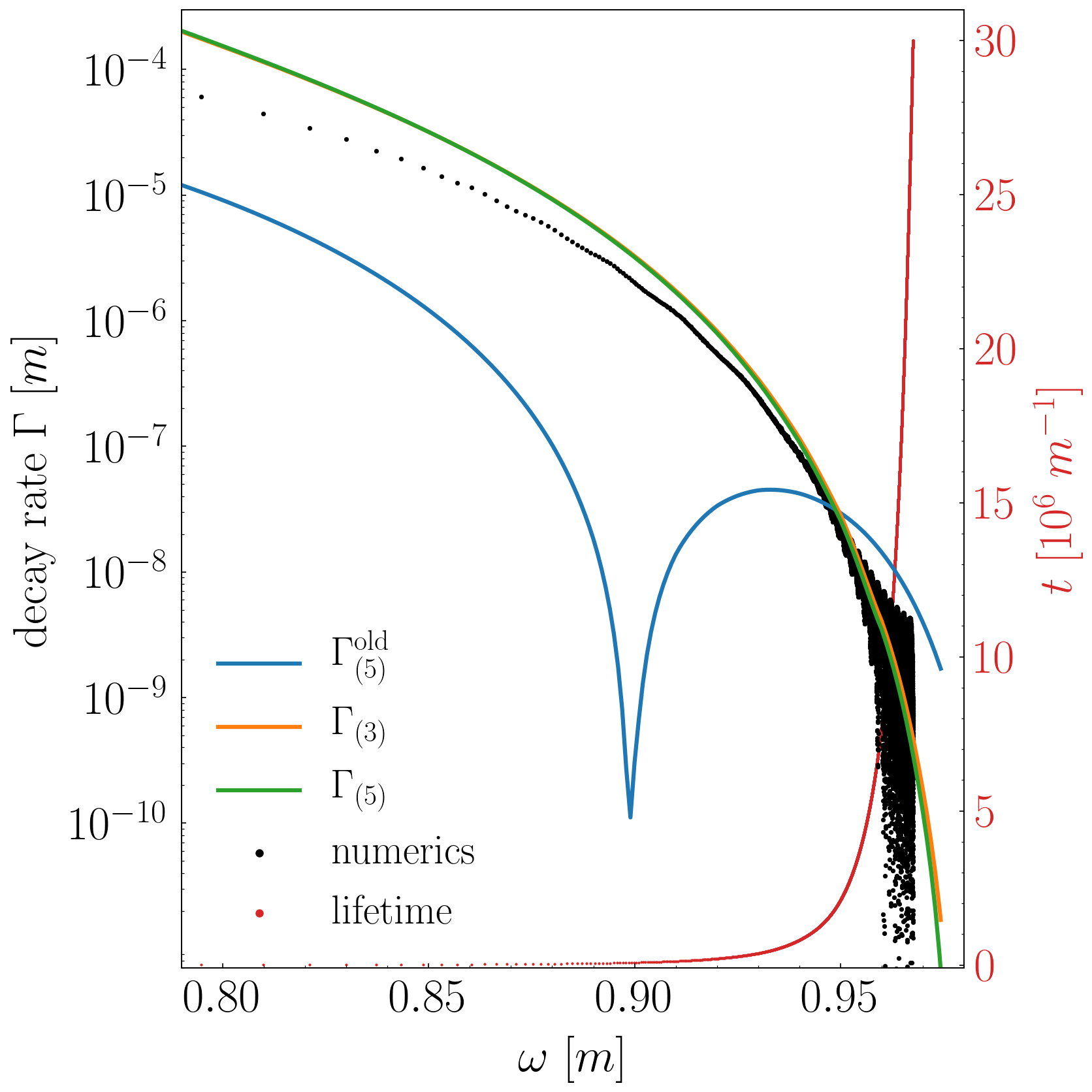}
	\caption{Comparison of decay rates computed in the quasi-breather approximation and fully numerical simulations for potential $V(\phi)=\f12 m^2 M^2 \ln \left(1+\phi^2 / M^2\right)$ at various base frequencies $\oi$. Here $\Gamma_{(3)}$ and $\Gamma_{(5)}$ represent truncations of the perturbation expansion at $3\oi$ and $5\oi$ respectively.  $\Gamma^{(\rm odd)}_{(5)}$ represents the case where the effective mass is not optimally chosen. The lifetime calculated from the $\Gamma_{(5)}$ decay rate is also plotted. This figure is from Ref \cite{Zhang:2020bec}.}
 \label{fig:oscillon_decayrate}
\end{figure}

Additionally, the $n\oi>m$ modes of the quasi-breather can be used to estimate the oscillon's radiation emission rate, by replacing the standing waves in the tail with the corresponding outgoing waves \cite{Fodor:2008du, Fodor:2009kf}.  As mentioned, the standing wave tail of a quasi-breather is sustained by balancing the outgoing radiation with ``unphysical" ingoing waves:
\bal
& ~~~~ \dfrac{\alpha_n}{r^\f{d-1}2} \sin (k_n r)\cos(n\oi t) ~ \Big\} \text{\footnotesize ~quasi-breather standing wave}\nonumber
\\
&= \dfrac{\alpha_n}{2r^\f{d-1}2}  \[  \underbrace{\sin (k_n r - n\oi t)}_\text{oscillon's radiation} + \underbrace{\sin (k_n r + n\oi t)}_\text{``unphysical" ingoing} \] .
\eal
Neglecting the ingoing waves of the quasi-breather turns out to be a good approximation to the radiation pattern of an oscillon, and the decay rate can be computed with the time-averaged outgoing flux divided by the oscillon energy. Ref \cite{Zhang:2020bec} systematically incorporates a spacetime-dependent effective mass term in the radiation equation, which significantly enhances the accuracy of the results. See Figure \ref{fig:oscillon_decayrate} for a comparison between the decay rates (and associated lifetimes) obtained with this semi-analytic method and the fully numerical results.

\subsubsection{Small amplitude limit}

In the small amplitude limit, the oscillon solution can be obtained perturbatively \cite{oscillon2, Segur:1987mg, Fodor:2008es} (see \cite{Fodor:2019ftc} for a review). To that end, one makes use of rescaled space and time coordinates
\be
\bsb\ri = \epi m\bfx, ~~~\ti = \oi(\epi) t .
\ee
The use of the $\bsb\rho$ coordinate takes into account the observation (based on numerical results) that the width of the oscillon shrinks as its amplitude diminishes. $\oi(\epi)$ is the dominant frequency of the oscillon, which is equal to the mass of the scalar in the free theory but is renormalized when interactions are included:
\be
\f{\oi(\epi)}{m} = 1 + \sum_I^\infty \epi^I \oi_I .
\ee
The equation of motion then becomes
\be
\label{qbeom0}
\f{\oi^2}{m^2} \pd_\ti^2\phi  - \epi^2 \pd_{\bsb\ri}^2  \phi +  \phi + \sum_{k=2} g_k \phi^k =0 .
\ee
Expanding the scalar field in terms of a small parameter 
\be
\phi(t,\bfx) = \sum_{I=1}^\infty \epi^I \phi_{(I)}  ,
\ee 
\eref{qbeom0} can then be decomposed into a set of equations for $\phi_{(I)}$, which can be solved perturbatively. 

For example, the first order equation is simply  $\pd_\ti^2{\phi}_{(1)}+\phi_{(1)}=0$ and gives the dominant harmonic oscillation of the quasi-breather $\phi_{(1)}=p_1(\bsb\ri) \cos \ti$, where $p_1$ will be determined by the higher order equations. Given the first order solution, the second order equation is simply a forced harmonic oscillator: $\pd_\ti^2{\phi}_{(2)}+\phi_{(2)}=-g_2 \phi_{(1)}^2-\omega_1 \pd_\ti^2{\phi}_{(1)}$; Requiring the absence of resonance in the ${\phi}_{(2)}$ solution gives $\oi_1=0$. Given the first and second order solutions, the third order equation again becomes a forced harmonic oscillator, and requiring the absence of resonance gives rise to an elliptic equation for $p_1$: 
\be
\label{p1equation}
\pd^2_{\bsb\ri} p_1+\omega_2 p_1+\(\frac{5}{6} g_2^2-\frac{3}{4} g_3\) p_1^3=0 .
\ee
For this equation to have an exponentially decaying solution at large $\rho$, $\oi_2$ must be negative and $\frac{5}{6} g_2^2-\frac{3}{4} g_3$ must be positive.  By scaling $p_1$ and $\epi$, $\oi_2$ can be chosen to be $\oi_2=-1$. To obtain a spherical solution, such as the quasi-breather, this equation simplifies to an ODE, which can be readily solved using, for example, a shooting method. Once $p_1$ is obtained numerically, the solution up to the first three orders is given by
\bal
\phi_1 & =p_1 \cos \tau , \\
\phi_2 & =\frac{1}{6} g_2 p_1^2(\cos (2 \tau)-3)  , \\
\phi_3 & =p_3 \cos \tau+\frac{1}{96}\left(2 g_2^2+3 g_3\right) p_1^3 \cos (3 \tau) ,
\eal
where $p_3$ can be determined by the higher order equations and also has a localized profile. Interestingly, it is always possible to re-arrange the solution such that $\oi/m=\sqrt{1-\epi^2}$ to all orders. If the potential is symmetric around $\phi=0$, the terms with even $k$ in \eref{qbeom0} vanish, which will result in the absence of all the harmonics with even multiples of $\oi$. Away from the small amplitude limit, the single frequency peaks will broaden, but a symmetric potential still results in the absence of broadened peaks around even multiples of $\oi$. From the Virial relations derived from \eref{p1equation} and the associated action, it follows that small amplitude oscillons only exist for $d\leq 3$ \cite{Fodor:2008es}, at least within the class of solutions explored using the method described above.

Note that the small amplitude expansion, as many perturbative expansions in physics, is actually an asymptotic expansion \cite{Segur:1987mg} and does not converge if summed to all orders. This is related to the fact that there does not exist strictly localized solutions. In other words, the small amplitude expansion only solves the perturbative equation order by order and is not an exact solution to the full equation of motion.  Nevertheless, it approximates the core of the oscillon increasingly well up to a certain order, but fails to accurately capture its tail --- the explicit solution listed above is essentially a quasi-breather. On the other hand, the radiation of the oscillon is determined by the tail. Matched asymptotic series expansion and Borel resummation techniques, which go beyond the perturbation method, have been used to obtain the radiation rates of oscillons in the small amplitude limit in 1+1D \cite{Segur:1987mg, Fodor:2008du} and higher dimensions \cite{Fodor:2009kf}. Indeed, with this treatment, the radiation rate of a small amplitude oscillon is found to be non-analytical in $\epi$ and of the form
\be
\label{oscDecayClass}
\text{(decay rate)} \sim \f{1}{\epi}\exp\(-\f{\mc{O}(1)}{\epi}\)  ,
\ee
which agrees quite well with the purely numerical results. (However, quantum mechanically, the decay rate is only power-law suppressed in terms of $\epi$ \cite{Hertzberg:2010yz}; see  \S\,\!  \ref{sec:osciQuantum} for more details.) For the 1+1D sine-Gordon model, which allows for true breather solutions, the overall proportional constant would be exactly zero. Alternatively, one can simply assume a small correction to the tail of the small amplitude solution, and the full equation of motion can be formulated as a linear wave equation for the correction, with the leading radiation mode of the small amplitude solution as the source. This equation can be easily solved with the Green's function method, which basically gives the quasi-breather's standing wave tail, and this allows one to capture the exponent of the above decay rate \cite{Hertzberg:2010yz}.

In 3+1D, small amplitude oscillons suffer from instabilities from modes whose wavelengths are comparable to the size of the oscillon \cite{Amin:2010jq, Fodor:2009kg}, depending on the specific models \cite{Fodor:2008es}. (These instabilities are absent in lower dimensions, where they can be long-lived, even in an expanding spacetime \cite{Graham:2006xs}.) To see these long wavelength instabilities, one can perform linear stability analysis on the oscillon background
\be
\phi= \phi_{\rm osc} + \chi .
\ee 
For short wavelengths, one can approximate the oscillon with a homogeneous background, for which the standard Floquet analysis can be applied. For a long-lived oscillon, the instability bands of the parametric resonance are very narrow. For the long wavelengths, one can introduce an additional (artificial) time variable
\be
\eta= \epi t, ~~|\epi|\ll 1  ,
\ee 
to capture the slow modes and apply a two variable expansion method to perturbatively solve the linear perturbation equations. For the specific model (\ref{oscphi6pot}) with $(\li/g)^2\ll 1$ considered in \cite{Amin:2010jq}, this perturbative expansion is effected by the smallness of $(\li/g)^2$, but the long wavelength instabilities seem to be prevalent beyond this limit. The leading order solution is a harmonic oscillator with slowly changing amplitudes
\be
\chi_0=u(\eta, r) \cos \oi t+v(\eta, r) \sin \oi t  .
\ee
The solutions to $u(\eta, r)$ and $v(\eta, r)$ are determined by the removal-of-resonance requirement at the next leading order equation for $\chi_1$. It turns out that absence of growing modes in $u(\eta, r)$ and $v(\eta, r)$ can be formulated as the Vakhitov-Kolokolov condition \cite{VKcondition}
\be
\label{NoscStab}
\frac{d {N}(\omega)}{d \omega}\leq 0, ~~~~~N \equiv \omega \int_0^{\infty} d r 4 \pi r^2 f(r)^2,
\ee
where $f(r)$ is the profile function of the leading harmonic of the oscillon background. For a small amplitude oscillon, this condition is often violated, leading to the long wavelength instabilities.

Note that the condition \eqref{NoscStab} is very similar to the stability condition of a Q-ball (see Eqs.~\eqref{dQdoiQball} and \eqref{Qdef1}). Indeed, the lifetime of an oscillon can be understood as an adiabatic evolution of its frequency and the approximate charge $N$ \cite{Kasuya:2002zs, Kawasaki:2015vga, Mukaida:2016hwd, Amin:2019ums, Levkov:2022egq, Zhang:2024bjo} until the oscillon, often referred to as I-ball \cite{Kasuya:2002zs} in this context, reaches the critical frequency $\oi_{\rm crit}$ such that ${d {N}(\omega_{\rm crit})}/{d \omega} = 0$, at which point the oscillon quickly decays. The role of the approximate charge becomes more prominent in the non-relativistic limit, where the particle number is conserved. In this case, the dynamics of the real oscillon field can be approximated by a nonlinear Schroedinger equation for a complex scalar field with U(1) symmetry. The longevity and instabilities of the oscillon can then be understood in terms of the U(1)-symmetric and symmetry-violating terms respectively \cite{Mukaida:2014oza}.

\subsection{Complex oscillons}
\label{sec:complexOsc}

While an oscillon on its intrinsic evolutionary track can be accurately approximated by a sequence of quasi-breathers, the transition to the oscillon track from generic initial profiles also reveals some intriguing features about the quasi-attractor behaviours and the complex structures of the oscillon.

\subsubsection{Resonant structure and critical scalings}
\label{sec:resonantS}

Starting the evolution from a lump of a Gaussian profile in 3+1D 
\be
\label{inigausslump}
\phi(t=0,r) = C e^{-r^2/r_0^2}, ~~~\dot\phi(t=0,r) =0 ,
\ee
which lies off the oscillon track, Refs.~\cite{Copeland:1995fq, Honda:2001xg} carefully mapped the lifetimes of the oscillon for a fixed $C$ (specifically $C=2v$) and various $r_0$ in the symmetric double well potential \eqref{DWpotential}. As the oscillon is long-lived, to uncover the ``fine structure'' of this spectrum in a finite-sized lattice, Ref \cite{Honda:2001xg} introduces a special spherical coordinate system to implement effective absorbing boundary conditions at large $r$. This efficiently eliminates the relatively large amount of radiation emitted by the initial Gaussian lump that would otherwise be reflected back from the boundaries of a periodic lattice, polluting the physical system of concern. The clever coordinate system, dubbed monotonically increasingly boosted (MIB) coordinate system, chooses a new radius coordinate $\tilde{r}$, which is related to the standard radius by
\be
\tilde{r} = r - f(r) t ,
\ee 
where $f(r)$ is chosen to smoothly interpolate from 0 to 1 at some sufficiently large $r_c$.  So for $r\ll r_c$, where the oscillon is located, the MIB coordinates reduce to the standard spherical coordinates; for $r>r_c$, $\tilde{r}$ becomes an almost null coordinate, resulting in both the outgoing and ingoing waves being effectively frozen (their propagation seeds dropping to zero) in the transition layer around $r_c$. Furthermore, as the outgoing waves are blue-shifted as they approach $r_c$ in the MIB coordinates, the Kreiss-Oliger dissipation term, which is explicitly added to the equation of motion, can effectively quench the outgoing waves. With this method, the computational resources required for a simulation scale almost linearly with the total time.

\begin{figure}
	\centering
		\includegraphics[height=6.7cm]{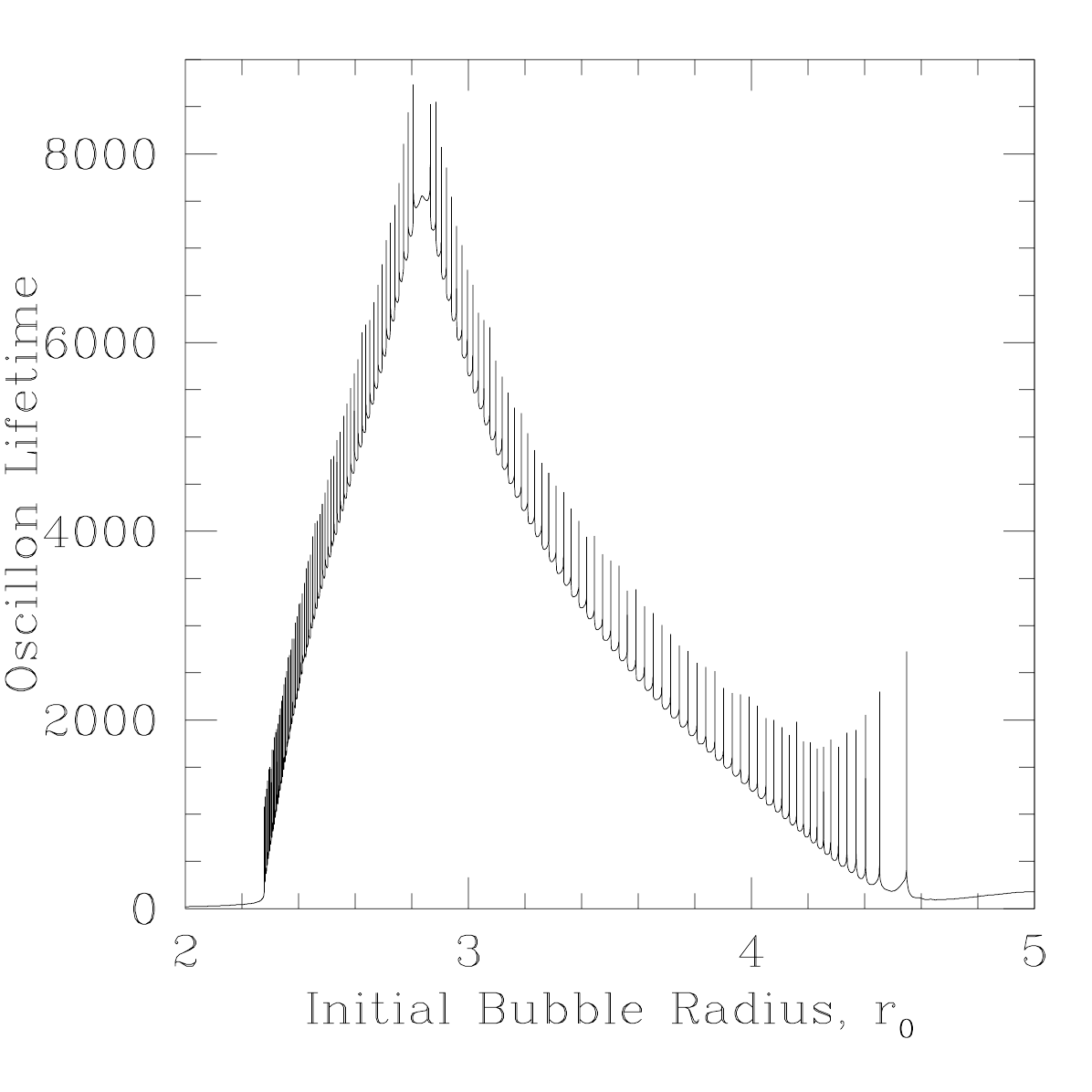}
	\caption{Fine resonant structure of the oscillon lifetime with a Gaussian lump (see \eref{inigausslump}) as the initial profile for $2\leq r_0\leq 5$. The potential is the symmetric double well potential \eqref{DWpotential}, and the units are such that the mass of the field is $m=\sqrt{2}$.  This figure is from Ref \cite{Honda:2001xg}.}
 \label{fig:HClifetime}
\end{figure}

\begin{figure}
	\centering
		\includegraphics[height=3.2cm]{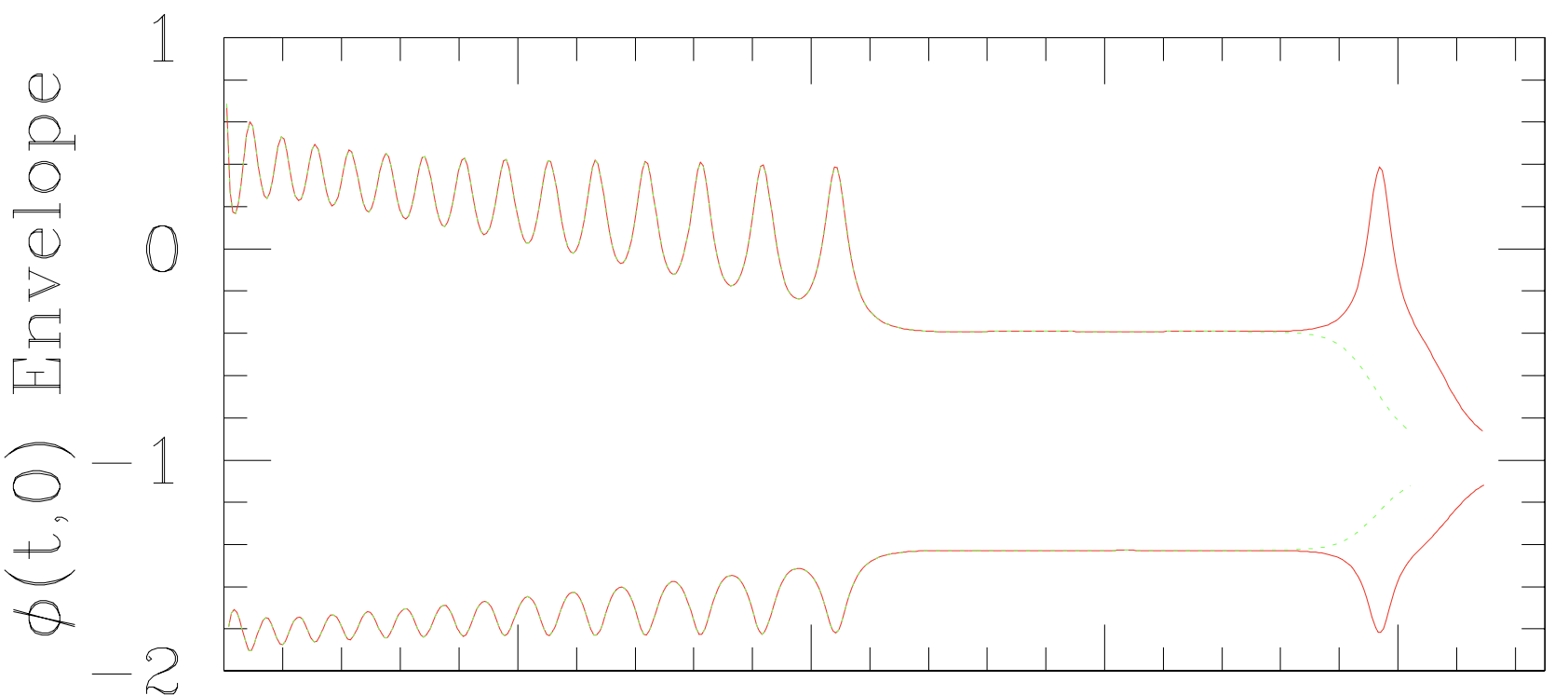}
	\caption{Evolution of $\phi(t,r=0)$ of the oscillon near a resonance peak of Figure \ref{fig:HClifetime}. This figure is from Ref \cite{Honda:2001xg}.}
 \label{fig:HCphenv_energy}
\end{figure}

Using the MIB coordinates and implementing extreme fine-tuning of $r_0$, Ref \cite{Honda:2001xg} was able to identify more than a hundred resonance peaks in the lifetime vs $r_0$ spectrum; see Figure \ref{fig:HClifetime}. (For $r_0\geq 5$, oscillons can also form, and their lifetimes are also charted in \cite{Honda:2001xg}. For large $r_0$, the Gaussian lump tends to experience more dramatic bounces in its initial evolution.) A typical initial Gaussian lump will gradually radiate away a significant amount of energy before decaying in a fast burst. During its lifetime of slow radiation, the amplitude envelope of the lump oscillates and slowly decreases before the fast decay. By tuning $r_0$ near a resonance peak, however, a constant plateau emerges in the amplitude envelope after a period of slow oscillations; see Figure \ref{fig:HCphenv_energy}. When the amplitude envelope stabilizes, the accumulated radiated energy also reaches a plateau. According to our terminology in \S\,\! \ref{sec:oscevolution}, the plateau period near the resonant peak is the oscillon track stage, and the initial period with an oscillating envelope is the initial relaxation/modulated amplitude stage.

Interestingly, the lifetime vs $r_0$ relation for an initial Gaussian lump follows a scaling law similar to the type I critical phenomena in gravitational collapse. Specifically, for each resonant peak $r_0^\star$, the lifetime of an initial Gaussian lump obeys the following power-law scaling
\be
(\text{lifetime})^{\rm 3+1D}_{\rm oscillon} \simeq c_r\left|r_0-r_0^{\star}\right|^{\gamma_r} ,
\ee
where $c_r$ and $\gamma_r$ are constants depending on the resonant peak. $c_r$ is also slightly different when approaching the peak from the left and the right. Although it varies for different peaks, the critical exponent $\gamma_r$ is around 40 for most peaks (see Figure 8 of Ref \cite{Honda:2001xg}).

The dependence of the lifetime on both $r_0$ and $C$ in \eref{inigausslump} has also been carefully mapped \cite{Andersen:2012wg}, which found that in this 2D parameter space the resonant spikes become thin bands. See \cite{Zhang:2020ntm} for gravitational effects on the lifetimes of the oscillons.

\subsubsection{Composite oscillons}

The radial profile of the lowest energy oscillon is nodeless at any given moment---its solution cannot be separated into temporal and spatial factors for all times. Similar to multi-nodal excited Q-balls, it is possible to set up initial spherical configurations where the spatial profile has multiple nodes in the case of a real scalar field. However, this type of excited oscillon is relatively short-lived. 

\begin{figure}
\center
\centering
\includegraphics[width=7.3cm]{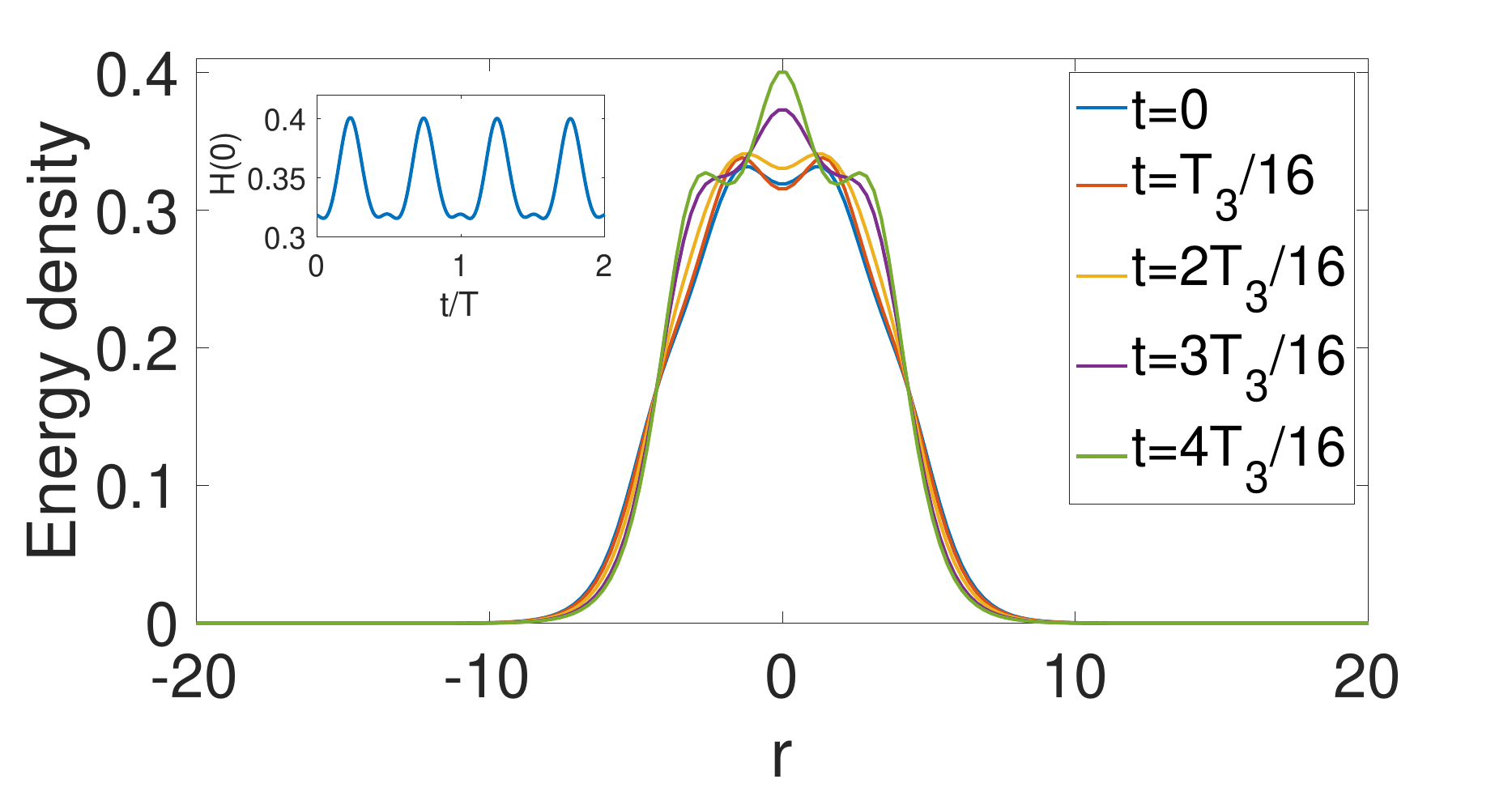}

\includegraphics[width=7cm]{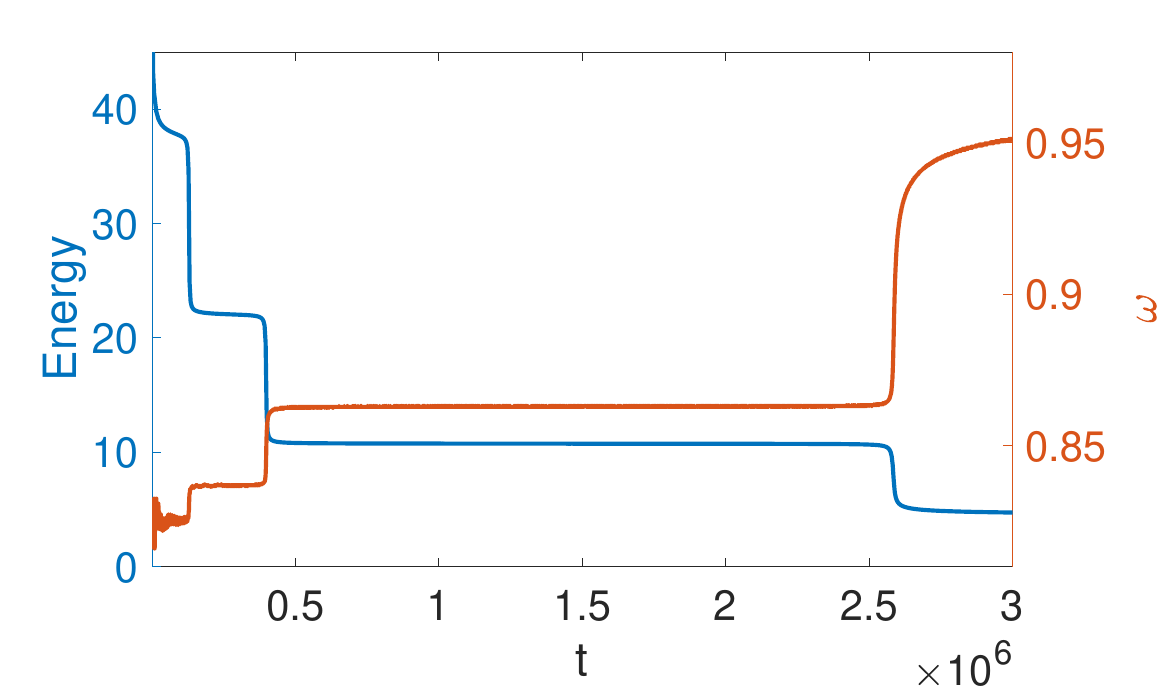}
	\caption{({\it top}) Evolution of the energy density of an excited oscillon in 2+1D (the 3rd lowest plateau of the bottom plot, $T_3$ being the oscillation period of the field $\phi$ at the center of the oscillon. The inset is the evolution of the energy density at the center. ({\it bottom}) Cascading energy levels of an excited spherical oscillon in 2+1D. The frequency $\oi$ is defined to be the (angular) frequency of a point in the center of the oscillon. Four distinct energy plateaus can be seen in this example, the last plateau being the nodeless oscillon. This figure is from Ref \cite{Wang:2022rhk}. }
	\label{fig:osciLevel3}
\end{figure}

Nevertheless, there exist another type of excited oscillon which has a dominant nodeless component plus an additional weakly multi-nodal structure \cite{Wang:2022rhk}. During the time evolution, the spatial profile of this composite oscillon exhibits approximate nodes; See Figure \ref{fig:osciLevel3} for an example of this kind of oscillon with 3 approximate nodes for a $\phi^6$ potential in 2+1D. These weakly multi-nodal oscillons are extremely long-lived in 2+1D, and they are local attractor solutions, similar to the basic oscillons.

Intriguingly, a composite oscillon with multiple approximate nodes will undergo a cascading decay, successively transitioning through configurations with decreasing numbers of nodes; see Figure \ref{fig:osciLevel3}. In this cascading process, the energy of the oscillon decreases and its frequency increases, both of which exhibit various correlated plateaus. Similar cascading behaviours in the energy have also been observed in 3+1D \cite{Honda:2001xg}. The lifespans of these composite oscillons can be relatively accurately estimated semi-analytically using a quasi-breather approximation \cite{Wang:2022rhk}. The existence of the cascading levels of the oscillon frequency is consistent with the fact that as the frequency changes, the decay rate will have multiple dips where the oscillon's radiation is extremely suppressed \cite{Ibe:2019vyo}. Although not yet explicitly confirmed, this also seems to be related to the fine resonant structure of the lifetime spectrum discussed in  \S\,\!  \ref{sec:resonantS}.

\begin{figure}
	\centering
	\includegraphics[width=9cm]{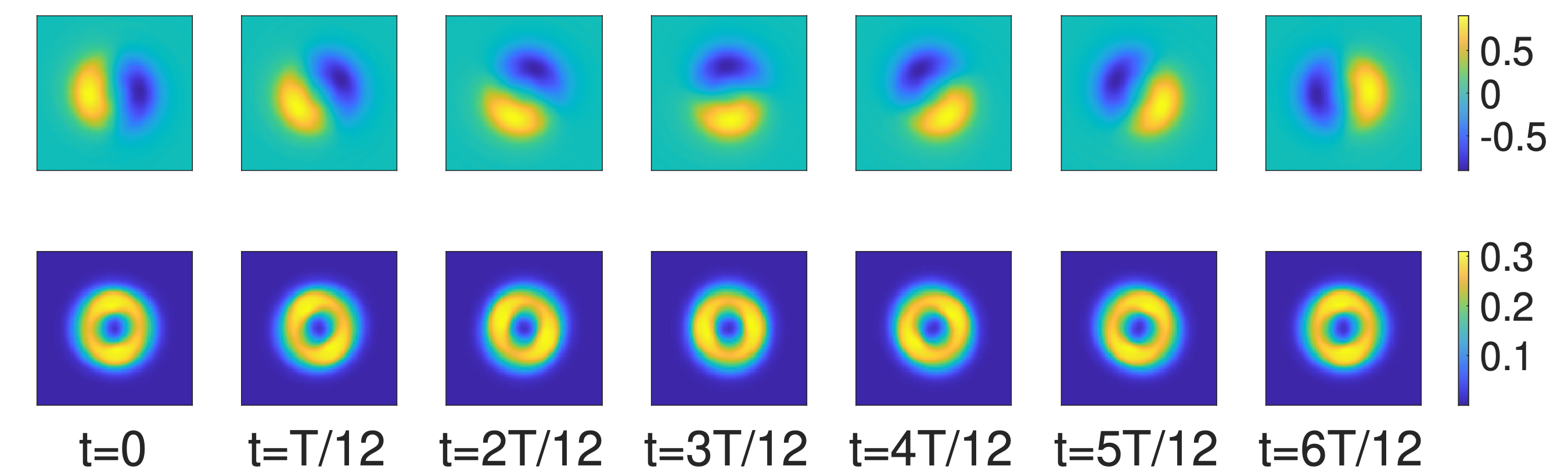}
	\caption{Time sequence of a dipolar oscillon. The $\phi$ evolution is shown in the top row, and the bottom row is for the energy density, with $T=2\pi/\oi$ being the oscilation period of the field. This figure is from Ref \cite{Wang:2022rhk}. }
	\label{fig:oscHandphi}
\end{figure}

The equivalent of spinning Q-balls (see  \S\,\!  \ref{sec:spinningQball}) can also be constructed for oscillons \cite{Wang:2022rhk}. Different from the Q-ball case, due to the restriction of a real scalar, such a composite oscillon necessarily contains multiple multipoles.  In 2+1D, a quasi-breather approximation for this kind of composite oscillon is given by:
\be
\label{spinningoscExp}
\phi =\sum_{n\geq 1, \tilde m\geq \tilde m_0}  f^{\tilde m}_n(r) \cos \left(n\omega t-\tilde m \theta\right) .
\ee
It is possible to construct a solution where the lowest multipole ($\tilde m_0$) dominates. Compared with a spherical oscillon, a spinning oscillon has a smaller attractor basin for the initial conditions. To numerically construct an $\tilde m_0$-multipole dominated oscillon, it is usually necessary to use the equation of motion to get a good estimate for $f^{\tilde m_0}_n(r)$. For example, to construct a dipolar oscillon, we can substitute the ansatz $\phi(t,r,\thi)=f^1_0(r) \cos \left(\omega t- \theta\right)$ into the equation of motion and solve for  $f^1_0(r)$, which can then be used as the initial configuration to relax to the spinning oscillon. See Figure \ref{fig:oscHandphi} for a typical evolution of a dipolar oscillon in 2+1D. For the $\phi^6$ model, the dipolar spinning oscillon can live for $\mc{O}(10^4/m)$, which is shorter than the lifetime of the level-4 spherical multi-nodal oscillon discussed above. The $\tilde m_0=2,3$ spinning oscillons last for $\mc{O}(10^3/m)$ and $\mc{O}(10^2/m)$ respectively. The decay rates of these spinning oscillons can also be accurately calculated by treating the higher order terms in \eref{spinningoscExp} as radiation fields. 

To construct more stable spinning oscillons, one can utilize vector fields with intrinsic polarisations \cite{Jain:2021pnk, Zhang:2021xxa, Jain:2022kwq, Zhang:2023ktk, Wang:2023tly}, removing the need for orbital angular momentum. In addition to circularly polarised or spinning oscillons, vector oscillons can also be linearly polarised. Both types maintain approximately spherically symmetric energy densities and can evolve from initial Gaussian configurations.

\subsection{Quantum corrections}
\label{sec:osciQuantum}

The quantum stability of small amplitude oscillons has been analyzed in \cite{Hertzberg:2010yz}. The quantization of the field is performed on top of the oscillon background $\phi_{\text {osc}}(t,\bfx)$:
\be
\phi(t,\mathbf{x})=\phi_{\text {osc}}(t,\bfx)+{\eta}(t,\mathbf{x}) ,
\ee
where the quantum field ${\eta}(t,\mathbf{x})$ is approximated to linear order in the Heisenberg equation of motion. For a small amplitude oscillon, its profile is dominated by small wavenumber modes with $k=\mc{O}(\epi)$, with $\epi$ charactering the small amplitude, so it is more convenient to solve the problem in the $k$ space. The $k$-mode equation of motion is given by
\be
\ddot\eta_k +\oi_k^2 \eta_k + \int \frac{d^d k^{\prime}}{(2 \pi)^d} B\left({k}-{k}^{\prime}\right) {\eta}_{k^{\prime}}=0 ,
\ee
where $\oi_k^2=k^2+m^2$ and $B\left({k}-{k}^{\prime}\right)$ is the contribution from the oscillon background. At linear order, the mode functions of ${\eta}_k(t)= \int \!\! \frac{d^d q}{(2 \pi)^d} {a}_q v_{q k}(t)+\text {h.c.}$ also satisfy the same equations of motion
\be
\label{oscQuantummodes}
\ddot v_{qk} +\oi_k^2 v_{qk} + \int \frac{d^d k^{\prime}}{(2 \pi)^d} B\left({k}-{k}^{\prime}\right) v_{qk^{\prime}}=0 .
\ee

For a potential containing some weak coupling, this set of coupled ODEs can be solved perturbatively. For example, for potential $V= \f12 m^2\phi^2-\lambda_4 \phi^4+ \lambda_5 \phi^5+\ldots$, the background oscillon can be approximated by $\phi_{\text {osc }}=\phi_\epsilon(r) \cos (\omega t)+\mathcal{O}\left(\epsilon^3\right)$, where $ \phi_\epsilon\propto \epi p_1(\epi r)/\sqrt{\li}$ with $p_1$ satisfying \eref{p1equation}. The mode functions $v_{qk}$ can be perturbatively solved in powers of $\li_5$. Specifically, the zero-th order solution is simply
\be
v_{q k}^{(0)}(t)=\frac{e^{-i \omega_k t}}{\sqrt{2 \omega_k}}(2 \pi)^d \delta^d(\mathbf{q}-\mathbf{k})+\mathcal{O}\left(\epsilon^2\right)
\ee
and the next order equation is given by
\be
\ddot{v}_{q k}^{(1)}+\omega_k^2 v_{q k}^{(1)}=-\frac{\tilde\Phi_\epi(|\bfq-\bfk|)}{48 \sqrt{2 \omega_k}}\left[e^{i \omega^-_q t}+e^{-i \omega^+_q t}\right] ,
\ee
where $\oi^\pm_q=3\oi\pm \oi_q$ and $\tilde\Phi_\epi(k)$ is the Fourier transform of $\phi_\epi^3(r)$. These are just forced harmonic oscillators, which can be easily solved, assuming the quantum modes start in the vacuum, with initial conditions $v_{q k}^{(1)}(0)=\dot{v}_{q k}^{(1)}(0)=0$. The energy of the quantum perturbations can be approximated with
\be
E(t)=\frac{1}{2} \int \frac{d^d q}{(2 \pi)^d} \frac{d^d k}{(2 \pi)^d}\left[\left|\dot{v}_{q k}\right|^2+\omega_k^2\left|v_{q k}\right|^2\right] ,
\ee
and subtracting the usual quantum zero-point energy gives the radiated energy of the oscillon.  To evaluate the radiated energy, one recognizes that $\tilde\Phi_\epi(|\bfq-\bfk|)$ is peaked near $\bfq-\bfk=0$ and makes use of Fermi's golden rule. The decay rate of the oscillon can then be computed by taking the time derivative of the radiated energy divided by the oscillon energy. The resulting decay rate turns out to be the same as the tree-level decay rate of the scattering process $3\phi\to2\phi$ if viewing the oscillon as a gas of particles. This is reasonable as the calculation above assumes the underlying theory is weakly coupled. Building on this correspondence, the results can be easily generalized to the case of a general potential.

Compared to the classical decay rate of a small amplitude oscillon, which is exponentially suppressed (see \eref{oscDecayClass}), the quantum decay rate generally has a power-law suppression. For example, if the interaction potential starts with $\phi^3$ (or $-\li_4\phi^4+\li_5\phi^5$), the decay rate is suppressed by $\epi^4$; if it starts with $-\li_4\phi^4+\li_6\phi^6$, the decay rate is suppressed by $\epi^6$. The decay rate is also suppressed by powers of the weak couplings, the explicit form of which depends on the sizes of the couplings.

The mode functions \eqref{oscQuantummodes} can be numerically solved nonperturbatively. With the numerical solutions, a Floquet analysis reveals that the growth of the perturbations on the oscillon is linear, suggesting the absence of the parametric resonance type of instabilities. The situation changes if $\phi$ is coupled to an extra scalar, where parametric resonance can occur for many choices of couplings, leading to the exponential growth of perturbations.

Beyond the small amplitude limit, the inhomogeneous Hartree approximation has been used to nonperturbatively simulate the quantum evolution of oscillons \cite{Saffin:2014yka}. These simulations are similar to those in the Q-ball case, which has been discussed in much detail in \S\,\!  \ref{sec:QballInhomo}). In particular, the stochastic ensemble method can be used to speed up the time consuming quantum simulations \cite{Borsanyi:2007wm, Borsanyi:2008eu}.  Ref \cite{Saffin:2014yka} focused on a specific $\phi^4$ model, and by scaling the mass and the quartic coupling to unity, it essentially restricts to a model that is relatively strongly coupled. As a result, the quantum corrections were found to significantly reduce the lifetime of oscillons even in 2+1D. The outcomes will differ if a smaller quartic coupling is chosen. Nevertheless, even in a relatively strongly coupled case, the attractor basin of the quantum oscillon is similar to that of the classical one.

\section{Applications in cosmology and particle physics}
\label{sec:applications}

In the preceding sections, we mainly focused on the properties and dynamics of non-topological solitons and quasi-solitons, examining them as nonperturbative field-theoretical structures in their own right. In this section, we shall turn to  applications of these objects and their significance in particles physics and cosmology.

\subsection{Existence of Q-balls in MSSM}
\label{sec:existMSSM}

For possible Q-ball-supporting global continuous symmetries in fundamental physics, two natural candidates to consider are the baryon number ($B$) and lepton number ($L$). In the Standard Model, the baryon and lepton U(1) symmetries are only violated nonperturbatively at the quantum level, which is very small except at high temperatures, although the $B-L$ symmetry is exactly conserved. Unfortunately, the Standard Model does not accommodate Q-balls. (In the hypothetical scenario without the gauge fields, the Higgs field can allow Q-ball to form if thermal effects are taken into account \cite{Pearce:2022ovj}.) However, as already envisaged by Coleman \cite{Coleman:1985ki}, Q-balls are easy to realize in extensions of the Standard Model, particularly in the supersymmetric (SUSY) models \cite{Kusenko:1997zq, Kusenko:1997si, Enqvist:1997si, Kasuya:1999wu}. The Q-balls in these contexts are often called B-balls or L-balls.

As experimental data continue to accumulate, the simplest versions of the Minimal Supersymmetric Standard Model (MSSM) at low scales have faced increasing tensions with measurements such as those from the Large Hadron Collider \cite{ParticleDataGroup:2022pth}. While the generic MSSM model, with its over a hundred free parameters, is far from being ruled out, a substantial space of phenomenological MSSM models remains consistent with the experimental data. In the absence of an experimentally verified concrete model, studies of Q-balls in the MSSM are typically restricted to simplified analyses of their dynamics, using well-motivated potentials and couplings.

\subsubsection{Flat directions in MSSM}

The reason why Q-balls naturally arise in SUSY models is that there are many flat directions in the scalar potential, {\it i.e.}, field-space directions where the scalar potential vanishes \cite{Dine:1995kz}. They are already ubiquitous in the MSSM, due to the large accidental degeneracies in the model. After these flat directions are lifted by the soft SUSY breaking terms, the potential often grows slower than the quadratic one, rendering them of the Q-ball-supporting type. 

In the MSSM, the scalar potential consists of the F-terms and D-terms: 
\be
V=\sum_I\left|F_I\right|^2+\frac{1}{2} \sum_A  D^A D^A, 
\ee
with $F_I = {\partial W_{\rm RN}}/{\partial \phi_I}$ and $D^A=g_A \phi^{\dagger} T^A \phi$, where scalar $\phi$ lives in the gauge group generated by generator $T^A$ with gauge coupling $g_A$ and $W_{\rm RN}$ is the renormalisable superpotential 
\be
W_{\rm RN} = \bar{u} {y}_{u} Q H_u-\bar{d} {y}_{{d}} Q H_d-\bar{e} {y}_{{e}} L H_d+\mu H_u H_d .
\ee
In the superpotential, $y_u$, $y_d$ and $y_e$ are the $3\times 3$ Yukawa couplings (due to the 3 generations), $H_u$ and $H_d$ are the two Higgs fields required in the MSSM, and $Q$, $L$, $\bar u$, $\bar d$ and $\bar e$ denote the corresponding squarks and sleptons, with $Q$ and $L$ being SU(2)$_L$ doublets and $\bar u$, $\bar d$ and $\bar e$ being SU(2)$_L$ singlets. In principle, the space of the flat directions, often referred to as the moduli space, can be determined by solving the following two sets of equations simultaneously 
\be
F_I =0~~~{\rm and}~~~D^A=0 .
\ee 
In practice, this space can be more readily characterized by searching for holomorphic, gauge-invariant polynomials of the chiral superfields \cite{Buccella:1982nx, Affleck:1983mk}, thanks to the condition $D^A=0$. (That is, the flat directions are necessarily gauge singlets.) Using this method, a systematic catalog of MSSM flat directions for the renormalisable and SUSY-preserving part was compiled in \cite{Gherghetta:1995dv}. Examples of these renormalisable flat directions include $LH_u$, $\bar u \bar d \bar d$,  $QL\bar d$ and so on.

Many of these flat directions carry a baryon or lepton number. (Since $B+L$ is destroyed by the sphaleron processes at high temperatures, these directions are often labeled by the  $B-L$ number.) For example, the following flat direction carries $B-L=-1$ \cite{Dine:1995kz}:
\be
Q_1^\alpha=\frac{1}{\sqrt{3}}\binom{\varphi}{0}, \quad L_1=\frac{1}{\sqrt{3}}\binom{0}{\varphi}, \quad \bar{d}_2^\alpha=\frac{1}{\sqrt{3}} \varphi ,
\ee
where $\ai$ is the color index, the subscripts label the generation and $\varphi$ is a complex scalar field parametrizing this flat direction. Of course, in phenomenological viable MSSM models, the flat directions are only approximately flat, lifted by soft SUSY breaking terms as well as by the non-renormalisable terms in the superpotential at high energies.

There are a number of proposals for spontaneously breaking SUSY, which reduce to the Standard Model under conventional circumstances. These proposals typically involve introducing new fields at higher energy scales, because neither D-term or F-term breaking works in the minimal setup (see, {\it e.g.,} \cite{Martin:1997ns}). They assume the spontaneous SUSY breaking occurs in this hidden sector of new fields, indirectly mediated to the visible sector ({\it i.e.,} the MSSM) via flavor-blind interactions. Two popular such proposals are the gravity-mediated SUSY breaking and the gauge-mediated SUSY breaking, which will be discussed momentarily, and we will see that they often lead to Q-ball-supporting potentials. 

It is useful to explicitly parameterize possible SUSY breaking at low energies. One of the main motivations to introduce SUSY is to address the hierarchy problem, reducing the sensitivity of the Higgs mass to the UV scales. Thus, for the broken SUSY model to remain a solution to the hierarchy problem, the explicit SUSY breaking must be soft, which can generally be achieved by Lagrangian terms whose coupling constants have positive mass dimensions and are around the weak scale. Explicitly, the general soft SUSY breaking Lagrangian terms are given by \cite{Martin:1997ns}
\bal
\mathcal{L}_{\text {soft }} &\!=-\frac{1}{2}\left(M_a \lambda^a \lambda^a+\text {c.c.}\right)-(m^2)_J^I \phi^{J\dagger} \phi_I
\nn
&~~  -\left(\frac{1}{2} b^{IJ} \phi_I \phi_J+\frac{1}{6} a^{I J K} \phi_I \phi_J \phi_K+\text {c.c.}\right) ,
\label{Lsoftterms}
\eal
where  $M_a$'s are gaugino masses, $(m^2)_J^I$ is the mass matrix of the squark, sleptons and Higgses, and $a^{IJK}$'s are the so-called A-term couplings. In particular, mass terms of the form $m_I^2|\phi_I|^2$ are not forbidden by the symmetries of the theory, so they are generically expected to be present, which weakly lift the flat directions.

Of course, quadratic mass terms alone do not support Q-balls. However, as we will see in the next subsections, in either the gauge- or gravity-mediated SUSY breaking scenario, quantum corrections typically cause the potential to grow more slowly than the mass terms \cite{Kusenko:1997zq, Kusenko:1997si, Enqvist:1997si}, thus allowing for Q-ball solutions.

Eventually, at large field values, we may expect the non-renormalisable terms, originated from integrating out higher energy degrees of freedom, to kick in, which strongly lift the flat directions. At the level of the superpotential, the dominant contributions can be schematically written as
\be
W\supset \sum_{n>3} \frac{\lambda_n}{\Lambda^{n-3}} \phi^n .
\ee
where the $\phi^n$ terms may be purely flat-direction fields or contain a non-flat-direction field \cite{Dine:1995kz}, $\Lambda$ is a heavy UV mass scale and $\lambda_n$ is dimensionless. The $\phi^n$ order term above leads to a potential term of the form
\be
\label{V2nm6}
V \supset \frac{|\lambda_n|^2}{\Lambda^{2 n-6}}(\phi^\dagger\phi)^{ n-1} ,
\ee
which conserves the global symmetries of the complex scalars.  Compared with the soft SUSY breaking mass term, the higher dimensional term (\ref{V2nm6}) dominates if $({m \Lambda^{n-3}}/{|\lambda_n|})^{1 /(n-2)}<|\phi|<\Lambda$, where $m$ is the soft mass scale. In fact, in the gravity-mediated SUSY breaking scenario, as we will see shortly, these terms are needed to stabilize the vacuum.  How the MSSM flat directions are lifted by these non-renormalisable terms has been cataloged in \cite{Gherghetta:1995dv}.  While most of the flat directions are lifted by terms with $n\leq 6$, a couple of ``very flat'' directions can only be fully lifted by higher dimensional operators, the ``flattest'' one being the $L\bar u\bar d$ direction, which can only be fully lifted by $n=9$ terms, corresponding to dimension-16 operators in the scalar potential.

There are also U(1) breaking potential terms but they turn out to be always subdominant in the MSSM, largely thanks to $R$-parity invariance \cite{Dine:1995kz, Gherghetta:1995dv}. In particular, there can be the following A-terms
\be
\label{Vsterms}
V \supset \frac{A}{\Lambda^{s-3}} \phi^s+{\rm c.c.} ,
\ee
where $A$ is expected to be around the weak scale in the vacuum so as not to spoil the solution to the hierarchy problem. These A-terms, though subdominant, play a key role in Affleck-Dine baryogenesis. They violate the baryon number (or $B-L$) symmetry and CP invariance, the latter due to the generally complex nature of the $A$ couplings.

In principle, when dealing with Q-balls in the MSSM, the treatment in \S\,\! \ref{sec:Q-balls}, which deals with a single complex scalar, should be generalised to accommodate multiple fields. In the multi-field case, an argument similar to the one around \eref{Efunctional} indicates that we should seek a Q-ball solution of the form \cite{Kusenko:1997zq}
\be
\phi_I(x) = f_I(r) e^{i  q_I \oi  t}   ,
\ee
and the existence condition now is that
\be
\min\( \f{V(|f_I|)}{\sum_I q_I f^2_I} \) \text{ is reached at }f_I \neq 0 .
\ee
As the minimum of the scalar energy functional, a Q-ball is stable against decay into scalar particles. However, carrying baryon or lepton numbers, Q-balls in the MSSM can decay into the Standard Model particles via fermionic evaporation on the surface (see \S\!   \ref{sec:fermionEvap}). For this reason, there is a significant difference between B-balls and L-balls. This is because the lightest leptons, neutrinos, are nearly massless, whereas the evaporation of a B-ball is often obstructed by the large mass of the proton. 

In practice, lack of confirmation of SUSY and concrete information about the theory parameters, Q-balls in the MSSM are usually modeled with one U(1) scalar field. This is often justified because the non-flat directions surrounding a flat one are much steeper as they move away from the minimum of the potential, a regime that is particularly pertinent to many properties of Q-balls.

\subsubsection{Gauge-mediated SUSY breaking}
\label{sec:gaugeSUSY}

The properties of Q-balls in the MSSM differ significantly regarding how the SUSY breaking is mediated to the visible sector. Let us first look at the Q-balls in the gauge-mediated SUSY breaking scenario \cite{Kusenko:1997si}.

In gauge-mediated SUSY breaking \cite{Dine:1981gu, Nappi:1982hm, Alvarez-Gaume:1981abe, Dine:1993yw, Dine:1994vc, Dine:1995ag}, the soft MSSM terms arise from loop-level interactions between the MSSM fields and some new messenger fields via gauge and gaugino couplings, the messengers transmitting the SUSY breaking from the hidden sector. In the minimal setup, the messenger fields are left-handed chiral superfields charged under the Standard Model gauge groups: $\mc{Q} \sim(\mathbf{3}, \mathbf{1},-\frac{1}{3}), \bar{\mc{Q}} \sim(\overline{\mathbf{3}}, \mathbf{1}, \frac{1}{3}), \mc{L} \sim(\mathbf{1}, \mathbf{2}, \frac{1}{2}), \bar{\mc{L}} \sim(\mathbf{1}, \mathbf{2},-\frac{1}{2})$.  They are assumed to be coupled to a hidden sector gauge-singlet chiral superfield, denoted as $S$, via the superpotential
\be
W_{\rm mgr} = y_2 S \mc{L} \bar{\mc{L}}+y_3 S \mc{Q} \bar{\mc{Q}}  .
\ee
The SUSY is broken by $S$ obtaining a VEV $\langle S \rangle$ from the F-term in its superpotential, and the MSSM sparticles can get soft masses at one-loop or two-loop order
\be
\label{msoftGauge}
m_{\rm soft} \sim \f{g^2_a}{(4\pi)^2}\f{\langle F \rangle}{M_{\mathrm{mgr}}} ,
\ee 
where $g_a$ is a relevant Standard Model gauge coupling and $M_{\mathrm{mgr}}\sim y_i\langle S \rangle$ is the characteristic mass scale of the messengers. Specifically, one-loop diagrams with $\mc{Q}$ and $\bar{\mc{Q}}$ ($\mc{L}$ and $\bar{\mc{L}}$) generate masses for the gluino and the bino (the wino and the bino), which are of the form $M_a={g^2_a\langle F \rangle}/((4\pi)^2\langle S\rangle)$, and the masses of the scalar sparticles come from two loops and depend on the quadratic Casimir invariants of the gauge groups. The soft-breaking $a^{IJK}$ and $b^{IJ}$ terms (see \eref{Lsoftterms}) are usually further suppressed at least at the messenger scale. 

For simplicity, let us focus on one particular flat direction parameterized by a complex scalar $\varphi$. 
While the minimum of the scalar potential is controlled by the soft mass term $m^2|\varphi|^2$, a typical feature in the gauge-mediated SUSY breaking is that the potential tends to be very flat moving away from the minimum \cite{Kolda:1997wt}. This is of course before the non-renomalisable terms \eqref{V2nm6} become important, but the non-renomalisable terms are usually negligible for the Q-ball solutions. For example, taking into account one-loop corrections, a typical effective potential for the lifted flat direction in this scenario is given by \cite{Dvali:1994ms}
\be
V(|\varphi|)=m^4 \ln \left(1+\frac{ |\varphi|^2}{m^2}\right) +\frac{\lambda_4^2}{M_P^2} |\varphi|^6  ,
\ee
where $\lambda_4$ is dimensionless and $m$ is the soft mass $m_{\rm soft}$ introduced above, which may be taken to be around the weak scale. In the context of the early universe, as will be discussed later, there will also be a soft mass term of the form $c H^2  |\varphi|^2$, where $H$ is the Hubble parameter and $c$ is a dimensionless constant. This term is induced by the finite density in the early universe, and $c$ can often be negative, which is important for the Q-ball generation via the Affleck-Dine condensate in the early universe (see \S\,\! \ref{sec:QballEarlyUni}). The SUSY breaking A-terms have been neglected in this potential, as they are important for baryogenesis but negligible for the Q-ball dynamics.

This kind of potential satisfies the existence conditions of Q-balls (see \S\,\! \ref{sec:Qballbasics}). In fact, due to the ultra flatness of the potential for large field values, $V(|\varphi|\gg m)\sim V_0=const$, the total energy of the Q-ball in this model scales with the total charge $Q$ as \cite{Dvali:1997qv} 
\be
\label{EQ3s4}
E_Q\propto Q^{3/4}   .
\ee
This scaling is different from the thin-wall Q-ball for which we have $E_Q \sim Q$ (see \eref{twEQ}). In fact, the standard thin-wall ansatz \eqref{thinwallansatz} does not apply to this kind of Q-ball. Instead, the following approximate ansatz is more appropriate
\be
f(r)= 
\begin{cases}
f_{\rm in}\f{\sin \omega r} { \omega r}, & r<R 
\\ 
f_{\rm out} e^{-m r}, & r \geq R,
\end{cases}
\ee
where $f(r)$ is continuous at $R$, and $f_{\rm in}$, $f_{\rm out}$, $R$ and  $\oi$ are determined by minimizing the energy for a given total charge. It can be obtained by solving the equation of motion (see \eref{fEoM}) asymptotically as $r\to 0$ (using the fact that $V\to V_0$ in this region) and as $r\to \infty$. From the shape of the ${\sin \omega r} /{ \omega r}$ function, we see that the size of this kind of Q-ball can be taken as $R\sim 1/\oi$. 

To understand the scaling \eqref{EQ3s4} intuitively, one can define dimensionless variables
\be
\hat{x}_\mu = \oi x_\mu,~~~\hat{f}=f/\oi .
\ee
For a large Q-ball, we can neglect the surface energy, thus neglecting the gradient terms in the energy functional. Also, by integrating within the radius $|\hat{\bfx}|<1$ ({\it i.e.,} $|\bfx|<R$) from the center of the Q-ball, we can re-write the energy as
\bal
E & \simeq  \int \d^3 x \[ \dot{\varphi}^{\dagger} \dot{\varphi} +V \] = \int_{|{\bfx}|<R}\!\! \d^3 x \[ \oi^2 f^2 +V_0 \] 
\\
\label{dropTermsQ34}
&  \simeq \f12 \omega Q+ \f1{\oi^3} \int_{|\hat{\bfx}|<1}\!\! \mathrm{d}^3 \hat{x} {V_0}  \simeq  \f12 \omega Q  +\f{V_0}{\oi^3} .
\eal
Note that this approximation relies on the fact that the potential is almost constant  inside the Q-ball where the field values are large. To get the Q-ball solution, we can minimize $E$ with respect to $\oi$, which gives $E_{Q}\simeq V_0^{1/4} Q^{3/4}$, obtained when $\oi\simeq (V_0/Q)^{1/4}$.

The intriguing aspect of the scaling \eqref{EQ3s4} is that the energy per particle for this kind of Q-ball is
\be
\label{EQQMSSM}
\f{E_Q}{Q}\simeq \(\f{V_0}{Q}\)^{1/4} \sim \f{m}{Q^{1/4}} ,
\ee
where $V_0$ has been approximated by the mass of the flat direction field, $V_0\sim m^4$. 
This means that, as the Q-ball grows, the binding energy of the ball increases and its energy per particle decreases. In the MSSM, B-balls can only evaporate to fermionic baryons. The lightest baryons are nucleons, whose masses are around 1\,GeV. So if the energy per particle of the B-ball is smaller than 1\,GeV, which, according to \eref{EQQMSSM}, requires
\be
Q \equiv B \gtrsim \left(\frac{m}{1\,{\rm GeV}}\right)^4  ,
\ee
then the B-ball is stable even against fermionic evaporation. For $m\simeq 1$TeV, this requires $B\gtrsim10^{12}$. Thus, in the absence of thermal effects, gauge-mediated MSSM naturally supports stable, large B-balls. The thermal effects of the hot Big Bang on the lifetime of a B-ball will be discussed in the next subsection.

For L-balls, they will evaporate via neutrinos which are almost massless. Using the evaporation bound \eqref{upperBEvap}, one can infer that the lifetime of an L-ball exceeds the age of the universe if $Q\equiv L \gtrsim 10^{32}$ \cite{Kusenko:1997si}. If L-balls decay after the Big Bang nucleosynthesis, it will lead to unwanted entropy increase or even distort the cosmic microwave background (CMB). Flipping the argument around, this fact can be used to place constraints on the MSSM models and the initial conditions from inflation. 

When the binding energy increases as the Q-ball becomes bigger, one may need to consider the gravitational corrections. As a rough estimate, these effects are important when the size of the Q-ball becomes comparable to its Schwarzschild radius. As mentioned above, the size of a Q-ball in the gauge-mediated scenario is $R \sim 1/\oi \simeq Q^{1/4}/m$, while the Schwarzschild radius of the Q-ball is $r_S\sim E_Q/m_P^2\simeq m Q^{3/4}/m^2_P$, where $m_P$ is the Planck mass. This means that gravitational effects are negligible if $Q\lesssim (m_P/m)^4$. For $m\simeq 1$TeV, this means $Q \lesssim  10^{64}$.

\subsubsection{Gravity-mediated SUSY breaking}
\label{sec:gravitySUSY}

Q-balls also exist in the gravity-mediated SUSY breaking scenario \cite{Enqvist:1997si, Enqvist:1998en}. Gravity-mediated SUSY breaking assumes that the SUSY breaking is passed to the visible sector by gravitational interactions \cite{Chamseddine:1982jx, Barbieri:1982eh, Ibanez:1982ee, Ellis:1982wr, Hall:1983iz, Alvarez-Gaume:1983drc}, which are famously ``universal'' rather than merely flavor-blind. This is also known as Planck-mediated SUSY breaking, as the gravitational coupling is of the order of $1/M_P$. 

Typically, one might expect that the characteristic mass scale of the soft breaking couplings (see \eref{Lsoftterms}) in this scenario is of the order
\be
\label{gravitymsoft}
 m_{\text {soft }} \sim \f{\langle F\rangle} {M_{\mathrm{P}} }  .
\ee
(The $b^{IJ}$ parameters may involve an additional mass scale: $b^{IJ}\sim m_{\text {soft }} \mu$.)
If $m_{\rm soft}$ is around or slightly above the weak scale, the SUSY breaking scale might be around $\langle F\rangle^{1/2}\sim 10^{11}$GeV or slightly higher. On the other hand, after the SUSY breaking, the gravitino ``eats'' the spin-$\f12$ goldstino and becomes massive.  By dimensional analysis, the gravitino mass $m_{3/2}$ is expected to be around 
\be
m_{3/2} \sim \f{\langle F\rangle}{M_{\mathrm{P}}} ,
\ee
which applies to both gauge-mediated and gravity-mediated SUSY breaking. So the gravitino mass is quite different in the two SUSY breaking scenarios, as the breaking scale $\langle F\rangle^{1/2}$ differs significantly in these two scenarios.  In gravity-mediated breaking, $m_{3/2}$ is around the soft mass scale, but the gravitino interacts with the gravitational coupling $1/M_P$, so it is essentially irrelevant in collider physics. Nevertheless, it can play interesting roles in the early universe. In contrast, in gauge-mediated breaking, the mass of the gravitino is much smaller than the soft masses of sparticles, assuming $M_{\rm mgr}\ll M_P$, and it is likely that the gravitino is the lightest SUSY particle, which is absolutely stable and a candidate of dark mater.

Although the gravity-mediation paradigm sets the typical mass scale of the soft couplings, it does not reduce the large number of parameters in the theory. Experimentally, as a starting point, the fiducial model to test is the so-called constrained MSSM, in which many of these masses and couplings are assumed to be equal, leaving only five parameters. More specific models also relate these parameters to the gravitino mass $m_{3/2}$. See \cite{Ellis:2022emx} for a recent comparison of the constrained MSSM with experimental data.

Let us look at the example of the $\bar u\bar u\bar d$ flat direction\,\footnote{As another example, the $\bar u\bar u\bar d \bar e$ flat direction would give a similar leading terms but with a $|\varphi|^{6}$ term, instead of $|\varphi|^{10}$.}, along which the lifted scalar effective potential is given by \cite{Enqvist:1997si, Enqvist:1998en}
\be
\label{VEnqvist}
V = m^2|\varphi|^2\[1+K \log \left(\f{|\varphi|^2}{M^2}\right)\]+\frac{\lambda^2|\varphi|^{10}}{M_P^6} ,
\ee
where the $K$ term takes into account the one-loop running effects for the scalar mass, arising mainly from the gauginos. Explicitly, the value of $K$ is given by \cite{Nilles:1983ge}
\be
K \simeq -\frac{1}{3} \sum_{a} \frac{\alpha_{g_a}}{8 \pi} \frac{M_a^2}{m^2}  <0 ,
\ee
where $\ai_{g_a}$'s are the gauge couplings and the sum is over the gauginos that become massive due to the VEV of $\varphi$ from the couplings to $\bar u$ and $\bar d$. The main contributions come from the gluinos, so choosing $\ai_{g_3}\sim 0.1$, we see that $K$ is naturally in the range of  $-0.01$ to $-0.1$. $M$ is the reference scale at which the scalar mass is $m$ and is often taken to be the GUT scale. Essentially, this effective potential is obtained by replacing the tree-level mass parameter with the one-loop corrected one and promoting the renormalisation scale $\mu$ to be the field $\varphi$ \cite{Einhorn:1982pp}.

Because of the negativity of $K$, the potential \eqref{VEnqvist} supports Q-balls. If the field values of the Q-balls are not too large, the effects of the $|\varphi|^{10}$ term are negligible. In the potential above, we have neglected the A-terms, ${A \varphi^6}/{M_P^3}+ {\rm c.c.}$, which are usually negligible for the Q-ball formation at late times. If we only keep the first term in \eref{VEnqvist}, an interesting aspect is that the exact Q-ball solution is analytic and has a Gaussian profile with radius $\si=(-Km^2)^{-1/2}$ (see \eref{gaussianQball}), which is not of the thin-wall type. In the presence of a sizable $|\varphi|^{10}$ term, $V/|\varphi|^2$ has a minimum and thin-wall Q-balls can form. The energy of a large Q-ball in this scenario scales with the charge as $E_Q\simeq \oi Q$. Therefore, different from the gauge-mediated scenario, even large B-balls can now decay into nucleons or lightest SUSY particles such as neutralinos. The exact lifetimes of these B-balls depend on the thermal effects in the early universe. 

In this scenario, the charge of a large Q-ball is roughly $Q \simeq \f43\pi R^3 f_0^2\oi \simeq E_Q/\oi$, where $R$ is the radius of the Q-ball. For the gravitational effects to be negligible, we again need $R$ to be smaller than the would-be Schwarzschild radius, which gives $Q\lesssim m^3_P/f_0\oi^2\sim m_P^3/m^3$. For $m\sim 1\,$TeV, this means that the gravitational effects are negligible for Q-balls with $Q\lesssim 10^{48}$.

Thick-wall Q-balls may also arise in this scenario. For example, the $H_uL$ flat direction carries lepton number, and the quantum-corrected effective potential can be roughly modeled as
\be
V(|\varphi|) = \frac{m^2}{2}\left(2 e^{-s |\varphi|}-1\right) |\varphi|^2 ,
\ee
where $s$ is a mass scale not far from $m$. This potential gives rise to thick-wall L-balls (see \S\,\! \ref{sec:nonAnalyPot}). The typical radius of such a L-ball is $R \sim m^{-1}$ and the total charge is about $L \sim (s m)^{-2}$ \cite{Enqvist:1997si}, so these are small Q-balls that couple to neutrinos, and thus will evaporate very quickly at high temperatures.

\subsubsection{Hybrid-mediated type}

Lastly, it is important to note that because gravity is universal, even in the gauge-mediated SUSY breaking scenario, the gravity-mediated effects may also be sizable. So, neglecting the higher order lifting terms, generally, one may consider the following potential 
\bal
V(\varphi) &=  m^4 \log \left(1+\frac{|\varphi|^2}{m^2}\right) 
\nn
&~~~~ +m_{3 / 2}^2|\varphi|^2\left[1+K \log \left(\frac{|\varphi|^2}{M^2}\right)\right] ,
\eal
where $m$ is around the electroweak scale and $m_{3/2}$ is the gravitino mass, which can now be as low as $10^4$GeV$-$1GeV. This gives rise to a new type of MSSM Q-balls \cite{Kasuya:2000sc, Kasuya:2014ofa}. The Q-ball solution in this potential extrapolates between the two types of Q-balls discussed above
\be
\frac{E_Q}{Q}\simeq \begin{cases}m_\phi Q^{-1 / 4} & \varphi \lesssim \varphi_{\mathrm{eq}} \\ m_{3 / 2} & \varphi \gtrsim \varphi_{\mathrm{eq}}\end{cases} ,
\ee
where $\varphi_{\mathrm{eq}} \sim {m_\phi^2}/{m_{3 / 2}}$. Note that in this case, although the energy of a large B-ball scales with the charge as $E_Q\simeq m_{3/2} Q$, $m_{3/2}$ can be smaller than the mass of a nucleon. So large B-balls are also stable in this scenario, and they can be a candidate of dark matter.  In the presence of sizable thermal effects, however, large Q-balls with $Q\sim10^{22}$$-$$10^{25}$ may evaporate to generate baryonic matter in the current universe.

\subsection{Q-balls from early universe}
\label{sec:QballEarlyUni}

In the early universe, Q-balls may be generated in a phase transition at high temperatures if there is a false vacuum in the potential that traps particles and a large charge density in the universe \cite{Frieman:1988ut} (see also \cite{Frieman:1989bx}). For the scenario of B-balls, the charge density or the matter asymmetry is likely to be very small, in which case the statistical fluctuations of the charge density might also source Q-ball generation \cite{Griest:1989cb}. However, the Q-ball production rate via this mechanism is exponentially suppressed by their charge, so only small Q-balls are efficiently generated, provided the charge exceeds the minimum required for Q-ball formation. It is only under certain special circumstances that large Q-balls may be synthesized from seed Q-balls \cite{Postma:2001ea}. Collisions, fusions and pair production are also generally inefficient to produce larger Q-balls, as the production of larger Q-balls is again suppressed by their total charge \cite{Griest:1989bq} (see also \cite{Kusenko:1997si}). However, a first-order phase transition precipitated by a period of Q-balls accumulating charge can result in a significantly higher nucleation rate of critical bubbles, compared to a thermal phase transition \cite{Croon:2019rqu} (see \cite{Krylov:2013qe} for Q-ball production as a result of a first-order phase transition). 

In the presence of light fermions charged under the same symmetry, small Q-balls can decay via evaporation and only exist for a relative short period of time. However, if B-balls persist until the temperature of the QCD phase transition, the trapped quarks inside them can survive as relics and facilitate the synthesis of heavy nuclei even before the standard Big Bang nucleosynthesis \cite{Kusenko:1997zq}. SUSY L-balls can also decay into pions towards the end of their lifetimes, which then decay into MeV gamma rays with distinctive spectra \cite{Kasuya:2024ldq}. Q-balls may also come from nonminimally coupled inflation models in Palatini gravity \cite{Lloyd-Stubbs:2021xlk}.

The most discussed and promising mechanism for producing large Q-balls in the early universe is the fragmentation of the Affleck-Dine (AD) condensate \cite{Kusenko:1997si}, which will be the focus of the rest of this subsection. This occurs when the Hubble rate drops below the mass of the AD field and the condensate starts to oscillate coherently.

\subsubsection{Affleck-Dine baryogenesis}

Before diving into Q-ball formation, let us first briefly review the AD mechanism  \cite{Affleck:1984fy, Dine:1995kz}, which is a widely studied framework for explaining the matter-antimatter asymmetry observed in the universe (see \cite{Dine:2003ax} for a review). This mechanism builds up on the fact that there are many $B$-carrying flat directions in the MSSM\;\footnote{For many purposes, $L$-carrying flat directions are equally valid, as high-temperature sphaleron processes can efficiently convert lepton number into baryon number.}, as reviewed above, along which VEVs or condensates of the flat direction fields (known as AD fields) can be developed during inflation. The AD field condensates are subdominant in the energy budget during inflation, but the subsequent evolution of the AD field condensates after inflation can lead to baryogenesis.

More concretely, in an expanding universe,  an AD/flat-direction field $\varphi$ can gain a mass term from non-minimal kinetic terms in the Kahler potential: $c H^2 |\varphi|^2$, where $H$ is the Hubble parameter and $c$ can be negative \cite{Dine:1995kz}. During inflation, the negative mass term drives the (homogenous) AD field away from $\varphi=0$. The VEV does not grow arbitrarily large because the potential is ultimately stabilized by the non-renormalisable terms (see \eref{V2nm6}). For example, if the potential is schematically $V\sim -H^2|\varphi|^2+|\varphi|^{2n-4}/\Lambda^{2 n-6}$, the VEV of $\varphi$ is around $\varphi_0\sim \Lambda (H/\Lambda)^{1/(n-2)}$.

After inflation, the VEV is adjusted almost adiabatically as $H$ decreases, which changes the instantaneous minimum of the potential. When $H$ drops below the mass of the AD field $m$, the Hubble-induced mass term becomes negligible compared to the soft mass term $m^2|\varphi|^2$, where $m\sim m_{\rm soft}$ (see Eqs.~\eqref{msoftGauge} and \eqref{gravitymsoft}), at which point the AD field starts to oscillate around $\varphi=0$. At the beginning of this period, the $B$-violating A-terms ${A} \phi^s/{\Lambda^{s-3}} + {A_H} \phi^{s'}/{\Lambda^{s'-3}}  +\, {\rm c.c.}$ (see \eref{Vsterms}) play a significant role, where $A_H\sim a H$ is induced by the finite energy density. Since $A$ and $A_H$ are generally complex, these A-terms violate CP, leading to generation of baryon number during this period. From the AD condensate's perspective, baryon number generation simply means that $\varphi(t)$ acquires a rotation in the internal field space, while initially before baryogenesis $\varphi(t)$ only moves along one direction (say the $\Re(\varphi)$ axis) in the field space. As the value of the AD field decreases, the A-terms become negligible and baryogenesis ceases.  

The AD condensate is a coherent collection of zero-momentum particles, which are usually quite stable as the particles they couple to have large masses, by the very nature of a flat direction. In the original AD scenario, the oscillating AD condensate will eventually decay and the baryon number is passed to the Standard Model quarks. One way for the AD condensate to decay is via interactions with the ambient thermal bath \cite{Dine:1995kz, Allahverdi:2000zd, Anisimov:2000wx}. The timing of the decay depends heavily on the amplitude of the AD condensate, the leading order ($n$) of the non-renormalisable terms, as well as the cosmic history including factors such as the expansion rate, the reheating temperature and the fractional energy density of the AD condensate. So the AD condensate can decay shortly after the baryon number is generated and frozen, or it can decay much later. 

Assuming the inflaton field has not decayed (and still dominates the energy of the universe) when the AD condensate decays, the ratio between the energy densities of the AD and the inflaton condensate at that time is given by ${\rho_{\varphi}}/{\rho_I} \sim\left({m_{\varphi}}/{M_P}\right)^{2 /(n-2)}$, where $n$ is the power of the monomial in \eref{V2nm6}. Using this together with $\rho_I \sim n_\gamma T_R$ and $\rho_\varphi \sim m n_B$, one can obtain \cite{Dine:1995kz, Dine:2003ax}:
\be
\frac{n_B}{n_\gamma}   \sim  \frac{T_R}{m_{\varphi}} \frac{\rho_{\varphi}}{\rho_I}
 \sim 10^{-10}\!\left[\frac{T_R}{10^9 \mathrm{GeV}}\right]\left[\frac{M_P}{m_{\varphi}}\right]^{\frac{(n-4)}{(n-2)}} ,
\ee
where $n_B$ and $n_\gamma$ are the baryon and photon number density and $T_R$ is the reheating temperature. We see that the resulting baryon number density is insensitive to the initial conditions of the AD condensate and is largely determined by the reheating temperature and the lowest order of the flat-lifting non-renormalisable terms as well as the mass of the AD field. For $n=4$, the lowest possible order for the flat-lifting terms in the MSSM, the AD mechanism leads to the precisely required matter-antimatter asymmetry. For a larger $n$, the generated asymmetry is much greater than the currently observed one. This overproduction of baryon number is often considered beneficial, as various processes in the early universe may release large amounts of entropy (see, {\it e.g.}, \cite{Coughlan:1983ci}), which dilute the baryon number.

\subsubsection{Q-ball formation from AD condensate}
\label{sec:QballFormedAD}

If the potential of the AD field supports Q-ball solutions, then the original AD mechanism discussed above needs to be revised in its final stage. Instead of homogeneous decay, the AD condensate fragments into Q-balls after the Hubble rate drops below the mass of the AD field \cite{Kusenko:1997si}. Essentially, the oscillating AD condensate induces parametric resonance that exponentially enhances minute fluctuations in the condensate, leading to large contrasts in field values. The Q-ball-supporting potential provides attractive forces between these field contrasts, causing them to coalesce into Q-balls. This is much like preheating, the initial phase of a special kind of reheating that can occur after inflation for a very flat inflaton potential. 

More concretely, when $H\lesssim m$, the homogeneous AD condensate begins to oscillate around the minimum $|\varphi|=0$. For a Q-ball-supporting potential, the oscillation of the AD condensate provides a periodic driving that induces parametric resonance in the perturbations. To see this, one can parametrize the AD field as 
\be
\varphi(t) \equiv R(t) e^{i \Omega(t)} ,
\ee
and derive the linear perturbation equations of motion for this system, which can be then decomposed into spatial Fourier modes $\delta R_\bfk(t)$ and $\delta \Omega_\bfk(t)$. These Fourier modes are decoupled, thanks to the homogeneity of the AD condensate. To capture the instability bands of wavenumbers in the adiabatic limit, one assumes $\delta R_\bfk(t), \delta \Omega_\bfk(t)\propto e^{S_\bfk (t)}$, and derives the dispersion relation from the linear perturbation equations of motion for $\delta R_\bfk$ and $\delta \Omega_\bfk$ \cite{Kusenko:1997si}:
\bal
&\left[\alpha_\bfk^2+3 H \alpha_\bfk+\frac{2 \dot{R}}{R} \alpha_\bfk+\frac{k^2}{a^2}\right]\bigg[\alpha_\bfk^2+3 H \alpha_\bfk
\nn
&~~~ +\frac{k^2}{a^2}-\dot{\Omega}^2+V^{\prime \prime}\bigg] +4 \dot{\Omega}^2\left[\alpha_\bfk-\frac{\dot{R}}{R}\right] \alpha_\bfk=0  ,
\eal
where  $\ai_\bfk(t)=\d S_\bfk(t)/\d t $ and $k^2=\bfk^2$. The transition from a stable $k$ to an unstable $k$ happens when $\ai_\bfk$ changes its sign. So the boundaries of the instability bands of $k$ are obtained by setting $\ai_\bfk$ to zero in the above dispersion relation. If $(\dot{\Omega}^2-V^{\prime \prime})>0$, there is an instability band close to the long wavelength limit:
\be
0<\f{k}{a}<\sqrt{\dot{\Omega}^2-V^{\prime \prime}(R)} .
\ee
Note that this does not contradict the fact that Q-matter is stable (see \S\,\! \ref{sec:thinwall}), in which case we have $V^{\prime \prime}>\dot{\Omega}^2$---an AD condensate is generally not Q-matter. To estimate the most amplified mode, we can look for the $\bfk$ mode that maximizes $S_\bfk = \int \ai_\bfk(t)\d t$.

It is instructive to examine the potential \eqref{VEnqvist} without the small $|\varphi|^{10}$ term, where a nice trick allows us to quickly view the AD condensate as a perfect fluid with a negative pressure \cite{Enqvist:1997si}. This is readily seen if, for a small logarithmic term, we approximate the truncated potential with $V= m^2|\varphi|^2[1+ \log ((|\varphi|/M)^{2K})]\simeq  m^2|\varphi|^2 \exp[\log(|\varphi|/M)^{2K})]\propto |\varphi|^{2+2K}$. Recall that an oscillating condensate dominated by a monomial potential $|\varphi|^n$ will have an effective equation of state: $p=\rho(n-2)/(n+2)$ \cite{Turner:1983he}. Thus, for the AD condensate with $|\varphi|^{2+2K}$, the pressure $p=\rho K/2$ is negative since $K<0$. For such a fluid, standard treatment tells us that its $\bfk$-wavenumber density contrast $\delta_{\mathbf{k}}=\delta \rho_{\mathbf{k}} / \rho$ will grow according to $\ddot{\delta}_{\mathbf{k}}=-{K \mathbf{k}^2} \delta_{\mathbf{k}}/2$. This approximation breaks down for large $|\bfk|$, where it leads to exponential growth on time scales shorter than the background oscillations.

When the AD condensate begins to oscillate, the initial field value of the homogeneous AD condensate is determined by the balance between the soft mass term and the high order lifting term. The initial inhomogeneities in the AD condensate mainly arise from quantum fluctuations that exit the horizon, become classical, and then re-enter the horizon after inflation ends. At the horizon exit, the fluctuations around the minimum of the $H^2$-mass-corrected potential are $\delta \varphi\simeq H_I/2\pi$. 
When the condensate begins to oscillate, the sizes of these initial fluctuations become roughly \cite{Enqvist:1997si}
\be
\label{dphiX8}
\delta \varphi_\bfk \sim \frac{|\bfk|^{5 / 2}}{(2 \pi)^{7/2} m H_I^{1 / 2}}  ,
\ee
where $2\pi/|\bfk|$ is the length scale at the time of $H\sim m$ and $H_I$ is the Hubble parameter during inflation. 

As the AD fragmentation process is highly nonlinearly after the initial parametric resonance, a full understanding of this process requires nonperturbative simulations, preferably taking into account the expansion of the universe. Typically, these simulations are performed with the finite difference method in a lattice large enough to resolve the excited modes in the process. In particular, the lattice spacing must be much smaller than the typical sizes of the Q-balls, and the simulation box, often equipped with periodic boundary conditions, must be large enough to encompass the relevant dynamics. Both the initial coherent AD condensate and the final Q-balls are highly classical, and the simulations are usually justifiably done with the classical approximation \cite{Khlebnikov:1996mc, Berges:2004yj}, for which accurate high order finite difference methods can be easily adopted.

\begin{figure}
	\centering
	\includegraphics[width=5.5cm]{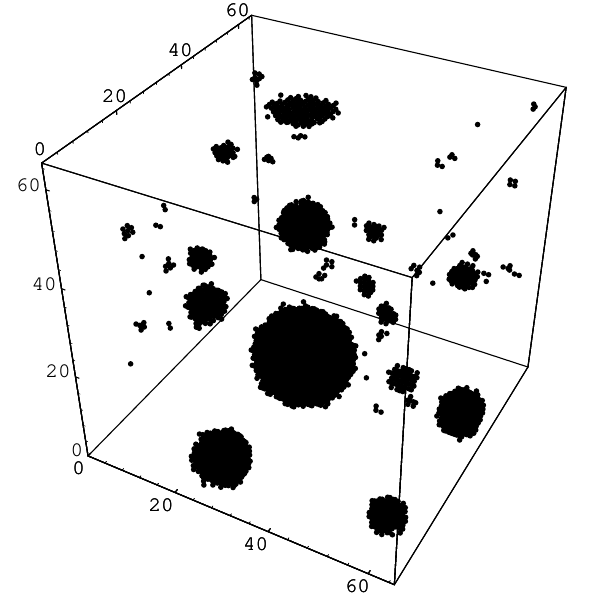}
	\caption{Q-balls generated from the AD condensate fragmentation. Most charges are absorbed into the Q-balls.	This figure is from \cite{Kasuya:1999wu}.}
	\label{fig:Qball3D}
\end{figure}

Toy model simulations in the 1+1D gauge-mediated scenario were conducted in \cite{Kusenko:1997si}, whereas spherically symmetric simulations in the gravity-mediated scenario were presented in \cite{Enqvist:1999mv}. Full 3+1D simulations for both scenarios have been detailed in \cite{Kasuya:1999wu, Kasuya:2000wx, Multamaki:2000ey, Kasuya:2001hg, Multamaki:2002hv, Tsumagari:2009na, Hiramatsu:2010dx, Chiba:2010ff, Zhou:2015yfa, Wang:2021rfk}. Indeed, starting from initial conditions that mimic the baryon asymmetry generated by the A-terms, the AD condensate typically fragments and Q-balls of varying sizes are formed as a consequence. The overall features of the AD fragmentation into Q-balls are similar in these two scenarios. The typical size of the Q-balls corresponds to the most amplified mode, more or less as predicted by linear analysis. When Q-balls are properly formed, the typical size of Q-balls is roughly one or two orders of magnitude smaller than the horizon scale. In the linear perturbative analysis of Fourier modes, mode mixing is neglected. When linear perturbations grow to the scale of the background AD field, the linear analysis breaks down and mode mixing becomes important. Nonlinear interactions then induce rescattering of the most amplified lower $k$ modes to the initially unamplified higher $k$ modes. It is also found that the Q-ball formation is so efficient that most of the charges in the AD condensate can be absorbed into the formed Q-balls; see, {\it e.g.,} Figure \ref{fig:Qball3D}. If the initial charge in the condensate is small, the naturally formed Q-balls are typically of the thick-wall type, while high initial charges lead to formation of large-charge/thin-wall Q-balls. 

In SUSY flat direction scenarios, the produced Q-balls will generally carry a nonzero $B-L$ charge, except in the case of a flat direction with $B=L$. Due to the smallness of the matter asymmetry, the net $B-L$ total charge for the fragmented condensate must be small. However, even in this case, the produced Q-balls can still dominate the energy budget of the universe by producing Q-balls and anti-Q-balls simultaneously \cite{Enqvist:2002si, Enqvist:2002rj}.

As discussed in \S\,\! \ref{sec:compositeQballs}, the logarithmic interaction $|\varphi|^2\log \left({|\varphi|^2}/{M^2}\right)$ in the gravity-mediated scenario is very weak, and thus the binding energies for these Q-balls are extremely small. Therefore, although not the minimum of the energy functional, non-spherical structures such as the charge-swapping Q-balls \cite{Copeland:2014qra} are very stable in this scenario. In fact, these composite states of Q-balls are naturally formed in the AD condensate fragmentation in the early universe \cite{Enqvist:1999mv, Hou:2022jcd}. Also, the generated Q-balls in this scenario seem to be more prone to moving and collisions \cite{Kasuya:2000wx}.

The thermal effects in the early universe can generate sizable quadratic or logarithmic potential terms. When such effects dominate, the density of these thermal Q-balls is diluted at least as fast as radiation as the universe expands, compared to the matter-like behaviour for the zero-temperature case. If the AD condensate also carries a nonzero $B-L$ charge, the viable parameter space in the gauge-mediated scenario significantly shrinks due to the constraint from the baryon asymmetry of the universe, unless the Q-ball formation is delayed \cite{Kasuya:2001hg}.

\subsubsection{Q-balls and dark matter}

The cosmological consequences following the formation of Q-balls from the AD condensate can be very diverse. Many of these scenarios are intimately related the dark matter problem. Depending on whether the lifetimes of Q-balls are longer than the age of the universe, Q-balls can be a candidate of dark matter or  decay into SUSY dark matter particles. One appealing aspect of the AD mechanism in general is that it can simultaneously generate both ordinary matter and dark matter from a common origin, which explains why ordinary matter and dark matter are of a similar amount in the current universe. This feature in one way or another persists in the scenarios where the AD condensate decays into Q-balls \cite{Enqvist:1998en, Laine:1998rg, Enqvist:1998ds, Banerjee:2000mb, Fujii:2002aj, Kasuya:2012mh}.

In the gauge-mediated SUSY breaking scenario, large Q-balls formed out of the AD condensate are absolutely stable in the absence of thermal effects (see \S\,\! \ref{sec:gaugeSUSY}). However, high temperature effects may alter the form of the scalar potential. If the thermally modified potential remains flat at large field values, the salient features of this scenario should be unaffected. Otherwise, one must estimate the thermal erosion of the Q-ball. If a particle $\psi$ couples to $\varphi$ via a Yukawa coupling $y$, then the particle acquires a thermal mass $m_\psi\sim T$ in an ambient bath at temperature $T$. Inside a Q-ball with a profile $\varphi(r)$, the particle's mass becomes $m_\psi(r) \sim y\varphi(r)$. Consequently, $\psi$ can penetrate the Q-ball only up to a depth $r_p$, where $\varphi(r_p)\sim T/g$. This information can be used to estimate the number of collisions during the lifetime of a cooling universe. Each collision knocks out  a small portion of energy $\delta E \sim T (m/Q^{1/4}/m_\psi(r))^2$, and the total erosion of the charge $\Delta Q\sim \Delta E/m$ is given by \cite{Kusenko:1997si, Laine:1998rg}
\be
\Delta Q \sim\left(\frac{T}{m}\right)^2 \frac{M_P /\big(1.66 n_{\rm eff}^{1 / 2}\big)}{m}  ,
\ee
where $n_{\rm eff}$ is the effective thermal degrees of freedom that interact with $\varphi$. So even in the case of adverse thermal corrections to the potential, Q-balls with $Q\gtrsim \Delta Q$ can survive to the present day. Since the field values inside a large Q-ball can be very high, one may imagine using this to probe high energy physics \cite{Dvali:1997qv}. The thermal effects of the charges escaping from the outer layer have also been considered in \cite{Enqvist:1998en}. For the gravity-mediated scenario, both of these two effects can be avoided if the reheating temperature is below $10^{3}-10^5$GeV.

If the high temperature effects do not destroy the Q-balls in the gauge-mediated scenario, they survive to this day as dark matter. Large B-balls are very massive, so if they are dark matter, their number density is extremely small in the current universe, making a direct detection quite challenging. See \cite{Arafune:2000yv, Kusenko:2005du, Super-Kamiokande:2006sdq, JacksonKimball:2017qgk, Hong:2016ict, Hong:2017qvx, Bakari:2000dq, Belolaptikov:1998mn, Kusenko:1997vp} for various observational bounds on Q-balls. Many of the experiments that search for monopole-catalyzed proton decay can also be used to constrain B-balls. For electrically neutral B-balls, these experimental bounds rely on the fact that B-balls carry baryon number and can interact with ordinary matter via the strong interaction. Specifically, the SU(3) gauge symmetry is broken inside a B-ball, inducing QCD deconfinement. As ordinary fermions can penetrate the B-ball's outer layer, nucleons entering this outer layer will dissolve into quarks and the energy is released as pions \cite{Kusenko:2005du, Kusenko:1997vp}. Also, see the Snowmass white paper \cite{Windchime:2022whs} for the Windchime project, which aims to use quantum enhanced mechanical sensors to detect dark matter including Q-balls solely through the gravitational coupling to ordinary matter.

However, the existence of large B-balls in the gauge-mediated scenario is strongly constrained by the observed abundance of neutron stars in the current universe \cite{Cotner:2016dhw, Kusenko:1997it, Hisano:2001dr}. As mentioned above, B-balls can absorb baryons and emit pions. Main sequence stars have low densities, so B-balls essentially pass through them with little change. However, the extreme density of a neutron star allows it to interact much more strongly with B-balls, making it capable of capturing and trapping them inside. Since these B-balls are stable, they can grow without bound inside a neutron star, either consuming it completely or causing it to collapse in about $10^8$ years. Then, neutron stars can be used to eliminate large regions of the parameter space in the gauge-mediated scenario \cite{Cotner:2016dhw}.  However, if the baryon number violating A-terms are taken into account in the potential, which become important when B-balls are sufficiently large, large B-balls become destabilized  \cite{Kawasaki:2005xc}, which alleviates many of the constraints \cite{Kawasaki:2019ywz}.

Q-balls in the gravity-mediated scenario are not absolutely stable, but they can still be dark matter if their lifetimes, after including the evaporation and thermal effects, are longer than the age of the universe \cite{Enqvist:1998en}. Otherwise, Q-balls in this scenario can decay into the lightest supersymmetric particles (LSPs), usually neutralinos, generating LSP dark matter non-thermally. If the LSPs do not annihilate, there will be typically much more dark matter than ordinary matter, which will contradict the fact that the universe is almost flat. One solution is to allow for a large annihilation cross section for the LSPs, in which case dark matter is more Higgsino-like or wino-like, rather than bino-like. Thus, this addresses one of the coincidence problem, even if the AD condensate fragments into Q-balls. Reversing the argument, it can be used to constrain the parameter space of the MSSM \cite{Fujii:2001xp, Fujii:2002kr, Fujii:2002aj}. The properties of these alternative LSPs and the fact that these LSPs can be generated non-thermally from Q-balls provide new paradigms for dark matter model-building and experimental searches.

An important application of B-balls produced in AD baryogenesis is that the B-balls can protect the generated baryon asymmetry from the $B+L$ violating processes above the electroweak scale \cite{Kusenko:1997si}. For example, in the  gravity-mediated SUSY breaking scenario, large B-balls with $B\gtrsim 10^{14}$ can survive well below the electroweak scale, and the baryon asymmetry included in them can be protected from being washed away by sphaleron processes \cite{Enqvist:1998en}. 
 
Q-balls may also be a candidate for self-interacting dark matter \cite{Kusenko:2001vu}. Specific self-interactions and radiation of Q-balls and oscillons can help differentiate whether a dark matter field is real or complex \cite{Zhang:2024bjo}. The possibility of supermassive Q-balls or charge-swapping Q-ball-like structures residing in the galactic centers is considered in \cite{Troitsky:2015mda, Jaramillo:2024smx}. An updated survey of astrophysical constraints on Q-balls as dark matter is presented in \cite{Ansari:2023cay}, which also establishes new theoretical limits on Q-ball sizes and bounds on the dark matter fraction from Q-balls through gravitational micro-lensing observations.

\subsubsection{GWs and primordial BHs from Q-balls}

Strictly spherical objects such as properly formed Q-balls do not emit gravitational waves, as a result of Birkhoff's theorem. However, the process of the AD condensate fragmenting into Q-balls during and after parametric resonance is highly nonperturbative and can emit sizable gravitational waves (GWs). Generally, more significant (stochastic) GW backgrounds are generated if the AD condensate fragmentation involves most of the energy within the horizon.

The characteristic frequency and energy density of GWs during Q-ball formation in both the gravity-mediated and gauge-mediated SUSY breaking scenarios were estimated in \cite{Kusenko:2008zm}. Analytical and numerical results on the AD condensate fragmentation suggest that the fastest growing mode has a frequency around $\oi_k\sim k\sim 10^2H_*$, where $k$ is the wavenumber and $H_*$ is the Hubble parameter during the fragmentation. This essentially determines the peak frequency of the GWs produced at the time, and to convert it to the frequency observed today, it must be red-shifted due to the expansion of the universe. Similarly, the GW energy density is also diluted by the expansion. Assuming instant reheating along with a couple of other assumptions (see \cite{Kusenko:2008zm} for caveats), the estimated peak GW frequency and the fractional energy density of GWs observed today are found to potentially fall within the sensitivity range of aLIGO or eLISA, depending on the reheating temperature.  A typical reheating temperature considered in scenarios with potential observable GW signals is about $1\!-\!100$TeV, which overlaps with the high energy frontier of particle physics beyond the Standard Model. This opens an exciting avenue to explore the synergies between collider physics and GW astronomy in such scenarios. 

The GW production from the AD condensate fragmentation in the gravity-mediated scenario has been investigated numerically with real-time lattice simulations \cite{Kusenko:2009cv} \cite{Zhou:2015yfa}. With these fully nonlinear computations, the fractional GW energy density is found to be a couple of orders smaller than the initial analytical estimates but still sizable. Interestingly, the resulting GW spectra seem to be quite insensitive to the sizes of the initial fluctuations. If the Q-balls do not decay \cite{Kawasaki:2012gk} soon after their formation, they can dominate the universe for a long period of time, in which case the GWs generated in the AD condensate fragmentation will be significantly diluted to unobservable levels \cite{Chiba:2009zu}. On the other hand, when (logarithmic) thermal effects are sizable, thermally supported Q-balls will not dominate the universe, and the GWs from the Q-ball formation will not be as diluted, resulting in potentially detectable GW spectra.

\begin{figure}
	\centering
	\includegraphics[width=7.5cm]{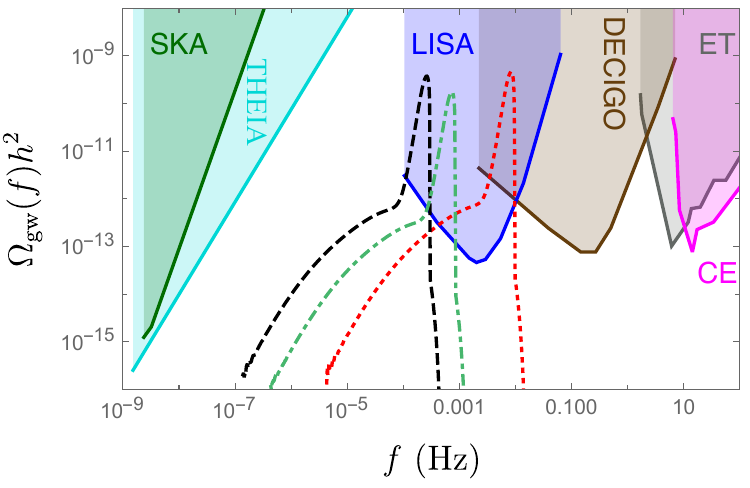}
	\caption{Stochastic GW backgrounds from Q-balls induced by a rapid transition from a Q-ball-dominated epoch to the radiation epoch. This figure is from Ref \cite{White:2021hwi}.}
	\label{fig:QballGWsp}
\end{figure}

Since Q-balls are extremely stable, another way Q-balls can generate a stochastic GW background in the early universe is through a prolonged Q-ball-dominated epoch, followed by a rapid decay of Q-balls \cite{White:2021hwi}. A Q-ball-dominated epoch is like a matter-dominated epoch where scalar perturbations from inflation can grow, and the enhanced scalar perturbations or overdensities can induce GWs during the transition to a radiation epoch when these overdensities are converted to relativistic sound waves \cite{Inomata:2019zqy}. For this to happen, it is necessary that the matter-radiation transition is rapid, which is natural in the AD mechanism setup as Q-balls can evaporate via coupling to fermions. The resulting GW spectrum will have a prominent sharp peak due to a resonance mode in this process, and the GW background can fall in the range of LISA, Einstein Telescope or DECIGO; see Figure \ref{fig:QballGWsp}. Moreover, Q-balls can also enhance second-order GWs produced during the matter-radiation transition as a result of their decay, and such a spectrum is consistent with the recent pulsar timing array observations \cite{Kawasaki:2023rfx}.

Since the number density of Q-balls formed from the AD condensate fragmentation can fluctuate, statistically, it is possible that some overdensities induced during a Q-ball-dominated epoch can exceed the critical (fractional) density $\delta_c\sim 1.7$. This can lead to the formation of primordial black holes (BHs) \cite{Sakai:2011wn, Cotner:2016cvr, Cotner:2019ykd}. Assuming a Poisson distribution for the probability to find equally charged Q-balls in a given spatial volume, Ref \cite{Cotner:2016cvr} estimated the mass function of the produced primordial BHs in this scenario. For a generic U(1) scalar, this mechanism can essentially produce primordial BHs across the entire mass range permitted by dark matter observations, and can account for the total dark matter budget in some cases. For models involving Q-balls arising from the MSSM scenarios, on the other hand, the masses of the produced primordial BHs are constrained to be smaller than about $10^{23}$g. See \cite{Cotner:2019ykd} for a general analytical framework for treating the fragmentation of a scalar condensate into lumps such as Q-balls and oscillons, as well as the subsequent formation of primordial BHs in the early universe. Interestingly, gravitational effects will limit the Q-ball size even in cases where a Q-ball is not dense enough to collapse into a BH \cite{Multamaki:2002wk}.

\subsection{Soliton bag models for hadrons}

Quantum Chromodynamics (QCD) is asymptotically free at high energies, but becomes strongly coupled at low energies, responsible for the strong interaction at subatomic scales. Due to the strong coupling at low energies, describing the structures and interactions of hadrons within QCD has been quite challenging. Nowadays, nonperturbative numerical techniques such as lattice QCD have enabled increasingly accurate descriptions of hadrons, thanks to advances in computational methods and resources. On the other hand, many phenomenological models for hadrons have been proposed, which include models involving non-topological solitons such as the SLAC bag model \cite{Bardeen:1974wr}, the MIT bag model \cite{Chodos:1974je}, and their generalization, the Friedberg-Lee model \cite{Friedberg:1977xf, Friedberg:1978sc} (see also \cite{Goldflam:1981tg, Cahill:1985mh}). Although still rudimentary in nature, these models can often yield surprisingly precise approximations of many parameters in hadronic physics, without needing substantial computational resources.

Hadrons can be characterized by their valence quarks, so fermionic fields are essential in these soliton models. In fact, they typically consist of quarks and some phenomenological, color-singlet scalar but neglect vector fields, making them similar to the fermionic non-topological soliton model already discussed in \ref{sec:fermiSoliton}. The phenomenological scalar $\si$ (sometimes called ``scalar gluon'') is not an elementary field in the Standard Model; rather, it is introduced to take into account the salient aspects of the nonperturbative interactions between gluons and quarks that bind the hadron together. 

Apart from the hadronic models, fermionic non-topological solitons also find their relevance in the various models of quark matter/nuggets \cite{Witten:1984rs, Terazawa:1989iw, Farhi:1984qu,  Buballa:1998pr, Fukushima:2004zq, Liang:2016tqc, Holdom:2017gdc, Bai:2018dxf}, and quark star \cite{Itoh:1970uw, Alcock:1986hz, Haensel:1986qb}, many of which are dark matter candidates. See \cite{Holdom:1987ep, Macpherson:1994wf, Hong:2020est, Kawana:2021tde, DelGrosso:2023trq} for other applications of solitons involving fermions.

\subsubsection{Friedberg-Lee model}
\label{sec:FriedLee}

Let us take the Friedberg-Lee model \cite{Friedberg:1977xf, Friedberg:1978sc, Lee:1991ax} as an example, which is given by the following phenomenological Lagrangian
\bal
\mc{L} &= \bar\psi (i \slashed{D} - m)\psi -\f14\kappa(\si) F_c^\mn F^c_\mn 
\nn
&~~~ - \f12 \pd^\mu \si \pd_\mu \si -V(\si) - f\si \bar\psi\psi  ,
\eal
where the quark field $\psi$ contains the color as well as the flavor degrees of freedom. Note that the original model \cite{Friedberg:1977xf} differs slightly from the above model in terms of the field content, while in \cite{Friedberg:1978sc} the scalar $\si$ is re-interpreted as a long-range order in the vacuum and the model is further justified and developed from the perspectives of QCD. Since this model has already been reviewed in some detail in \cite{Lee:1991ax}, our discussion here will be brief. 

In the Friedberg-Lee model, a quartic potential $V(\si)$ is chosen such that there is a local minimum/metastable vacuum at $\si=0$ and an absolute minimum/physical vacuum at $\si=\si_v\neq 0$. Around the physical vacuum $\si_v$, the Yukawa coupling $f\si \bar\psi\psi$ provides a large mass $f\si_v$ for the quarks (if $f\si_v\gg m$), while the quark mass is small around the metastable vacuum, making the quarks energetically more favourable to stay in the metastable vacuum. On the other hand, when the quarks congregate around the metastable vacuum, a nonzero quark density provides a linear $\si$ term to the scalar potential, which raises the energy of the physical vacuum and hence renders the metastable vacuum at $\si=0$ very stable. Therefore, quarks are confined to a finite region around the  $\si=0$ vacuum where $\si$ is close to zero. Since the scalar stays at the physical vacuum $\si=\si_v$ away from the soliton, gluon confinement is also achieved if we choose the function $\kappa(\si)$ to be 
\be
\kappa(0)=1,~~\kappa(\si_v)=0,~~\f{\d\kappa(\si_v)}{\d \si} =0 .
\ee
Numerical classical solutions of the Friedberg-Lee model without the gauge fields have been extensively investigated \cite{Saly:1983fbs, Koeppel:1985tt, Horn:1986qj}, and an example of the profile functions is given in Figure 15 of Ref \cite{Lee:1991ax}. The Green's functions in the one gluon exchange approximation are obtained in \cite{Lee:1978mf, Tang:1990hd}.

One benefit of the soliton bag models is that they can be straightforwardly quantized on top of the classical soliton background by the mean field method, as briefly mentioned in \ref{sec:fermiSoliton}. However, in the mean field approximation, since the classical solution is spatially localized, the quantum expectation value of the momentum operator squared does not vanish $\langle P^2\rangle \neq 0$. This leads to unphysical center-of-mass fluctuations. To get rid of this unphysical contribution to the soliton energy, one can adopt the recoil-corrected mass \cite{Dethier:1982ax}
\be
m=\sqrt{\< H\>^2-\<P\>^2} ,
\ee
where $H$ is the Hamiltonian. To go beyond this {\it ad hoc} fix, it is necessary to separate the collective motion of the system from the true vibrational modes, which is however difficult to compute in practice. 

Alternatively, one can construct the $\si$ part of the soliton state as a coherent state and make the Peierls-Yoccoz projection \cite{Peierls:1957er} to compute the energies of the solitons \cite{Lubeck:1986if, Lubeck:1987pj}. To construct the coherent state, we expand the $\si$ field around the vacuum
\be
\si = \si_v +  \int\! \f{\d^3k}{(2\pi)^3 2\oi_\bfk} (d_\bfk(t) e^{i\bfk\cdot \bfx} + d^\dagger_\bfk(t) e^{-i\bfk\cdot \bfx}) ,
\ee
where $d_\bfk$ and $d^\dagger_\bfk$ satisfy the standard commutation relations.  A coherent state\;\footnote{For a given mode $[a,a^\dagger]=1$, the coherent state or Glauber state $|\ai\rangle$ is defined to be the eigenstate of the annihilation operator $a |\ai\rangle = \ai |\ai\rangle$, with $\ai$ being a complex number. Explicitly, we have $|\ai\rangle =e^{-\f12 |\ai|^2} e^{\alpha a^{\dagger}} |0\rangle$.} corresponding to the $\si_c$ configuration can be defined as follows
\be
|\si_c\rangle \propto {\rm exp}\( \int \!\f{\d^3 k}{(2\pi)^3}  \tilde\si_c(\bfk) d_\bfk^\dagger \) |0\rangle ,
\ee
with $\tilde\si_c(\bfk)$ being the Fourier transform of $\si_c(\bfx)$, which ensures $\<\si_c|\si|\si_c\>=\si_c$. The soliton state can then be obtained by acting on  $|\si_c\rangle$ with creation operators of the other fields. The Peierls-Yoccoz projection is based on the fact that given a localized soliton state $ |\bfX\rangle$ at $\bfX$, the momentum eigenstate can be obtained by 
\be
|\bfP\rangle = \int \d^3 \bfX e^{i \bfP\cdot \bfX} |\bfX\rangle .
\ee
Thus, the energy of the soliton can be computed by taking the zero momentum state $|\bfP=0\rangle = \int \d^3 \bfX |\bfX\rangle$ as the soliton state. To get a better fit with the experimental data, one can minimize the expectation value of the projected Hamiltonian by varying the profile functions of the scalar and fermions, as well as $\oi_\bfk$. 

A comparison between the experimental data of the hadrons and the predictions of the Friedberg-Lee model for a given set of theory parameters is summarized in Table 3 of \cite{Lee:1991ax}. One can see that, although understandably not a perfect match, the theoretical predictions obtained from the variation after the Peierls-Yoccoz projection are very close to the experimental values for many quantities. Similar phenomenological hadronic models can be found in \cite{Nielsen:1981fi, Kahana:1984dx, Kahana:1984be, Birse:1984js, Williams:1985ku,  Dodd:1987pw}.

\subsection{Oscillons from early universe}
\label{sec:oscUniverse}

Like Q-balls, the study of oscillons is, to a large part, motivated by the possibility of their formation in the early universe, such as in phase transitions, and their potential implications for cosmic evolution \cite{Copeland:1995fq, Gleiser:1993pt}. 

Oscillons can be found in the Abelian-Higgs model \cite{Witten:1976ck}, and can arise as remnants in low-momentum vortex and anti-vortex scatterings \cite{Gleiser:2007te}. The axion field may support oscillons, dubbed axitons, due to the axion's attractive self-interactions \cite{Kolb:1993hw}. The role of axion-like oscillons as a candidate for ultra-light dark matter has been explored in \cite{Olle:2019kbo}. Collisions and mergers of axion-like oscillons can generate electromagnetic bursts \cite{Amin:2020vja, Amin:2021tnq} (see \cite{Amin:2023imi} for a similar phenomenon for vector oscillons). Intriguingly, oscillons could exist within the bosonic sector of the electroweak Standard Model if the Higgs mass were exactly twice the W boson mass  \cite{Graham:2006vy, Graham:2007ds} (see \cite{Farhi:2005rz} for an early work which studied a SU(2) gauge theory spontaneously broken by a Higgs doublet, and \cite{Correa:2015kha} for oscillons in Lorentz-violating Standard Model extensions). However, as long as the Higgs and W boson masses differ from the 2-to-1 ratio by as little as 0.5\%, the lifetime of the resulting oscillon is reduced to about a few hundreds of oscillations. Therefore, oscillons unfortunately do not exist in our Standard Model.

Given the relatively loose conditions on the scalar potential that can accommodate oscillons, it is conceivable that new fields beyond the Standard Model could support their existence. However, even in the presence of such fields, oscillons would still need to be generated efficiently in the early universe for them to be impactful to the cosmic history. One natural possibility is that they could be thermally produced in the thermal bath of the hot Big Bang. While small-amplitude oscillons can be abundantly generated by thermal fluctuations in an expanding 1+1D spacetime \cite{Farhi:2007wj}, this mechanism is much less effective in 3+1D \cite{Riotto:1995yy}, where oscillons are only locally stable if their amplitudes are sufficiently large. Indeed, it is found that only a couple of percentages of the energy budget can be stored in the produced oscillons in this way \cite{Gleiser:2010qt}. Oscillons can also serve as precursors to bubble nucleation events during first-order phase transitions, as has been demonstrated in 1+1D \cite{Pirvu:2023plk}.

Another mechanism for generating oscillons, as demonstrated in 2+1D, is to have a scalar field that begins with a single-well potential, which, after reaching thermal equilibrium with an external heat bath, is then ``spontaneously'' broken to a double-well potential \cite{Gleiser:2002pw}. If the heat bath is also switched off when this happens, oscillons can be nucleated during the quench. Subject to thermal fluctuations, the formed oscillons will eventually decay when the system approaches equipartition, though their presence delays thermalization. On the other hand, the formation of oscillons can accelerate the decay of a thermally metastable vacuum when the system undergoes a similar spontaneous symmetry breaking of the potential followed by a fast quench \cite{Gleiser:2004iy}. In this case, after the single-well potential transitions to a double-well, oscillons again form due to parametric resonance and then coalesce into critical nuclei. Interestingly, this process results in a power-law decay rate, instead of an exponential one typically observed in homogeneous nucleation. Additionally, oscillons can be generated from an evolving network of domain walls \cite{Hindmarsh:2007jb}, collapsing asymmetric bubbles \cite{Adib:2002ff} or kinetic relaxation \cite{Jain:2023tsr, Jain:2023ojg}

\subsubsection{Oscillons from preheating}
\label{sec:oscillonForm}

A very promising scenario for oscillons to form in the early universe is via parametric resonance in the preheating period after inflation \cite{Amin:2011hj, Gleiser:2006te,  Amin:2010dc, McDonald:2001iv, Gleiser:2011xj, Lozanov:2014zfa, Gleiser:2014ipa, Adshead:2015pva, Antusch:2017flz, Lozanov:2017hjm, Sang:2020kpd, Mahbub:2023faw, Jia:2024fmo, Shafi:2024jig} (see \cite{Amin:2014eta} for a review). This is because current cosmological observations suggest an inflationary model with a potential that supports oscillon formation after inflation. The absence of isocurvature perturbations from inflation implies that the inflaton is essentially a single field, and the current constraints on the spectrum of the curvature perturbations suggest that the part of the inflaton potential that supports the quasi-de Sitter expansion is flatter than $\phi^2$ \cite{Planck:2018vyg}. Assuming the inflaton is massive, {\it i.e.}, equipped with a quadratic minimum, the inflaton potential is of the ``open-up'' type that allows oscillons to form. In other words, the current observations favor a scenario where the inflaton field exhibits attractive force, which can fragment the homogeneous inflaton condensate into oscillons.

Indeed, quite generically, oscillons can be copiously generated in the reheating period, thanks to the coherent oscillations of the inflaton condensate after inflation ends. This can happen in the generalised axion monodromy model \eqref{oscGmono} \cite{Amin:2011hj}, a popular model still consistent with the cosmological observations \cite{Planck:2018vyg}. 
From a theoretical perspective, the motivation for the axion monodromy model (with $\alpha = 1/2$) stems from the fact that it can be obtained by a braneworld construction where a space-filling 5-brane wraps a 2-cycle of the compact internal space in a manner that is consistent with moduli stabilization \cite{Silverstein:2008sg, McAllister:2008hb}, as well as from its ability to exceed the Lyth bound $\Delta \phi \gtrsim M_P$, allowing for potentially observational tensor modes ({\it i.e.,} a primordial GW background) from inflation itself. For a generic $\ai$, to fit the power spectrum of the primordial curvature perturbations from the CMB, one of the three parameters, $m$, $\ai$ and $M$, in the theory is fixed by \cite{Amin:2011hj}
\be
\label{oscCP109}
\Delta_R^2=\frac{(4 \alpha N)^{1+\alpha}}{96 \pi^2 \alpha^3}\left(\frac{m}{M_{\mathrm{P}}}\right)^2\left(\frac{M}{M_{\mathrm{P}}}\right)^{2-2 \alpha}\sim 10^{-9}  ,
\ee
if a fiducial value of the e-folds before the end of inflation $N$ is chosen. In this model with quite generic choices of the theory parameters, the beginning of reheating is proceeded with preheating.

\begin{figure}
	\centering
	\includegraphics[width=5.8cm]{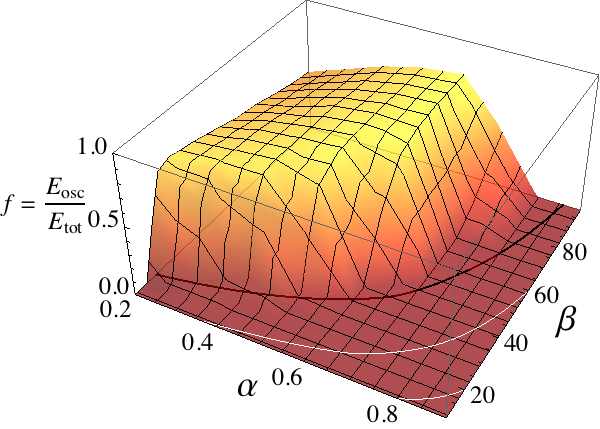}
	\caption{Fraction of the total energy stored in oscillons after preheating in the generalised axion monodromy model \eqref{oscGmono}. $\bi=M_P/M$ and the black ($\max (|\Re (\mu_k)| / H)=7$) and white ($\max (|\Re (\mu_k)| / H)=3,1$) contours correspond to the maximum values of $|\Re (\mu_k)| / H$ for given $\ai$ and $\bi$, where $H$ is the Hubble constant and $\Re\left(\mu_k\right)$ is the real part of the Floquet exponent $\mu_k$ for the perturbation field with the $k$ comoving wavenumber $\delta\phi_k\propto e^{\mu_k t}$. This figure is from Ref \cite{Amin:2011hj}. }
	\label{fig:oscFraction}
\end{figure}

Preheating refers to a process in which particles with non-zero momenta are explosively produced, in contrast to the scenario of gradual perturbative decay of the inflaton condensate. In this stage, much like the AD condensate fragmenting into Q-balls, the oscillating inflaton condensate induces parametric resonance. Essentially, every perturbative $\bfk$ mode $\delta\phi_\bfk(t)$  of the inflaton field behaves like a damped Mathieu's oscillator
\be
\ddot{X} +\li \dot {X} + (a+q \cos \oi t) X= 0 ,
\ee
where $X\sim \delta\phi_\bfk(t)$, $\cos \oi t$ is provided by the oscillating inflaton background, the detuning term $\li\dot {X}$ comes from the cosmic expansion, and $a$ and $q$ also weakly depend on time due to the cosmic expansion. As oscillons are generated well within the cosmic horizon by causal interactions, the expansion effects are usually small. Thus, as a reasonable approximation, a flat-space Floquet analysis can be applied to find approximate parameters for the model to preheat (see, {\it e.g.,} \cite{Amin:2010xe} for details). The linear parametric resonance exponentially enhances the initially minute quantum fluctuations from inflation, and when perturbations become nonlinear, rescattering transfers energy between the different modes, resulting in the copious production of oscillons. The parametric resonance can be so efficient that most oscillons are formed within a few dozens of oscillations of the inflaton field. 

Numerical lattice simulations in the FRW background have shown that the produced oscillons can often dominate the energy budget of the universe \cite{Amin:2011hj}; see Figure \ref{fig:oscFraction}. As we might expect, strong parametric resonance ($\max (|\Re (\mu_k)| / H) \gtrsim 10$, $\Re (\mu_k)$ being the real part of the Floquet exponent) is necessary for inducing an oscillon-dominated universe. As these oscillons are long-lived, an oscillon-dominated period will follow after preheating, in which the scale factor scales as in a matter-dominated epoch. As oscillons are quasi-stable, they will eventually decay. However, in some cases, this leads to a prolonged period, thereby delaying thermalization compared to the scenario without oscillon formation. 

Oscillon formation and dynamics  have also been studied in other preheating or preheating-like scenarios, such as in the logarithmic potential \cite{Kawasaki:2013hka}, the string moduli scenario \cite{Antusch:2017flz} and the $\alpha$-attractor inflation model \cite{Jia:2024fmo}. Oscillons can also act as an engine for generating massive fermions when the cosmic temperature has dropped far below the natural mass scale of the fermions \cite{Saffin:2016kof}.

\subsubsection{GWs and primordial BHs from oscillons}

An important consequence of the inflaton condensate fragmenting into oscillons is the generation of a stochastic GW background \cite{Zhou:2013tsa, Antusch:2017vga,  Amin:2018xfe, Lozanov:2019ylm, Hiramatsu:2020obh}. Similar to the case of the AD condensate fragmentation, significant GWs can be generated during the formation of oscillons. 

To compute the GW production from oscillon formation, one can nonlinearly evolve the inflaton field after inflation ends, calculate the energy-momentum tensor of the inflaton field $T_{ij}$, and use it as the source for the perturbative metric equation of motion
\be
\left(\frac{\mathrm{d}^2}{\mathrm{~d} t^2}+3 H \frac{\mathrm{~d}}{\mathrm{~d} t}-\frac{1}{a^2} \nabla^2\right) h_{i j}=\frac{2}{M_P^2} T_{i j} ,
\ee
where $a$ is the scalar factor, $H$ is the Hubble parameter and $h_{ij}$'s are the spatial components of the perturbative metric. The GW energy density is then given by
\be
 \rho_{\mathrm{gw}}  =\frac{M_P^2}{4}\left\langle\dot{h}^{TT}_{i j} \dot{h}^{TT}_{i j}\right\rangle  ,
\ee
where $h^{TT}_{i j}$ is the transverse and traceless projection of $h_{ij}$ and the average is over a spatial volume that is sufficiently large to include all relevant GW wavelengths. The GW spectra can again be computed with real-time lattice simulations. The GW spectrum from oscillons can exhibit a unique multi-peak structure, with troughs imprinted by the suppression of certain frequency bands due to the formation of spherically symmetric oscillons \cite{Zhou:2013tsa}. 

The GWs from oscillons are mostly generated in a short period after inflation ends. Due to spherical symmetry, when oscillons are properly formed, neither the oscillon itself nor the superposition of well-separated oscillons can produce sizable GWs, which can be seen analytically \cite{Zhou:2013tsa, Antusch:2017vga} or numerically \cite{Amin:2018xfe}. The latter is challenging as it requires running the lattice simulation for an extended period in an expanding box. As time progresses in an oscillon-dominated universe, the lattice separation $a(t)\Delta x_i$ can increase to a point where the lattice can no longer adequately resolve the relevant modes for computing the GW spectrum, resulting in numerical artifacts in the high-frequency region \cite{Amin:2018xfe, Antusch:2016con}. For such a simulation, empirically, numerical artefects appear when the lattice separation $a(t)\Delta x_i$ exceeds $m^{-1}$, with $m$ being the mass of the scalar \cite{Amin:2018xfe}.

\begin{figure}
	\centering
	\includegraphics[width=9.5cm]{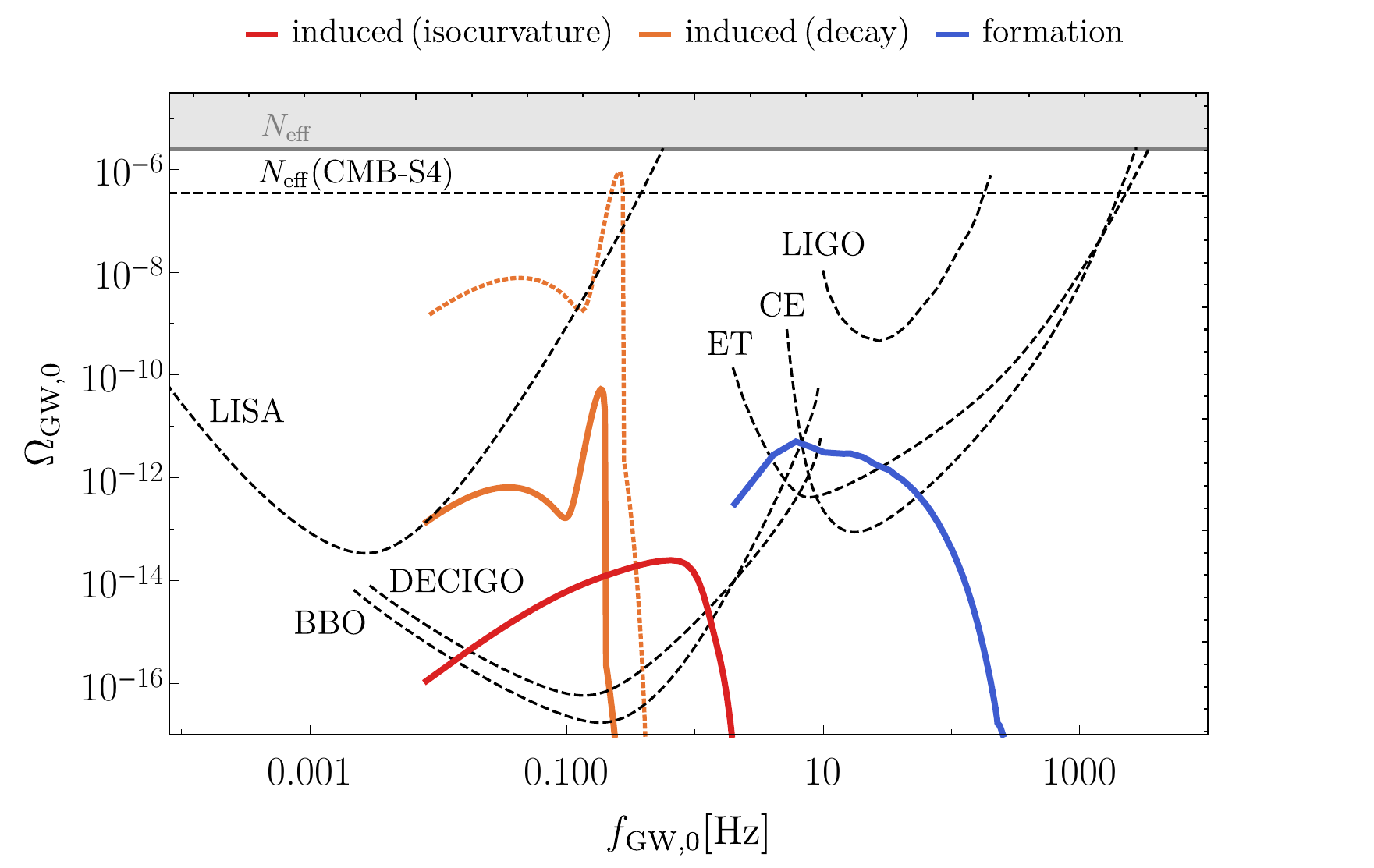}
	\caption{Stochastic GW backgrounds from oscillons in the early universe. The blue line is the GW spectrum directly from oscillon formation, and the red and orange lines are the GW spectra induced by 2nd order curvature perturbations originated from fast oscillon decay and oscillon isocurvature perturbations respectively. This figure is from Ref \cite{Lozanov:2023aez}.}
	\label{fig:oscGWsp}
\end{figure}

The generated GWs are redshifted and diluted by the subsequent cosmic expansion until today, which simply amounts to scale the GW spectrum by suitable numerical factors \cite{Dufaux:2007pt}. As mentioned, this however depends on the reheating temperature.
In the generalised axion monodromy model, it is assumed that the oscillon field is the inflaton field that generates the primordial curvature perturbations observed in the CMB. In that case, the generated GW spectra, as observed today, are peaked around extremely high frequencies $f\sim 10^9$Hz, which is beyond the reach of current and foreseeable future observational capabilities. This is because the scale of the inflation in this model is fixed by the CMB constraint \eqref{oscCP109} to be very high. If we relax this by assuming the inflaton field is distinct from the oscillon field and the oscillon field obtains a VEV nonetheless (like the AD condensate), then the frequencies of the GW spectrum can fall within the observable bands of the current or upcoming GW experiments \cite{Liu:2017hua} (see, {\it e.g.,} Figure \ref{fig:oscGWsp}). GW spectra of such a preheating-like scenario have been computed for cuspy potentials $|\phi|^{p}$ with $p\leq 1$ \cite{Liu:2017hua, Liu:2018rrt, Sang:2019ndv} or the generalised axion monodromy potential \cite{Sang:2019ndv}---the cuspy potentials might be viewed as approximations of the generalised axion monodromy potentials. For some choices of the parameters, the peaks of the GW spectra can lie above the sensitivity curves of aLIGO, making these models particularly testable with ongoing GW experiments.

\begin{figure}
	\centering
	\includegraphics[width=8.2cm]{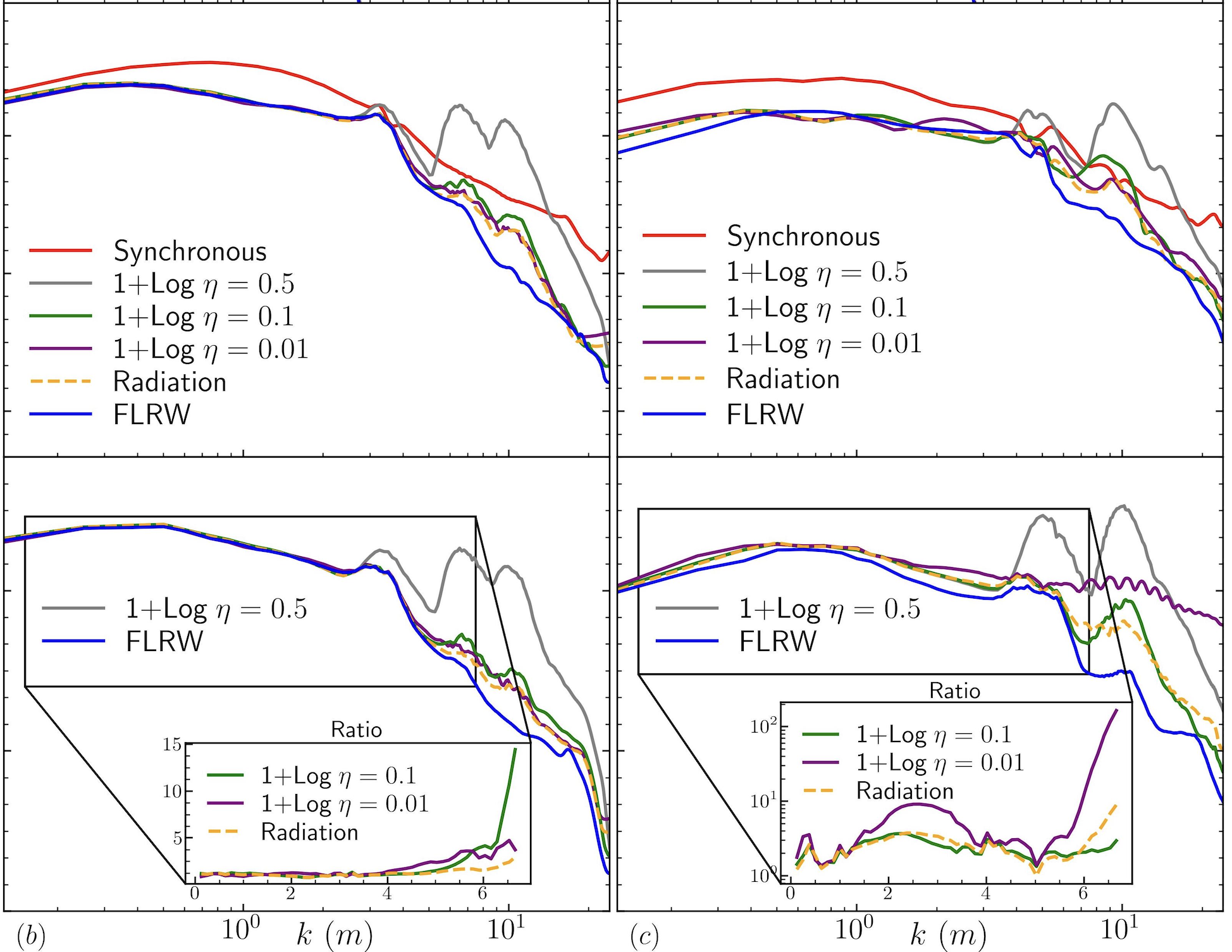}
	\caption{GW spectra $\Omega_{\rm GW}(k)$ from oscillon formation using full numerical relativity. The left column is for relative strong scalar (parametric) resonance and the right column is for weak scalar resonance but strong gravitational effects; For each column, the 2 different panels are for 2 different times of oscillon preheating: top $t\simeq 94m^{-1}$ and bottom $t\simeq 112m^{-1}$. Different gauge choices are compared with the rigid FRW case, and the often used synchronous gauge is found to break down very quickly. This figure is from Ref \cite{Kou:2021bij}.}
	\label{fig:oscGWNR}
\end{figure}

Computations of oscillon preheating are usually performed in a rigid FRW background sourced by the average energy density of the matter fields. As oscillons are highly dense, nonlinear objects that can form at high energies, it is natural to ask whether, and under what conditions, neglecting inhomogeneous gravitational back-reactions is a valid approximation. Refs \cite{Kou:2019bbc, Aurrekoetxea:2023jwd} simulated oscillon preheating in full general relativity using the BSSN formalism \cite{Baumgarte:1998te, Shibata:1995we} (see \cite{Aurrekoetxea:2024mdy} for a review of the applications of numerical relativity in cosmology). For the model \eqref{oscGmono}, it is found that the FRW simulations are sufficient for the parameter region where (flat-space) scalar parametric resonance is strong. However, when inhomogeneous gravitational back-reactions are included, BSSN simulations reveal that in regions where parametric resonance is less prominent, a larger portion of the inflaton condensate is transferred to oscillons. In extreme cases, oscillons can collapse into BHs, where the BH apparent horizons can be resolved using adaptive mesh refinement in lattice simulations \cite{Kou:2019bbc} or through the collapse of the lapse function \cite{Nazari:2020fmk}. The GW production in full general relativity has also been computed, which entails clarification of the gauge ambiguities around an expanding background \cite{Kou:2021bij}. It is found that synchronous gauge tends to generate large artificial effects in the GW spectrum if gravitational effects are significant in the formation of oscillons, while radiation gauge or  suitably chosen 1+log gauges can efficiently reduce the gauge effects. The GW spectra obtained in the simulations with a rigid FRW background are found to be fairly accurate unless strong gravitational effects are involved in the formation of oscillons, which can result in $\mc{O}(1)$ enhancement in the GW spectrum; see Figure \ref{fig:oscGWNR}. The mass spectrum of the produced primordial BHs have also been estimated, and it is found that these primordial BHs can potentially account for all dark matter \cite{Cotner:2018vug, Cotner:2019ykd}.

Apart from the stochastic GW background produced directly from nonlinear causal interactions during oscillon formation, there are also a couple of stochastic GW backgrounds indirectly associated with the oscillon formation in the early universe \cite{Lozanov:2022yoy, Lozanov:2023aez, Lozanov:2023knf}; see Figure \ref{fig:oscGWsp}. These GW backgrounds are induced by enhanced second order primordial curvature perturbations \cite{Ananda:2006af}, parallel to the scenarios of Q-ball formation from the AD condensate discussed previously.

First, an oscillon-dominated universe behaves just as an effective matter-dominated epoch, which, due to the longevity of oscillons, can be particularly prolonged for certain potentials. This phase can be followed by a rapid transition to the radiation-dominated phase, as oscillons are quasi-stable and can decay very fast at the end of their lifetimes,  compared to the Hubble scale at the time. Such a rapid transition can amplify primordial second-order curvature perturbations \cite{Inomata:2019ivs}, which in turn can induce significant GW backgrounds within the detectable frequency and amplitude range of low-frequency GW detectors \cite{Lozanov:2022yoy}. Secondly, using the axion-like-particle oscillon as an example, Ref \cite{Lozanov:2023aez} pointed out that soliton formation in the early universe is universally accompanied by another GW background from the soliton isocurvature perturbations. Specifically, the Poissonian long-wavelength tail of the density power spectrum leads to isocurvature perturbations towards the end of the soliton formation. The conversion of the isocurvature perturbations to curvature perturbations can then induce a detectable stochastic GW background \cite{Lozanov:2023aez}. This universal GW background can be significantly enhanced by soliton interactions. Particularly, gravitational clustering of oscillons can induce sizable correlations on large scales  \cite{Lozanov:2023knf}. Both the decay and isocurvature induced GW backgrounds are a few orders lower than the GW background from oscillon formation, as we can see in Figure \ref{fig:oscGWsp}. This opens up a new window to detect oscillons and probe the early universe physics with upcoming GW experiments.\\
~\\

\begin{acknowledgments}

We thank Mustafa Amin, Fu-Ming Chang, Ed Copeland, Hong-Yi Liu, Paul Saffin, Qi-Xin Xie, Guo-Dong Zhang and Hong-Yi Zhang for helpful discussions. We extend special thanks to Qi-Xin Xie and Guo-Dong Zhang for their assistance in producing a couple of illustrative figures. We acknowledge support from the National Natural Science Foundation of China under grant No.~12075233 and No.~12247103 and from the National Key R\&~D Program of China under grant No.~2022YFC2204603.

\end{acknowledgments}

\appendix

\section{Derrick's theorem}
\label{sec:Derrick}

Here we review Derrick's simple argument on the non-existence of static solitons in canonical scalar theories in higher than 1+1D spacetimes. Consider a set of (real) scalar fields $\phi=(\phi_1,\phi_2,...)$ in $d+1$ dimensions whose dynamics are described by the Lagrangian
\be
\label{LagDerrick}
\mc{L} = -\frac12 \pd_\mu \phi \cdot \pd^\mu \phi - V(\phi) .
\ee
We shall assume that potential $V(\phi)$ is bounded from below and, without loss of generality, choose its minimum to be $\min(V(\phi))=0$. Consider a finite-energy/localized, static configuration $\phi(\bfx)$, and its rescaling deformation $\phi_\li\equiv \phi(\li\bfx)$, whose energy is given by
\be
E_\li \!=\!\!\int\!\! \d^d x \!  \[ \frac12 \pd_i \phi_\li \cdot \pd^i \phi_\li + V(\phi_\li)\] = \li^{2-d}G+\f{P}{\li^d} ,
\ee
where we have defined 
\bal
G&= \int\! \d^d x \, \f12 \pd_i \phi(\bfx) \cdot \pd^i \phi(\bfx)\geq 0,
\\
P &= \int\! \d^d x  V(\phi(\bfx))\geq 0 .
\eal
 If $\phi(\bfx)$ is a localized soliton solution to the equation of motion, it must be a stationary point of $E_\li$: ${\d E_{\li=1}}/{\d \li} = 0$. However, for $d\geq 2$, it is clear that $E_\li$ cannot be stationary at $\li=1$. Thus, static solitons do not exist for the Lagrangian \eqref{LagDerrick} in a higher than 1+1D spacetime\,\footnote{It is actually not necessary to assume that the potential $V$ is positive or bounded from below. One can just additionally evaluate ${\d^2 E_{\li=1}}/{\d \li^2}$ and find that it can not be positive \cite{Derrick:1964ww}, which means that the localized solution, if exists, can not be stable.}.
For $d=1$, it is possible to have a stationary point at $\li=1$, consistent with the existence of static solitons in 1+1D. In fact, this scaling argument in 1+1D provides a Virial theorem that relates the gradient and potential energies. On the other hand, by extending the  Lagrangian \eqref{LagDerrick} with higher derivative terms and/or nonlinear sigma-like constraints, it is actually possible to have scalar topological solitons in 2+1D \cite{Manton:2004tk}.

\bibliographystyle{apsrev4-2}
\bibliography{nontopRefs}

\begin{thebibliography}{397}%
\makeatletter
\providecommand \@ifxundefined [1]{%
 \@ifx{#1\undefined}
}%
\providecommand \@ifnum [1]{%
 \ifnum #1\expandafter \@firstoftwo
 \else \expandafter \@secondoftwo
 \fi
}%
\providecommand \@ifx [1]{%
 \ifx #1\expandafter \@firstoftwo
 \else \expandafter \@secondoftwo
 \fi
}%
\providecommand \natexlab [1]{#1}%
\providecommand \enquote  [1]{``#1''}%
\providecommand \bibnamefont  [1]{#1}%
\providecommand \bibfnamefont [1]{#1}%
\providecommand \citenamefont [1]{#1}%
\providecommand \href@noop [0]{\@secondoftwo}%
\providecommand \href [0]{\begingroup \@sanitize@url \@href}%
\providecommand \@href[1]{\@@startlink{#1}\@@href}%
\providecommand \@@href[1]{\endgroup#1\@@endlink}%
\providecommand \@sanitize@url [0]{\catcode `\\12\catcode `\$12\catcode
  `\&12\catcode `\#12\catcode `\^12\catcode `\_12\catcode `\%12\relax}%
\providecommand \@@startlink[1]{}%
\providecommand \@@endlink[0]{}%
\providecommand \url  [0]{\begingroup\@sanitize@url \@url }%
\providecommand \@url [1]{\endgroup\@href {#1}{\urlprefix }}%
\providecommand \urlprefix  [0]{URL }%
\providecommand \Eprint [0]{\href }%
\providecommand \doibase [0]{https://doi.org/}%
\providecommand \selectlanguage [0]{\@gobble}%
\providecommand \bibinfo  [0]{\@secondoftwo}%
\providecommand \bibfield  [0]{\@secondoftwo}%
\providecommand \translation [1]{[#1]}%
\providecommand \BibitemOpen [0]{}%
\providecommand \bibitemStop [0]{}%
\providecommand \bibitemNoStop [0]{.\EOS\space}%
\providecommand \EOS [0]{\spacefactor3000\relax}%
\providecommand \BibitemShut  [1]{\csname bibitem#1\endcsname}%
\let\auto@bib@innerbib\@empty
%</preamble>
\bibitem [{\citenamefont {Gardner}\ \emph {et~al.}(1967)\citenamefont
  {Gardner}, \citenamefont {Greene}, \citenamefont {Kruskal},\ and\
  \citenamefont {Miura}}]{Gardner:1967wc}%
  \BibitemOpen
  \bibfield  {author} {\bibinfo {author} {\bibfnamefont {C.~S.}\ \bibnamefont
  {Gardner}}, \bibinfo {author} {\bibfnamefont {J.~M.}\ \bibnamefont {Greene}},
  \bibinfo {author} {\bibfnamefont {M.~D.}\ \bibnamefont {Kruskal}},\ and\
  \bibinfo {author} {\bibfnamefont {R.~M.}\ \bibnamefont {Miura}},\ }\href
  {https://doi.org/10.1103/PhysRevLett.19.1095} {\bibfield  {journal} {\bibinfo
   {journal} {Phys. Rev. Lett.}\ }\textbf {\bibinfo {volume} {19}},\ \bibinfo
  {pages} {1095} (\bibinfo {year} {1967})}\BibitemShut {NoStop}%
\bibitem [{\citenamefont {Manton}\ and\ \citenamefont
  {Sutcliffe}(2004)}]{Manton:2004tk}%
  \BibitemOpen
  \bibfield  {author} {\bibinfo {author} {\bibfnamefont {N.~S.}\ \bibnamefont
  {Manton}}\ and\ \bibinfo {author} {\bibfnamefont {P.}~\bibnamefont
  {Sutcliffe}},\ }\href {https://doi.org/10.1017/CBO9780511617034} {\emph
  {\bibinfo {title} {{Topological solitons}}}},\ Cambridge Monographs on
  Mathematical Physics\ (\bibinfo  {publisher} {Cambridge University Press},\
  \bibinfo {year} {2004})\BibitemShut {NoStop}%
\bibitem [{\citenamefont {Ryder}(1996)}]{Ryder:1985wq}%
  \BibitemOpen
  \bibfield  {author} {\bibinfo {author} {\bibfnamefont {L.~H.}\ \bibnamefont
  {Ryder}},\ }\href {https://doi.org/10.1017/CBO9780511813900} {\emph {\bibinfo
  {title} {{Quantum Field Theory}}}}\ (\bibinfo  {publisher} {Cambridge
  University Press},\ \bibinfo {year} {1996})\BibitemShut {NoStop}%
\bibitem [{\citenamefont {Derrick}(1964)}]{Derrick:1964ww}%
  \BibitemOpen
  \bibfield  {author} {\bibinfo {author} {\bibfnamefont {G.~H.}\ \bibnamefont
  {Derrick}},\ }\href {https://doi.org/10.1063/1.1704233} {\bibfield  {journal}
  {\bibinfo  {journal} {J. Math. Phys.}\ }\textbf {\bibinfo {volume} {5}},\
  \bibinfo {pages} {1252} (\bibinfo {year} {1964})}\BibitemShut {NoStop}%
\bibitem [{\citenamefont {Nielsen}\ and\ \citenamefont
  {Olesen}(1973)}]{Nielsen:1973cs}%
  \BibitemOpen
  \bibfield  {author} {\bibinfo {author} {\bibfnamefont {H.~B.}\ \bibnamefont
  {Nielsen}}\ and\ \bibinfo {author} {\bibfnamefont {P.}~\bibnamefont
  {Olesen}},\ }\href {https://doi.org/10.1016/0550-3213(73)90350-7} {\bibfield
  {journal} {\bibinfo  {journal} {Nucl. Phys. B}\ }\textbf {\bibinfo {volume}
  {61}},\ \bibinfo {pages} {45} (\bibinfo {year} {1973})}\BibitemShut {NoStop}%
\bibitem [{\citenamefont {'t~Hooft}(1974)}]{tHooft:1974kcl}%
  \BibitemOpen
  \bibfield  {author} {\bibinfo {author} {\bibfnamefont {G.}~\bibnamefont
  {'t~Hooft}},\ }\href {https://doi.org/10.1016/0550-3213(74)90486-6}
  {\bibfield  {journal} {\bibinfo  {journal} {Nucl. Phys. B}\ }\textbf
  {\bibinfo {volume} {79}},\ \bibinfo {pages} {276} (\bibinfo {year}
  {1974})}\BibitemShut {NoStop}%
\bibitem [{\citenamefont {Polyakov}(1974)}]{Polyakov:1974ek}%
  \BibitemOpen
  \bibfield  {author} {\bibinfo {author} {\bibfnamefont {A.~M.}\ \bibnamefont
  {Polyakov}},\ }\href@noop {} {\bibfield  {journal} {\bibinfo  {journal} {JETP
  Lett.}\ }\textbf {\bibinfo {volume} {20}},\ \bibinfo {pages} {194} (\bibinfo
  {year} {1974})}\BibitemShut {NoStop}%
\bibitem [{\citenamefont {Friedberg}\ \emph {et~al.}(1976)\citenamefont
  {Friedberg}, \citenamefont {Lee},\ and\ \citenamefont
  {Sirlin}}]{Friedberg:1976me}%
  \BibitemOpen
  \bibfield  {author} {\bibinfo {author} {\bibfnamefont {R.}~\bibnamefont
  {Friedberg}}, \bibinfo {author} {\bibfnamefont {T.~D.}\ \bibnamefont {Lee}},\
  and\ \bibinfo {author} {\bibfnamefont {A.}~\bibnamefont {Sirlin}},\ }\href
  {https://doi.org/10.1103/PhysRevD.13.2739} {\bibfield  {journal} {\bibinfo
  {journal} {Phys. Rev. D}\ }\textbf {\bibinfo {volume} {13}},\ \bibinfo
  {pages} {2739} (\bibinfo {year} {1976})}\BibitemShut {NoStop}%
\bibitem [{\citenamefont {Friedberg}\ and\ \citenamefont
  {Lee}(1977{\natexlab{a}})}]{Friedberg:1977xf}%
  \BibitemOpen
  \bibfield  {author} {\bibinfo {author} {\bibfnamefont {R.}~\bibnamefont
  {Friedberg}}\ and\ \bibinfo {author} {\bibfnamefont {T.~D.}\ \bibnamefont
  {Lee}},\ }\href {https://doi.org/10.1103/PhysRevD.16.1096} {\bibfield
  {journal} {\bibinfo  {journal} {Phys. Rev. D}\ }\textbf {\bibinfo {volume}
  {16}},\ \bibinfo {pages} {1096} (\bibinfo {year}
  {1977}{\natexlab{a}})}\BibitemShut {NoStop}%
\bibitem [{\citenamefont {Coleman}(1985)}]{Coleman:1985ki}%
  \BibitemOpen
  \bibfield  {author} {\bibinfo {author} {\bibfnamefont {S.~R.}\ \bibnamefont
  {Coleman}},\ }\href {https://doi.org/10.1016/0550-3213(86)90520-1} {\bibfield
   {journal} {\bibinfo  {journal} {Nucl. Phys. B}\ }\textbf {\bibinfo {volume}
  {262}},\ \bibinfo {pages} {263} (\bibinfo {year} {1985})},\ \bibinfo {note}
  {[Addendum: Nucl.Phys.B 269, 744 (1986)]}\BibitemShut {NoStop}%
\bibitem [{\citenamefont {Lee}\ and\ \citenamefont {Pang}(1992)}]{Lee:1991ax}%
  \BibitemOpen
  \bibfield  {author} {\bibinfo {author} {\bibfnamefont {T.~D.}\ \bibnamefont
  {Lee}}\ and\ \bibinfo {author} {\bibfnamefont {Y.}~\bibnamefont {Pang}},\
  }\href {https://doi.org/10.1016/0370-1573(92)90064-7} {\bibfield  {journal}
  {\bibinfo  {journal} {Phys. Rept.}\ }\textbf {\bibinfo {volume} {221}},\
  \bibinfo {pages} {251} (\bibinfo {year} {1992})}\BibitemShut {NoStop}%
\bibitem [{\citenamefont {Rosen}(1968{\natexlab{a}})}]{Rosen:1968mfz}%
  \BibitemOpen
  \bibfield  {author} {\bibinfo {author} {\bibfnamefont {G.}~\bibnamefont
  {Rosen}},\ }\href {https://doi.org/10.1063/1.1664693} {\bibfield  {journal}
  {\bibinfo  {journal} {J. Math. Phys.}\ }\textbf {\bibinfo {volume} {9}},\
  \bibinfo {pages} {996} (\bibinfo {year} {1968}{\natexlab{a}})}\BibitemShut
  {NoStop}%
\bibitem [{\citenamefont {Kusenko}(1997{\natexlab{a}})}]{Kusenko:1997zq}%
  \BibitemOpen
  \bibfield  {author} {\bibinfo {author} {\bibfnamefont {A.}~\bibnamefont
  {Kusenko}},\ }\href {https://doi.org/10.1016/S0370-2693(97)00584-4}
  {\bibfield  {journal} {\bibinfo  {journal} {Phys. Lett. B}\ }\textbf
  {\bibinfo {volume} {405}},\ \bibinfo {pages} {108} (\bibinfo {year}
  {1997}{\natexlab{a}})},\ \Eprint {https://arxiv.org/abs/hep-ph/9704273}
  {arXiv:hep-ph/9704273} \BibitemShut {NoStop}%
\bibitem [{\citenamefont {Kusenko}\ and\ \citenamefont
  {Shaposhnikov}(1998)}]{Kusenko:1997si}%
  \BibitemOpen
  \bibfield  {author} {\bibinfo {author} {\bibfnamefont {A.}~\bibnamefont
  {Kusenko}}\ and\ \bibinfo {author} {\bibfnamefont {M.~E.}\ \bibnamefont
  {Shaposhnikov}},\ }\href {https://doi.org/10.1016/S0370-2693(97)01375-0}
  {\bibfield  {journal} {\bibinfo  {journal} {Phys. Lett. B}\ }\textbf
  {\bibinfo {volume} {418}},\ \bibinfo {pages} {46} (\bibinfo {year} {1998})},\
  \Eprint {https://arxiv.org/abs/hep-ph/9709492} {arXiv:hep-ph/9709492}
  \BibitemShut {NoStop}%
\bibitem [{\citenamefont {Enqvist}\ and\ \citenamefont
  {McDonald}(1998{\natexlab{a}})}]{Enqvist:1997si}%
  \BibitemOpen
  \bibfield  {author} {\bibinfo {author} {\bibfnamefont {K.}~\bibnamefont
  {Enqvist}}\ and\ \bibinfo {author} {\bibfnamefont {J.}~\bibnamefont
  {McDonald}},\ }\href {https://doi.org/10.1016/S0370-2693(98)00271-8}
  {\bibfield  {journal} {\bibinfo  {journal} {Phys. Lett. B}\ }\textbf
  {\bibinfo {volume} {425}},\ \bibinfo {pages} {309} (\bibinfo {year}
  {1998}{\natexlab{a}})},\ \Eprint {https://arxiv.org/abs/hep-ph/9711514}
  {arXiv:hep-ph/9711514} \BibitemShut {NoStop}%
\bibitem [{\citenamefont {Kasuya}\ and\ \citenamefont
  {Kawasaki}(2000{\natexlab{a}})}]{Kasuya:1999wu}%
  \BibitemOpen
  \bibfield  {author} {\bibinfo {author} {\bibfnamefont {S.}~\bibnamefont
  {Kasuya}}\ and\ \bibinfo {author} {\bibfnamefont {M.}~\bibnamefont
  {Kawasaki}},\ }\href {https://doi.org/10.1103/PhysRevD.61.041301} {\bibfield
  {journal} {\bibinfo  {journal} {Phys. Rev. D}\ }\textbf {\bibinfo {volume}
  {61}},\ \bibinfo {pages} {041301} (\bibinfo {year} {2000}{\natexlab{a}})},\
  \Eprint {https://arxiv.org/abs/hep-ph/9909509} {arXiv:hep-ph/9909509}
  \BibitemShut {NoStop}%
\bibitem [{\citenamefont {Kasuya}\ and\ \citenamefont
  {Kawasaki}(2000{\natexlab{b}})}]{Kasuya:2000wx}%
  \BibitemOpen
  \bibfield  {author} {\bibinfo {author} {\bibfnamefont {S.}~\bibnamefont
  {Kasuya}}\ and\ \bibinfo {author} {\bibfnamefont {M.}~\bibnamefont
  {Kawasaki}},\ }\href {https://doi.org/10.1103/PhysRevD.62.023512} {\bibfield
  {journal} {\bibinfo  {journal} {Phys. Rev. D}\ }\textbf {\bibinfo {volume}
  {62}},\ \bibinfo {pages} {023512} (\bibinfo {year} {2000}{\natexlab{b}})},\
  \Eprint {https://arxiv.org/abs/hep-ph/0002285} {arXiv:hep-ph/0002285}
  \BibitemShut {NoStop}%
\bibitem [{\citenamefont {Bogolyubsky}(1977)}]{Bogolyubsky:1977tc}%
  \BibitemOpen
  \bibfield  {author} {\bibinfo {author} {\bibfnamefont {I.~L.}\ \bibnamefont
  {Bogolyubsky}},\ }\href {https://doi.org/10.1016/0375-9601(77)90138-4}
  {\bibfield  {journal} {\bibinfo  {journal} {Phys. Lett. A}\ }\textbf
  {\bibinfo {volume} {61}},\ \bibinfo {pages} {205} (\bibinfo {year}
  {1977})}\BibitemShut {NoStop}%
\bibitem [{\citenamefont {Bogolyubsky}\ and\ \citenamefont
  {Makhankov}(1977)}]{oscillon2}%
  \BibitemOpen
  \bibfield  {author} {\bibinfo {author} {\bibfnamefont {I.~L.}\ \bibnamefont
  {Bogolyubsky}}\ and\ \bibinfo {author} {\bibfnamefont {V.~G.}\ \bibnamefont
  {Makhankov}},\ }\href@noop {} {\bibfield  {journal} {\bibinfo  {journal}
  {JETP Letters}\ }\textbf {\bibinfo {volume} {25}},\ \bibinfo {pages} {107}
  (\bibinfo {year} {1977})}\BibitemShut {NoStop}%
\bibitem [{\citenamefont {Gleiser}(1994)}]{Gleiser:1993pt}%
  \BibitemOpen
  \bibfield  {author} {\bibinfo {author} {\bibfnamefont {M.}~\bibnamefont
  {Gleiser}},\ }\href {https://doi.org/10.1103/PhysRevD.49.2978} {\bibfield
  {journal} {\bibinfo  {journal} {Phys. Rev. D}\ }\textbf {\bibinfo {volume}
  {49}},\ \bibinfo {pages} {2978} (\bibinfo {year} {1994})},\ \Eprint
  {https://arxiv.org/abs/hep-ph/9308279} {arXiv:hep-ph/9308279} \BibitemShut
  {NoStop}%
\bibitem [{\citenamefont {Copeland}\ \emph {et~al.}(1995)\citenamefont
  {Copeland}, \citenamefont {Gleiser},\ and\ \citenamefont
  {Muller}}]{Copeland:1995fq}%
  \BibitemOpen
  \bibfield  {author} {\bibinfo {author} {\bibfnamefont {E.~J.}\ \bibnamefont
  {Copeland}}, \bibinfo {author} {\bibfnamefont {M.}~\bibnamefont {Gleiser}},\
  and\ \bibinfo {author} {\bibfnamefont {H.~R.}\ \bibnamefont {Muller}},\
  }\href {https://doi.org/10.1103/PhysRevD.52.1920} {\bibfield  {journal}
  {\bibinfo  {journal} {Phys. Rev. D}\ }\textbf {\bibinfo {volume} {52}},\
  \bibinfo {pages} {1920} (\bibinfo {year} {1995})},\ \Eprint
  {https://arxiv.org/abs/hep-ph/9503217} {arXiv:hep-ph/9503217} \BibitemShut
  {NoStop}%
\bibitem [{\citenamefont {Khlebnikov}\ and\ \citenamefont
  {Tkachev}(1996)}]{Khlebnikov:1996mc}%
  \BibitemOpen
  \bibfield  {author} {\bibinfo {author} {\bibfnamefont {S.~Y.}\ \bibnamefont
  {Khlebnikov}}\ and\ \bibinfo {author} {\bibfnamefont {I.~I.}\ \bibnamefont
  {Tkachev}},\ }\href {https://doi.org/10.1103/PhysRevLett.77.219} {\bibfield
  {journal} {\bibinfo  {journal} {Phys. Rev. Lett.}\ }\textbf {\bibinfo
  {volume} {77}},\ \bibinfo {pages} {219} (\bibinfo {year} {1996})},\ \Eprint
  {https://arxiv.org/abs/hep-ph/9603378} {arXiv:hep-ph/9603378} \BibitemShut
  {NoStop}%
\bibitem [{\citenamefont {Berges}(2004)}]{Berges:2004yj}%
  \BibitemOpen
  \bibfield  {author} {\bibinfo {author} {\bibfnamefont {J.}~\bibnamefont
  {Berges}},\ }\href {https://doi.org/10.1063/1.1843591} {\bibfield  {journal}
  {\bibinfo  {journal} {AIP Conf. Proc.}\ }\textbf {\bibinfo {volume} {739}},\
  \bibinfo {pages} {3} (\bibinfo {year} {2004})},\ \Eprint
  {https://arxiv.org/abs/hep-ph/0409233} {arXiv:hep-ph/0409233} \BibitemShut
  {NoStop}%
\bibitem [{\citenamefont {Dashen}\ \emph {et~al.}(1974)\citenamefont {Dashen},
  \citenamefont {Hasslacher},\ and\ \citenamefont {Neveu}}]{Dashen:1974ci}%
  \BibitemOpen
  \bibfield  {author} {\bibinfo {author} {\bibfnamefont {R.~F.}\ \bibnamefont
  {Dashen}}, \bibinfo {author} {\bibfnamefont {B.}~\bibnamefont {Hasslacher}},\
  and\ \bibinfo {author} {\bibfnamefont {A.}~\bibnamefont {Neveu}},\ }\href
  {https://doi.org/10.1103/PhysRevD.10.4114} {\bibfield  {journal} {\bibinfo
  {journal} {Phys. Rev. D}\ }\textbf {\bibinfo {volume} {10}},\ \bibinfo
  {pages} {4114} (\bibinfo {year} {1974})}\BibitemShut {NoStop}%
\bibitem [{\citenamefont {Gervais}\ and\ \citenamefont
  {Sakita}(1975)}]{Gervais:1974dc}%
  \BibitemOpen
  \bibfield  {author} {\bibinfo {author} {\bibfnamefont {J.-L.}\ \bibnamefont
  {Gervais}}\ and\ \bibinfo {author} {\bibfnamefont {B.}~\bibnamefont
  {Sakita}},\ }\href {https://doi.org/10.1103/PhysRevD.11.2943} {\bibfield
  {journal} {\bibinfo  {journal} {Phys. Rev. D}\ }\textbf {\bibinfo {volume}
  {11}},\ \bibinfo {pages} {2943} (\bibinfo {year} {1975})}\BibitemShut
  {NoStop}%
\bibitem [{\citenamefont {Goldstone}\ and\ \citenamefont
  {Jackiw}(1975)}]{Goldstone:1974gf}%
  \BibitemOpen
  \bibfield  {author} {\bibinfo {author} {\bibfnamefont {J.}~\bibnamefont
  {Goldstone}}\ and\ \bibinfo {author} {\bibfnamefont {R.}~\bibnamefont
  {Jackiw}},\ }\href {https://doi.org/10.1103/PhysRevD.11.1486} {\bibfield
  {journal} {\bibinfo  {journal} {Phys. Rev. D}\ }\textbf {\bibinfo {volume}
  {11}},\ \bibinfo {pages} {1486} (\bibinfo {year} {1975})}\BibitemShut
  {NoStop}%
\bibitem [{\citenamefont {Christ}\ and\ \citenamefont
  {Lee}(1975)}]{Christ:1975wt}%
  \BibitemOpen
  \bibfield  {author} {\bibinfo {author} {\bibfnamefont {N.~H.}\ \bibnamefont
  {Christ}}\ and\ \bibinfo {author} {\bibfnamefont {T.~D.}\ \bibnamefont
  {Lee}},\ }\href {https://doi.org/10.1103/PhysRevD.12.1606} {\bibfield
  {journal} {\bibinfo  {journal} {Phys. Rev. D}\ }\textbf {\bibinfo {volume}
  {12}},\ \bibinfo {pages} {1606} (\bibinfo {year} {1975})}\BibitemShut
  {NoStop}%
\bibitem [{\citenamefont {Tomboulis}(1975)}]{Tomboulis:1975gf}%
  \BibitemOpen
  \bibfield  {author} {\bibinfo {author} {\bibfnamefont {E.}~\bibnamefont
  {Tomboulis}},\ }\href {https://doi.org/10.1103/PhysRevD.12.1678} {\bibfield
  {journal} {\bibinfo  {journal} {Phys. Rev. D}\ }\textbf {\bibinfo {volume}
  {12}},\ \bibinfo {pages} {1678} (\bibinfo {year} {1975})}\BibitemShut
  {NoStop}%
\bibitem [{\citenamefont {Creutz}(1975)}]{Creutz:1975qt}%
  \BibitemOpen
  \bibfield  {author} {\bibinfo {author} {\bibfnamefont {M.}~\bibnamefont
  {Creutz}},\ }\href {https://doi.org/10.1103/PhysRevD.12.3126} {\bibfield
  {journal} {\bibinfo  {journal} {Phys. Rev. D}\ }\textbf {\bibinfo {volume}
  {12}},\ \bibinfo {pages} {3126} (\bibinfo {year} {1975})}\BibitemShut
  {NoStop}%
\bibitem [{\citenamefont {Salle}(2004)}]{Salle:2003ju}%
  \BibitemOpen
  \bibfield  {author} {\bibinfo {author} {\bibfnamefont {M.}~\bibnamefont
  {Salle}},\ }\href {https://doi.org/10.1103/PhysRevD.69.025005} {\bibfield
  {journal} {\bibinfo  {journal} {Phys. Rev. D}\ }\textbf {\bibinfo {volume}
  {69}},\ \bibinfo {pages} {025005} (\bibinfo {year} {2004})},\ \Eprint
  {https://arxiv.org/abs/hep-ph/0307080} {arXiv:hep-ph/0307080} \BibitemShut
  {NoStop}%
\bibitem [{\citenamefont {Borsanyi}\ and\ \citenamefont
  {Hindmarsh}(2008)}]{Borsanyi:2007wm}%
  \BibitemOpen
  \bibfield  {author} {\bibinfo {author} {\bibfnamefont {S.}~\bibnamefont
  {Borsanyi}}\ and\ \bibinfo {author} {\bibfnamefont {M.}~\bibnamefont
  {Hindmarsh}},\ }\href {https://doi.org/10.1103/PhysRevD.77.045022} {\bibfield
   {journal} {\bibinfo  {journal} {Phys. Rev. D}\ }\textbf {\bibinfo {volume}
  {77}},\ \bibinfo {pages} {045022} (\bibinfo {year} {2008})},\ \Eprint
  {https://arxiv.org/abs/0712.0300} {arXiv:0712.0300 [hep-ph]} \BibitemShut
  {NoStop}%
\bibitem [{\citenamefont {Tranberg}\ and\ \citenamefont
  {Weir}(2014)}]{Tranberg:2013cka}%
  \BibitemOpen
  \bibfield  {author} {\bibinfo {author} {\bibfnamefont {A.}~\bibnamefont
  {Tranberg}}\ and\ \bibinfo {author} {\bibfnamefont {D.~J.}\ \bibnamefont
  {Weir}},\ }\href {https://doi.org/10.1007/JHEP04(2014)184} {\bibfield
  {journal} {\bibinfo  {journal} {JHEP}\ }\textbf {\bibinfo {volume} {04}},\
  \bibinfo {pages} {184}},\ \Eprint {https://arxiv.org/abs/1310.7487}
  {arXiv:1310.7487 [hep-ph]} \BibitemShut {NoStop}%
\bibitem [{\citenamefont {Xie}\ \emph {et~al.}(2024)\citenamefont {Xie},
  \citenamefont {Saffin}, \citenamefont {Tranberg},\ and\ \citenamefont
  {Zhou}}]{Xie:2023psz}%
  \BibitemOpen
  \bibfield  {author} {\bibinfo {author} {\bibfnamefont {Q.-X.}\ \bibnamefont
  {Xie}}, \bibinfo {author} {\bibfnamefont {P.~M.}\ \bibnamefont {Saffin}},
  \bibinfo {author} {\bibfnamefont {A.}~\bibnamefont {Tranberg}},\ and\
  \bibinfo {author} {\bibfnamefont {S.-Y.}\ \bibnamefont {Zhou}},\ }\href
  {https://doi.org/10.1007/JHEP01(2024)165} {\bibfield  {journal} {\bibinfo
  {journal} {JHEP}\ }\textbf {\bibinfo {volume} {01}},\ \bibinfo {pages}
  {165}},\ \Eprint {https://arxiv.org/abs/2312.01139} {arXiv:2312.01139
  [hep-th]} \BibitemShut {NoStop}%
\bibitem [{\citenamefont {Jetzer}(1992)}]{Jetzer:1991jr}%
  \BibitemOpen
  \bibfield  {author} {\bibinfo {author} {\bibfnamefont {P.}~\bibnamefont
  {Jetzer}},\ }\href {https://doi.org/10.1016/0370-1573(92)90123-H} {\bibfield
  {journal} {\bibinfo  {journal} {Phys. Rept.}\ }\textbf {\bibinfo {volume}
  {220}},\ \bibinfo {pages} {163} (\bibinfo {year} {1992})}\BibitemShut
  {NoStop}%
\bibitem [{\citenamefont {Liebling}\ and\ \citenamefont
  {Palenzuela}(2023)}]{Liebling:2012fv}%
  \BibitemOpen
  \bibfield  {author} {\bibinfo {author} {\bibfnamefont {S.~L.}\ \bibnamefont
  {Liebling}}\ and\ \bibinfo {author} {\bibfnamefont {C.}~\bibnamefont
  {Palenzuela}},\ }\href {https://doi.org/10.1007/s41114-023-00043-4}
  {\bibfield  {journal} {\bibinfo  {journal} {Living Rev. Rel.}\ }\textbf
  {\bibinfo {volume} {26}},\ \bibinfo {pages} {1} (\bibinfo {year} {2023})},\
  \Eprint {https://arxiv.org/abs/1202.5809} {arXiv:1202.5809 [gr-qc]}
  \BibitemShut {NoStop}%
\bibitem [{\citenamefont {Visinelli}(2021)}]{Visinelli:2021uve}%
  \BibitemOpen
  \bibfield  {author} {\bibinfo {author} {\bibfnamefont {L.}~\bibnamefont
  {Visinelli}},\ }\href {https://doi.org/10.1142/S0218271821300068} {\bibfield
  {journal} {\bibinfo  {journal} {Int. J. Mod. Phys. D}\ }\textbf {\bibinfo
  {volume} {30}},\ \bibinfo {pages} {2130006} (\bibinfo {year} {2021})},\
  \Eprint {https://arxiv.org/abs/2109.05481} {arXiv:2109.05481 [gr-qc]}
  \BibitemShut {NoStop}%
\bibitem [{\citenamefont {Nugaev}\ and\ \citenamefont
  {Shkerin}(2020)}]{Nugaev:2019vru}%
  \BibitemOpen
  \bibfield  {author} {\bibinfo {author} {\bibfnamefont {E.~Y.}\ \bibnamefont
  {Nugaev}}\ and\ \bibinfo {author} {\bibfnamefont {A.~V.}\ \bibnamefont
  {Shkerin}},\ }\href {https://doi.org/10.1134/S1063776120020077} {\bibfield
  {journal} {\bibinfo  {journal} {J. Exp. Theor. Phys.}\ }\textbf {\bibinfo
  {volume} {130}},\ \bibinfo {pages} {301} (\bibinfo {year} {2020})},\ \Eprint
  {https://arxiv.org/abs/1905.05146} {arXiv:1905.05146 [hep-th]} \BibitemShut
  {NoStop}%
\bibitem [{\citenamefont {Mai}\ and\ \citenamefont
  {Schweitzer}(2012)}]{Mai:2012cx}%
  \BibitemOpen
  \bibfield  {author} {\bibinfo {author} {\bibfnamefont {M.}~\bibnamefont
  {Mai}}\ and\ \bibinfo {author} {\bibfnamefont {P.}~\bibnamefont
  {Schweitzer}},\ }\href {https://doi.org/10.1103/PhysRevD.86.096002}
  {\bibfield  {journal} {\bibinfo  {journal} {Phys. Rev. D}\ }\textbf {\bibinfo
  {volume} {86}},\ \bibinfo {pages} {096002} (\bibinfo {year} {2012})},\
  \Eprint {https://arxiv.org/abs/1206.2930} {arXiv:1206.2930 [hep-ph]}
  \BibitemShut {NoStop}%
\bibitem [{\citenamefont {Almumin}\ \emph {et~al.}(2022)\citenamefont
  {Almumin}, \citenamefont {Heeck}, \citenamefont {Rajaraman},\ and\
  \citenamefont {Verhaaren}}]{Almumin:2021gax}%
  \BibitemOpen
  \bibfield  {author} {\bibinfo {author} {\bibfnamefont {Y.}~\bibnamefont
  {Almumin}}, \bibinfo {author} {\bibfnamefont {J.}~\bibnamefont {Heeck}},
  \bibinfo {author} {\bibfnamefont {A.}~\bibnamefont {Rajaraman}},\ and\
  \bibinfo {author} {\bibfnamefont {C.~B.}\ \bibnamefont {Verhaaren}},\ }\href
  {https://doi.org/10.1140/epjc/s10052-022-10772-5} {\bibfield  {journal}
  {\bibinfo  {journal} {Eur. Phys. J. C}\ }\textbf {\bibinfo {volume} {82}},\
  \bibinfo {pages} {801} (\bibinfo {year} {2022})},\ \Eprint
  {https://arxiv.org/abs/2112.00657} {arXiv:2112.00657 [hep-th]} \BibitemShut
  {NoStop}%
\bibitem [{\citenamefont {Kusenko}(1997{\natexlab{b}})}]{Kusenko:1997ad}%
  \BibitemOpen
  \bibfield  {author} {\bibinfo {author} {\bibfnamefont {A.}~\bibnamefont
  {Kusenko}},\ }\href {https://doi.org/10.1016/S0370-2693(97)00582-0}
  {\bibfield  {journal} {\bibinfo  {journal} {Phys. Lett. B}\ }\textbf
  {\bibinfo {volume} {404}},\ \bibinfo {pages} {285} (\bibinfo {year}
  {1997}{\natexlab{b}})},\ \Eprint {https://arxiv.org/abs/hep-th/9704073}
  {arXiv:hep-th/9704073} \BibitemShut {NoStop}%
\bibitem [{\citenamefont {Sakai}\ and\ \citenamefont
  {Sasaki}(2008)}]{Sakai:2007ft}%
  \BibitemOpen
  \bibfield  {author} {\bibinfo {author} {\bibfnamefont {N.}~\bibnamefont
  {Sakai}}\ and\ \bibinfo {author} {\bibfnamefont {M.}~\bibnamefont {Sasaki}},\
  }\href {https://doi.org/10.1143/PTP.119.929} {\bibfield  {journal} {\bibinfo
  {journal} {Prog. Theor. Phys.}\ }\textbf {\bibinfo {volume} {119}},\ \bibinfo
  {pages} {929} (\bibinfo {year} {2008})},\ \Eprint
  {https://arxiv.org/abs/0712.1450} {arXiv:0712.1450 [hep-ph]} \BibitemShut
  {NoStop}%
\bibitem [{\citenamefont {Smolyakov}(2019)}]{Smolyakov:2019cld}%
  \BibitemOpen
  \bibfield  {author} {\bibinfo {author} {\bibfnamefont {M.~N.}\ \bibnamefont
  {Smolyakov}},\ }\href {https://doi.org/10.1103/PhysRevD.100.045002}
  {\bibfield  {journal} {\bibinfo  {journal} {Phys. Rev. D}\ }\textbf {\bibinfo
  {volume} {100}},\ \bibinfo {pages} {045002} (\bibinfo {year} {2019})},\
  \Eprint {https://arxiv.org/abs/1906.02117} {arXiv:1906.02117 [hep-th]}
  \BibitemShut {NoStop}%
\bibitem [{\citenamefont {Battye}\ and\ \citenamefont
  {Sutcliffe}(2000)}]{Battye:2000qj}%
  \BibitemOpen
  \bibfield  {author} {\bibinfo {author} {\bibfnamefont {R.}~\bibnamefont
  {Battye}}\ and\ \bibinfo {author} {\bibfnamefont {P.}~\bibnamefont
  {Sutcliffe}},\ }\href {https://doi.org/10.1016/S0550-3213(00)00506-X}
  {\bibfield  {journal} {\bibinfo  {journal} {Nucl. Phys. B}\ }\textbf
  {\bibinfo {volume} {590}},\ \bibinfo {pages} {329} (\bibinfo {year}
  {2000})},\ \Eprint {https://arxiv.org/abs/hep-th/0003252}
  {arXiv:hep-th/0003252} \BibitemShut {NoStop}%
\bibitem [{\citenamefont {Campanelli}\ and\ \citenamefont
  {Ruggieri}(2008)}]{Campanelli:2007um}%
  \BibitemOpen
  \bibfield  {author} {\bibinfo {author} {\bibfnamefont {L.}~\bibnamefont
  {Campanelli}}\ and\ \bibinfo {author} {\bibfnamefont {M.}~\bibnamefont
  {Ruggieri}},\ }\href {https://doi.org/10.1103/PhysRevD.77.043504} {\bibfield
  {journal} {\bibinfo  {journal} {Phys. Rev. D}\ }\textbf {\bibinfo {volume}
  {77}},\ \bibinfo {pages} {043504} (\bibinfo {year} {2008})},\ \Eprint
  {https://arxiv.org/abs/0712.3669} {arXiv:0712.3669 [hep-th]} \BibitemShut
  {NoStop}%
\bibitem [{\citenamefont {Tsumagari}\ \emph {et~al.}(2008)\citenamefont
  {Tsumagari}, \citenamefont {Copeland},\ and\ \citenamefont
  {Saffin}}]{Tsumagari:2008bv}%
  \BibitemOpen
  \bibfield  {author} {\bibinfo {author} {\bibfnamefont {M.~I.}\ \bibnamefont
  {Tsumagari}}, \bibinfo {author} {\bibfnamefont {E.~J.}\ \bibnamefont
  {Copeland}},\ and\ \bibinfo {author} {\bibfnamefont {P.~M.}\ \bibnamefont
  {Saffin}},\ }\href {https://doi.org/10.1103/PhysRevD.78.065021} {\bibfield
  {journal} {\bibinfo  {journal} {Phys. Rev. D}\ }\textbf {\bibinfo {volume}
  {78}},\ \bibinfo {pages} {065021} (\bibinfo {year} {2008})},\ \Eprint
  {https://arxiv.org/abs/0805.3233} {arXiv:0805.3233 [hep-th]} \BibitemShut
  {NoStop}%
\bibitem [{\citenamefont {Copeland}\ and\ \citenamefont
  {Tsumagari}(2009)}]{Copeland:2009as}%
  \BibitemOpen
  \bibfield  {author} {\bibinfo {author} {\bibfnamefont {E.~J.}\ \bibnamefont
  {Copeland}}\ and\ \bibinfo {author} {\bibfnamefont {M.~I.}\ \bibnamefont
  {Tsumagari}},\ }\href {https://doi.org/10.1103/PhysRevD.80.025016} {\bibfield
   {journal} {\bibinfo  {journal} {Phys. Rev. D}\ }\textbf {\bibinfo {volume}
  {80}},\ \bibinfo {pages} {025016} (\bibinfo {year} {2009})},\ \Eprint
  {https://arxiv.org/abs/0905.0125} {arXiv:0905.0125 [hep-th]} \BibitemShut
  {NoStop}%
\bibitem [{\citenamefont {Gleiser}\ and\ \citenamefont
  {Thorarinson}(2006)}]{Gleiser:2005iq}%
  \BibitemOpen
  \bibfield  {author} {\bibinfo {author} {\bibfnamefont {M.}~\bibnamefont
  {Gleiser}}\ and\ \bibinfo {author} {\bibfnamefont {J.}~\bibnamefont
  {Thorarinson}},\ }\href {https://doi.org/10.1103/PhysRevD.73.065008}
  {\bibfield  {journal} {\bibinfo  {journal} {Phys. Rev. D}\ }\textbf {\bibinfo
  {volume} {73}},\ \bibinfo {pages} {065008} (\bibinfo {year} {2006})},\
  \Eprint {https://arxiv.org/abs/hep-th/0505251} {arXiv:hep-th/0505251}
  \BibitemShut {NoStop}%
\bibitem [{\citenamefont {Bazeia}\ \emph {et~al.}(2017)\citenamefont {Bazeia},
  \citenamefont {Losano}, \citenamefont {Marques},\ and\ \citenamefont
  {Menezes}}]{Bazeia:2016xrf}%
  \BibitemOpen
  \bibfield  {author} {\bibinfo {author} {\bibfnamefont {D.}~\bibnamefont
  {Bazeia}}, \bibinfo {author} {\bibfnamefont {L.}~\bibnamefont {Losano}},
  \bibinfo {author} {\bibfnamefont {M.~A.}\ \bibnamefont {Marques}},\ and\
  \bibinfo {author} {\bibfnamefont {R.}~\bibnamefont {Menezes}},\ }\href
  {https://doi.org/10.1016/j.physletb.2016.12.033} {\bibfield  {journal}
  {\bibinfo  {journal} {Phys. Lett. B}\ }\textbf {\bibinfo {volume} {765}},\
  \bibinfo {pages} {359} (\bibinfo {year} {2017})},\ \Eprint
  {https://arxiv.org/abs/1612.04442} {arXiv:1612.04442 [hep-th]} \BibitemShut
  {NoStop}%
\bibitem [{\citenamefont {Bazeia}\ \emph {et~al.}(2016)\citenamefont {Bazeia},
  \citenamefont {Losano}, \citenamefont {Marques}, \citenamefont {Menezes},\
  and\ \citenamefont {da~Rocha}}]{Bazeia:2016wco}%
  \BibitemOpen
  \bibfield  {author} {\bibinfo {author} {\bibfnamefont {D.}~\bibnamefont
  {Bazeia}}, \bibinfo {author} {\bibfnamefont {L.}~\bibnamefont {Losano}},
  \bibinfo {author} {\bibfnamefont {M.~A.}\ \bibnamefont {Marques}}, \bibinfo
  {author} {\bibfnamefont {R.}~\bibnamefont {Menezes}},\ and\ \bibinfo {author}
  {\bibfnamefont {R.}~\bibnamefont {da~Rocha}},\ }\href
  {https://doi.org/10.1016/j.physletb.2016.04.060} {\bibfield  {journal}
  {\bibinfo  {journal} {Phys. Lett. B}\ }\textbf {\bibinfo {volume} {758}},\
  \bibinfo {pages} {146} (\bibinfo {year} {2016})},\ \Eprint
  {https://arxiv.org/abs/1604.08871} {arXiv:1604.08871 [hep-th]} \BibitemShut
  {NoStop}%
\bibitem [{\citenamefont {Bazeia}\ \emph {et~al.}(2019)\citenamefont {Bazeia},
  \citenamefont {Marques},\ and\ \citenamefont {Menezes}}]{Bazeia:2019ymk}%
  \BibitemOpen
  \bibfield  {author} {\bibinfo {author} {\bibfnamefont {D.}~\bibnamefont
  {Bazeia}}, \bibinfo {author} {\bibfnamefont {M.~A.}\ \bibnamefont
  {Marques}},\ and\ \bibinfo {author} {\bibfnamefont {R.}~\bibnamefont
  {Menezes}},\ }\href {https://doi.org/10.1209/0295-5075/127/21001} {\bibfield
  {journal} {\bibinfo  {journal} {EPL}\ }\textbf {\bibinfo {volume} {127}},\
  \bibinfo {pages} {21001} (\bibinfo {year} {2019})},\ \Eprint
  {https://arxiv.org/abs/1909.01163} {arXiv:1909.01163 [hep-th]} \BibitemShut
  {NoStop}%
\bibitem [{\citenamefont {Theodorakis}(2000)}]{Theodorakis:2000bz}%
  \BibitemOpen
  \bibfield  {author} {\bibinfo {author} {\bibfnamefont {S.}~\bibnamefont
  {Theodorakis}},\ }\href {https://doi.org/10.1103/PhysRevD.61.047701}
  {\bibfield  {journal} {\bibinfo  {journal} {Phys. Rev. D}\ }\textbf {\bibinfo
  {volume} {61}},\ \bibinfo {pages} {047701} (\bibinfo {year}
  {2000})}\BibitemShut {NoStop}%
\bibitem [{\citenamefont {Gulamov}\ \emph {et~al.}(2013)\citenamefont
  {Gulamov}, \citenamefont {Nugaev},\ and\ \citenamefont
  {Smolyakov}}]{Gulamov:2013ema}%
  \BibitemOpen
  \bibfield  {author} {\bibinfo {author} {\bibfnamefont {I.~E.}\ \bibnamefont
  {Gulamov}}, \bibinfo {author} {\bibfnamefont {E.~Y.}\ \bibnamefont
  {Nugaev}},\ and\ \bibinfo {author} {\bibfnamefont {M.~N.}\ \bibnamefont
  {Smolyakov}},\ }\href {https://doi.org/10.1103/PhysRevD.87.085043} {\bibfield
   {journal} {\bibinfo  {journal} {Phys. Rev. D}\ }\textbf {\bibinfo {volume}
  {87}},\ \bibinfo {pages} {085043} (\bibinfo {year} {2013})},\ \Eprint
  {https://arxiv.org/abs/1303.1173} {arXiv:1303.1173 [hep-th]} \BibitemShut
  {NoStop}%
\bibitem [{\citenamefont {Enqvist}\ and\ \citenamefont
  {McDonald}(1999)}]{Enqvist:1998en}%
  \BibitemOpen
  \bibfield  {author} {\bibinfo {author} {\bibfnamefont {K.}~\bibnamefont
  {Enqvist}}\ and\ \bibinfo {author} {\bibfnamefont {J.}~\bibnamefont
  {McDonald}},\ }\href {https://doi.org/10.1016/S0550-3213(98)00695-6}
  {\bibfield  {journal} {\bibinfo  {journal} {Nucl. Phys. B}\ }\textbf
  {\bibinfo {volume} {538}},\ \bibinfo {pages} {321} (\bibinfo {year}
  {1999})},\ \Eprint {https://arxiv.org/abs/hep-ph/9803380}
  {arXiv:hep-ph/9803380} \BibitemShut {NoStop}%
\bibitem [{\citenamefont {Dine}\ and\ \citenamefont
  {Kusenko}(2003)}]{Dine:2003ax}%
  \BibitemOpen
  \bibfield  {author} {\bibinfo {author} {\bibfnamefont {M.}~\bibnamefont
  {Dine}}\ and\ \bibinfo {author} {\bibfnamefont {A.}~\bibnamefont {Kusenko}},\
  }\href {https://doi.org/10.1103/RevModPhys.76.1} {\bibfield  {journal}
  {\bibinfo  {journal} {Rev. Mod. Phys.}\ }\textbf {\bibinfo {volume} {76}},\
  \bibinfo {pages} {1} (\bibinfo {year} {2003})},\ \Eprint
  {https://arxiv.org/abs/hep-ph/0303065} {arXiv:hep-ph/0303065} \BibitemShut
  {NoStop}%
\bibitem [{\citenamefont {Bialynicki-Birula}\ and\ \citenamefont
  {Mycielski}(1975)}]{Bialynicki-Birula:1975nws}%
  \BibitemOpen
  \bibfield  {author} {\bibinfo {author} {\bibfnamefont {I.}~\bibnamefont
  {Bialynicki-Birula}}\ and\ \bibinfo {author} {\bibfnamefont {J.}~\bibnamefont
  {Mycielski}},\ }\href@noop {} {\  (\bibinfo {year} {1975})}\BibitemShut
  {NoStop}%
\bibitem [{\citenamefont {Kovtun}\ \emph {et~al.}(2018)\citenamefont {Kovtun},
  \citenamefont {Nugaev},\ and\ \citenamefont {Shkerin}}]{Kovtun:2018jae}%
  \BibitemOpen
  \bibfield  {author} {\bibinfo {author} {\bibfnamefont {A.}~\bibnamefont
  {Kovtun}}, \bibinfo {author} {\bibfnamefont {E.}~\bibnamefont {Nugaev}},\
  and\ \bibinfo {author} {\bibfnamefont {A.}~\bibnamefont {Shkerin}},\ }\href
  {https://doi.org/10.1103/PhysRevD.98.096016} {\bibfield  {journal} {\bibinfo
  {journal} {Phys. Rev. D}\ }\textbf {\bibinfo {volume} {98}},\ \bibinfo
  {pages} {096016} (\bibinfo {year} {2018})},\ \Eprint
  {https://arxiv.org/abs/1805.03518} {arXiv:1805.03518 [hep-th]} \BibitemShut
  {NoStop}%
\bibitem [{\citenamefont {Heeck}\ \emph
  {et~al.}(2021{\natexlab{a}})\citenamefont {Heeck}, \citenamefont {Rajaraman},
  \citenamefont {Riley},\ and\ \citenamefont {Verhaaren}}]{Heeck:2020bau}%
  \BibitemOpen
  \bibfield  {author} {\bibinfo {author} {\bibfnamefont {J.}~\bibnamefont
  {Heeck}}, \bibinfo {author} {\bibfnamefont {A.}~\bibnamefont {Rajaraman}},
  \bibinfo {author} {\bibfnamefont {R.}~\bibnamefont {Riley}},\ and\ \bibinfo
  {author} {\bibfnamefont {C.~B.}\ \bibnamefont {Verhaaren}},\ }\href
  {https://doi.org/10.1103/PhysRevD.103.045008} {\bibfield  {journal} {\bibinfo
   {journal} {Phys. Rev. D}\ }\textbf {\bibinfo {volume} {103}},\ \bibinfo
  {pages} {045008} (\bibinfo {year} {2021}{\natexlab{a}})},\ \Eprint
  {https://arxiv.org/abs/2009.08462} {arXiv:2009.08462 [hep-th]} \BibitemShut
  {NoStop}%
\bibitem [{\citenamefont {Heeck}\ and\ \citenamefont
  {Sokhashvili}(2023{\natexlab{a}})}]{Heeck:2022iky}%
  \BibitemOpen
  \bibfield  {author} {\bibinfo {author} {\bibfnamefont {J.}~\bibnamefont
  {Heeck}}\ and\ \bibinfo {author} {\bibfnamefont {M.}~\bibnamefont
  {Sokhashvili}},\ }\href {https://doi.org/10.1103/PhysRevD.107.016006}
  {\bibfield  {journal} {\bibinfo  {journal} {Phys. Rev. D}\ }\textbf {\bibinfo
  {volume} {107}},\ \bibinfo {pages} {016006} (\bibinfo {year}
  {2023}{\natexlab{a}})},\ \Eprint {https://arxiv.org/abs/2211.00021}
  {arXiv:2211.00021 [hep-ph]} \BibitemShut {NoStop}%
\bibitem [{\citenamefont {Paccetti~Correia}\ and\ \citenamefont
  {Schmidt}(2001)}]{PaccettiCorreia:2001wtt}%
  \BibitemOpen
  \bibfield  {author} {\bibinfo {author} {\bibfnamefont {F.}~\bibnamefont
  {Paccetti~Correia}}\ and\ \bibinfo {author} {\bibfnamefont {M.~G.}\
  \bibnamefont {Schmidt}},\ }\href {https://doi.org/10.1007/s100520100710}
  {\bibfield  {journal} {\bibinfo  {journal} {Eur. Phys. J. C}\ }\textbf
  {\bibinfo {volume} {21}},\ \bibinfo {pages} {181} (\bibinfo {year} {2001})},\
  \Eprint {https://arxiv.org/abs/hep-th/0103189} {arXiv:hep-th/0103189}
  \BibitemShut {NoStop}%
\bibitem [{\citenamefont {Lennon}(2021{\natexlab{a}})}]{Lennon:2021zzx}%
  \BibitemOpen
  \bibfield  {author} {\bibinfo {author} {\bibfnamefont {O.}~\bibnamefont
  {Lennon}},\ }\href@noop {} {\  (\bibinfo {year} {2021}{\natexlab{a}})},\
  \Eprint {https://arxiv.org/abs/2112.14263} {arXiv:2112.14263 [hep-ph]}
  \BibitemShut {NoStop}%
\bibitem [{\citenamefont {Lennon}(2021{\natexlab{b}})}]{Lennon:2021fde}%
  \BibitemOpen
  \bibfield  {author} {\bibinfo {author} {\bibfnamefont {O.}~\bibnamefont
  {Lennon}},\ }\href@noop {} {\  (\bibinfo {year} {2021}{\natexlab{b}})},\
  \Eprint {https://arxiv.org/abs/2201.00024} {arXiv:2201.00024 [hep-ph]}
  \BibitemShut {NoStop}%
\bibitem [{\citenamefont {Lennon}(2021{\natexlab{c}})}]{Lennon:2021uqu}%
  \BibitemOpen
  \bibfield  {author} {\bibinfo {author} {\bibfnamefont {O.}~\bibnamefont
  {Lennon}},\ }\href@noop {} {\  (\bibinfo {year} {2021}{\natexlab{c}})},\
  \Eprint {https://arxiv.org/abs/2112.12547} {arXiv:2112.12547 [hep-ph]}
  \BibitemShut {NoStop}%
\bibitem [{\citenamefont {Fa\'undez}\ and\ \citenamefont
  {Gannouji}(2023)}]{Faundez:2023svu}%
  \BibitemOpen
  \bibfield  {author} {\bibinfo {author} {\bibfnamefont {A.}~\bibnamefont
  {Fa\'undez}}\ and\ \bibinfo {author} {\bibfnamefont {R.}~\bibnamefont
  {Gannouji}},\ }\href {https://doi.org/10.1103/PhysRevD.107.104058} {\bibfield
   {journal} {\bibinfo  {journal} {Phys. Rev. D}\ }\textbf {\bibinfo {volume}
  {107}},\ \bibinfo {pages} {104058} (\bibinfo {year} {2023})},\ \Eprint
  {https://arxiv.org/abs/2301.05890} {arXiv:2301.05890 [hep-th]} \BibitemShut
  {NoStop}%
\bibitem [{\citenamefont {Kuniyasu}\ \emph {et~al.}(2016)\citenamefont
  {Kuniyasu}, \citenamefont {Sakai},\ and\ \citenamefont
  {Shiraishi}}]{Kuniyasu:2016tse}%
  \BibitemOpen
  \bibfield  {author} {\bibinfo {author} {\bibfnamefont {M.}~\bibnamefont
  {Kuniyasu}}, \bibinfo {author} {\bibfnamefont {N.}~\bibnamefont {Sakai}},\
  and\ \bibinfo {author} {\bibfnamefont {K.}~\bibnamefont {Shiraishi}},\ }\href
  {https://doi.org/10.1103/PhysRevD.94.116001} {\bibfield  {journal} {\bibinfo
  {journal} {Phys. Rev. D}\ }\textbf {\bibinfo {volume} {94}},\ \bibinfo
  {pages} {116001} (\bibinfo {year} {2016})}\BibitemShut {NoStop}%
\bibitem [{\citenamefont {Pearce}\ \emph {et~al.}(2022)\citenamefont {Pearce},
  \citenamefont {White},\ and\ \citenamefont {Kusenko}}]{Pearce:2022ovj}%
  \BibitemOpen
  \bibfield  {author} {\bibinfo {author} {\bibfnamefont {L.}~\bibnamefont
  {Pearce}}, \bibinfo {author} {\bibfnamefont {G.}~\bibnamefont {White}},\ and\
  \bibinfo {author} {\bibfnamefont {A.}~\bibnamefont {Kusenko}},\ }\href
  {https://doi.org/10.1007/JHEP08(2022)033} {\bibfield  {journal} {\bibinfo
  {journal} {JHEP}\ }\textbf {\bibinfo {volume} {08}},\ \bibinfo {pages}
  {033}},\ \Eprint {https://arxiv.org/abs/2205.13557} {arXiv:2205.13557
  [hep-ph]} \BibitemShut {NoStop}%
\bibitem [{\citenamefont {Linde}(1983)}]{Linde:1981zj}%
  \BibitemOpen
  \bibfield  {author} {\bibinfo {author} {\bibfnamefont {A.~D.}\ \bibnamefont
  {Linde}},\ }\href {https://doi.org/10.1016/0550-3213(83)90072-X} {\bibfield
  {journal} {\bibinfo  {journal} {Nucl. Phys. B}\ }\textbf {\bibinfo {volume}
  {216}},\ \bibinfo {pages} {421} (\bibinfo {year} {1983})},\ \bibinfo {note}
  {[Erratum: Nucl.Phys.B 223, 544 (1983)]}\BibitemShut {NoStop}%
\bibitem [{\citenamefont {Espinosa}\ \emph {et~al.}(2023)\citenamefont
  {Espinosa}, \citenamefont {Heeck},\ and\ \citenamefont
  {Sokhashvili}}]{Espinosa:2023osv}%
  \BibitemOpen
  \bibfield  {author} {\bibinfo {author} {\bibfnamefont {J.~R.}\ \bibnamefont
  {Espinosa}}, \bibinfo {author} {\bibfnamefont {J.}~\bibnamefont {Heeck}},\
  and\ \bibinfo {author} {\bibfnamefont {M.}~\bibnamefont {Sokhashvili}},\
  }\href {https://doi.org/10.1103/PhysRevD.108.056019} {\bibfield  {journal}
  {\bibinfo  {journal} {Phys. Rev. D}\ }\textbf {\bibinfo {volume} {108}},\
  \bibinfo {pages} {056019} (\bibinfo {year} {2023})},\ \Eprint
  {https://arxiv.org/abs/2307.05667} {arXiv:2307.05667 [hep-ph]} \BibitemShut
  {NoStop}%
\bibitem [{\citenamefont {Graham}(2001)}]{Graham:2001hr}%
  \BibitemOpen
  \bibfield  {author} {\bibinfo {author} {\bibfnamefont {N.}~\bibnamefont
  {Graham}},\ }\href {https://doi.org/10.1016/S0370-2693(01)00669-4} {\bibfield
   {journal} {\bibinfo  {journal} {Phys. Lett. B}\ }\textbf {\bibinfo {volume}
  {513}},\ \bibinfo {pages} {112} (\bibinfo {year} {2001})},\ \Eprint
  {https://arxiv.org/abs/hep-th/0105009} {arXiv:hep-th/0105009} \BibitemShut
  {NoStop}%
\bibitem [{\citenamefont {Levkov}\ \emph {et~al.}(2017)\citenamefont {Levkov},
  \citenamefont {Nugaev},\ and\ \citenamefont {Popescu}}]{Levkov:2017paj}%
  \BibitemOpen
  \bibfield  {author} {\bibinfo {author} {\bibfnamefont {D.}~\bibnamefont
  {Levkov}}, \bibinfo {author} {\bibfnamefont {E.}~\bibnamefont {Nugaev}},\
  and\ \bibinfo {author} {\bibfnamefont {A.}~\bibnamefont {Popescu}},\ }\href
  {https://doi.org/10.1007/JHEP12(2017)131} {\bibfield  {journal} {\bibinfo
  {journal} {JHEP}\ }\textbf {\bibinfo {volume} {12}},\ \bibinfo {pages}
  {131}},\ \Eprint {https://arxiv.org/abs/1711.05279} {arXiv:1711.05279
  [hep-ph]} \BibitemShut {NoStop}%
\bibitem [{\citenamefont {Nugaev}\ and\ \citenamefont
  {Smolyakov}(2014)}]{Nugaev:2013poa}%
  \BibitemOpen
  \bibfield  {author} {\bibinfo {author} {\bibfnamefont {E.~Y.}\ \bibnamefont
  {Nugaev}}\ and\ \bibinfo {author} {\bibfnamefont {M.~N.}\ \bibnamefont
  {Smolyakov}},\ }\href {https://doi.org/10.1007/JHEP07(2014)009} {\bibfield
  {journal} {\bibinfo  {journal} {JHEP}\ }\textbf {\bibinfo {volume} {07}},\
  \bibinfo {pages} {009}},\ \Eprint {https://arxiv.org/abs/1311.3418}
  {arXiv:1311.3418 [hep-th]} \BibitemShut {NoStop}%
\bibitem [{\citenamefont {Klimas}\ and\ \citenamefont
  {Livramento}(2017)}]{Klimas:2017eft}%
  \BibitemOpen
  \bibfield  {author} {\bibinfo {author} {\bibfnamefont {P.}~\bibnamefont
  {Klimas}}\ and\ \bibinfo {author} {\bibfnamefont {L.~R.}\ \bibnamefont
  {Livramento}},\ }\href {https://doi.org/10.1103/PhysRevD.96.016001}
  {\bibfield  {journal} {\bibinfo  {journal} {Phys. Rev. D}\ }\textbf {\bibinfo
  {volume} {96}},\ \bibinfo {pages} {016001} (\bibinfo {year} {2017})},\
  \Eprint {https://arxiv.org/abs/1704.01132} {arXiv:1704.01132 [hep-th]}
  \BibitemShut {NoStop}%
\bibitem [{\citenamefont {Klimas}\ \emph {et~al.}(2021)\citenamefont {Klimas},
  \citenamefont {Kubaski}, \citenamefont {Sawado},\ and\ \citenamefont
  {Yanai}}]{Klimas:2021eue}%
  \BibitemOpen
  \bibfield  {author} {\bibinfo {author} {\bibfnamefont {P.}~\bibnamefont
  {Klimas}}, \bibinfo {author} {\bibfnamefont {L.~C.}\ \bibnamefont {Kubaski}},
  \bibinfo {author} {\bibfnamefont {N.}~\bibnamefont {Sawado}},\ and\ \bibinfo
  {author} {\bibfnamefont {S.}~\bibnamefont {Yanai}},\ }\href
  {https://doi.org/10.1007/JHEP09(2021)084} {\bibfield  {journal} {\bibinfo
  {journal} {JHEP}\ }\textbf {\bibinfo {volume} {09}},\ \bibinfo {pages}
  {084}},\ \Eprint {https://arxiv.org/abs/2107.09831} {arXiv:2107.09831
  [hep-th]} \BibitemShut {NoStop}%
\bibitem [{\citenamefont {Klimas}\ \emph {et~al.}(2022)\citenamefont {Klimas},
  \citenamefont {Sawado},\ and\ \citenamefont {Yanai}}]{Klimas:2022ghu}%
  \BibitemOpen
  \bibfield  {author} {\bibinfo {author} {\bibfnamefont {P.}~\bibnamefont
  {Klimas}}, \bibinfo {author} {\bibfnamefont {N.}~\bibnamefont {Sawado}},\
  and\ \bibinfo {author} {\bibfnamefont {S.}~\bibnamefont {Yanai}},\ }\href
  {https://doi.org/10.1103/PhysRevD.105.085004} {\bibfield  {journal} {\bibinfo
   {journal} {Phys. Rev. D}\ }\textbf {\bibinfo {volume} {105}},\ \bibinfo
  {pages} {085004} (\bibinfo {year} {2022})},\ \Eprint
  {https://arxiv.org/abs/2201.09239} {arXiv:2201.09239 [hep-th]} \BibitemShut
  {NoStop}%
\bibitem [{\citenamefont {Klimas}\ \emph {et~al.}(2023)\citenamefont {Klimas},
  \citenamefont {Kubaski}, \citenamefont {Sawado},\ and\ \citenamefont
  {Yanai}}]{Klimas:2023zxm}%
  \BibitemOpen
  \bibfield  {author} {\bibinfo {author} {\bibfnamefont {P.}~\bibnamefont
  {Klimas}}, \bibinfo {author} {\bibfnamefont {L.~C.}\ \bibnamefont {Kubaski}},
  \bibinfo {author} {\bibfnamefont {N.}~\bibnamefont {Sawado}},\ and\ \bibinfo
  {author} {\bibfnamefont {S.}~\bibnamefont {Yanai}},\ }\href@noop {} {\
  (\bibinfo {year} {2023})},\ \Eprint {https://arxiv.org/abs/2311.13076}
  {arXiv:2311.13076 [hep-th]} \BibitemShut {NoStop}%
\bibitem [{\citenamefont {Bishara}\ and\ \citenamefont
  {Lennon}(2022)}]{Bishara:2021fag}%
  \BibitemOpen
  \bibfield  {author} {\bibinfo {author} {\bibfnamefont {F.}~\bibnamefont
  {Bishara}}\ and\ \bibinfo {author} {\bibfnamefont {O.}~\bibnamefont
  {Lennon}},\ }\href {https://doi.org/10.1007/JHEP10(2022)079} {\bibfield
  {journal} {\bibinfo  {journal} {JHEP}\ }\textbf {\bibinfo {volume} {10}},\
  \bibinfo {pages} {079}},\ \Eprint {https://arxiv.org/abs/2110.02236}
  {arXiv:2110.02236 [hep-ph]} \BibitemShut {NoStop}%
\bibitem [{\citenamefont {Sutcliffe}(2023)}]{Sutcliffe:2023xau}%
  \BibitemOpen
  \bibfield  {author} {\bibinfo {author} {\bibfnamefont {P.}~\bibnamefont
  {Sutcliffe}},\ }\href {https://doi.org/10.1007/JHEP06(2023)162} {\bibfield
  {journal} {\bibinfo  {journal} {JHEP}\ }\textbf {\bibinfo {volume} {06}},\
  \bibinfo {pages} {162}},\ \Eprint {https://arxiv.org/abs/2304.05521}
  {arXiv:2304.05521 [hep-th]} \BibitemShut {NoStop}%
\bibitem [{\citenamefont {Heusler}\ \emph {et~al.}(1998)\citenamefont
  {Heusler}, \citenamefont {Straumann},\ and\ \citenamefont
  {Volkov}}]{Heusler:1998ec}%
  \BibitemOpen
  \bibfield  {author} {\bibinfo {author} {\bibfnamefont {M.}~\bibnamefont
  {Heusler}}, \bibinfo {author} {\bibfnamefont {N.}~\bibnamefont {Straumann}},\
  and\ \bibinfo {author} {\bibfnamefont {M.~S.}\ \bibnamefont {Volkov}},\
  }\href {https://doi.org/10.1103/PhysRevD.58.105021} {\bibfield  {journal}
  {\bibinfo  {journal} {Phys. Rev. D}\ }\textbf {\bibinfo {volume} {58}},\
  \bibinfo {pages} {105021} (\bibinfo {year} {1998})},\ \Eprint
  {https://arxiv.org/abs/gr-qc/9805061} {arXiv:gr-qc/9805061} \BibitemShut
  {NoStop}%
\bibitem [{\citenamefont {Volkov}\ and\ \citenamefont
  {Wohnert}(2002)}]{Volkov:2002aj}%
  \BibitemOpen
  \bibfield  {author} {\bibinfo {author} {\bibfnamefont {M.~S.}\ \bibnamefont
  {Volkov}}\ and\ \bibinfo {author} {\bibfnamefont {E.}~\bibnamefont
  {Wohnert}},\ }\href {https://doi.org/10.1103/PhysRevD.66.085003} {\bibfield
  {journal} {\bibinfo  {journal} {Phys. Rev. D}\ }\textbf {\bibinfo {volume}
  {66}},\ \bibinfo {pages} {085003} (\bibinfo {year} {2002})},\ \Eprint
  {https://arxiv.org/abs/hep-th/0205157} {arXiv:hep-th/0205157} \BibitemShut
  {NoStop}%
\bibitem [{\citenamefont {Kleihaus}\ \emph {et~al.}(2005)\citenamefont
  {Kleihaus}, \citenamefont {Kunz},\ and\ \citenamefont
  {List}}]{Kleihaus:2005me}%
  \BibitemOpen
  \bibfield  {author} {\bibinfo {author} {\bibfnamefont {B.}~\bibnamefont
  {Kleihaus}}, \bibinfo {author} {\bibfnamefont {J.}~\bibnamefont {Kunz}},\
  and\ \bibinfo {author} {\bibfnamefont {M.}~\bibnamefont {List}},\ }\href
  {https://doi.org/10.1103/PhysRevD.72.064002} {\bibfield  {journal} {\bibinfo
  {journal} {Phys. Rev. D}\ }\textbf {\bibinfo {volume} {72}},\ \bibinfo
  {pages} {064002} (\bibinfo {year} {2005})},\ \Eprint
  {https://arxiv.org/abs/gr-qc/0505143} {arXiv:gr-qc/0505143} \BibitemShut
  {NoStop}%
\bibitem [{\citenamefont {Kleihaus}\ \emph {et~al.}(2008)\citenamefont
  {Kleihaus}, \citenamefont {Kunz}, \citenamefont {List},\ and\ \citenamefont
  {Schaffer}}]{Kleihaus:2007vk}%
  \BibitemOpen
  \bibfield  {author} {\bibinfo {author} {\bibfnamefont {B.}~\bibnamefont
  {Kleihaus}}, \bibinfo {author} {\bibfnamefont {J.}~\bibnamefont {Kunz}},
  \bibinfo {author} {\bibfnamefont {M.}~\bibnamefont {List}},\ and\ \bibinfo
  {author} {\bibfnamefont {I.}~\bibnamefont {Schaffer}},\ }\href
  {https://doi.org/10.1103/PhysRevD.77.064025} {\bibfield  {journal} {\bibinfo
  {journal} {Phys. Rev. D}\ }\textbf {\bibinfo {volume} {77}},\ \bibinfo
  {pages} {064025} (\bibinfo {year} {2008})},\ \Eprint
  {https://arxiv.org/abs/0712.3742} {arXiv:0712.3742 [gr-qc]} \BibitemShut
  {NoStop}%
\bibitem [{\citenamefont {Campanelli}\ and\ \citenamefont
  {Ruggieri}(2009)}]{Campanelli:2009su}%
  \BibitemOpen
  \bibfield  {author} {\bibinfo {author} {\bibfnamefont {L.}~\bibnamefont
  {Campanelli}}\ and\ \bibinfo {author} {\bibfnamefont {M.}~\bibnamefont
  {Ruggieri}},\ }\href {https://doi.org/10.1103/PhysRevD.80.036006} {\bibfield
  {journal} {\bibinfo  {journal} {Phys. Rev. D}\ }\textbf {\bibinfo {volume}
  {80}},\ \bibinfo {pages} {036006} (\bibinfo {year} {2009})},\ \Eprint
  {https://arxiv.org/abs/0904.4802} {arXiv:0904.4802 [hep-th]} \BibitemShut
  {NoStop}%
\bibitem [{\citenamefont {Almumin}\ \emph {et~al.}(2024)\citenamefont
  {Almumin}, \citenamefont {Heeck}, \citenamefont {Rajaraman},\ and\
  \citenamefont {Verhaaren}}]{Almumin:2023wwi}%
  \BibitemOpen
  \bibfield  {author} {\bibinfo {author} {\bibfnamefont {Y.}~\bibnamefont
  {Almumin}}, \bibinfo {author} {\bibfnamefont {J.}~\bibnamefont {Heeck}},
  \bibinfo {author} {\bibfnamefont {A.}~\bibnamefont {Rajaraman}},\ and\
  \bibinfo {author} {\bibfnamefont {C.~B.}\ \bibnamefont {Verhaaren}},\ }\href
  {https://doi.org/10.1140/epjc/s10052-024-12712-x} {\bibfield  {journal}
  {\bibinfo  {journal} {Eur. Phys. J. C}\ }\textbf {\bibinfo {volume} {84}},\
  \bibinfo {pages} {364} (\bibinfo {year} {2024})},\ \Eprint
  {https://arxiv.org/abs/2302.11589} {arXiv:2302.11589 [hep-th]} \BibitemShut
  {NoStop}%
\bibitem [{\citenamefont {Yoshida}\ and\ \citenamefont
  {Eriguchi}(1997)}]{Yoshida:1997qf}%
  \BibitemOpen
  \bibfield  {author} {\bibinfo {author} {\bibfnamefont {S.}~\bibnamefont
  {Yoshida}}\ and\ \bibinfo {author} {\bibfnamefont {Y.}~\bibnamefont
  {Eriguchi}},\ }\href {https://doi.org/10.1103/PhysRevD.56.762} {\bibfield
  {journal} {\bibinfo  {journal} {Phys. Rev. D}\ }\textbf {\bibinfo {volume}
  {56}},\ \bibinfo {pages} {762} (\bibinfo {year} {1997})}\BibitemShut
  {NoStop}%
\bibitem [{\citenamefont {Copeland}\ \emph {et~al.}(2014)\citenamefont
  {Copeland}, \citenamefont {Saffin},\ and\ \citenamefont
  {Zhou}}]{Copeland:2014qra}%
  \BibitemOpen
  \bibfield  {author} {\bibinfo {author} {\bibfnamefont {E.~J.}\ \bibnamefont
  {Copeland}}, \bibinfo {author} {\bibfnamefont {P.~M.}\ \bibnamefont
  {Saffin}},\ and\ \bibinfo {author} {\bibfnamefont {S.-Y.}\ \bibnamefont
  {Zhou}},\ }\href {https://doi.org/10.1103/PhysRevLett.113.231603} {\bibfield
  {journal} {\bibinfo  {journal} {Phys. Rev. Lett.}\ }\textbf {\bibinfo
  {volume} {113}},\ \bibinfo {pages} {231603} (\bibinfo {year} {2014})},\
  \Eprint {https://arxiv.org/abs/1409.3232} {arXiv:1409.3232 [hep-th]}
  \BibitemShut {NoStop}%
\bibitem [{\citenamefont {Xie}\ \emph {et~al.}(2021)\citenamefont {Xie},
  \citenamefont {Saffin},\ and\ \citenamefont {Zhou}}]{Xie:2021glp}%
  \BibitemOpen
  \bibfield  {author} {\bibinfo {author} {\bibfnamefont {Q.-X.}\ \bibnamefont
  {Xie}}, \bibinfo {author} {\bibfnamefont {P.~M.}\ \bibnamefont {Saffin}},\
  and\ \bibinfo {author} {\bibfnamefont {S.-Y.}\ \bibnamefont {Zhou}},\ }\href
  {https://doi.org/10.1007/JHEP07(2021)062} {\bibfield  {journal} {\bibinfo
  {journal} {JHEP}\ }\textbf {\bibinfo {volume} {07}},\ \bibinfo {pages}
  {062}},\ \Eprint {https://arxiv.org/abs/2101.06988} {arXiv:2101.06988
  [hep-th]} \BibitemShut {NoStop}%
\bibitem [{\citenamefont {Hou}\ \emph {et~al.}(2022)\citenamefont {Hou},
  \citenamefont {Saffin}, \citenamefont {Xie},\ and\ \citenamefont
  {Zhou}}]{Hou:2022jcd}%
  \BibitemOpen
  \bibfield  {author} {\bibinfo {author} {\bibfnamefont {S.-Y.}\ \bibnamefont
  {Hou}}, \bibinfo {author} {\bibfnamefont {P.~M.}\ \bibnamefont {Saffin}},
  \bibinfo {author} {\bibfnamefont {Q.-X.}\ \bibnamefont {Xie}},\ and\ \bibinfo
  {author} {\bibfnamefont {S.-Y.}\ \bibnamefont {Zhou}},\ }\href
  {https://doi.org/10.1007/JHEP07(2022)060} {\bibfield  {journal} {\bibinfo
  {journal} {JHEP}\ }\textbf {\bibinfo {volume} {07}},\ \bibinfo {pages}
  {060}},\ \Eprint {https://arxiv.org/abs/2202.08392} {arXiv:2202.08392
  [hep-ph]} \BibitemShut {NoStop}%
\bibitem [{\citenamefont {Jaramillo}\ and\ \citenamefont
  {Zhou}(2024{\natexlab{a}})}]{Jaramillo:2024smx}%
  \BibitemOpen
  \bibfield  {author} {\bibinfo {author} {\bibfnamefont {V.}~\bibnamefont
  {Jaramillo}}\ and\ \bibinfo {author} {\bibfnamefont {S.-Y.}\ \bibnamefont
  {Zhou}},\ }\href {https://doi.org/10.1103/PhysRevD.110.084069} {\bibfield
  {journal} {\bibinfo  {journal} {Phys. Rev. D}\ }\textbf {\bibinfo {volume}
  {110}},\ \bibinfo {pages} {084069} (\bibinfo {year} {2024}{\natexlab{a}})},\
  \Eprint {https://arxiv.org/abs/2407.12084} {arXiv:2407.12084 [gr-qc]}
  \BibitemShut {NoStop}%
\bibitem [{\citenamefont {Hasegawa}\ \emph {et~al.}(2019)\citenamefont
  {Hasegawa}, \citenamefont {Hong},\ and\ \citenamefont
  {Suzuki}}]{Hasegawa:2019bbo}%
  \BibitemOpen
  \bibfield  {author} {\bibinfo {author} {\bibfnamefont {F.}~\bibnamefont
  {Hasegawa}}, \bibinfo {author} {\bibfnamefont {J.-P.}\ \bibnamefont {Hong}},\
  and\ \bibinfo {author} {\bibfnamefont {M.}~\bibnamefont {Suzuki}},\ }\href
  {https://doi.org/10.1016/j.physletb.2019.135001} {\bibfield  {journal}
  {\bibinfo  {journal} {Phys. Lett. B}\ }\textbf {\bibinfo {volume} {798}},\
  \bibinfo {pages} {135001} (\bibinfo {year} {2019})},\ \Eprint
  {https://arxiv.org/abs/1903.07281} {arXiv:1903.07281 [hep-ph]} \BibitemShut
  {NoStop}%
\bibitem [{\citenamefont {Jaramillo}\ and\ \citenamefont
  {Zhou}(2024{\natexlab{b}})}]{Jaramillo:2024cus}%
  \BibitemOpen
  \bibfield  {author} {\bibinfo {author} {\bibfnamefont {V.}~\bibnamefont
  {Jaramillo}}\ and\ \bibinfo {author} {\bibfnamefont {S.-Y.}\ \bibnamefont
  {Zhou}},\ }\href@noop {} {\  (\bibinfo {year} {2024}{\natexlab{b}})},\
  \Eprint {https://arxiv.org/abs/2411.08985} {arXiv:2411.08985 [gr-qc]}
  \BibitemShut {NoStop}%
\bibitem [{\citenamefont {Loiko}\ \emph {et~al.}(2021)\citenamefont {Loiko},
  \citenamefont {Perapechka},\ and\ \citenamefont {Shnir}}]{Loiko:2020htk}%
  \BibitemOpen
  \bibfield  {author} {\bibinfo {author} {\bibfnamefont {V.}~\bibnamefont
  {Loiko}}, \bibinfo {author} {\bibfnamefont {I.}~\bibnamefont {Perapechka}},\
  and\ \bibinfo {author} {\bibfnamefont {Y.}~\bibnamefont {Shnir}},\ }\href
  {https://doi.org/10.1209/0295-5075/133/41001} {\bibfield  {journal} {\bibinfo
   {journal} {EPL}\ }\textbf {\bibinfo {volume} {133}},\ \bibinfo {pages}
  {41001} (\bibinfo {year} {2021})},\ \Eprint
  {https://arxiv.org/abs/2012.01052} {arXiv:2012.01052 [hep-th]} \BibitemShut
  {NoStop}%
\bibitem [{\citenamefont {Ward}(2003)}]{Ward:2003un}%
  \BibitemOpen
  \bibfield  {author} {\bibinfo {author} {\bibfnamefont {R.~S.}\ \bibnamefont
  {Ward}},\ }\href {https://doi.org/10.1063/1.1584527} {\bibfield  {journal}
  {\bibinfo  {journal} {J. Math. Phys.}\ }\textbf {\bibinfo {volume} {44}},\
  \bibinfo {pages} {3555} (\bibinfo {year} {2003})},\ \Eprint
  {https://arxiv.org/abs/hep-th/0302045} {arXiv:hep-th/0302045} \BibitemShut
  {NoStop}%
\bibitem [{\citenamefont {Shnir}(2011)}]{Shnir:2011gr}%
  \BibitemOpen
  \bibfield  {author} {\bibinfo {author} {\bibfnamefont {Y.}~\bibnamefont
  {Shnir}},\ }\href {https://doi.org/10.1088/1751-8113/44/42/425202} {\bibfield
   {journal} {\bibinfo  {journal} {J. Phys. A}\ }\textbf {\bibinfo {volume}
  {44}},\ \bibinfo {pages} {425202} (\bibinfo {year} {2011})},\ \Eprint
  {https://arxiv.org/abs/1101.5366} {arXiv:1101.5366 [hep-th]} \BibitemShut
  {NoStop}%
\bibitem [{\citenamefont {Loginov}\ and\ \citenamefont
  {Gauzshtein}(2019)}]{Loginov:2019rwz}%
  \BibitemOpen
  \bibfield  {author} {\bibinfo {author} {\bibfnamefont {A.~Y.}\ \bibnamefont
  {Loginov}}\ and\ \bibinfo {author} {\bibfnamefont {V.~V.}\ \bibnamefont
  {Gauzshtein}},\ }\href {https://doi.org/10.1140/epjc/s10052-019-7302-6}
  {\bibfield  {journal} {\bibinfo  {journal} {Eur. Phys. J. C}\ }\textbf
  {\bibinfo {volume} {79}},\ \bibinfo {pages} {780} (\bibinfo {year} {2019})},\
  \Eprint {https://arxiv.org/abs/1906.02447} {arXiv:1906.02447 [hep-th]}
  \BibitemShut {NoStop}%
\bibitem [{\citenamefont {Bai}\ \emph {et~al.}(2022)\citenamefont {Bai},
  \citenamefont {Lu},\ and\ \citenamefont {Orlofsky}}]{Bai:2021mzu}%
  \BibitemOpen
  \bibfield  {author} {\bibinfo {author} {\bibfnamefont {Y.}~\bibnamefont
  {Bai}}, \bibinfo {author} {\bibfnamefont {S.}~\bibnamefont {Lu}},\ and\
  \bibinfo {author} {\bibfnamefont {N.}~\bibnamefont {Orlofsky}},\ }\href
  {https://doi.org/10.1007/JHEP01(2022)109} {\bibfield  {journal} {\bibinfo
  {journal} {JHEP}\ }\textbf {\bibinfo {volume} {01}},\ \bibinfo {pages}
  {109}},\ \Eprint {https://arxiv.org/abs/2111.10360} {arXiv:2111.10360
  [hep-ph]} \BibitemShut {NoStop}%
\bibitem [{\citenamefont {Alonso-Izquierdo}\ and\ \citenamefont
  {Sanchez}(2023)}]{Alonso-Izquierdo:2023xni}%
  \BibitemOpen
  \bibfield  {author} {\bibinfo {author} {\bibfnamefont {A.}~\bibnamefont
  {Alonso-Izquierdo}}\ and\ \bibinfo {author} {\bibfnamefont {C.~G.}\
  \bibnamefont {Sanchez}},\ }\href
  {https://doi.org/10.1103/PhysRevD.107.125004} {\bibfield  {journal} {\bibinfo
   {journal} {Phys. Rev. D}\ }\textbf {\bibinfo {volume} {107}},\ \bibinfo
  {pages} {125004} (\bibinfo {year} {2023})},\ \Eprint
  {https://arxiv.org/abs/2303.01537} {arXiv:2303.01537 [hep-th]} \BibitemShut
  {NoStop}%
\bibitem [{\citenamefont {Garc\'\i{}a}\ \emph {et~al.}(2023)\citenamefont
  {Garc\'\i{}a}, \citenamefont {del Moral}, \citenamefont {Pe\~na},\ and\
  \citenamefont {Prado-Fuentes}}]{Garcia:2023hrn}%
  \BibitemOpen
  \bibfield  {author} {\bibinfo {author} {\bibfnamefont {P.}~\bibnamefont
  {Garc\'\i{}a}}, \bibinfo {author} {\bibfnamefont {M.~P.~G.}\ \bibnamefont
  {del Moral}}, \bibinfo {author} {\bibfnamefont {J.~M.}\ \bibnamefont
  {Pe\~na}},\ and\ \bibinfo {author} {\bibfnamefont {R.}~\bibnamefont
  {Prado-Fuentes}},\ }\href@noop {} {\  (\bibinfo {year} {2023})},\ \Eprint
  {https://arxiv.org/abs/2302.12373} {arXiv:2302.12373 [hep-th]} \BibitemShut
  {NoStop}%
\bibitem [{\citenamefont {Nugaev}\ \emph {et~al.}(2016)\citenamefont {Nugaev},
  \citenamefont {Shkerin},\ and\ \citenamefont {Smolyakov}}]{Nugaev:2016wyt}%
  \BibitemOpen
  \bibfield  {author} {\bibinfo {author} {\bibfnamefont {E.}~\bibnamefont
  {Nugaev}}, \bibinfo {author} {\bibfnamefont {A.}~\bibnamefont {Shkerin}},\
  and\ \bibinfo {author} {\bibfnamefont {M.}~\bibnamefont {Smolyakov}},\ }\href
  {https://doi.org/10.1007/JHEP12(2016)032} {\bibfield  {journal} {\bibinfo
  {journal} {JHEP}\ }\textbf {\bibinfo {volume} {12}},\ \bibinfo {pages}
  {032}},\ \Eprint {https://arxiv.org/abs/1609.05568} {arXiv:1609.05568
  [hep-th]} \BibitemShut {NoStop}%
\bibitem [{\citenamefont {Axenides}\ \emph {et~al.}(2000)\citenamefont
  {Axenides}, \citenamefont {Komineas}, \citenamefont {Perivolaropoulos},\ and\
  \citenamefont {Floratos}}]{Axenides:1999hs}%
  \BibitemOpen
  \bibfield  {author} {\bibinfo {author} {\bibfnamefont {M.}~\bibnamefont
  {Axenides}}, \bibinfo {author} {\bibfnamefont {S.}~\bibnamefont {Komineas}},
  \bibinfo {author} {\bibfnamefont {L.}~\bibnamefont {Perivolaropoulos}},\ and\
  \bibinfo {author} {\bibfnamefont {M.}~\bibnamefont {Floratos}},\ }\href
  {https://doi.org/10.1103/PhysRevD.61.085006} {\bibfield  {journal} {\bibinfo
  {journal} {Phys. Rev. D}\ }\textbf {\bibinfo {volume} {61}},\ \bibinfo
  {pages} {085006} (\bibinfo {year} {2000})},\ \Eprint
  {https://arxiv.org/abs/hep-ph/9910388} {arXiv:hep-ph/9910388} \BibitemShut
  {NoStop}%
\bibitem [{\citenamefont {Ruback}(1988)}]{Ruback:1988ba}%
  \BibitemOpen
  \bibfield  {author} {\bibinfo {author} {\bibfnamefont {P.~J.}\ \bibnamefont
  {Ruback}},\ }\href {https://doi.org/10.1016/0550-3213(88)90038-7} {\bibfield
  {journal} {\bibinfo  {journal} {Nucl. Phys. B}\ }\textbf {\bibinfo {volume}
  {296}},\ \bibinfo {pages} {669} (\bibinfo {year} {1988})}\BibitemShut
  {NoStop}%
\bibitem [{\citenamefont {Bowcock}\ \emph {et~al.}(2009)\citenamefont
  {Bowcock}, \citenamefont {Foster},\ and\ \citenamefont
  {Sutcliffe}}]{Bowcock:2008dn}%
  \BibitemOpen
  \bibfield  {author} {\bibinfo {author} {\bibfnamefont {P.}~\bibnamefont
  {Bowcock}}, \bibinfo {author} {\bibfnamefont {D.}~\bibnamefont {Foster}},\
  and\ \bibinfo {author} {\bibfnamefont {P.}~\bibnamefont {Sutcliffe}},\ }\href
  {https://doi.org/10.1088/1751-8113/42/8/085403} {\bibfield  {journal}
  {\bibinfo  {journal} {J. Phys. A}\ }\textbf {\bibinfo {volume} {42}},\
  \bibinfo {pages} {085403} (\bibinfo {year} {2009})},\ \Eprint
  {https://arxiv.org/abs/0809.3895} {arXiv:0809.3895 [hep-th]} \BibitemShut
  {NoStop}%
\bibitem [{\citenamefont {Gorshkov}\ and\ \citenamefont
  {Ostrovsky}(1981)}]{Gorshkov:1981ke}%
  \BibitemOpen
  \bibfield  {author} {\bibinfo {author} {\bibfnamefont {K.~A.}\ \bibnamefont
  {Gorshkov}}\ and\ \bibinfo {author} {\bibfnamefont {L.~A.}\ \bibnamefont
  {Ostrovsky}},\ }\href@noop {} {\bibfield  {journal} {\bibinfo  {journal}
  {Physica D}\ }\textbf {\bibinfo {volume} {3}},\ \bibinfo {pages} {428}
  (\bibinfo {year} {1981})}\BibitemShut {NoStop}%
\bibitem [{\citenamefont {Kinach}\ and\ \citenamefont
  {Choptuik}(2024{\natexlab{a}})}]{Kinach:2024hfa}%
  \BibitemOpen
  \bibfield  {author} {\bibinfo {author} {\bibfnamefont {M.~P.}\ \bibnamefont
  {Kinach}}\ and\ \bibinfo {author} {\bibfnamefont {M.~W.}\ \bibnamefont
  {Choptuik}},\ }\href {https://doi.org/10.1103/PhysRevD.110.015012} {\bibfield
   {journal} {\bibinfo  {journal} {Phys. Rev. D}\ }\textbf {\bibinfo {volume}
  {110}},\ \bibinfo {pages} {015012} (\bibinfo {year} {2024}{\natexlab{a}})},\
  \Eprint {https://arxiv.org/abs/2404.04323} {arXiv:2404.04323 [hep-th]}
  \BibitemShut {NoStop}%
\bibitem [{\citenamefont {Kinach}\ and\ \citenamefont
  {Choptuik}(2024{\natexlab{b}})}]{Kinach:2024qzc}%
  \BibitemOpen
  \bibfield  {author} {\bibinfo {author} {\bibfnamefont {M.~P.}\ \bibnamefont
  {Kinach}}\ and\ \bibinfo {author} {\bibfnamefont {M.~W.}\ \bibnamefont
  {Choptuik}},\ }\href {https://doi.org/10.1103/PhysRevD.110.075033} {\bibfield
   {journal} {\bibinfo  {journal} {Phys. Rev. D}\ }\textbf {\bibinfo {volume}
  {110}},\ \bibinfo {pages} {075033} (\bibinfo {year} {2024}{\natexlab{b}})},\
  \Eprint {https://arxiv.org/abs/2408.07561} {arXiv:2408.07561 [hep-th]}
  \BibitemShut {NoStop}%
\bibitem [{\citenamefont {Hong}\ and\ \citenamefont
  {Lonsdale}(2024)}]{Hong:2024uxl}%
  \BibitemOpen
  \bibfield  {author} {\bibinfo {author} {\bibfnamefont {D.~K.}\ \bibnamefont
  {Hong}}\ and\ \bibinfo {author} {\bibfnamefont {S.~J.}\ \bibnamefont
  {Lonsdale}},\ }\href@noop {} {\  (\bibinfo {year} {2024})},\ \Eprint
  {https://arxiv.org/abs/2408.12342} {arXiv:2408.12342 [hep-ph]} \BibitemShut
  {NoStop}%
\bibitem [{\citenamefont {Dicke}(1954)}]{Dicke:1954zz}%
  \BibitemOpen
  \bibfield  {author} {\bibinfo {author} {\bibfnamefont {R.~H.}\ \bibnamefont
  {Dicke}},\ }\href {https://doi.org/10.1103/PhysRev.93.99} {\bibfield
  {journal} {\bibinfo  {journal} {Phys. Rev.}\ }\textbf {\bibinfo {volume}
  {93}},\ \bibinfo {pages} {99} (\bibinfo {year} {1954})}\BibitemShut {NoStop}%
\bibitem [{\citenamefont {Masson}\ and\ \citenamefont
  {Asenjo-Garcia}(2022)}]{MassonSuperrad}%
  \BibitemOpen
  \bibfield  {author} {\bibinfo {author} {\bibfnamefont {S.~J.}\ \bibnamefont
  {Masson}}\ and\ \bibinfo {author} {\bibfnamefont {A.}~\bibnamefont
  {Asenjo-Garcia}},\ }\href {https://doi.org/10.1038/s41467-022-29805-4}
  {\bibfield  {journal} {\bibinfo  {journal} {Nature Communications}\ }\textbf
  {\bibinfo {volume} {13}},\ \bibinfo {pages} {2285} (\bibinfo {year}
  {2022})}\BibitemShut {NoStop}%
\bibitem [{\citenamefont {Bekenstein}\ and\ \citenamefont
  {Schiffer}(1998)}]{Bekenstein:1998nt}%
  \BibitemOpen
  \bibfield  {author} {\bibinfo {author} {\bibfnamefont {J.~D.}\ \bibnamefont
  {Bekenstein}}\ and\ \bibinfo {author} {\bibfnamefont {M.}~\bibnamefont
  {Schiffer}},\ }\href {https://doi.org/10.1103/PhysRevD.58.064014} {\bibfield
  {journal} {\bibinfo  {journal} {Phys. Rev. D}\ }\textbf {\bibinfo {volume}
  {58}},\ \bibinfo {pages} {064014} (\bibinfo {year} {1998})},\ \Eprint
  {https://arxiv.org/abs/gr-qc/9803033} {arXiv:gr-qc/9803033} \BibitemShut
  {NoStop}%
\bibitem [{\citenamefont {Brito}\ \emph {et~al.}(2015)\citenamefont {Brito},
  \citenamefont {Cardoso},\ and\ \citenamefont {Pani}}]{Brito:2015oca}%
  \BibitemOpen
  \bibfield  {author} {\bibinfo {author} {\bibfnamefont {R.}~\bibnamefont
  {Brito}}, \bibinfo {author} {\bibfnamefont {V.}~\bibnamefont {Cardoso}},\
  and\ \bibinfo {author} {\bibfnamefont {P.}~\bibnamefont {Pani}},\ }\href
  {https://doi.org/10.1007/978-3-319-19000-6} {\bibfield  {journal} {\bibinfo
  {journal} {Lect. Notes Phys.}\ }\textbf {\bibinfo {volume} {906}},\ \bibinfo
  {pages} {pp.1} (\bibinfo {year} {2015})},\ \Eprint
  {https://arxiv.org/abs/1501.06570} {arXiv:1501.06570 [gr-qc]} \BibitemShut
  {NoStop}%
\bibitem [{\citenamefont {Saffin}\ \emph {et~al.}(2023)\citenamefont {Saffin},
  \citenamefont {Xie},\ and\ \citenamefont {Zhou}}]{Saffin:2022tub}%
  \BibitemOpen
  \bibfield  {author} {\bibinfo {author} {\bibfnamefont {P.~M.}\ \bibnamefont
  {Saffin}}, \bibinfo {author} {\bibfnamefont {Q.-X.}\ \bibnamefont {Xie}},\
  and\ \bibinfo {author} {\bibfnamefont {S.-Y.}\ \bibnamefont {Zhou}},\ }\href
  {https://doi.org/10.1103/PhysRevLett.131.111601} {\bibfield  {journal}
  {\bibinfo  {journal} {Phys. Rev. Lett.}\ }\textbf {\bibinfo {volume} {131}},\
  \bibinfo {pages} {111601} (\bibinfo {year} {2023})},\ \Eprint
  {https://arxiv.org/abs/2212.03269} {arXiv:2212.03269 [hep-th]} \BibitemShut
  {NoStop}%
\bibitem [{\citenamefont {Cardoso}\ \emph {et~al.}(2023)\citenamefont
  {Cardoso}, \citenamefont {Vicente},\ and\ \citenamefont
  {Zhong}}]{Cardoso:2023dtm}%
  \BibitemOpen
  \bibfield  {author} {\bibinfo {author} {\bibfnamefont {V.}~\bibnamefont
  {Cardoso}}, \bibinfo {author} {\bibfnamefont {R.}~\bibnamefont {Vicente}},\
  and\ \bibinfo {author} {\bibfnamefont {Z.}~\bibnamefont {Zhong}},\ }\href
  {https://doi.org/10.1103/PhysRevLett.131.111602} {\bibfield  {journal}
  {\bibinfo  {journal} {Phys. Rev. Lett.}\ }\textbf {\bibinfo {volume} {131}},\
  \bibinfo {pages} {111602} (\bibinfo {year} {2023})},\ \Eprint
  {https://arxiv.org/abs/2307.13734} {arXiv:2307.13734 [hep-th]} \BibitemShut
  {NoStop}%
\bibitem [{\citenamefont {Zhang}\ \emph {et~al.}(2024)\citenamefont {Zhang},
  \citenamefont {Chang}, \citenamefont {Saffin}, \citenamefont {Xie},\ and\
  \citenamefont {Zhou}}]{Zhang:2024ufh}%
  \BibitemOpen
  \bibfield  {author} {\bibinfo {author} {\bibfnamefont {G.-D.}\ \bibnamefont
  {Zhang}}, \bibinfo {author} {\bibfnamefont {F.-M.}\ \bibnamefont {Chang}},
  \bibinfo {author} {\bibfnamefont {P.~M.}\ \bibnamefont {Saffin}}, \bibinfo
  {author} {\bibfnamefont {Q.-X.}\ \bibnamefont {Xie}},\ and\ \bibinfo {author}
  {\bibfnamefont {S.-Y.}\ \bibnamefont {Zhou}},\ }\href
  {https://doi.org/10.1103/PhysRevD.110.043504} {\bibfield  {journal} {\bibinfo
   {journal} {Phys. Rev. D}\ }\textbf {\bibinfo {volume} {110}},\ \bibinfo
  {pages} {043504} (\bibinfo {year} {2024})},\ \Eprint
  {https://arxiv.org/abs/2402.03193} {arXiv:2402.03193 [hep-th]} \BibitemShut
  {NoStop}%
\bibitem [{\citenamefont {Gao}\ \emph {et~al.}(2024)\citenamefont {Gao},
  \citenamefont {Saffin}, \citenamefont {Wang}, \citenamefont {Xie},\ and\
  \citenamefont {Zhou}}]{Gao:2023gof}%
  \BibitemOpen
  \bibfield  {author} {\bibinfo {author} {\bibfnamefont {H.-Y.}\ \bibnamefont
  {Gao}}, \bibinfo {author} {\bibfnamefont {P.~M.}\ \bibnamefont {Saffin}},
  \bibinfo {author} {\bibfnamefont {Y.-J.}\ \bibnamefont {Wang}}, \bibinfo
  {author} {\bibfnamefont {Q.-X.}\ \bibnamefont {Xie}},\ and\ \bibinfo {author}
  {\bibfnamefont {S.-Y.}\ \bibnamefont {Zhou}},\ }\href
  {https://doi.org/10.1007/s11433-023-2357-4} {\bibfield  {journal} {\bibinfo
  {journal} {Sci. China Phys. Mech. Astron.}\ }\textbf {\bibinfo {volume}
  {67}},\ \bibinfo {pages} {260413} (\bibinfo {year} {2024})},\ \Eprint
  {https://arxiv.org/abs/2306.01868} {arXiv:2306.01868 [gr-qc]} \BibitemShut
  {NoStop}%
\bibitem [{\citenamefont {Chang}\ \emph {et~al.}()\citenamefont {Chang},
  \citenamefont {Gao}, \citenamefont {Jaramillo},\ and\ \citenamefont
  {Meng}}]{BSStime}%
  \BibitemOpen
  \bibfield  {author} {\bibinfo {author} {\bibfnamefont {F.-M.}\ \bibnamefont
  {Chang}}, \bibinfo {author} {\bibfnamefont {H.-Y.}\ \bibnamefont {Gao}},
  \bibinfo {author} {\bibfnamefont {V.}~\bibnamefont {Jaramillo}},\ and\
  \bibinfo {author} {\bibfnamefont {X.}~\bibnamefont {Meng}},\ }\href@noop {}
  {\bibinfo  {journal} {To appear}\ }\BibitemShut {NoStop}%
\bibitem [{\citenamefont {Borsanyi}\ and\ \citenamefont
  {Hindmarsh}(2009)}]{Borsanyi:2008eu}%
  \BibitemOpen
\bibfield  {journal} {  }\bibfield  {author} {\bibinfo {author} {\bibfnamefont
  {S.}~\bibnamefont {Borsanyi}}\ and\ \bibinfo {author} {\bibfnamefont
  {M.}~\bibnamefont {Hindmarsh}},\ }\href
  {https://doi.org/10.1103/PhysRevD.79.065010} {\bibfield  {journal} {\bibinfo
  {journal} {Phys. Rev. D}\ }\textbf {\bibinfo {volume} {79}},\ \bibinfo
  {pages} {065010} (\bibinfo {year} {2009})},\ \Eprint
  {https://arxiv.org/abs/0809.4711} {arXiv:0809.4711 [hep-ph]} \BibitemShut
  {NoStop}%
\bibitem [{\citenamefont {Coleman}(1977)}]{Coleman:1977py}%
  \BibitemOpen
  \bibfield  {author} {\bibinfo {author} {\bibfnamefont {S.~R.}\ \bibnamefont
  {Coleman}},\ }\href {https://doi.org/10.1103/PhysRevD.16.1248} {\bibfield
  {journal} {\bibinfo  {journal} {Phys. Rev. D}\ }\textbf {\bibinfo {volume}
  {15}},\ \bibinfo {pages} {2929} (\bibinfo {year} {1977})},\ \bibinfo {note}
  {[Erratum: Phys.Rev.D 16, 1248 (1977)]}\BibitemShut {NoStop}%
\bibitem [{\citenamefont {Cohen}\ \emph {et~al.}(1986)\citenamefont {Cohen},
  \citenamefont {Coleman}, \citenamefont {Georgi},\ and\ \citenamefont
  {Manohar}}]{Cohen:1986ct}%
  \BibitemOpen
  \bibfield  {author} {\bibinfo {author} {\bibfnamefont {A.~G.}\ \bibnamefont
  {Cohen}}, \bibinfo {author} {\bibfnamefont {S.~R.}\ \bibnamefont {Coleman}},
  \bibinfo {author} {\bibfnamefont {H.}~\bibnamefont {Georgi}},\ and\ \bibinfo
  {author} {\bibfnamefont {A.}~\bibnamefont {Manohar}},\ }\href
  {https://doi.org/10.1016/0550-3213(86)90004-0} {\bibfield  {journal}
  {\bibinfo  {journal} {Nucl. Phys. B}\ }\textbf {\bibinfo {volume} {272}},\
  \bibinfo {pages} {301} (\bibinfo {year} {1986})}\BibitemShut {NoStop}%
\bibitem [{\citenamefont {Bardeen}\ \emph {et~al.}(1975)\citenamefont
  {Bardeen}, \citenamefont {Chanowitz}, \citenamefont {Drell}, \citenamefont
  {Weinstein},\ and\ \citenamefont {Yan}}]{Bardeen:1974wr}%
  \BibitemOpen
  \bibfield  {author} {\bibinfo {author} {\bibfnamefont {W.~A.}\ \bibnamefont
  {Bardeen}}, \bibinfo {author} {\bibfnamefont {M.~S.}\ \bibnamefont
  {Chanowitz}}, \bibinfo {author} {\bibfnamefont {S.~D.}\ \bibnamefont
  {Drell}}, \bibinfo {author} {\bibfnamefont {M.}~\bibnamefont {Weinstein}},\
  and\ \bibinfo {author} {\bibfnamefont {T.-M.}\ \bibnamefont {Yan}},\ }\href
  {https://doi.org/10.1103/PhysRevD.11.1094} {\bibfield  {journal} {\bibinfo
  {journal} {Phys. Rev. D}\ }\textbf {\bibinfo {volume} {11}},\ \bibinfo
  {pages} {1094} (\bibinfo {year} {1975})}\BibitemShut {NoStop}%
\bibitem [{\citenamefont {Friedberg}\ and\ \citenamefont
  {Lee}(1977{\natexlab{b}})}]{Friedberg:1976eg}%
  \BibitemOpen
  \bibfield  {author} {\bibinfo {author} {\bibfnamefont {R.}~\bibnamefont
  {Friedberg}}\ and\ \bibinfo {author} {\bibfnamefont {T.~D.}\ \bibnamefont
  {Lee}},\ }\href {https://doi.org/10.1103/PhysRevD.15.1694} {\bibfield
  {journal} {\bibinfo  {journal} {Phys. Rev. D}\ }\textbf {\bibinfo {volume}
  {15}},\ \bibinfo {pages} {1694} (\bibinfo {year}
  {1977}{\natexlab{b}})}\BibitemShut {NoStop}%
\bibitem [{\citenamefont {Demir}(2000)}]{Demir:2000gj}%
  \BibitemOpen
  \bibfield  {author} {\bibinfo {author} {\bibfnamefont {D.~A.}\ \bibnamefont
  {Demir}},\ }\href {https://doi.org/10.1016/S0370-2693(00)01262-4} {\bibfield
  {journal} {\bibinfo  {journal} {Phys. Lett. B}\ }\textbf {\bibinfo {volume}
  {495}},\ \bibinfo {pages} {357} (\bibinfo {year} {2000})},\ \Eprint
  {https://arxiv.org/abs/hep-ph/0006344} {arXiv:hep-ph/0006344} \BibitemShut
  {NoStop}%
\bibitem [{\citenamefont {Abel}\ and\ \citenamefont
  {Kehagias}(2015)}]{Abel:2015tca}%
  \BibitemOpen
  \bibfield  {author} {\bibinfo {author} {\bibfnamefont {S.}~\bibnamefont
  {Abel}}\ and\ \bibinfo {author} {\bibfnamefont {A.}~\bibnamefont
  {Kehagias}},\ }\href {https://doi.org/10.1007/JHEP11(2015)096} {\bibfield
  {journal} {\bibinfo  {journal} {JHEP}\ }\textbf {\bibinfo {volume} {11}},\
  \bibinfo {pages} {096}},\ \Eprint {https://arxiv.org/abs/1507.04557}
  {arXiv:1507.04557 [hep-th]} \BibitemShut {NoStop}%
\bibitem [{\citenamefont {Bl\'azquez-Salcedo}\ \emph
  {et~al.}(2022)\citenamefont {Bl\'azquez-Salcedo}, \citenamefont {Dariescu},
  \citenamefont {Dariescu}, \citenamefont {Radu},\ and\ \citenamefont
  {Stelea}}]{Blazquez-Salcedo:2022kaw}%
  \BibitemOpen
  \bibfield  {author} {\bibinfo {author} {\bibfnamefont {J.~L.}\ \bibnamefont
  {Bl\'azquez-Salcedo}}, \bibinfo {author} {\bibfnamefont {M.-A.}\ \bibnamefont
  {Dariescu}}, \bibinfo {author} {\bibfnamefont {C.}~\bibnamefont {Dariescu}},
  \bibinfo {author} {\bibfnamefont {E.}~\bibnamefont {Radu}},\ and\ \bibinfo
  {author} {\bibfnamefont {C.}~\bibnamefont {Stelea}},\ }\href
  {https://doi.org/10.1016/j.physletb.2022.136993} {\bibfield  {journal}
  {\bibinfo  {journal} {Phys. Lett. B}\ }\textbf {\bibinfo {volume} {827}},\
  \bibinfo {pages} {136993} (\bibinfo {year} {2022})},\ \Eprint
  {https://arxiv.org/abs/2204.05244} {arXiv:2204.05244 [gr-qc]} \BibitemShut
  {NoStop}%
\bibitem [{\citenamefont {Rajaraman}\ \emph {et~al.}(2024)\citenamefont
  {Rajaraman}, \citenamefont {Stewart},\ and\ \citenamefont
  {Verhaaren}}]{Rajaraman:2023ygy}%
  \BibitemOpen
  \bibfield  {author} {\bibinfo {author} {\bibfnamefont {A.}~\bibnamefont
  {Rajaraman}}, \bibinfo {author} {\bibfnamefont {A.}~\bibnamefont {Stewart}},\
  and\ \bibinfo {author} {\bibfnamefont {C.~B.}\ \bibnamefont {Verhaaren}},\
  }\href {https://doi.org/10.1103/PhysRevD.109.086003} {\bibfield  {journal}
  {\bibinfo  {journal} {Phys. Rev. D}\ }\textbf {\bibinfo {volume} {109}},\
  \bibinfo {pages} {086003} (\bibinfo {year} {2024})},\ \Eprint
  {https://arxiv.org/abs/2310.13660} {arXiv:2310.13660 [hep-th]} \BibitemShut
  {NoStop}%
\bibitem [{\citenamefont {Clark}(2006)}]{Clark:2005zc}%
  \BibitemOpen
  \bibfield  {author} {\bibinfo {author} {\bibfnamefont {S.~S.}\ \bibnamefont
  {Clark}},\ }\href {https://doi.org/10.1016/j.nuclphysb.2006.08.019}
  {\bibfield  {journal} {\bibinfo  {journal} {Nucl. Phys. B}\ }\textbf
  {\bibinfo {volume} {756}},\ \bibinfo {pages} {38} (\bibinfo {year} {2006})},\
  \Eprint {https://arxiv.org/abs/hep-ph/0510078} {arXiv:hep-ph/0510078}
  \BibitemShut {NoStop}%
\bibitem [{\citenamefont {Multamaki}\ and\ \citenamefont
  {Vilja}(2000{\natexlab{a}})}]{Multamaki:1999an}%
  \BibitemOpen
  \bibfield  {author} {\bibinfo {author} {\bibfnamefont {T.}~\bibnamefont
  {Multamaki}}\ and\ \bibinfo {author} {\bibfnamefont {I.}~\bibnamefont
  {Vilja}},\ }\href {https://doi.org/10.1016/S0550-3213(99)00827-5} {\bibfield
  {journal} {\bibinfo  {journal} {Nucl. Phys. B}\ }\textbf {\bibinfo {volume}
  {574}},\ \bibinfo {pages} {130} (\bibinfo {year} {2000}{\natexlab{a}})},\
  \Eprint {https://arxiv.org/abs/hep-ph/9908446} {arXiv:hep-ph/9908446}
  \BibitemShut {NoStop}%
\bibitem [{\citenamefont {Friedberg}\ and\ \citenamefont
  {Lee}(1978)}]{Friedberg:1978sc}%
  \BibitemOpen
  \bibfield  {author} {\bibinfo {author} {\bibfnamefont {R.}~\bibnamefont
  {Friedberg}}\ and\ \bibinfo {author} {\bibfnamefont {T.~D.}\ \bibnamefont
  {Lee}},\ }\href {https://doi.org/10.1103/PhysRevD.18.2623} {\bibfield
  {journal} {\bibinfo  {journal} {Phys. Rev. D}\ }\textbf {\bibinfo {volume}
  {18}},\ \bibinfo {pages} {2623} (\bibinfo {year} {1978})}\BibitemShut
  {NoStop}%
\bibitem [{\citenamefont {Goldflam}\ and\ \citenamefont
  {Wilets}(1982)}]{Goldflam:1981tg}%
  \BibitemOpen
  \bibfield  {author} {\bibinfo {author} {\bibfnamefont {R.}~\bibnamefont
  {Goldflam}}\ and\ \bibinfo {author} {\bibfnamefont {L.}~\bibnamefont
  {Wilets}},\ }\href {https://doi.org/10.1103/PhysRevD.25.1951} {\bibfield
  {journal} {\bibinfo  {journal} {Phys. Rev. D}\ }\textbf {\bibinfo {volume}
  {25}},\ \bibinfo {pages} {1951} (\bibinfo {year} {1982})}\BibitemShut
  {NoStop}%
\bibitem [{\citenamefont {Cahill}\ and\ \citenamefont
  {Roberts}(1985)}]{Cahill:1985mh}%
  \BibitemOpen
  \bibfield  {author} {\bibinfo {author} {\bibfnamefont {R.~T.}\ \bibnamefont
  {Cahill}}\ and\ \bibinfo {author} {\bibfnamefont {C.~D.}\ \bibnamefont
  {Roberts}},\ }\href {https://doi.org/10.1103/PhysRevD.32.2419} {\bibfield
  {journal} {\bibinfo  {journal} {Phys. Rev. D}\ }\textbf {\bibinfo {volume}
  {32}},\ \bibinfo {pages} {2419} (\bibinfo {year} {1985})}\BibitemShut
  {NoStop}%
\bibitem [{\citenamefont {Xie}(2024)}]{Xie:2024mxr}%
  \BibitemOpen
  \bibfield  {author} {\bibinfo {author} {\bibfnamefont {K.-P.}\ \bibnamefont
  {Xie}},\ }\href {https://doi.org/10.1007/JHEP09(2024)077} {\bibfield
  {journal} {\bibinfo  {journal} {JHEP}\ }\textbf {\bibinfo {volume} {09}},\
  \bibinfo {pages} {077}},\ \Eprint {https://arxiv.org/abs/2405.01227}
  {arXiv:2405.01227 [hep-ph]} \BibitemShut {NoStop}%
\bibitem [{\citenamefont {Lee}\ \emph {et~al.}(1989)\citenamefont {Lee},
  \citenamefont {Stein-Schabes}, \citenamefont {Watkins},\ and\ \citenamefont
  {Widrow}}]{Lee:1988ag}%
  \BibitemOpen
  \bibfield  {author} {\bibinfo {author} {\bibfnamefont {K.-M.}\ \bibnamefont
  {Lee}}, \bibinfo {author} {\bibfnamefont {J.~A.}\ \bibnamefont
  {Stein-Schabes}}, \bibinfo {author} {\bibfnamefont {R.}~\bibnamefont
  {Watkins}},\ and\ \bibinfo {author} {\bibfnamefont {L.~M.}\ \bibnamefont
  {Widrow}},\ }\href {https://doi.org/10.1103/PhysRevD.39.1665} {\bibfield
  {journal} {\bibinfo  {journal} {Phys. Rev. D}\ }\textbf {\bibinfo {volume}
  {39}},\ \bibinfo {pages} {1665} (\bibinfo {year} {1989})}\BibitemShut
  {NoStop}%
\bibitem [{\citenamefont {Levi}\ and\ \citenamefont
  {Gleiser}(2002)}]{Levi:2001aw}%
  \BibitemOpen
  \bibfield  {author} {\bibinfo {author} {\bibfnamefont {T.~S.}\ \bibnamefont
  {Levi}}\ and\ \bibinfo {author} {\bibfnamefont {M.}~\bibnamefont {Gleiser}},\
  }\href {https://doi.org/10.1103/PhysRevD.66.087701} {\bibfield  {journal}
  {\bibinfo  {journal} {Phys. Rev. D}\ }\textbf {\bibinfo {volume} {66}},\
  \bibinfo {pages} {087701} (\bibinfo {year} {2002})},\ \Eprint
  {https://arxiv.org/abs/hep-ph/0110395} {arXiv:hep-ph/0110395} \BibitemShut
  {NoStop}%
\bibitem [{\citenamefont {Kusenko}\ \emph
  {et~al.}(1998{\natexlab{a}})\citenamefont {Kusenko}, \citenamefont
  {Shaposhnikov},\ and\ \citenamefont {Tinyakov}}]{Kusenko:1997vi}%
  \BibitemOpen
  \bibfield  {author} {\bibinfo {author} {\bibfnamefont {A.}~\bibnamefont
  {Kusenko}}, \bibinfo {author} {\bibfnamefont {M.~E.}\ \bibnamefont
  {Shaposhnikov}},\ and\ \bibinfo {author} {\bibfnamefont {P.~G.}\ \bibnamefont
  {Tinyakov}},\ }\href {https://doi.org/10.1134/1.567658} {\bibfield  {journal}
  {\bibinfo  {journal} {Pisma Zh. Eksp. Teor. Fiz.}\ }\textbf {\bibinfo
  {volume} {67}},\ \bibinfo {pages} {229} (\bibinfo {year}
  {1998}{\natexlab{a}})},\ \Eprint {https://arxiv.org/abs/hep-th/9801041}
  {arXiv:hep-th/9801041} \BibitemShut {NoStop}%
\bibitem [{\citenamefont {Benci}\ and\ \citenamefont
  {Fortunato}(2011)}]{Benci:2010cs}%
  \BibitemOpen
  \bibfield  {author} {\bibinfo {author} {\bibfnamefont {V.}~\bibnamefont
  {Benci}}\ and\ \bibinfo {author} {\bibfnamefont {D.}~\bibnamefont
  {Fortunato}},\ }\href {https://doi.org/10.1063/1.3629848} {\bibfield
  {journal} {\bibinfo  {journal} {J. Math. Phys.}\ }\textbf {\bibinfo {volume}
  {52}},\ \bibinfo {pages} {093701} (\bibinfo {year} {2011})},\ \Eprint
  {https://arxiv.org/abs/1011.5044} {arXiv:1011.5044 [math-ph]} \BibitemShut
  {NoStop}%
\bibitem [{\citenamefont {Rosen}(1968{\natexlab{b}})}]{Rosen:1968zwl}%
  \BibitemOpen
  \bibfield  {author} {\bibinfo {author} {\bibfnamefont {G.}~\bibnamefont
  {Rosen}},\ }\href {https://doi.org/10.1063/1.1664694} {\bibfield  {journal}
  {\bibinfo  {journal} {J. Math. Phys.}\ }\textbf {\bibinfo {volume} {9}},\
  \bibinfo {pages} {999} (\bibinfo {year} {1968}{\natexlab{b}})}\BibitemShut
  {NoStop}%
\bibitem [{\citenamefont {Gulamov}\ \emph {et~al.}(2014)\citenamefont
  {Gulamov}, \citenamefont {Nugaev},\ and\ \citenamefont
  {Smolyakov}}]{Gulamov:2013cra}%
  \BibitemOpen
  \bibfield  {author} {\bibinfo {author} {\bibfnamefont {I.~E.}\ \bibnamefont
  {Gulamov}}, \bibinfo {author} {\bibfnamefont {E.~Y.}\ \bibnamefont
  {Nugaev}},\ and\ \bibinfo {author} {\bibfnamefont {M.~N.}\ \bibnamefont
  {Smolyakov}},\ }\href {https://doi.org/10.1103/PhysRevD.89.085006} {\bibfield
   {journal} {\bibinfo  {journal} {Phys. Rev. D}\ }\textbf {\bibinfo {volume}
  {89}},\ \bibinfo {pages} {085006} (\bibinfo {year} {2014})},\ \Eprint
  {https://arxiv.org/abs/1311.0325} {arXiv:1311.0325 [hep-th]} \BibitemShut
  {NoStop}%
\bibitem [{\citenamefont {Gulamov}\ \emph {et~al.}(2015)\citenamefont
  {Gulamov}, \citenamefont {Nugaev}, \citenamefont {Panin},\ and\ \citenamefont
  {Smolyakov}}]{Gulamov:2015fya}%
  \BibitemOpen
  \bibfield  {author} {\bibinfo {author} {\bibfnamefont {I.~E.}\ \bibnamefont
  {Gulamov}}, \bibinfo {author} {\bibfnamefont {E.~Y.}\ \bibnamefont {Nugaev}},
  \bibinfo {author} {\bibfnamefont {A.~G.}\ \bibnamefont {Panin}},\ and\
  \bibinfo {author} {\bibfnamefont {M.~N.}\ \bibnamefont {Smolyakov}},\ }\href
  {https://doi.org/10.1103/PhysRevD.92.045011} {\bibfield  {journal} {\bibinfo
  {journal} {Phys. Rev. D}\ }\textbf {\bibinfo {volume} {92}},\ \bibinfo
  {pages} {045011} (\bibinfo {year} {2015})},\ \Eprint
  {https://arxiv.org/abs/1506.05786} {arXiv:1506.05786 [hep-th]} \BibitemShut
  {NoStop}%
\bibitem [{\citenamefont {Panin}\ and\ \citenamefont
  {Smolyakov}(2017)}]{Panin:2016ooo}%
  \BibitemOpen
  \bibfield  {author} {\bibinfo {author} {\bibfnamefont {A.~G.}\ \bibnamefont
  {Panin}}\ and\ \bibinfo {author} {\bibfnamefont {M.~N.}\ \bibnamefont
  {Smolyakov}},\ }\href {https://doi.org/10.1103/PhysRevD.95.065006} {\bibfield
   {journal} {\bibinfo  {journal} {Phys. Rev. D}\ }\textbf {\bibinfo {volume}
  {95}},\ \bibinfo {pages} {065006} (\bibinfo {year} {2017})},\ \Eprint
  {https://arxiv.org/abs/1612.00737} {arXiv:1612.00737 [hep-th]} \BibitemShut
  {NoStop}%
\bibitem [{\citenamefont {Heeck}\ \emph
  {et~al.}(2021{\natexlab{b}})\citenamefont {Heeck}, \citenamefont
  {Rajaraman},\ and\ \citenamefont {Verhaaren}}]{Heeck:2021gam}%
  \BibitemOpen
  \bibfield  {author} {\bibinfo {author} {\bibfnamefont {J.}~\bibnamefont
  {Heeck}}, \bibinfo {author} {\bibfnamefont {A.}~\bibnamefont {Rajaraman}},\
  and\ \bibinfo {author} {\bibfnamefont {C.~B.}\ \bibnamefont {Verhaaren}},\
  }\href {https://doi.org/10.1103/PhysRevD.104.016030} {\bibfield  {journal}
  {\bibinfo  {journal} {Phys. Rev. D}\ }\textbf {\bibinfo {volume} {104}},\
  \bibinfo {pages} {016030} (\bibinfo {year} {2021}{\natexlab{b}})},\ \Eprint
  {https://arxiv.org/abs/2105.02893} {arXiv:2105.02893 [hep-th]} \BibitemShut
  {NoStop}%
\bibitem [{\citenamefont {Heeck}\ \emph
  {et~al.}(2021{\natexlab{c}})\citenamefont {Heeck}, \citenamefont {Rajaraman},
  \citenamefont {Riley},\ and\ \citenamefont {Verhaaren}}]{Heeck:2021zvk}%
  \BibitemOpen
  \bibfield  {author} {\bibinfo {author} {\bibfnamefont {J.}~\bibnamefont
  {Heeck}}, \bibinfo {author} {\bibfnamefont {A.}~\bibnamefont {Rajaraman}},
  \bibinfo {author} {\bibfnamefont {R.}~\bibnamefont {Riley}},\ and\ \bibinfo
  {author} {\bibfnamefont {C.~B.}\ \bibnamefont {Verhaaren}},\ }\href
  {https://doi.org/10.1103/PhysRevD.103.116004} {\bibfield  {journal} {\bibinfo
   {journal} {Phys. Rev. D}\ }\textbf {\bibinfo {volume} {103}},\ \bibinfo
  {pages} {116004} (\bibinfo {year} {2021}{\natexlab{c}})},\ \Eprint
  {https://arxiv.org/abs/2103.06905} {arXiv:2103.06905 [hep-th]} \BibitemShut
  {NoStop}%
\bibitem [{\citenamefont {Heeck}\ \emph
  {et~al.}(2021{\natexlab{d}})\citenamefont {Heeck}, \citenamefont {Rajaraman},
  \citenamefont {Riley},\ and\ \citenamefont {Verhaaren}}]{Heeck:2021bce}%
  \BibitemOpen
  \bibfield  {author} {\bibinfo {author} {\bibfnamefont {J.}~\bibnamefont
  {Heeck}}, \bibinfo {author} {\bibfnamefont {A.}~\bibnamefont {Rajaraman}},
  \bibinfo {author} {\bibfnamefont {R.}~\bibnamefont {Riley}},\ and\ \bibinfo
  {author} {\bibfnamefont {C.~B.}\ \bibnamefont {Verhaaren}},\ }\href
  {https://doi.org/10.1007/JHEP10(2021)103} {\bibfield  {journal} {\bibinfo
  {journal} {JHEP}\ }\textbf {\bibinfo {volume} {10}},\ \bibinfo {pages}
  {103}},\ \Eprint {https://arxiv.org/abs/2107.10280} {arXiv:2107.10280
  [hep-th]} \BibitemShut {NoStop}%
\bibitem [{\citenamefont {Loiko}\ and\ \citenamefont
  {Shnir}(2022)}]{Loiko:2022noq}%
  \BibitemOpen
  \bibfield  {author} {\bibinfo {author} {\bibfnamefont {V.}~\bibnamefont
  {Loiko}}\ and\ \bibinfo {author} {\bibfnamefont {Y.}~\bibnamefont {Shnir}},\
  }\href {https://doi.org/10.1103/PhysRevD.106.045021} {\bibfield  {journal}
  {\bibinfo  {journal} {Phys. Rev. D}\ }\textbf {\bibinfo {volume} {106}},\
  \bibinfo {pages} {045021} (\bibinfo {year} {2022})},\ \Eprint
  {https://arxiv.org/abs/2207.02646} {arXiv:2207.02646 [hep-th]} \BibitemShut
  {NoStop}%
\bibitem [{\citenamefont {Loiko}\ and\ \citenamefont
  {Shnir}(2019)}]{Loiko:2019gwk}%
  \BibitemOpen
  \bibfield  {author} {\bibinfo {author} {\bibfnamefont {V.}~\bibnamefont
  {Loiko}}\ and\ \bibinfo {author} {\bibfnamefont {Y.}~\bibnamefont {Shnir}},\
  }\href {https://doi.org/10.1016/j.physletb.2019.134810} {\bibfield  {journal}
  {\bibinfo  {journal} {Phys. Lett. B}\ }\textbf {\bibinfo {volume} {797}},\
  \bibinfo {pages} {134810} (\bibinfo {year} {2019})},\ \Eprint
  {https://arxiv.org/abs/1906.01943} {arXiv:1906.01943 [hep-th]} \BibitemShut
  {NoStop}%
\bibitem [{\citenamefont {Kinach}\ and\ \citenamefont
  {Choptuik}(2023)}]{Kinach:2022jdx}%
  \BibitemOpen
  \bibfield  {author} {\bibinfo {author} {\bibfnamefont {M.~P.}\ \bibnamefont
  {Kinach}}\ and\ \bibinfo {author} {\bibfnamefont {M.~W.}\ \bibnamefont
  {Choptuik}},\ }\href {https://doi.org/10.1103/PhysRevD.107.035022} {\bibfield
   {journal} {\bibinfo  {journal} {Phys. Rev. D}\ }\textbf {\bibinfo {volume}
  {107}},\ \bibinfo {pages} {035022} (\bibinfo {year} {2023})},\ \Eprint
  {https://arxiv.org/abs/2211.11198} {arXiv:2211.11198 [hep-th]} \BibitemShut
  {NoStop}%
\bibitem [{\citenamefont {Rajaraman}(2024)}]{Rajaraman:2024erp}%
  \BibitemOpen
  \bibfield  {author} {\bibinfo {author} {\bibfnamefont {A.}~\bibnamefont
  {Rajaraman}},\ }\href@noop {} {\  (\bibinfo {year} {2024})},\ \Eprint
  {https://arxiv.org/abs/2406.02817} {arXiv:2406.02817 [hep-th]} \BibitemShut
  {NoStop}%
\bibitem [{\citenamefont {Loginov}\ and\ \citenamefont
  {Gauzshtein}(2020)}]{Loginov:2020xoj}%
  \BibitemOpen
  \bibfield  {author} {\bibinfo {author} {\bibfnamefont {A.~Y.}\ \bibnamefont
  {Loginov}}\ and\ \bibinfo {author} {\bibfnamefont {V.~V.}\ \bibnamefont
  {Gauzshtein}},\ }\href {https://doi.org/10.1103/PhysRevD.102.025010}
  {\bibfield  {journal} {\bibinfo  {journal} {Phys. Rev. D}\ }\textbf {\bibinfo
  {volume} {102}},\ \bibinfo {pages} {025010} (\bibinfo {year} {2020})},\
  \Eprint {https://arxiv.org/abs/2004.03446} {arXiv:2004.03446 [hep-th]}
  \BibitemShut {NoStop}%
\bibitem [{\citenamefont {Almumin}(2024)}]{Almumin:2024dem}%
  \BibitemOpen
  \bibfield  {author} {\bibinfo {author} {\bibfnamefont {Y.}~\bibnamefont
  {Almumin}},\ }\href@noop {} {\  (\bibinfo {year} {2024})},\ \Eprint
  {https://arxiv.org/abs/2404.03053} {arXiv:2404.03053 [hep-th]} \BibitemShut
  {NoStop}%
\bibitem [{\citenamefont {Arodz}\ and\ \citenamefont
  {Lis}(2009)}]{Arodz:2008nm}%
  \BibitemOpen
  \bibfield  {author} {\bibinfo {author} {\bibfnamefont {H.}~\bibnamefont
  {Arodz}}\ and\ \bibinfo {author} {\bibfnamefont {J.}~\bibnamefont {Lis}},\
  }\href {https://doi.org/10.1103/PhysRevD.79.045002} {\bibfield  {journal}
  {\bibinfo  {journal} {Phys. Rev. D}\ }\textbf {\bibinfo {volume} {79}},\
  \bibinfo {pages} {045002} (\bibinfo {year} {2009})},\ \Eprint
  {https://arxiv.org/abs/0812.3284} {arXiv:0812.3284 [hep-th]} \BibitemShut
  {NoStop}%
\bibitem [{\citenamefont {Han}\ and\ \citenamefont {Su}(2023)}]{Han:2023uyo}%
  \BibitemOpen
  \bibfield  {author} {\bibinfo {author} {\bibfnamefont {X.}~\bibnamefont
  {Han}}\ and\ \bibinfo {author} {\bibfnamefont {G.}~\bibnamefont {Su}},\
  }\href {https://doi.org/10.1063/5.0151329} {\bibfield  {journal} {\bibinfo
  {journal} {J. Math. Phys.}\ }\textbf {\bibinfo {volume} {64}},\ \bibinfo
  {pages} {111505} (\bibinfo {year} {2023})},\ \Eprint
  {https://arxiv.org/abs/2308.06458} {arXiv:2308.06458 [math-ph]} \BibitemShut
  {NoStop}%
\bibitem [{\citenamefont {Loginov}(2015)}]{Loginov:2015rya}%
  \BibitemOpen
  \bibfield  {author} {\bibinfo {author} {\bibfnamefont {A.~Y.}\ \bibnamefont
  {Loginov}},\ }\href {https://doi.org/10.1103/PhysRevD.91.105028} {\bibfield
  {journal} {\bibinfo  {journal} {Phys. Rev. D}\ }\textbf {\bibinfo {volume}
  {91}},\ \bibinfo {pages} {105028} (\bibinfo {year} {2015})}\BibitemShut
  {NoStop}%
\bibitem [{\citenamefont {Ishihara}\ and\ \citenamefont
  {Ogawa}(2019)}]{Ishihara:2018rxg}%
  \BibitemOpen
  \bibfield  {author} {\bibinfo {author} {\bibfnamefont {H.}~\bibnamefont
  {Ishihara}}\ and\ \bibinfo {author} {\bibfnamefont {T.}~\bibnamefont
  {Ogawa}},\ }\href {https://doi.org/10.1093/ptep/ptz005} {\bibfield  {journal}
  {\bibinfo  {journal} {PTEP}\ }\textbf {\bibinfo {volume} {2019}},\ \bibinfo
  {pages} {021B01} (\bibinfo {year} {2019})},\ \Eprint
  {https://arxiv.org/abs/1811.10894} {arXiv:1811.10894 [hep-th]} \BibitemShut
  {NoStop}%
\bibitem [{\citenamefont {Forg\'acs}\ and\ \citenamefont
  {Luk\'acs}(2020)}]{Forgacs:2020vcy}%
  \BibitemOpen
  \bibfield  {author} {\bibinfo {author} {\bibfnamefont {P.}~\bibnamefont
  {Forg\'acs}}\ and\ \bibinfo {author} {\bibfnamefont {A.}~\bibnamefont
  {Luk\'acs}},\ }\href {https://doi.org/10.1103/PhysRevD.102.076017} {\bibfield
   {journal} {\bibinfo  {journal} {Phys. Rev. D}\ }\textbf {\bibinfo {volume}
  {102}},\ \bibinfo {pages} {076017} (\bibinfo {year} {2020})},\ \Eprint
  {https://arxiv.org/abs/2008.09844} {arXiv:2008.09844 [hep-th]} \BibitemShut
  {NoStop}%
\bibitem [{\citenamefont {Loginov}(2024)}]{Loginov:2023wmh}%
  \BibitemOpen
  \bibfield  {author} {\bibinfo {author} {\bibfnamefont {A.~Y.}\ \bibnamefont
  {Loginov}},\ }\href {https://doi.org/10.1016/j.physletb.2023.138336}
  {\bibfield  {journal} {\bibinfo  {journal} {Phys. Lett. B}\ }\textbf
  {\bibinfo {volume} {848}},\ \bibinfo {pages} {138336} (\bibinfo {year}
  {2024})},\ \Eprint {https://arxiv.org/abs/2307.00282} {arXiv:2307.00282
  [hep-th]} \BibitemShut {NoStop}%
\bibitem [{\citenamefont {Hong}\ and\ \citenamefont
  {Kawasaki}(2017{\natexlab{a}})}]{Hong:2017uhi}%
  \BibitemOpen
  \bibfield  {author} {\bibinfo {author} {\bibfnamefont {J.-P.}\ \bibnamefont
  {Hong}}\ and\ \bibinfo {author} {\bibfnamefont {M.}~\bibnamefont
  {Kawasaki}},\ }\href {https://doi.org/10.1103/PhysRevD.96.103526} {\bibfield
  {journal} {\bibinfo  {journal} {Phys. Rev. D}\ }\textbf {\bibinfo {volume}
  {96}},\ \bibinfo {pages} {103526} (\bibinfo {year} {2017}{\natexlab{a}})},\
  \Eprint {https://arxiv.org/abs/1706.01651} {arXiv:1706.01651 [hep-ph]}
  \BibitemShut {NoStop}%
\bibitem [{\citenamefont {Anagnostopoulos}\ \emph {et~al.}(2001)\citenamefont
  {Anagnostopoulos}, \citenamefont {Axenides}, \citenamefont {Floratos},\ and\
  \citenamefont {Tetradis}}]{Anagnostopoulos:2001dh}%
  \BibitemOpen
  \bibfield  {author} {\bibinfo {author} {\bibfnamefont {K.~N.}\ \bibnamefont
  {Anagnostopoulos}}, \bibinfo {author} {\bibfnamefont {M.}~\bibnamefont
  {Axenides}}, \bibinfo {author} {\bibfnamefont {E.~G.}\ \bibnamefont
  {Floratos}},\ and\ \bibinfo {author} {\bibfnamefont {N.}~\bibnamefont
  {Tetradis}},\ }\href {https://doi.org/10.1103/PhysRevD.64.125006} {\bibfield
  {journal} {\bibinfo  {journal} {Phys. Rev. D}\ }\textbf {\bibinfo {volume}
  {64}},\ \bibinfo {pages} {125006} (\bibinfo {year} {2001})},\ \Eprint
  {https://arxiv.org/abs/hep-ph/0109080} {arXiv:hep-ph/0109080} \BibitemShut
  {NoStop}%
\bibitem [{\citenamefont {Deshaies-Jacques}\ and\ \citenamefont
  {MacKenzie}(2006)}]{Deshaies-Jacques:2006clf}%
  \BibitemOpen
  \bibfield  {author} {\bibinfo {author} {\bibfnamefont {M.}~\bibnamefont
  {Deshaies-Jacques}}\ and\ \bibinfo {author} {\bibfnamefont {R.}~\bibnamefont
  {MacKenzie}},\ }\href {https://doi.org/10.1103/PhysRevD.74.025006} {\bibfield
   {journal} {\bibinfo  {journal} {Phys. Rev. D}\ }\textbf {\bibinfo {volume}
  {74}},\ \bibinfo {pages} {025006} (\bibinfo {year} {2006})},\ \Eprint
  {https://arxiv.org/abs/hep-th/0604036} {arXiv:hep-th/0604036} \BibitemShut
  {NoStop}%
\bibitem [{\citenamefont {Wilson}(1971)}]{Wilson:1971bg}%
  \BibitemOpen
  \bibfield  {author} {\bibinfo {author} {\bibfnamefont {K.~G.}\ \bibnamefont
  {Wilson}},\ }\href {https://doi.org/10.1103/PhysRevB.4.3174} {\bibfield
  {journal} {\bibinfo  {journal} {Phys. Rev. B}\ }\textbf {\bibinfo {volume}
  {4}},\ \bibinfo {pages} {3174} (\bibinfo {year} {1971})}\BibitemShut
  {NoStop}%
\bibitem [{\citenamefont {Weinberg}(1979)}]{Weinberg:1978kz}%
  \BibitemOpen
  \bibfield  {author} {\bibinfo {author} {\bibfnamefont {S.}~\bibnamefont
  {Weinberg}},\ }\href {https://doi.org/10.1016/0378-4371(79)90223-1}
  {\bibfield  {journal} {\bibinfo  {journal} {Physica A}\ }\textbf {\bibinfo
  {volume} {96}},\ \bibinfo {pages} {327} (\bibinfo {year} {1979})}\BibitemShut
  {NoStop}%
\bibitem [{\citenamefont {Burgess}(2020)}]{Burgess:2020tbq}%
  \BibitemOpen
  \bibfield  {author} {\bibinfo {author} {\bibfnamefont {C.~P.}\ \bibnamefont
  {Burgess}},\ }\href {https://doi.org/10.1017/9781139048040} {\emph {\bibinfo
  {title} {{Introduction to Effective Field Theory}}}}\ (\bibinfo  {publisher}
  {Cambridge University Press},\ \bibinfo {year} {2020})\BibitemShut {NoStop}%
\bibitem [{\citenamefont {Adams}\ \emph {et~al.}(2006)\citenamefont {Adams},
  \citenamefont {Arkani-Hamed}, \citenamefont {Dubovsky}, \citenamefont
  {Nicolis},\ and\ \citenamefont {Rattazzi}}]{Adams:2006sv}%
  \BibitemOpen
  \bibfield  {author} {\bibinfo {author} {\bibfnamefont {A.}~\bibnamefont
  {Adams}}, \bibinfo {author} {\bibfnamefont {N.}~\bibnamefont {Arkani-Hamed}},
  \bibinfo {author} {\bibfnamefont {S.}~\bibnamefont {Dubovsky}}, \bibinfo
  {author} {\bibfnamefont {A.}~\bibnamefont {Nicolis}},\ and\ \bibinfo {author}
  {\bibfnamefont {R.}~\bibnamefont {Rattazzi}},\ }\href
  {https://doi.org/10.1088/1126-6708/2006/10/014} {\bibfield  {journal}
  {\bibinfo  {journal} {JHEP}\ }\textbf {\bibinfo {volume} {10}},\ \bibinfo
  {pages} {014}},\ \Eprint {https://arxiv.org/abs/hep-th/0602178}
  {arXiv:hep-th/0602178} \BibitemShut {NoStop}%
\bibitem [{\citenamefont {Tolley}\ \emph {et~al.}(2021)\citenamefont {Tolley},
  \citenamefont {Wang},\ and\ \citenamefont {Zhou}}]{Tolley:2020gtv}%
  \BibitemOpen
  \bibfield  {author} {\bibinfo {author} {\bibfnamefont {A.~J.}\ \bibnamefont
  {Tolley}}, \bibinfo {author} {\bibfnamefont {Z.-Y.}\ \bibnamefont {Wang}},\
  and\ \bibinfo {author} {\bibfnamefont {S.-Y.}\ \bibnamefont {Zhou}},\ }\href
  {https://doi.org/10.1007/JHEP05(2021)255} {\bibfield  {journal} {\bibinfo
  {journal} {JHEP}\ }\textbf {\bibinfo {volume} {05}},\ \bibinfo {pages}
  {255}},\ \Eprint {https://arxiv.org/abs/2011.02400} {arXiv:2011.02400
  [hep-th]} \BibitemShut {NoStop}%
\bibitem [{\citenamefont {de~Rham}\ \emph {et~al.}(2022)\citenamefont
  {de~Rham}, \citenamefont {Kundu}, \citenamefont {Reece}, \citenamefont
  {Tolley},\ and\ \citenamefont {Zhou}}]{deRham:2022hpx}%
  \BibitemOpen
  \bibfield  {author} {\bibinfo {author} {\bibfnamefont {C.}~\bibnamefont
  {de~Rham}}, \bibinfo {author} {\bibfnamefont {S.}~\bibnamefont {Kundu}},
  \bibinfo {author} {\bibfnamefont {M.}~\bibnamefont {Reece}}, \bibinfo
  {author} {\bibfnamefont {A.~J.}\ \bibnamefont {Tolley}},\ and\ \bibinfo
  {author} {\bibfnamefont {S.-Y.}\ \bibnamefont {Zhou}},\ }in\ \href@noop {}
  {\emph {\bibinfo {booktitle} {{Snowmass 2021}}}}\ (\bibinfo {year} {2022})\
  \Eprint {https://arxiv.org/abs/2203.06805} {arXiv:2203.06805 [hep-th]}
  \BibitemShut {NoStop}%
\bibitem [{\citenamefont {Hamada}\ \emph {et~al.}(2024)\citenamefont {Hamada},
  \citenamefont {Kawana}, \citenamefont {Kim},\ and\ \citenamefont
  {Lu}}]{Hamada:2024pbs}%
  \BibitemOpen
  \bibfield  {author} {\bibinfo {author} {\bibfnamefont {Y.}~\bibnamefont
  {Hamada}}, \bibinfo {author} {\bibfnamefont {K.}~\bibnamefont {Kawana}},
  \bibinfo {author} {\bibfnamefont {T.}~\bibnamefont {Kim}},\ and\ \bibinfo
  {author} {\bibfnamefont {P.}~\bibnamefont {Lu}},\ }\href
  {https://doi.org/10.1007/JHEP08(2024)242} {\bibfield  {journal} {\bibinfo
  {journal} {JHEP}\ }\textbf {\bibinfo {volume} {08}},\ \bibinfo {pages}
  {242}},\ \Eprint {https://arxiv.org/abs/2407.11115} {arXiv:2407.11115
  [hep-ph]} \BibitemShut {NoStop}%
\bibitem [{\citenamefont {Kim}\ and\ \citenamefont
  {Nugaev}(2024)}]{Kim:2023zvf}%
  \BibitemOpen
  \bibfield  {author} {\bibinfo {author} {\bibfnamefont {E.}~\bibnamefont
  {Kim}}\ and\ \bibinfo {author} {\bibfnamefont {E.}~\bibnamefont {Nugaev}},\
  }\href {https://doi.org/10.1140/epjc/s10052-024-13167-w} {\bibfield
  {journal} {\bibinfo  {journal} {Eur. Phys. J. C}\ }\textbf {\bibinfo {volume}
  {84}},\ \bibinfo {pages} {797} (\bibinfo {year} {2024})},\ \Eprint
  {https://arxiv.org/abs/2309.09661} {arXiv:2309.09661 [hep-ph]} \BibitemShut
  {NoStop}%
\bibitem [{\citenamefont {Kim}\ \emph {et~al.}(2024)\citenamefont {Kim},
  \citenamefont {Nugaev},\ and\ \citenamefont {Shnir}}]{Kim:2024vam}%
  \BibitemOpen
  \bibfield  {author} {\bibinfo {author} {\bibfnamefont {E.}~\bibnamefont
  {Kim}}, \bibinfo {author} {\bibfnamefont {E.}~\bibnamefont {Nugaev}},\ and\
  \bibinfo {author} {\bibfnamefont {Y.}~\bibnamefont {Shnir}},\ }\href
  {https://doi.org/10.1016/j.physletb.2024.138881} {\bibfield  {journal}
  {\bibinfo  {journal} {Phys. Lett. B}\ }\textbf {\bibinfo {volume} {856}},\
  \bibinfo {pages} {138881} (\bibinfo {year} {2024})},\ \Eprint
  {https://arxiv.org/abs/2405.09262} {arXiv:2405.09262 [hep-ph]} \BibitemShut
  {NoStop}%
\bibitem [{\citenamefont {Heeck}\ and\ \citenamefont
  {Sokhashvili}(2023{\natexlab{b}})}]{Heeck:2023idx}%
  \BibitemOpen
  \bibfield  {author} {\bibinfo {author} {\bibfnamefont {J.}~\bibnamefont
  {Heeck}}\ and\ \bibinfo {author} {\bibfnamefont {M.}~\bibnamefont
  {Sokhashvili}},\ }\href {https://doi.org/10.1140/epjc/s10052-023-11710-9}
  {\bibfield  {journal} {\bibinfo  {journal} {Eur. Phys. J. C}\ }\textbf
  {\bibinfo {volume} {83}},\ \bibinfo {pages} {526} (\bibinfo {year}
  {2023}{\natexlab{b}})},\ \Eprint {https://arxiv.org/abs/2303.09566}
  {arXiv:2303.09566 [hep-ph]} \BibitemShut {NoStop}%
\bibitem [{\citenamefont {Zhong}\ and\ \citenamefont
  {Cheng}(2019)}]{Zhong:2018hwm}%
  \BibitemOpen
  \bibfield  {author} {\bibinfo {author} {\bibfnamefont {Y.}~\bibnamefont
  {Zhong}}\ and\ \bibinfo {author} {\bibfnamefont {H.}~\bibnamefont {Cheng}},\
  }\href {https://doi.org/10.1007/s10773-019-04117-4} {\bibfield  {journal}
  {\bibinfo  {journal} {Int. J. Theor. Phys.}\ }\textbf {\bibinfo {volume}
  {58}},\ \bibinfo {pages} {2251} (\bibinfo {year} {2019})},\ \Eprint
  {https://arxiv.org/abs/1807.03695} {arXiv:1807.03695 [hep-th]} \BibitemShut
  {NoStop}%
\bibitem [{\citenamefont {Loiko}\ \emph {et~al.}(2018)\citenamefont {Loiko},
  \citenamefont {Perapechka},\ and\ \citenamefont {Shnir}}]{Loiko:2018mhb}%
  \BibitemOpen
  \bibfield  {author} {\bibinfo {author} {\bibfnamefont {V.}~\bibnamefont
  {Loiko}}, \bibinfo {author} {\bibfnamefont {I.}~\bibnamefont {Perapechka}},\
  and\ \bibinfo {author} {\bibfnamefont {Y.}~\bibnamefont {Shnir}},\ }\href
  {https://doi.org/10.1103/PhysRevD.98.045018} {\bibfield  {journal} {\bibinfo
  {journal} {Phys. Rev. D}\ }\textbf {\bibinfo {volume} {98}},\ \bibinfo
  {pages} {045018} (\bibinfo {year} {2018})},\ \Eprint
  {https://arxiv.org/abs/1805.11929} {arXiv:1805.11929 [hep-th]} \BibitemShut
  {NoStop}%
\bibitem [{\citenamefont {Brihaye}\ and\ \citenamefont
  {Buisseret}(2024)}]{Brihaye:2024oji}%
  \BibitemOpen
  \bibfield  {author} {\bibinfo {author} {\bibfnamefont {Y.}~\bibnamefont
  {Brihaye}}\ and\ \bibinfo {author} {\bibfnamefont {F.}~\bibnamefont
  {Buisseret}},\ }\href {https://doi.org/10.1103/PhysRevD.109.076029}
  {\bibfield  {journal} {\bibinfo  {journal} {Phys. Rev. D}\ }\textbf {\bibinfo
  {volume} {109}},\ \bibinfo {pages} {076029} (\bibinfo {year} {2024})},\
  \Eprint {https://arxiv.org/abs/2402.15396} {arXiv:2402.15396 [hep-th]}
  \BibitemShut {NoStop}%
\bibitem [{\citenamefont {Safian}\ \emph {et~al.}(1988)\citenamefont {Safian},
  \citenamefont {Coleman},\ and\ \citenamefont {Axenides}}]{Safian:1987pr}%
  \BibitemOpen
  \bibfield  {author} {\bibinfo {author} {\bibfnamefont {A.~M.}\ \bibnamefont
  {Safian}}, \bibinfo {author} {\bibfnamefont {S.~R.}\ \bibnamefont
  {Coleman}},\ and\ \bibinfo {author} {\bibfnamefont {M.}~\bibnamefont
  {Axenides}},\ }\href {https://doi.org/10.1016/0550-3213(88)90315-X}
  {\bibfield  {journal} {\bibinfo  {journal} {Nucl. Phys. B}\ }\textbf
  {\bibinfo {volume} {297}},\ \bibinfo {pages} {498} (\bibinfo {year}
  {1988})}\BibitemShut {NoStop}%
\bibitem [{\citenamefont {Safian}(1988)}]{Safian:1988cz}%
  \BibitemOpen
  \bibfield  {author} {\bibinfo {author} {\bibfnamefont {A.~M.}\ \bibnamefont
  {Safian}},\ }\href {https://doi.org/10.1016/0550-3213(88)90633-5} {\bibfield
  {journal} {\bibinfo  {journal} {Nucl. Phys. B}\ }\textbf {\bibinfo {volume}
  {304}},\ \bibinfo {pages} {392} (\bibinfo {year} {1988})}\BibitemShut
  {NoStop}%
\bibitem [{\citenamefont {Amin}\ and\ \citenamefont
  {Shirokoff}(2010)}]{Amin:2010jq}%
  \BibitemOpen
  \bibfield  {author} {\bibinfo {author} {\bibfnamefont {M.~A.}\ \bibnamefont
  {Amin}}\ and\ \bibinfo {author} {\bibfnamefont {D.}~\bibnamefont
  {Shirokoff}},\ }\href {https://doi.org/10.1103/PhysRevD.81.085045} {\bibfield
   {journal} {\bibinfo  {journal} {Phys. Rev. D}\ }\textbf {\bibinfo {volume}
  {81}},\ \bibinfo {pages} {085045} (\bibinfo {year} {2010})},\ \Eprint
  {https://arxiv.org/abs/1002.3380} {arXiv:1002.3380 [astro-ph.CO]}
  \BibitemShut {NoStop}%
\bibitem [{\citenamefont {Amin}(2010)}]{Amin:2010xe}%
  \BibitemOpen
  \bibfield  {author} {\bibinfo {author} {\bibfnamefont {M.~A.}\ \bibnamefont
  {Amin}},\ }\href@noop {} {\  (\bibinfo {year} {2010})},\ \Eprint
  {https://arxiv.org/abs/1006.3075} {arXiv:1006.3075 [astro-ph.CO]}
  \BibitemShut {NoStop}%
\bibitem [{\citenamefont {Silverstein}\ and\ \citenamefont
  {Westphal}(2008)}]{Silverstein:2008sg}%
  \BibitemOpen
  \bibfield  {author} {\bibinfo {author} {\bibfnamefont {E.}~\bibnamefont
  {Silverstein}}\ and\ \bibinfo {author} {\bibfnamefont {A.}~\bibnamefont
  {Westphal}},\ }\href {https://doi.org/10.1103/PhysRevD.78.106003} {\bibfield
  {journal} {\bibinfo  {journal} {Phys. Rev. D}\ }\textbf {\bibinfo {volume}
  {78}},\ \bibinfo {pages} {106003} (\bibinfo {year} {2008})},\ \Eprint
  {https://arxiv.org/abs/0803.3085} {arXiv:0803.3085 [hep-th]} \BibitemShut
  {NoStop}%
\bibitem [{\citenamefont {McAllister}\ \emph {et~al.}(2010)\citenamefont
  {McAllister}, \citenamefont {Silverstein},\ and\ \citenamefont
  {Westphal}}]{McAllister:2008hb}%
  \BibitemOpen
  \bibfield  {author} {\bibinfo {author} {\bibfnamefont {L.}~\bibnamefont
  {McAllister}}, \bibinfo {author} {\bibfnamefont {E.}~\bibnamefont
  {Silverstein}},\ and\ \bibinfo {author} {\bibfnamefont {A.}~\bibnamefont
  {Westphal}},\ }\href {https://doi.org/10.1103/PhysRevD.82.046003} {\bibfield
  {journal} {\bibinfo  {journal} {Phys. Rev. D}\ }\textbf {\bibinfo {volume}
  {82}},\ \bibinfo {pages} {046003} (\bibinfo {year} {2010})},\ \Eprint
  {https://arxiv.org/abs/0808.0706} {arXiv:0808.0706 [hep-th]} \BibitemShut
  {NoStop}%
\bibitem [{\citenamefont {Amin}\ \emph {et~al.}(2012)\citenamefont {Amin},
  \citenamefont {Easther}, \citenamefont {Finkel}, \citenamefont {Flauger},\
  and\ \citenamefont {Hertzberg}}]{Amin:2011hj}%
  \BibitemOpen
  \bibfield  {author} {\bibinfo {author} {\bibfnamefont {M.~A.}\ \bibnamefont
  {Amin}}, \bibinfo {author} {\bibfnamefont {R.}~\bibnamefont {Easther}},
  \bibinfo {author} {\bibfnamefont {H.}~\bibnamefont {Finkel}}, \bibinfo
  {author} {\bibfnamefont {R.}~\bibnamefont {Flauger}},\ and\ \bibinfo {author}
  {\bibfnamefont {M.~P.}\ \bibnamefont {Hertzberg}},\ }\href
  {https://doi.org/10.1103/PhysRevLett.108.241302} {\bibfield  {journal}
  {\bibinfo  {journal} {Phys. Rev. Lett.}\ }\textbf {\bibinfo {volume} {108}},\
  \bibinfo {pages} {241302} (\bibinfo {year} {2012})},\ \Eprint
  {https://arxiv.org/abs/1106.3335} {arXiv:1106.3335 [astro-ph.CO]}
  \BibitemShut {NoStop}%
\bibitem [{\citenamefont {Aghanim}\ \emph {et~al.}(2020)\citenamefont {Aghanim}
  \emph {et~al.}}]{Planck:2018vyg}%
  \BibitemOpen
  \bibfield  {author} {\bibinfo {author} {\bibfnamefont {N.}~\bibnamefont
  {Aghanim}} \emph {et~al.} (\bibinfo {collaboration} {Planck}),\ }\href
  {https://doi.org/10.1051/0004-6361/201833910} {\bibfield  {journal} {\bibinfo
   {journal} {Astron. Astrophys.}\ }\textbf {\bibinfo {volume} {641}},\
  \bibinfo {pages} {A6} (\bibinfo {year} {2020})},\ \bibinfo {note} {[Erratum:
  Astron.Astrophys. 652, C4 (2021)]},\ \Eprint
  {https://arxiv.org/abs/1807.06209} {arXiv:1807.06209 [astro-ph.CO]}
  \BibitemShut {NoStop}%
\bibitem [{\citenamefont {Amin}(2013)}]{Amin:2013ika}%
  \BibitemOpen
  \bibfield  {author} {\bibinfo {author} {\bibfnamefont {M.~A.}\ \bibnamefont
  {Amin}},\ }\href {https://doi.org/10.1103/PhysRevD.87.123505} {\bibfield
  {journal} {\bibinfo  {journal} {Phys. Rev. D}\ }\textbf {\bibinfo {volume}
  {87}},\ \bibinfo {pages} {123505} (\bibinfo {year} {2013})},\ \Eprint
  {https://arxiv.org/abs/1303.1102} {arXiv:1303.1102 [astro-ph.CO]}
  \BibitemShut {NoStop}%
\bibitem [{\citenamefont {Sakstein}\ and\ \citenamefont
  {Trodden}(2018)}]{Sakstein:2018pfd}%
  \BibitemOpen
  \bibfield  {author} {\bibinfo {author} {\bibfnamefont {J.}~\bibnamefont
  {Sakstein}}\ and\ \bibinfo {author} {\bibfnamefont {M.}~\bibnamefont
  {Trodden}},\ }\href {https://doi.org/10.1103/PhysRevD.98.123512} {\bibfield
  {journal} {\bibinfo  {journal} {Phys. Rev. D}\ }\textbf {\bibinfo {volume}
  {98}},\ \bibinfo {pages} {123512} (\bibinfo {year} {2018})},\ \Eprint
  {https://arxiv.org/abs/1809.07724} {arXiv:1809.07724 [hep-th]} \BibitemShut
  {NoStop}%
\bibitem [{\citenamefont {Van~Dissel}\ and\ \citenamefont
  {Sfakianakis}(2022)}]{VanDissel:2020umg}%
  \BibitemOpen
  \bibfield  {author} {\bibinfo {author} {\bibfnamefont {F.}~\bibnamefont
  {Van~Dissel}}\ and\ \bibinfo {author} {\bibfnamefont {E.~I.}\ \bibnamefont
  {Sfakianakis}},\ }\href {https://doi.org/10.1103/PhysRevD.106.096018}
  {\bibfield  {journal} {\bibinfo  {journal} {Phys. Rev. D}\ }\textbf {\bibinfo
  {volume} {106}},\ \bibinfo {pages} {096018} (\bibinfo {year} {2022})},\
  \Eprint {https://arxiv.org/abs/2010.07789} {arXiv:2010.07789 [hep-th]}
  \BibitemShut {NoStop}%
\bibitem [{\citenamefont {Jain}\ and\ \citenamefont
  {Amin}(2022)}]{Jain:2021pnk}%
  \BibitemOpen
  \bibfield  {author} {\bibinfo {author} {\bibfnamefont {M.}~\bibnamefont
  {Jain}}\ and\ \bibinfo {author} {\bibfnamefont {M.~A.}\ \bibnamefont
  {Amin}},\ }\href {https://doi.org/10.1103/PhysRevD.105.056019} {\bibfield
  {journal} {\bibinfo  {journal} {Phys. Rev. D}\ }\textbf {\bibinfo {volume}
  {105}},\ \bibinfo {pages} {056019} (\bibinfo {year} {2022})},\ \Eprint
  {https://arxiv.org/abs/2109.04892} {arXiv:2109.04892 [hep-th]} \BibitemShut
  {NoStop}%
\bibitem [{\citenamefont {Zhang}\ \emph {et~al.}(2022)\citenamefont {Zhang},
  \citenamefont {Jain},\ and\ \citenamefont {Amin}}]{Zhang:2021xxa}%
  \BibitemOpen
  \bibfield  {author} {\bibinfo {author} {\bibfnamefont {H.-Y.}\ \bibnamefont
  {Zhang}}, \bibinfo {author} {\bibfnamefont {M.}~\bibnamefont {Jain}},\ and\
  \bibinfo {author} {\bibfnamefont {M.~A.}\ \bibnamefont {Amin}},\ }\href
  {https://doi.org/10.1103/PhysRevD.105.096037} {\bibfield  {journal} {\bibinfo
   {journal} {Phys. Rev. D}\ }\textbf {\bibinfo {volume} {105}},\ \bibinfo
  {pages} {096037} (\bibinfo {year} {2022})},\ \Eprint
  {https://arxiv.org/abs/2111.08700} {arXiv:2111.08700 [astro-ph.CO]}
  \BibitemShut {NoStop}%
\bibitem [{\citenamefont {Jain}(2022)}]{Jain:2022kwq}%
  \BibitemOpen
  \bibfield  {author} {\bibinfo {author} {\bibfnamefont {M.}~\bibnamefont
  {Jain}},\ }\href {https://doi.org/10.1103/PhysRevD.106.085011} {\bibfield
  {journal} {\bibinfo  {journal} {Phys. Rev. D}\ }\textbf {\bibinfo {volume}
  {106}},\ \bibinfo {pages} {085011} (\bibinfo {year} {2022})},\ \Eprint
  {https://arxiv.org/abs/2205.03418} {arXiv:2205.03418 [hep-ph]} \BibitemShut
  {NoStop}%
\bibitem [{\citenamefont {Zhang}()}]{Zhang:2023ktk}%
  \BibitemOpen
  \bibfield  {author} {\bibinfo {author} {\bibfnamefont {H.-Y.}\ \bibnamefont
  {Zhang}},\ }\href@noop {} {\ }\Eprint {https://arxiv.org/abs/2401.00043}
  {arXiv:2401.00043 [hep-ph]} \BibitemShut {NoStop}%
\bibitem [{\citenamefont {Wang}\ \emph {et~al.}(2024)\citenamefont {Wang},
  \citenamefont {Helfer},\ and\ \citenamefont {Amin}}]{Wang:2023tly}%
  \BibitemOpen
  \bibfield  {author} {\bibinfo {author} {\bibfnamefont {Z.}~\bibnamefont
  {Wang}}, \bibinfo {author} {\bibfnamefont {T.}~\bibnamefont {Helfer}},\ and\
  \bibinfo {author} {\bibfnamefont {M.~A.}\ \bibnamefont {Amin}},\ }\href
  {https://doi.org/10.1103/PhysRevD.109.024019} {\bibfield  {journal} {\bibinfo
   {journal} {Phys. Rev. D}\ }\textbf {\bibinfo {volume} {109}},\ \bibinfo
  {pages} {024019} (\bibinfo {year} {2024})},\ \Eprint
  {https://arxiv.org/abs/2309.04345} {arXiv:2309.04345 [gr-qc]} \BibitemShut
  {NoStop}%
\bibitem [{\citenamefont {Amin}\ \emph
  {et~al.}(2014{\natexlab{a}})\citenamefont {Amin}, \citenamefont {Banik},
  \citenamefont {Negreanu},\ and\ \citenamefont {Yang}}]{Amin:2014fua}%
  \BibitemOpen
  \bibfield  {author} {\bibinfo {author} {\bibfnamefont {M.~A.}\ \bibnamefont
  {Amin}}, \bibinfo {author} {\bibfnamefont {I.}~\bibnamefont {Banik}},
  \bibinfo {author} {\bibfnamefont {C.}~\bibnamefont {Negreanu}},\ and\
  \bibinfo {author} {\bibfnamefont {I.-S.}\ \bibnamefont {Yang}},\ }\href
  {https://doi.org/10.1103/PhysRevD.90.085024} {\bibfield  {journal} {\bibinfo
  {journal} {Phys. Rev. D}\ }\textbf {\bibinfo {volume} {90}},\ \bibinfo
  {pages} {085024} (\bibinfo {year} {2014}{\natexlab{a}})},\ \Eprint
  {https://arxiv.org/abs/1410.1822} {arXiv:1410.1822 [hep-th]} \BibitemShut
  {NoStop}%
\bibitem [{\citenamefont {Levkov}\ and\ \citenamefont
  {Maslov}(2023)}]{Levkov:2023ncb}%
  \BibitemOpen
  \bibfield  {author} {\bibinfo {author} {\bibfnamefont {D.~G.}\ \bibnamefont
  {Levkov}}\ and\ \bibinfo {author} {\bibfnamefont {V.~E.}\ \bibnamefont
  {Maslov}},\ }\href {https://doi.org/10.1103/PhysRevD.108.063514} {\bibfield
  {journal} {\bibinfo  {journal} {Phys. Rev. D}\ }\textbf {\bibinfo {volume}
  {108}},\ \bibinfo {pages} {063514} (\bibinfo {year} {2023})},\ \Eprint
  {https://arxiv.org/abs/2306.06171} {arXiv:2306.06171 [hep-th]} \BibitemShut
  {NoStop}%
\bibitem [{\citenamefont {Gleiser}\ and\ \citenamefont
  {Sornborger}(2000)}]{Gleiser:1999tj}%
  \BibitemOpen
  \bibfield  {author} {\bibinfo {author} {\bibfnamefont {M.}~\bibnamefont
  {Gleiser}}\ and\ \bibinfo {author} {\bibfnamefont {A.}~\bibnamefont
  {Sornborger}},\ }\href {https://doi.org/10.1103/PhysRevE.62.1368} {\bibfield
  {journal} {\bibinfo  {journal} {Phys. Rev. E}\ }\textbf {\bibinfo {volume}
  {62}},\ \bibinfo {pages} {1368} (\bibinfo {year} {2000})},\ \Eprint
  {https://arxiv.org/abs/patt-sol/9909002} {arXiv:patt-sol/9909002}
  \BibitemShut {NoStop}%
\bibitem [{\citenamefont {Fodor}\ \emph {et~al.}(2006)\citenamefont {Fodor},
  \citenamefont {Forgacs}, \citenamefont {Grandclement},\ and\ \citenamefont
  {Racz}}]{Fodor:2006zs}%
  \BibitemOpen
  \bibfield  {author} {\bibinfo {author} {\bibfnamefont {G.}~\bibnamefont
  {Fodor}}, \bibinfo {author} {\bibfnamefont {P.}~\bibnamefont {Forgacs}},
  \bibinfo {author} {\bibfnamefont {P.}~\bibnamefont {Grandclement}},\ and\
  \bibinfo {author} {\bibfnamefont {I.}~\bibnamefont {Racz}},\ }\href
  {https://doi.org/10.1103/PhysRevD.74.124003} {\bibfield  {journal} {\bibinfo
  {journal} {Phys. Rev. D}\ }\textbf {\bibinfo {volume} {74}},\ \bibinfo
  {pages} {124003} (\bibinfo {year} {2006})},\ \Eprint
  {https://arxiv.org/abs/hep-th/0609023} {arXiv:hep-th/0609023} \BibitemShut
  {NoStop}%
\bibitem [{\citenamefont {Hindmarsh}\ and\ \citenamefont
  {Salmi}(2006)}]{Hindmarsh:2006ur}%
  \BibitemOpen
  \bibfield  {author} {\bibinfo {author} {\bibfnamefont {M.}~\bibnamefont
  {Hindmarsh}}\ and\ \bibinfo {author} {\bibfnamefont {P.}~\bibnamefont
  {Salmi}},\ }\href {https://doi.org/10.1103/PhysRevD.74.105005} {\bibfield
  {journal} {\bibinfo  {journal} {Phys. Rev. D}\ }\textbf {\bibinfo {volume}
  {74}},\ \bibinfo {pages} {105005} (\bibinfo {year} {2006})},\ \Eprint
  {https://arxiv.org/abs/hep-th/0606016} {arXiv:hep-th/0606016} \BibitemShut
  {NoStop}%
\bibitem [{\citenamefont {Gleiser}\ and\ \citenamefont
  {Sicilia}(2009)}]{Gleiser:2009ys}%
  \BibitemOpen
  \bibfield  {author} {\bibinfo {author} {\bibfnamefont {M.}~\bibnamefont
  {Gleiser}}\ and\ \bibinfo {author} {\bibfnamefont {D.}~\bibnamefont
  {Sicilia}},\ }\href {https://doi.org/10.1103/PhysRevD.80.125037} {\bibfield
  {journal} {\bibinfo  {journal} {Phys. Rev. D}\ }\textbf {\bibinfo {volume}
  {80}},\ \bibinfo {pages} {125037} (\bibinfo {year} {2009})},\ \Eprint
  {https://arxiv.org/abs/0910.5922} {arXiv:0910.5922 [hep-th]} \BibitemShut
  {NoStop}%
\bibitem [{\citenamefont {Amin}\ \emph {et~al.}(2010)\citenamefont {Amin},
  \citenamefont {Easther},\ and\ \citenamefont {Finkel}}]{Amin:2010dc}%
  \BibitemOpen
  \bibfield  {author} {\bibinfo {author} {\bibfnamefont {M.~A.}\ \bibnamefont
  {Amin}}, \bibinfo {author} {\bibfnamefont {R.}~\bibnamefont {Easther}},\ and\
  \bibinfo {author} {\bibfnamefont {H.}~\bibnamefont {Finkel}},\ }\href
  {https://doi.org/10.1088/1475-7516/2010/12/001} {\bibfield  {journal}
  {\bibinfo  {journal} {JCAP}\ }\textbf {\bibinfo {volume} {12}},\ \bibinfo
  {pages} {001}},\ \Eprint {https://arxiv.org/abs/1009.2505} {arXiv:1009.2505
  [astro-ph.CO]} \BibitemShut {NoStop}%
\bibitem [{\citenamefont {Andersen}\ and\ \citenamefont
  {Tranberg}(2012)}]{Andersen:2012wg}%
  \BibitemOpen
  \bibfield  {author} {\bibinfo {author} {\bibfnamefont {E.~A.}\ \bibnamefont
  {Andersen}}\ and\ \bibinfo {author} {\bibfnamefont {A.}~\bibnamefont
  {Tranberg}},\ }\href {https://doi.org/10.1007/JHEP12(2012)016} {\bibfield
  {journal} {\bibinfo  {journal} {JHEP}\ }\textbf {\bibinfo {volume} {12}},\
  \bibinfo {pages} {016}},\ \Eprint {https://arxiv.org/abs/1210.2227}
  {arXiv:1210.2227 [hep-ph]} \BibitemShut {NoStop}%
\bibitem [{\citenamefont {Gleiser}\ and\ \citenamefont
  {Stamatopoulos}(2012)}]{Gleiser:2012tu}%
  \BibitemOpen
  \bibfield  {author} {\bibinfo {author} {\bibfnamefont {M.}~\bibnamefont
  {Gleiser}}\ and\ \bibinfo {author} {\bibfnamefont {N.}~\bibnamefont
  {Stamatopoulos}},\ }\href {https://doi.org/10.1103/PhysRevD.86.045004}
  {\bibfield  {journal} {\bibinfo  {journal} {Phys. Rev. D}\ }\textbf {\bibinfo
  {volume} {86}},\ \bibinfo {pages} {045004} (\bibinfo {year} {2012})},\
  \Eprint {https://arxiv.org/abs/1205.3061} {arXiv:1205.3061 [hep-th]}
  \BibitemShut {NoStop}%
\bibitem [{\citenamefont {Salmi}\ and\ \citenamefont
  {Hindmarsh}(2012)}]{Salmi:2012ta}%
  \BibitemOpen
  \bibfield  {author} {\bibinfo {author} {\bibfnamefont {P.}~\bibnamefont
  {Salmi}}\ and\ \bibinfo {author} {\bibfnamefont {M.}~\bibnamefont
  {Hindmarsh}},\ }\href {https://doi.org/10.1103/PhysRevD.85.085033} {\bibfield
   {journal} {\bibinfo  {journal} {Phys. Rev. D}\ }\textbf {\bibinfo {volume}
  {85}},\ \bibinfo {pages} {085033} (\bibinfo {year} {2012})},\ \Eprint
  {https://arxiv.org/abs/1201.1934} {arXiv:1201.1934 [hep-th]} \BibitemShut
  {NoStop}%
\bibitem [{\citenamefont {Honda}\ and\ \citenamefont
  {Choptuik}(2002)}]{Honda:2001xg}%
  \BibitemOpen
  \bibfield  {author} {\bibinfo {author} {\bibfnamefont {E.~P.}\ \bibnamefont
  {Honda}}\ and\ \bibinfo {author} {\bibfnamefont {M.~W.}\ \bibnamefont
  {Choptuik}},\ }\href {https://doi.org/10.1103/PhysRevD.65.084037} {\bibfield
  {journal} {\bibinfo  {journal} {Phys. Rev. D}\ }\textbf {\bibinfo {volume}
  {65}},\ \bibinfo {pages} {084037} (\bibinfo {year} {2002})},\ \Eprint
  {https://arxiv.org/abs/hep-ph/0110065} {arXiv:hep-ph/0110065} \BibitemShut
  {NoStop}%
\bibitem [{\citenamefont {Wang}\ \emph {et~al.}(2023)\citenamefont {Wang},
  \citenamefont {Xie},\ and\ \citenamefont {Zhou}}]{Wang:2022rhk}%
  \BibitemOpen
  \bibfield  {author} {\bibinfo {author} {\bibfnamefont {Y.-J.}\ \bibnamefont
  {Wang}}, \bibinfo {author} {\bibfnamefont {Q.-X.}\ \bibnamefont {Xie}},\ and\
  \bibinfo {author} {\bibfnamefont {S.-Y.}\ \bibnamefont {Zhou}},\ }\href
  {https://doi.org/10.1103/PhysRevD.108.025006} {\bibfield  {journal} {\bibinfo
   {journal} {Phys. Rev. D}\ }\textbf {\bibinfo {volume} {108}},\ \bibinfo
  {pages} {025006} (\bibinfo {year} {2023})},\ \Eprint
  {https://arxiv.org/abs/2210.04969} {arXiv:2210.04969 [hep-th]} \BibitemShut
  {NoStop}%
\bibitem [{\citenamefont {Saffin}\ and\ \citenamefont
  {Tranberg}(2007)}]{Saffin:2006yk}%
  \BibitemOpen
  \bibfield  {author} {\bibinfo {author} {\bibfnamefont {P.~M.}\ \bibnamefont
  {Saffin}}\ and\ \bibinfo {author} {\bibfnamefont {A.}~\bibnamefont
  {Tranberg}},\ }\href {https://doi.org/10.1088/1126-6708/2007/01/030}
  {\bibfield  {journal} {\bibinfo  {journal} {JHEP}\ }\textbf {\bibinfo
  {volume} {01}},\ \bibinfo {pages} {030}},\ \Eprint
  {https://arxiv.org/abs/hep-th/0610191} {arXiv:hep-th/0610191} \BibitemShut
  {NoStop}%
\bibitem [{\citenamefont {Zhang}\ \emph {et~al.}(2020)\citenamefont {Zhang},
  \citenamefont {Amin}, \citenamefont {Copeland}, \citenamefont {Saffin},\ and\
  \citenamefont {Lozanov}}]{Zhang:2020bec}%
  \BibitemOpen
  \bibfield  {author} {\bibinfo {author} {\bibfnamefont {H.-Y.}\ \bibnamefont
  {Zhang}}, \bibinfo {author} {\bibfnamefont {M.~A.}\ \bibnamefont {Amin}},
  \bibinfo {author} {\bibfnamefont {E.~J.}\ \bibnamefont {Copeland}}, \bibinfo
  {author} {\bibfnamefont {P.~M.}\ \bibnamefont {Saffin}},\ and\ \bibinfo
  {author} {\bibfnamefont {K.~D.}\ \bibnamefont {Lozanov}},\ }\href
  {https://doi.org/10.1088/1475-7516/2020/07/055} {\bibfield  {journal}
  {\bibinfo  {journal} {JCAP}\ }\textbf {\bibinfo {volume} {07}},\ \bibinfo
  {pages} {055}},\ \Eprint {https://arxiv.org/abs/2004.01202} {arXiv:2004.01202
  [hep-th]} \BibitemShut {NoStop}%
\bibitem [{\citenamefont {Cyncynates}\ and\ \citenamefont
  {Giurgica-Tiron}(2021)}]{Cyncynates:2021rtf}%
  \BibitemOpen
  \bibfield  {author} {\bibinfo {author} {\bibfnamefont {D.}~\bibnamefont
  {Cyncynates}}\ and\ \bibinfo {author} {\bibfnamefont {T.}~\bibnamefont
  {Giurgica-Tiron}},\ }\href {https://doi.org/10.1103/PhysRevD.103.116011}
  {\bibfield  {journal} {\bibinfo  {journal} {Phys. Rev. D}\ }\textbf {\bibinfo
  {volume} {103}},\ \bibinfo {pages} {116011} (\bibinfo {year} {2021})},\
  \Eprint {https://arxiv.org/abs/2104.02069} {arXiv:2104.02069 [hep-ph]}
  \BibitemShut {NoStop}%
\bibitem [{\citenamefont {Nagy}\ and\ \citenamefont
  {Takacs}(2021)}]{Nagy:2021plv}%
  \BibitemOpen
  \bibfield  {author} {\bibinfo {author} {\bibfnamefont {B.~C.}\ \bibnamefont
  {Nagy}}\ and\ \bibinfo {author} {\bibfnamefont {G.}~\bibnamefont {Takacs}},\
  }\href {https://doi.org/10.1103/PhysRevD.104.056033} {\bibfield  {journal}
  {\bibinfo  {journal} {Phys. Rev. D}\ }\textbf {\bibinfo {volume} {104}},\
  \bibinfo {pages} {056033} (\bibinfo {year} {2021})},\ \Eprint
  {https://arxiv.org/abs/2105.01089} {arXiv:2105.01089 [hep-th]} \BibitemShut
  {NoStop}%
\bibitem [{\citenamefont {Fodor}()}]{Fodor:2019ftc}%
  \BibitemOpen
  \bibfield  {author} {\bibinfo {author} {\bibfnamefont {G.}~\bibnamefont
  {Fodor}},\ }\href@noop {} {\ }\Eprint {https://arxiv.org/abs/1911.03340}
  {arXiv:1911.03340 [hep-th]} \BibitemShut {NoStop}%
\bibitem [{\citenamefont {Flach}\ and\ \citenamefont
  {Willis}(1998)}]{FLACH1998181}%
  \BibitemOpen
  \bibfield  {author} {\bibinfo {author} {\bibfnamefont {S.}~\bibnamefont
  {Flach}}\ and\ \bibinfo {author} {\bibfnamefont {C.}~\bibnamefont {Willis}},\
  }\href {https://doi.org/https://doi.org/10.1016/S0370-1573(97)00068-9}
  {\bibfield  {journal} {\bibinfo  {journal} {Physics Reports}\ }\textbf
  {\bibinfo {volume} {295}},\ \bibinfo {pages} {181} (\bibinfo {year}
  {1998})}\BibitemShut {NoStop}%
\bibitem [{\citenamefont {Fodor}\ \emph
  {et~al.}(2009{\natexlab{a}})\citenamefont {Fodor}, \citenamefont {Forgacs},
  \citenamefont {Horvath},\ and\ \citenamefont {Mezei}}]{Fodor:2008du}%
  \BibitemOpen
  \bibfield  {author} {\bibinfo {author} {\bibfnamefont {G.}~\bibnamefont
  {Fodor}}, \bibinfo {author} {\bibfnamefont {P.}~\bibnamefont {Forgacs}},
  \bibinfo {author} {\bibfnamefont {Z.}~\bibnamefont {Horvath}},\ and\ \bibinfo
  {author} {\bibfnamefont {M.}~\bibnamefont {Mezei}},\ }\href
  {https://doi.org/10.1103/PhysRevD.79.065002} {\bibfield  {journal} {\bibinfo
  {journal} {Phys. Rev. D}\ }\textbf {\bibinfo {volume} {79}},\ \bibinfo
  {pages} {065002} (\bibinfo {year} {2009}{\natexlab{a}})},\ \Eprint
  {https://arxiv.org/abs/0812.1919} {arXiv:0812.1919 [hep-th]} \BibitemShut
  {NoStop}%
\bibitem [{\citenamefont {Fodor}\ \emph
  {et~al.}(2009{\natexlab{b}})\citenamefont {Fodor}, \citenamefont {Forgacs},
  \citenamefont {Horvath},\ and\ \citenamefont {Mezei}}]{Fodor:2009kf}%
  \BibitemOpen
  \bibfield  {author} {\bibinfo {author} {\bibfnamefont {G.}~\bibnamefont
  {Fodor}}, \bibinfo {author} {\bibfnamefont {P.}~\bibnamefont {Forgacs}},
  \bibinfo {author} {\bibfnamefont {Z.}~\bibnamefont {Horvath}},\ and\ \bibinfo
  {author} {\bibfnamefont {M.}~\bibnamefont {Mezei}},\ }\href
  {https://doi.org/10.1016/j.physletb.2009.03.054} {\bibfield  {journal}
  {\bibinfo  {journal} {Phys. Lett. B}\ }\textbf {\bibinfo {volume} {674}},\
  \bibinfo {pages} {319} (\bibinfo {year} {2009}{\natexlab{b}})},\ \Eprint
  {https://arxiv.org/abs/0903.0953} {arXiv:0903.0953 [hep-th]} \BibitemShut
  {NoStop}%
\bibitem [{\citenamefont {Segur}\ and\ \citenamefont
  {Kruskal}(1987)}]{Segur:1987mg}%
  \BibitemOpen
  \bibfield  {author} {\bibinfo {author} {\bibfnamefont {H.}~\bibnamefont
  {Segur}}\ and\ \bibinfo {author} {\bibfnamefont {M.~D.}\ \bibnamefont
  {Kruskal}},\ }\href {https://doi.org/10.1103/PhysRevLett.58.747} {\bibfield
  {journal} {\bibinfo  {journal} {Phys. Rev. Lett.}\ }\textbf {\bibinfo
  {volume} {58}},\ \bibinfo {pages} {747} (\bibinfo {year} {1987})}\BibitemShut
  {NoStop}%
\bibitem [{\citenamefont {Fodor}\ \emph {et~al.}(2008)\citenamefont {Fodor},
  \citenamefont {Forgacs}, \citenamefont {Horvath},\ and\ \citenamefont
  {Lukacs}}]{Fodor:2008es}%
  \BibitemOpen
  \bibfield  {author} {\bibinfo {author} {\bibfnamefont {G.}~\bibnamefont
  {Fodor}}, \bibinfo {author} {\bibfnamefont {P.}~\bibnamefont {Forgacs}},
  \bibinfo {author} {\bibfnamefont {Z.}~\bibnamefont {Horvath}},\ and\ \bibinfo
  {author} {\bibfnamefont {A.}~\bibnamefont {Lukacs}},\ }\href
  {https://doi.org/10.1103/PhysRevD.78.025003} {\bibfield  {journal} {\bibinfo
  {journal} {Phys. Rev. D}\ }\textbf {\bibinfo {volume} {78}},\ \bibinfo
  {pages} {025003} (\bibinfo {year} {2008})},\ \Eprint
  {https://arxiv.org/abs/0802.3525} {arXiv:0802.3525 [hep-th]} \BibitemShut
  {NoStop}%
\bibitem [{\citenamefont {Hertzberg}(2010)}]{Hertzberg:2010yz}%
  \BibitemOpen
  \bibfield  {author} {\bibinfo {author} {\bibfnamefont {M.~P.}\ \bibnamefont
  {Hertzberg}},\ }\href {https://doi.org/10.1103/PhysRevD.82.045022} {\bibfield
   {journal} {\bibinfo  {journal} {Phys. Rev. D}\ }\textbf {\bibinfo {volume}
  {82}},\ \bibinfo {pages} {045022} (\bibinfo {year} {2010})},\ \Eprint
  {https://arxiv.org/abs/1003.3459} {arXiv:1003.3459 [hep-th]} \BibitemShut
  {NoStop}%
\bibitem [{\citenamefont {Fodor}\ \emph {et~al.}(2010)\citenamefont {Fodor},
  \citenamefont {Forgacs},\ and\ \citenamefont {Mezei}}]{Fodor:2009kg}%
  \BibitemOpen
  \bibfield  {author} {\bibinfo {author} {\bibfnamefont {G.}~\bibnamefont
  {Fodor}}, \bibinfo {author} {\bibfnamefont {P.}~\bibnamefont {Forgacs}},\
  and\ \bibinfo {author} {\bibfnamefont {M.}~\bibnamefont {Mezei}},\ }\href
  {https://doi.org/10.1103/PhysRevD.81.064029} {\bibfield  {journal} {\bibinfo
  {journal} {Phys. Rev. D}\ }\textbf {\bibinfo {volume} {81}},\ \bibinfo
  {pages} {064029} (\bibinfo {year} {2010})},\ \Eprint
  {https://arxiv.org/abs/0912.5351} {arXiv:0912.5351 [gr-qc]} \BibitemShut
  {NoStop}%
\bibitem [{\citenamefont {Graham}\ and\ \citenamefont
  {Stamatopoulos}(2006)}]{Graham:2006xs}%
  \BibitemOpen
  \bibfield  {author} {\bibinfo {author} {\bibfnamefont {N.}~\bibnamefont
  {Graham}}\ and\ \bibinfo {author} {\bibfnamefont {N.}~\bibnamefont
  {Stamatopoulos}},\ }\href {https://doi.org/10.1016/j.physletb.2006.06.070}
  {\bibfield  {journal} {\bibinfo  {journal} {Phys. Lett. B}\ }\textbf
  {\bibinfo {volume} {639}},\ \bibinfo {pages} {541} (\bibinfo {year}
  {2006})},\ \Eprint {https://arxiv.org/abs/hep-th/0604134}
  {arXiv:hep-th/0604134} \BibitemShut {NoStop}%
\bibitem [{\citenamefont {Vakhitov}\ and\ \citenamefont
  {Kolokolov}(1973)}]{VKcondition}%
  \BibitemOpen
  \bibfield  {author} {\bibinfo {author} {\bibfnamefont {N.~G.}\ \bibnamefont
  {Vakhitov}}\ and\ \bibinfo {author} {\bibfnamefont {A.~A.}\ \bibnamefont
  {Kolokolov}},\ }\href {https://doi.org/10.1007/BF01031343} {\bibfield
  {journal} {\bibinfo  {journal} {Radiophysics and Quantum Electronics}\
  }\textbf {\bibinfo {volume} {16}},\ \bibinfo {pages} {783} (\bibinfo {year}
  {1973})}\BibitemShut {NoStop}%
\bibitem [{\citenamefont {Kasuya}\ \emph {et~al.}(2003)\citenamefont {Kasuya},
  \citenamefont {Kawasaki},\ and\ \citenamefont {Takahashi}}]{Kasuya:2002zs}%
  \BibitemOpen
  \bibfield  {author} {\bibinfo {author} {\bibfnamefont {S.}~\bibnamefont
  {Kasuya}}, \bibinfo {author} {\bibfnamefont {M.}~\bibnamefont {Kawasaki}},\
  and\ \bibinfo {author} {\bibfnamefont {F.}~\bibnamefont {Takahashi}},\ }\href
  {https://doi.org/10.1016/S0370-2693(03)00344-7} {\bibfield  {journal}
  {\bibinfo  {journal} {Phys. Lett. B}\ }\textbf {\bibinfo {volume} {559}},\
  \bibinfo {pages} {99} (\bibinfo {year} {2003})},\ \Eprint
  {https://arxiv.org/abs/hep-ph/0209358} {arXiv:hep-ph/0209358} \BibitemShut
  {NoStop}%
\bibitem [{\citenamefont {Kawasaki}\ \emph {et~al.}(2015)\citenamefont
  {Kawasaki}, \citenamefont {Takahashi},\ and\ \citenamefont
  {Takeda}}]{Kawasaki:2015vga}%
  \BibitemOpen
  \bibfield  {author} {\bibinfo {author} {\bibfnamefont {M.}~\bibnamefont
  {Kawasaki}}, \bibinfo {author} {\bibfnamefont {F.}~\bibnamefont
  {Takahashi}},\ and\ \bibinfo {author} {\bibfnamefont {N.}~\bibnamefont
  {Takeda}},\ }\href {https://doi.org/10.1103/PhysRevD.92.105024} {\bibfield
  {journal} {\bibinfo  {journal} {Phys. Rev. D}\ }\textbf {\bibinfo {volume}
  {92}},\ \bibinfo {pages} {105024} (\bibinfo {year} {2015})},\ \Eprint
  {https://arxiv.org/abs/1508.01028} {arXiv:1508.01028 [hep-th]} \BibitemShut
  {NoStop}%
\bibitem [{\citenamefont {Mukaida}\ \emph {et~al.}(2017)\citenamefont
  {Mukaida}, \citenamefont {Takimoto},\ and\ \citenamefont
  {Yamada}}]{Mukaida:2016hwd}%
  \BibitemOpen
  \bibfield  {author} {\bibinfo {author} {\bibfnamefont {K.}~\bibnamefont
  {Mukaida}}, \bibinfo {author} {\bibfnamefont {M.}~\bibnamefont {Takimoto}},\
  and\ \bibinfo {author} {\bibfnamefont {M.}~\bibnamefont {Yamada}},\ }\href
  {https://doi.org/10.1007/JHEP03(2017)122} {\bibfield  {journal} {\bibinfo
  {journal} {JHEP}\ }\textbf {\bibinfo {volume} {03}},\ \bibinfo {pages}
  {122}},\ \Eprint {https://arxiv.org/abs/1612.07750} {arXiv:1612.07750
  [hep-ph]} \BibitemShut {NoStop}%
\bibitem [{\citenamefont {Amin}\ and\ \citenamefont
  {Mocz}(2019)}]{Amin:2019ums}%
  \BibitemOpen
  \bibfield  {author} {\bibinfo {author} {\bibfnamefont {M.~A.}\ \bibnamefont
  {Amin}}\ and\ \bibinfo {author} {\bibfnamefont {P.}~\bibnamefont {Mocz}},\
  }\href {https://doi.org/10.1103/PhysRevD.100.063507} {\bibfield  {journal}
  {\bibinfo  {journal} {Phys. Rev. D}\ }\textbf {\bibinfo {volume} {100}},\
  \bibinfo {pages} {063507} (\bibinfo {year} {2019})},\ \Eprint
  {https://arxiv.org/abs/1902.07261} {arXiv:1902.07261 [astro-ph.CO]}
  \BibitemShut {NoStop}%
\bibitem [{\citenamefont {Levkov}\ \emph {et~al.}(2022)\citenamefont {Levkov},
  \citenamefont {Maslov}, \citenamefont {Nugaev},\ and\ \citenamefont
  {Panin}}]{Levkov:2022egq}%
  \BibitemOpen
  \bibfield  {author} {\bibinfo {author} {\bibfnamefont {D.~G.}\ \bibnamefont
  {Levkov}}, \bibinfo {author} {\bibfnamefont {V.~E.}\ \bibnamefont {Maslov}},
  \bibinfo {author} {\bibfnamefont {E.~Y.}\ \bibnamefont {Nugaev}},\ and\
  \bibinfo {author} {\bibfnamefont {A.~G.}\ \bibnamefont {Panin}},\ }\href
  {https://doi.org/10.1007/JHEP12(2022)079} {\bibfield  {journal} {\bibinfo
  {journal} {JHEP}\ }\textbf {\bibinfo {volume} {12}},\ \bibinfo {pages}
  {079}},\ \Eprint {https://arxiv.org/abs/2208.04334} {arXiv:2208.04334
  [hep-th]} \BibitemShut {NoStop}%
\bibitem [{\citenamefont {Zhang}(2024)}]{Zhang:2024bjo}%
  \BibitemOpen
  \bibfield  {author} {\bibinfo {author} {\bibfnamefont {H.-Y.}\ \bibnamefont
  {Zhang}},\ }\href@noop {} {\  (\bibinfo {year} {2024})},\ \Eprint
  {https://arxiv.org/abs/2406.05031} {arXiv:2406.05031 [hep-ph]} \BibitemShut
  {NoStop}%
\bibitem [{\citenamefont {Mukaida}\ and\ \citenamefont
  {Takimoto}(2014)}]{Mukaida:2014oza}%
  \BibitemOpen
  \bibfield  {author} {\bibinfo {author} {\bibfnamefont {K.}~\bibnamefont
  {Mukaida}}\ and\ \bibinfo {author} {\bibfnamefont {M.}~\bibnamefont
  {Takimoto}},\ }\href {https://doi.org/10.1088/1475-7516/2014/08/051}
  {\bibfield  {journal} {\bibinfo  {journal} {JCAP}\ }\textbf {\bibinfo
  {volume} {08}},\ \bibinfo {pages} {051}},\ \Eprint
  {https://arxiv.org/abs/1405.3233} {arXiv:1405.3233 [hep-ph]} \BibitemShut
  {NoStop}%
\bibitem [{\citenamefont {Zhang}(2021)}]{Zhang:2020ntm}%
  \BibitemOpen
  \bibfield  {author} {\bibinfo {author} {\bibfnamefont {H.-Y.}\ \bibnamefont
  {Zhang}},\ }\href {https://doi.org/10.1088/1475-7516/2021/03/102} {\bibfield
  {journal} {\bibinfo  {journal} {JCAP}\ }\textbf {\bibinfo {volume} {03}},\
  \bibinfo {pages} {102}},\ \Eprint {https://arxiv.org/abs/2011.11720}
  {arXiv:2011.11720 [hep-th]} \BibitemShut {NoStop}%
\bibitem [{\citenamefont {Ibe}\ \emph {et~al.}(2019)\citenamefont {Ibe},
  \citenamefont {Kawasaki}, \citenamefont {Nakano},\ and\ \citenamefont
  {Sonomoto}}]{Ibe:2019vyo}%
  \BibitemOpen
  \bibfield  {author} {\bibinfo {author} {\bibfnamefont {M.}~\bibnamefont
  {Ibe}}, \bibinfo {author} {\bibfnamefont {M.}~\bibnamefont {Kawasaki}},
  \bibinfo {author} {\bibfnamefont {W.}~\bibnamefont {Nakano}},\ and\ \bibinfo
  {author} {\bibfnamefont {E.}~\bibnamefont {Sonomoto}},\ }\href
  {https://doi.org/10.1007/JHEP04(2019)030} {\bibfield  {journal} {\bibinfo
  {journal} {JHEP}\ }\textbf {\bibinfo {volume} {04}},\ \bibinfo {pages}
  {030}},\ \Eprint {https://arxiv.org/abs/1901.06130} {arXiv:1901.06130
  [hep-ph]} \BibitemShut {NoStop}%
\bibitem [{\citenamefont {Saffin}\ \emph {et~al.}(2014)\citenamefont {Saffin},
  \citenamefont {Tognarelli},\ and\ \citenamefont {Tranberg}}]{Saffin:2014yka}%
  \BibitemOpen
  \bibfield  {author} {\bibinfo {author} {\bibfnamefont {P.~M.}\ \bibnamefont
  {Saffin}}, \bibinfo {author} {\bibfnamefont {P.}~\bibnamefont {Tognarelli}},\
  and\ \bibinfo {author} {\bibfnamefont {A.}~\bibnamefont {Tranberg}},\ }\href
  {https://doi.org/10.1007/JHEP08(2014)125} {\bibfield  {journal} {\bibinfo
  {journal} {JHEP}\ }\textbf {\bibinfo {volume} {08}},\ \bibinfo {pages}
  {125}},\ \Eprint {https://arxiv.org/abs/1401.6168} {arXiv:1401.6168 [hep-ph]}
  \BibitemShut {NoStop}%
\bibitem [{\citenamefont {Workman}\ \emph {et~al.}(2022)\citenamefont {Workman}
  \emph {et~al.}}]{ParticleDataGroup:2022pth}%
  \BibitemOpen
  \bibfield  {author} {\bibinfo {author} {\bibfnamefont {R.~L.}\ \bibnamefont
  {Workman}} \emph {et~al.} (\bibinfo {collaboration} {Particle Data Group}),\
  }\href {https://doi.org/10.1093/ptep/ptac097} {\bibfield  {journal} {\bibinfo
   {journal} {PTEP}\ }\textbf {\bibinfo {volume} {2022}},\ \bibinfo {pages}
  {083C01} (\bibinfo {year} {2022})}\BibitemShut {NoStop}%
\bibitem [{\citenamefont {Dine}\ \emph
  {et~al.}(1996{\natexlab{a}})\citenamefont {Dine}, \citenamefont {Randall},\
  and\ \citenamefont {Thomas}}]{Dine:1995kz}%
  \BibitemOpen
  \bibfield  {author} {\bibinfo {author} {\bibfnamefont {M.}~\bibnamefont
  {Dine}}, \bibinfo {author} {\bibfnamefont {L.}~\bibnamefont {Randall}},\ and\
  \bibinfo {author} {\bibfnamefont {S.~D.}\ \bibnamefont {Thomas}},\ }\href
  {https://doi.org/10.1016/0550-3213(95)00538-2} {\bibfield  {journal}
  {\bibinfo  {journal} {Nucl. Phys. B}\ }\textbf {\bibinfo {volume} {458}},\
  \bibinfo {pages} {291} (\bibinfo {year} {1996}{\natexlab{a}})},\ \Eprint
  {https://arxiv.org/abs/hep-ph/9507453} {arXiv:hep-ph/9507453} \BibitemShut
  {NoStop}%
\bibitem [{\citenamefont {Buccella}\ \emph {et~al.}(1982)\citenamefont
  {Buccella}, \citenamefont {Derendinger}, \citenamefont {Ferrara},\ and\
  \citenamefont {Savoy}}]{Buccella:1982nx}%
  \BibitemOpen
  \bibfield  {author} {\bibinfo {author} {\bibfnamefont {F.}~\bibnamefont
  {Buccella}}, \bibinfo {author} {\bibfnamefont {J.~P.}\ \bibnamefont
  {Derendinger}}, \bibinfo {author} {\bibfnamefont {S.}~\bibnamefont
  {Ferrara}},\ and\ \bibinfo {author} {\bibfnamefont {C.~A.}\ \bibnamefont
  {Savoy}},\ }\href {https://doi.org/10.1016/0370-2693(82)90521-4} {\bibfield
  {journal} {\bibinfo  {journal} {Phys. Lett. B}\ }\textbf {\bibinfo {volume}
  {115}},\ \bibinfo {pages} {375} (\bibinfo {year} {1982})}\BibitemShut
  {NoStop}%
\bibitem [{\citenamefont {Affleck}\ \emph {et~al.}(1984)\citenamefont
  {Affleck}, \citenamefont {Dine},\ and\ \citenamefont
  {Seiberg}}]{Affleck:1983mk}%
  \BibitemOpen
  \bibfield  {author} {\bibinfo {author} {\bibfnamefont {I.}~\bibnamefont
  {Affleck}}, \bibinfo {author} {\bibfnamefont {M.}~\bibnamefont {Dine}},\ and\
  \bibinfo {author} {\bibfnamefont {N.}~\bibnamefont {Seiberg}},\ }\href
  {https://doi.org/10.1016/0550-3213(84)90058-0} {\bibfield  {journal}
  {\bibinfo  {journal} {Nucl. Phys. B}\ }\textbf {\bibinfo {volume} {241}},\
  \bibinfo {pages} {493} (\bibinfo {year} {1984})}\BibitemShut {NoStop}%
\bibitem [{\citenamefont {Gherghetta}\ \emph {et~al.}(1996)\citenamefont
  {Gherghetta}, \citenamefont {Kolda},\ and\ \citenamefont
  {Martin}}]{Gherghetta:1995dv}%
  \BibitemOpen
  \bibfield  {author} {\bibinfo {author} {\bibfnamefont {T.}~\bibnamefont
  {Gherghetta}}, \bibinfo {author} {\bibfnamefont {C.~F.}\ \bibnamefont
  {Kolda}},\ and\ \bibinfo {author} {\bibfnamefont {S.~P.}\ \bibnamefont
  {Martin}},\ }\href {https://doi.org/10.1016/0550-3213(96)00095-8} {\bibfield
  {journal} {\bibinfo  {journal} {Nucl. Phys. B}\ }\textbf {\bibinfo {volume}
  {468}},\ \bibinfo {pages} {37} (\bibinfo {year} {1996})},\ \Eprint
  {https://arxiv.org/abs/hep-ph/9510370} {arXiv:hep-ph/9510370} \BibitemShut
  {NoStop}%
\bibitem [{\citenamefont {Martin}(1998)}]{Martin:1997ns}%
  \BibitemOpen
  \bibfield  {author} {\bibinfo {author} {\bibfnamefont {S.~P.}\ \bibnamefont
  {Martin}},\ }\href {https://doi.org/10.1142/9789812839657_0001} {\bibfield
  {journal} {\bibinfo  {journal} {Adv. Ser. Direct. High Energy Phys.}\
  }\textbf {\bibinfo {volume} {18}},\ \bibinfo {pages} {1} (\bibinfo {year}
  {1998})},\ \Eprint {https://arxiv.org/abs/hep-ph/9709356}
  {arXiv:hep-ph/9709356} \BibitemShut {NoStop}%
\bibitem [{\citenamefont {Dine}\ and\ \citenamefont
  {Fischler}(1982)}]{Dine:1981gu}%
  \BibitemOpen
  \bibfield  {author} {\bibinfo {author} {\bibfnamefont {M.}~\bibnamefont
  {Dine}}\ and\ \bibinfo {author} {\bibfnamefont {W.}~\bibnamefont
  {Fischler}},\ }\href {https://doi.org/10.1016/0370-2693(82)91241-2}
  {\bibfield  {journal} {\bibinfo  {journal} {Phys. Lett. B}\ }\textbf
  {\bibinfo {volume} {110}},\ \bibinfo {pages} {227} (\bibinfo {year}
  {1982})}\BibitemShut {NoStop}%
\bibitem [{\citenamefont {Nappi}\ and\ \citenamefont
  {Ovrut}(1982)}]{Nappi:1982hm}%
  \BibitemOpen
  \bibfield  {author} {\bibinfo {author} {\bibfnamefont {C.~R.}\ \bibnamefont
  {Nappi}}\ and\ \bibinfo {author} {\bibfnamefont {B.~A.}\ \bibnamefont
  {Ovrut}},\ }\href {https://doi.org/10.1016/0370-2693(82)90418-X} {\bibfield
  {journal} {\bibinfo  {journal} {Phys. Lett. B}\ }\textbf {\bibinfo {volume}
  {113}},\ \bibinfo {pages} {175} (\bibinfo {year} {1982})}\BibitemShut
  {NoStop}%
\bibitem [{\citenamefont {Alvarez-Gaume}\ \emph {et~al.}(1982)\citenamefont
  {Alvarez-Gaume}, \citenamefont {Claudson},\ and\ \citenamefont
  {Wise}}]{Alvarez-Gaume:1981abe}%
  \BibitemOpen
  \bibfield  {author} {\bibinfo {author} {\bibfnamefont {L.}~\bibnamefont
  {Alvarez-Gaume}}, \bibinfo {author} {\bibfnamefont {M.}~\bibnamefont
  {Claudson}},\ and\ \bibinfo {author} {\bibfnamefont {M.~B.}\ \bibnamefont
  {Wise}},\ }\href {https://doi.org/10.1016/0550-3213(82)90138-9} {\bibfield
  {journal} {\bibinfo  {journal} {Nucl. Phys. B}\ }\textbf {\bibinfo {volume}
  {207}},\ \bibinfo {pages} {96} (\bibinfo {year} {1982})}\BibitemShut
  {NoStop}%
\bibitem [{\citenamefont {Dine}\ and\ \citenamefont
  {Nelson}(1993)}]{Dine:1993yw}%
  \BibitemOpen
  \bibfield  {author} {\bibinfo {author} {\bibfnamefont {M.}~\bibnamefont
  {Dine}}\ and\ \bibinfo {author} {\bibfnamefont {A.~E.}\ \bibnamefont
  {Nelson}},\ }\href {https://doi.org/10.1103/PhysRevD.48.1277} {\bibfield
  {journal} {\bibinfo  {journal} {Phys. Rev. D}\ }\textbf {\bibinfo {volume}
  {48}},\ \bibinfo {pages} {1277} (\bibinfo {year} {1993})},\ \Eprint
  {https://arxiv.org/abs/hep-ph/9303230} {arXiv:hep-ph/9303230} \BibitemShut
  {NoStop}%
\bibitem [{\citenamefont {Dine}\ \emph {et~al.}(1995)\citenamefont {Dine},
  \citenamefont {Nelson},\ and\ \citenamefont {Shirman}}]{Dine:1994vc}%
  \BibitemOpen
  \bibfield  {author} {\bibinfo {author} {\bibfnamefont {M.}~\bibnamefont
  {Dine}}, \bibinfo {author} {\bibfnamefont {A.~E.}\ \bibnamefont {Nelson}},\
  and\ \bibinfo {author} {\bibfnamefont {Y.}~\bibnamefont {Shirman}},\ }\href
  {https://doi.org/10.1103/PhysRevD.51.1362} {\bibfield  {journal} {\bibinfo
  {journal} {Phys. Rev. D}\ }\textbf {\bibinfo {volume} {51}},\ \bibinfo
  {pages} {1362} (\bibinfo {year} {1995})},\ \Eprint
  {https://arxiv.org/abs/hep-ph/9408384} {arXiv:hep-ph/9408384} \BibitemShut
  {NoStop}%
\bibitem [{\citenamefont {Dine}\ \emph
  {et~al.}(1996{\natexlab{b}})\citenamefont {Dine}, \citenamefont {Nelson},
  \citenamefont {Nir},\ and\ \citenamefont {Shirman}}]{Dine:1995ag}%
  \BibitemOpen
  \bibfield  {author} {\bibinfo {author} {\bibfnamefont {M.}~\bibnamefont
  {Dine}}, \bibinfo {author} {\bibfnamefont {A.~E.}\ \bibnamefont {Nelson}},
  \bibinfo {author} {\bibfnamefont {Y.}~\bibnamefont {Nir}},\ and\ \bibinfo
  {author} {\bibfnamefont {Y.}~\bibnamefont {Shirman}},\ }\href
  {https://doi.org/10.1103/PhysRevD.53.2658} {\bibfield  {journal} {\bibinfo
  {journal} {Phys. Rev. D}\ }\textbf {\bibinfo {volume} {53}},\ \bibinfo
  {pages} {2658} (\bibinfo {year} {1996}{\natexlab{b}})},\ \Eprint
  {https://arxiv.org/abs/hep-ph/9507378} {arXiv:hep-ph/9507378} \BibitemShut
  {NoStop}%
\bibitem [{\citenamefont {Kolda}(1998)}]{Kolda:1997wt}%
  \BibitemOpen
  \bibfield  {author} {\bibinfo {author} {\bibfnamefont {C.~F.}\ \bibnamefont
  {Kolda}},\ }\href {https://doi.org/10.1016/S0920-5632(97)00667-1} {\bibfield
  {journal} {\bibinfo  {journal} {Nucl. Phys. B Proc. Suppl.}\ }\textbf
  {\bibinfo {volume} {62}},\ \bibinfo {pages} {266} (\bibinfo {year} {1998})},\
  \Eprint {https://arxiv.org/abs/hep-ph/9707450} {arXiv:hep-ph/9707450}
  \BibitemShut {NoStop}%
\bibitem [{\citenamefont {Dvali}\ \emph {et~al.}(1994)\citenamefont {Dvali},
  \citenamefont {Shafi},\ and\ \citenamefont {Schaefer}}]{Dvali:1994ms}%
  \BibitemOpen
  \bibfield  {author} {\bibinfo {author} {\bibfnamefont {G.~R.}\ \bibnamefont
  {Dvali}}, \bibinfo {author} {\bibfnamefont {Q.}~\bibnamefont {Shafi}},\ and\
  \bibinfo {author} {\bibfnamefont {R.~K.}\ \bibnamefont {Schaefer}},\ }\href
  {https://doi.org/10.1103/PhysRevLett.73.1886} {\bibfield  {journal} {\bibinfo
   {journal} {Phys. Rev. Lett.}\ }\textbf {\bibinfo {volume} {73}},\ \bibinfo
  {pages} {1886} (\bibinfo {year} {1994})},\ \Eprint
  {https://arxiv.org/abs/hep-ph/9406319} {arXiv:hep-ph/9406319} \BibitemShut
  {NoStop}%
\bibitem [{\citenamefont {Dvali}\ \emph {et~al.}(1998)\citenamefont {Dvali},
  \citenamefont {Kusenko},\ and\ \citenamefont {Shaposhnikov}}]{Dvali:1997qv}%
  \BibitemOpen
  \bibfield  {author} {\bibinfo {author} {\bibfnamefont {G.~R.}\ \bibnamefont
  {Dvali}}, \bibinfo {author} {\bibfnamefont {A.}~\bibnamefont {Kusenko}},\
  and\ \bibinfo {author} {\bibfnamefont {M.~E.}\ \bibnamefont {Shaposhnikov}},\
  }\href {https://doi.org/10.1016/S0370-2693(97)01378-6} {\bibfield  {journal}
  {\bibinfo  {journal} {Phys. Lett. B}\ }\textbf {\bibinfo {volume} {417}},\
  \bibinfo {pages} {99} (\bibinfo {year} {1998})},\ \Eprint
  {https://arxiv.org/abs/hep-ph/9707423} {arXiv:hep-ph/9707423} \BibitemShut
  {NoStop}%
\bibitem [{\citenamefont {Chamseddine}\ \emph {et~al.}(1982)\citenamefont
  {Chamseddine}, \citenamefont {Arnowitt},\ and\ \citenamefont
  {Nath}}]{Chamseddine:1982jx}%
  \BibitemOpen
  \bibfield  {author} {\bibinfo {author} {\bibfnamefont {A.~H.}\ \bibnamefont
  {Chamseddine}}, \bibinfo {author} {\bibfnamefont {R.~L.}\ \bibnamefont
  {Arnowitt}},\ and\ \bibinfo {author} {\bibfnamefont {P.}~\bibnamefont
  {Nath}},\ }\href {https://doi.org/10.1103/PhysRevLett.49.970} {\bibfield
  {journal} {\bibinfo  {journal} {Phys. Rev. Lett.}\ }\textbf {\bibinfo
  {volume} {49}},\ \bibinfo {pages} {970} (\bibinfo {year} {1982})}\BibitemShut
  {NoStop}%
\bibitem [{\citenamefont {Barbieri}\ \emph {et~al.}(1982)\citenamefont
  {Barbieri}, \citenamefont {Ferrara},\ and\ \citenamefont
  {Savoy}}]{Barbieri:1982eh}%
  \BibitemOpen
  \bibfield  {author} {\bibinfo {author} {\bibfnamefont {R.}~\bibnamefont
  {Barbieri}}, \bibinfo {author} {\bibfnamefont {S.}~\bibnamefont {Ferrara}},\
  and\ \bibinfo {author} {\bibfnamefont {C.~A.}\ \bibnamefont {Savoy}},\ }\href
  {https://doi.org/10.1016/0370-2693(82)90685-2} {\bibfield  {journal}
  {\bibinfo  {journal} {Phys. Lett. B}\ }\textbf {\bibinfo {volume} {119}},\
  \bibinfo {pages} {343} (\bibinfo {year} {1982})}\BibitemShut {NoStop}%
\bibitem [{\citenamefont {Ibanez}(1982)}]{Ibanez:1982ee}%
  \BibitemOpen
  \bibfield  {author} {\bibinfo {author} {\bibfnamefont {L.~E.}\ \bibnamefont
  {Ibanez}},\ }\href {https://doi.org/10.1016/0370-2693(82)90604-9} {\bibfield
  {journal} {\bibinfo  {journal} {Phys. Lett. B}\ }\textbf {\bibinfo {volume}
  {118}},\ \bibinfo {pages} {73} (\bibinfo {year} {1982})}\BibitemShut
  {NoStop}%
\bibitem [{\citenamefont {Ellis}\ \emph {et~al.}(1983)\citenamefont {Ellis},
  \citenamefont {Nanopoulos},\ and\ \citenamefont {Tamvakis}}]{Ellis:1982wr}%
  \BibitemOpen
  \bibfield  {author} {\bibinfo {author} {\bibfnamefont {J.~R.}\ \bibnamefont
  {Ellis}}, \bibinfo {author} {\bibfnamefont {D.~V.}\ \bibnamefont
  {Nanopoulos}},\ and\ \bibinfo {author} {\bibfnamefont {K.}~\bibnamefont
  {Tamvakis}},\ }\href {https://doi.org/10.1016/0370-2693(83)90900-0}
  {\bibfield  {journal} {\bibinfo  {journal} {Phys. Lett. B}\ }\textbf
  {\bibinfo {volume} {121}},\ \bibinfo {pages} {123} (\bibinfo {year}
  {1983})}\BibitemShut {NoStop}%
\bibitem [{\citenamefont {Hall}\ \emph {et~al.}(1983)\citenamefont {Hall},
  \citenamefont {Lykken},\ and\ \citenamefont {Weinberg}}]{Hall:1983iz}%
  \BibitemOpen
  \bibfield  {author} {\bibinfo {author} {\bibfnamefont {L.~J.}\ \bibnamefont
  {Hall}}, \bibinfo {author} {\bibfnamefont {J.~D.}\ \bibnamefont {Lykken}},\
  and\ \bibinfo {author} {\bibfnamefont {S.}~\bibnamefont {Weinberg}},\ }\href
  {https://doi.org/10.1103/PhysRevD.27.2359} {\bibfield  {journal} {\bibinfo
  {journal} {Phys. Rev. D}\ }\textbf {\bibinfo {volume} {27}},\ \bibinfo
  {pages} {2359} (\bibinfo {year} {1983})}\BibitemShut {NoStop}%
\bibitem [{\citenamefont {Alvarez-Gaume}\ \emph {et~al.}(1983)\citenamefont
  {Alvarez-Gaume}, \citenamefont {Polchinski},\ and\ \citenamefont
  {Wise}}]{Alvarez-Gaume:1983drc}%
  \BibitemOpen
  \bibfield  {author} {\bibinfo {author} {\bibfnamefont {L.}~\bibnamefont
  {Alvarez-Gaume}}, \bibinfo {author} {\bibfnamefont {J.}~\bibnamefont
  {Polchinski}},\ and\ \bibinfo {author} {\bibfnamefont {M.~B.}\ \bibnamefont
  {Wise}},\ }\href {https://doi.org/10.1016/0550-3213(83)90591-6} {\bibfield
  {journal} {\bibinfo  {journal} {Nucl. Phys. B}\ }\textbf {\bibinfo {volume}
  {221}},\ \bibinfo {pages} {495} (\bibinfo {year} {1983})}\BibitemShut
  {NoStop}%
\bibitem [{\citenamefont {Ellis}\ \emph {et~al.}(2023)\citenamefont {Ellis},
  \citenamefont {Olive}, \citenamefont {Spanos},\ and\ \citenamefont
  {Stamou}}]{Ellis:2022emx}%
  \BibitemOpen
  \bibfield  {author} {\bibinfo {author} {\bibfnamefont {J.}~\bibnamefont
  {Ellis}}, \bibinfo {author} {\bibfnamefont {K.~A.}\ \bibnamefont {Olive}},
  \bibinfo {author} {\bibfnamefont {V.~C.}\ \bibnamefont {Spanos}},\ and\
  \bibinfo {author} {\bibfnamefont {I.~D.}\ \bibnamefont {Stamou}},\ }\href
  {https://doi.org/10.1140/epjc/s10052-023-11405-1} {\bibfield  {journal}
  {\bibinfo  {journal} {Eur. Phys. J. C}\ }\textbf {\bibinfo {volume} {83}},\
  \bibinfo {pages} {246} (\bibinfo {year} {2023})},\ \Eprint
  {https://arxiv.org/abs/2210.16337} {arXiv:2210.16337 [hep-ph]} \BibitemShut
  {NoStop}%
\bibitem [{\citenamefont {Nilles}(1984)}]{Nilles:1983ge}%
  \BibitemOpen
  \bibfield  {author} {\bibinfo {author} {\bibfnamefont {H.~P.}\ \bibnamefont
  {Nilles}},\ }\href {https://doi.org/10.1016/0370-1573(84)90008-5} {\bibfield
  {journal} {\bibinfo  {journal} {Phys. Rept.}\ }\textbf {\bibinfo {volume}
  {110}},\ \bibinfo {pages} {1} (\bibinfo {year} {1984})}\BibitemShut {NoStop}%
\bibitem [{\citenamefont {Einhorn}\ and\ \citenamefont
  {Jones}(1983)}]{Einhorn:1982pp}%
  \BibitemOpen
  \bibfield  {author} {\bibinfo {author} {\bibfnamefont {M.~B.}\ \bibnamefont
  {Einhorn}}\ and\ \bibinfo {author} {\bibfnamefont {D.~R.~T.}\ \bibnamefont
  {Jones}},\ }\href {https://doi.org/10.1016/0550-3213(83)90184-0} {\bibfield
  {journal} {\bibinfo  {journal} {Nucl. Phys. B}\ }\textbf {\bibinfo {volume}
  {211}},\ \bibinfo {pages} {29} (\bibinfo {year} {1983})}\BibitemShut
  {NoStop}%
\bibitem [{\citenamefont {Kasuya}\ and\ \citenamefont
  {Kawasaki}(2000{\natexlab{c}})}]{Kasuya:2000sc}%
  \BibitemOpen
  \bibfield  {author} {\bibinfo {author} {\bibfnamefont {S.}~\bibnamefont
  {Kasuya}}\ and\ \bibinfo {author} {\bibfnamefont {M.}~\bibnamefont
  {Kawasaki}},\ }\href {https://doi.org/10.1103/PhysRevLett.85.2677} {\bibfield
   {journal} {\bibinfo  {journal} {Phys. Rev. Lett.}\ }\textbf {\bibinfo
  {volume} {85}},\ \bibinfo {pages} {2677} (\bibinfo {year}
  {2000}{\natexlab{c}})},\ \Eprint {https://arxiv.org/abs/hep-ph/0006128}
  {arXiv:hep-ph/0006128} \BibitemShut {NoStop}%
\bibitem [{\citenamefont {Kasuya}\ and\ \citenamefont
  {Kawasaki}(2014)}]{Kasuya:2014ofa}%
  \BibitemOpen
  \bibfield  {author} {\bibinfo {author} {\bibfnamefont {S.}~\bibnamefont
  {Kasuya}}\ and\ \bibinfo {author} {\bibfnamefont {M.}~\bibnamefont
  {Kawasaki}},\ }\href {https://doi.org/10.1103/PhysRevD.89.103534} {\bibfield
  {journal} {\bibinfo  {journal} {Phys. Rev. D}\ }\textbf {\bibinfo {volume}
  {89}},\ \bibinfo {pages} {103534} (\bibinfo {year} {2014})},\ \Eprint
  {https://arxiv.org/abs/1402.4546} {arXiv:1402.4546 [hep-ph]} \BibitemShut
  {NoStop}%
\bibitem [{\citenamefont {Frieman}\ \emph {et~al.}(1988)\citenamefont
  {Frieman}, \citenamefont {Gelmini}, \citenamefont {Gleiser},\ and\
  \citenamefont {Kolb}}]{Frieman:1988ut}%
  \BibitemOpen
  \bibfield  {author} {\bibinfo {author} {\bibfnamefont {J.~A.}\ \bibnamefont
  {Frieman}}, \bibinfo {author} {\bibfnamefont {G.~B.}\ \bibnamefont
  {Gelmini}}, \bibinfo {author} {\bibfnamefont {M.}~\bibnamefont {Gleiser}},\
  and\ \bibinfo {author} {\bibfnamefont {E.~W.}\ \bibnamefont {Kolb}},\ }\href
  {https://doi.org/10.1103/PhysRevLett.60.2101} {\bibfield  {journal} {\bibinfo
   {journal} {Phys. Rev. Lett.}\ }\textbf {\bibinfo {volume} {60}},\ \bibinfo
  {pages} {2101} (\bibinfo {year} {1988})}\BibitemShut {NoStop}%
\bibitem [{\citenamefont {Frieman}\ \emph {et~al.}(1989)\citenamefont
  {Frieman}, \citenamefont {Olinto}, \citenamefont {Gleiser},\ and\
  \citenamefont {Alcock}}]{Frieman:1989bx}%
  \BibitemOpen
  \bibfield  {author} {\bibinfo {author} {\bibfnamefont {J.~A.}\ \bibnamefont
  {Frieman}}, \bibinfo {author} {\bibfnamefont {A.~V.}\ \bibnamefont {Olinto}},
  \bibinfo {author} {\bibfnamefont {M.}~\bibnamefont {Gleiser}},\ and\ \bibinfo
  {author} {\bibfnamefont {C.}~\bibnamefont {Alcock}},\ }\href
  {https://doi.org/10.1103/PhysRevD.40.3241} {\bibfield  {journal} {\bibinfo
  {journal} {Phys. Rev. D}\ }\textbf {\bibinfo {volume} {40}},\ \bibinfo
  {pages} {3241} (\bibinfo {year} {1989})}\BibitemShut {NoStop}%
\bibitem [{\citenamefont {Griest}\ \emph {et~al.}(1989)\citenamefont {Griest},
  \citenamefont {Kolb},\ and\ \citenamefont {Massarotti}}]{Griest:1989cb}%
  \BibitemOpen
  \bibfield  {author} {\bibinfo {author} {\bibfnamefont {K.}~\bibnamefont
  {Griest}}, \bibinfo {author} {\bibfnamefont {E.~W.}\ \bibnamefont {Kolb}},\
  and\ \bibinfo {author} {\bibfnamefont {A.}~\bibnamefont {Massarotti}},\
  }\href {https://doi.org/10.1103/PhysRevD.40.3529} {\bibfield  {journal}
  {\bibinfo  {journal} {Phys. Rev. D}\ }\textbf {\bibinfo {volume} {40}},\
  \bibinfo {pages} {3529} (\bibinfo {year} {1989})}\BibitemShut {NoStop}%
\bibitem [{\citenamefont {Postma}(2002)}]{Postma:2001ea}%
  \BibitemOpen
  \bibfield  {author} {\bibinfo {author} {\bibfnamefont {M.}~\bibnamefont
  {Postma}},\ }\href {https://doi.org/10.1103/PhysRevD.65.085035} {\bibfield
  {journal} {\bibinfo  {journal} {Phys. Rev. D}\ }\textbf {\bibinfo {volume}
  {65}},\ \bibinfo {pages} {085035} (\bibinfo {year} {2002})},\ \Eprint
  {https://arxiv.org/abs/hep-ph/0110199} {arXiv:hep-ph/0110199} \BibitemShut
  {NoStop}%
\bibitem [{\citenamefont {Griest}\ and\ \citenamefont
  {Kolb}(1989)}]{Griest:1989bq}%
  \BibitemOpen
  \bibfield  {author} {\bibinfo {author} {\bibfnamefont {K.}~\bibnamefont
  {Griest}}\ and\ \bibinfo {author} {\bibfnamefont {E.~W.}\ \bibnamefont
  {Kolb}},\ }\href {https://doi.org/10.1103/PhysRevD.40.3231} {\bibfield
  {journal} {\bibinfo  {journal} {Phys. Rev. D}\ }\textbf {\bibinfo {volume}
  {40}},\ \bibinfo {pages} {3231} (\bibinfo {year} {1989})}\BibitemShut
  {NoStop}%
\bibitem [{\citenamefont {Croon}\ \emph {et~al.}(2020)\citenamefont {Croon},
  \citenamefont {Kusenko}, \citenamefont {Mazumdar},\ and\ \citenamefont
  {White}}]{Croon:2019rqu}%
  \BibitemOpen
  \bibfield  {author} {\bibinfo {author} {\bibfnamefont {D.}~\bibnamefont
  {Croon}}, \bibinfo {author} {\bibfnamefont {A.}~\bibnamefont {Kusenko}},
  \bibinfo {author} {\bibfnamefont {A.}~\bibnamefont {Mazumdar}},\ and\
  \bibinfo {author} {\bibfnamefont {G.}~\bibnamefont {White}},\ }\href
  {https://doi.org/10.1103/PhysRevD.101.085010} {\bibfield  {journal} {\bibinfo
   {journal} {Phys. Rev. D}\ }\textbf {\bibinfo {volume} {101}},\ \bibinfo
  {pages} {085010} (\bibinfo {year} {2020})},\ \Eprint
  {https://arxiv.org/abs/1910.09562} {arXiv:1910.09562 [hep-ph]} \BibitemShut
  {NoStop}%
\bibitem [{\citenamefont {Krylov}\ \emph {et~al.}(2013)\citenamefont {Krylov},
  \citenamefont {Levin},\ and\ \citenamefont {Rubakov}}]{Krylov:2013qe}%
  \BibitemOpen
  \bibfield  {author} {\bibinfo {author} {\bibfnamefont {E.}~\bibnamefont
  {Krylov}}, \bibinfo {author} {\bibfnamefont {A.}~\bibnamefont {Levin}},\ and\
  \bibinfo {author} {\bibfnamefont {V.}~\bibnamefont {Rubakov}},\ }\href
  {https://doi.org/10.1103/PhysRevD.87.083528} {\bibfield  {journal} {\bibinfo
  {journal} {Phys. Rev. D}\ }\textbf {\bibinfo {volume} {87}},\ \bibinfo
  {pages} {083528} (\bibinfo {year} {2013})},\ \Eprint
  {https://arxiv.org/abs/1301.0354} {arXiv:1301.0354 [hep-ph]} \BibitemShut
  {NoStop}%
\bibitem [{\citenamefont {Kasuya}\ \emph {et~al.}(2024)\citenamefont {Kasuya},
  \citenamefont {Kawasaki},\ and\ \citenamefont {Tsuji}}]{Kasuya:2024ldq}%
  \BibitemOpen
  \bibfield  {author} {\bibinfo {author} {\bibfnamefont {S.}~\bibnamefont
  {Kasuya}}, \bibinfo {author} {\bibfnamefont {M.}~\bibnamefont {Kawasaki}},\
  and\ \bibinfo {author} {\bibfnamefont {N.}~\bibnamefont {Tsuji}},\ }\href
  {https://doi.org/10.1103/PhysRevD.109.083039} {\bibfield  {journal} {\bibinfo
   {journal} {Phys. Rev. D}\ }\textbf {\bibinfo {volume} {109}},\ \bibinfo
  {pages} {083039} (\bibinfo {year} {2024})},\ \Eprint
  {https://arxiv.org/abs/2403.01675} {arXiv:2403.01675 [hep-ph]} \BibitemShut
  {NoStop}%
\bibitem [{\citenamefont {Lloyd-Stubbs}\ and\ \citenamefont
  {McDonald}(2022)}]{Lloyd-Stubbs:2021xlk}%
  \BibitemOpen
  \bibfield  {author} {\bibinfo {author} {\bibfnamefont {A.~K.}\ \bibnamefont
  {Lloyd-Stubbs}}\ and\ \bibinfo {author} {\bibfnamefont {J.}~\bibnamefont
  {McDonald}},\ }\href {https://doi.org/10.1103/PhysRevD.105.103532} {\bibfield
   {journal} {\bibinfo  {journal} {Phys. Rev. D}\ }\textbf {\bibinfo {volume}
  {105}},\ \bibinfo {pages} {103532} (\bibinfo {year} {2022})},\ \Eprint
  {https://arxiv.org/abs/2112.09121} {arXiv:2112.09121 [hep-th]} \BibitemShut
  {NoStop}%
\bibitem [{\citenamefont {Affleck}\ and\ \citenamefont
  {Dine}(1985)}]{Affleck:1984fy}%
  \BibitemOpen
  \bibfield  {author} {\bibinfo {author} {\bibfnamefont {I.}~\bibnamefont
  {Affleck}}\ and\ \bibinfo {author} {\bibfnamefont {M.}~\bibnamefont {Dine}},\
  }\href {https://doi.org/10.1016/0550-3213(85)90021-5} {\bibfield  {journal}
  {\bibinfo  {journal} {Nucl. Phys. B}\ }\textbf {\bibinfo {volume} {249}},\
  \bibinfo {pages} {361} (\bibinfo {year} {1985})}\BibitemShut {NoStop}%
\bibitem [{\citenamefont {Allahverdi}\ \emph {et~al.}(2000)\citenamefont
  {Allahverdi}, \citenamefont {Campbell},\ and\ \citenamefont
  {Ellis}}]{Allahverdi:2000zd}%
  \BibitemOpen
  \bibfield  {author} {\bibinfo {author} {\bibfnamefont {R.}~\bibnamefont
  {Allahverdi}}, \bibinfo {author} {\bibfnamefont {B.~A.}\ \bibnamefont
  {Campbell}},\ and\ \bibinfo {author} {\bibfnamefont {J.~R.}\ \bibnamefont
  {Ellis}},\ }\href {https://doi.org/10.1016/S0550-3213(00)00124-3} {\bibfield
  {journal} {\bibinfo  {journal} {Nucl. Phys. B}\ }\textbf {\bibinfo {volume}
  {579}},\ \bibinfo {pages} {355} (\bibinfo {year} {2000})},\ \Eprint
  {https://arxiv.org/abs/hep-ph/0001122} {arXiv:hep-ph/0001122} \BibitemShut
  {NoStop}%
\bibitem [{\citenamefont {Anisimov}\ and\ \citenamefont
  {Dine}(2001)}]{Anisimov:2000wx}%
  \BibitemOpen
  \bibfield  {author} {\bibinfo {author} {\bibfnamefont {A.}~\bibnamefont
  {Anisimov}}\ and\ \bibinfo {author} {\bibfnamefont {M.}~\bibnamefont
  {Dine}},\ }\href {https://doi.org/10.1016/S0550-3213(01)00550-8} {\bibfield
  {journal} {\bibinfo  {journal} {Nucl. Phys. B}\ }\textbf {\bibinfo {volume}
  {619}},\ \bibinfo {pages} {729} (\bibinfo {year} {2001})},\ \Eprint
  {https://arxiv.org/abs/hep-ph/0008058} {arXiv:hep-ph/0008058} \BibitemShut
  {NoStop}%
\bibitem [{\citenamefont {Coughlan}\ \emph {et~al.}(1983)\citenamefont
  {Coughlan}, \citenamefont {Fischler}, \citenamefont {Kolb}, \citenamefont
  {Raby},\ and\ \citenamefont {Ross}}]{Coughlan:1983ci}%
  \BibitemOpen
  \bibfield  {author} {\bibinfo {author} {\bibfnamefont {G.~D.}\ \bibnamefont
  {Coughlan}}, \bibinfo {author} {\bibfnamefont {W.}~\bibnamefont {Fischler}},
  \bibinfo {author} {\bibfnamefont {E.~W.}\ \bibnamefont {Kolb}}, \bibinfo
  {author} {\bibfnamefont {S.}~\bibnamefont {Raby}},\ and\ \bibinfo {author}
  {\bibfnamefont {G.~G.}\ \bibnamefont {Ross}},\ }\href
  {https://doi.org/10.1016/0370-2693(83)91091-2} {\bibfield  {journal}
  {\bibinfo  {journal} {Phys. Lett. B}\ }\textbf {\bibinfo {volume} {131}},\
  \bibinfo {pages} {59} (\bibinfo {year} {1983})}\BibitemShut {NoStop}%
\bibitem [{\citenamefont {Turner}(1983)}]{Turner:1983he}%
  \BibitemOpen
  \bibfield  {author} {\bibinfo {author} {\bibfnamefont {M.~S.}\ \bibnamefont
  {Turner}},\ }\href {https://doi.org/10.1103/PhysRevD.28.1243} {\bibfield
  {journal} {\bibinfo  {journal} {Phys. Rev. D}\ }\textbf {\bibinfo {volume}
  {28}},\ \bibinfo {pages} {1243} (\bibinfo {year} {1983})}\BibitemShut
  {NoStop}%
\bibitem [{\citenamefont {Enqvist}\ and\ \citenamefont
  {McDonald}(2000)}]{Enqvist:1999mv}%
  \BibitemOpen
  \bibfield  {author} {\bibinfo {author} {\bibfnamefont {K.}~\bibnamefont
  {Enqvist}}\ and\ \bibinfo {author} {\bibfnamefont {J.}~\bibnamefont
  {McDonald}},\ }\href {https://doi.org/10.1016/S0550-3213(99)00776-2}
  {\bibfield  {journal} {\bibinfo  {journal} {Nucl. Phys. B}\ }\textbf
  {\bibinfo {volume} {570}},\ \bibinfo {pages} {407} (\bibinfo {year}
  {2000})},\ \bibinfo {note} {[Erratum: Nucl.Phys.B 582, 763--763 (2000)]},\
  \Eprint {https://arxiv.org/abs/hep-ph/9908316} {arXiv:hep-ph/9908316}
  \BibitemShut {NoStop}%
\bibitem [{\citenamefont {Multamaki}\ and\ \citenamefont
  {Vilja}(2000{\natexlab{b}})}]{Multamaki:2000ey}%
  \BibitemOpen
  \bibfield  {author} {\bibinfo {author} {\bibfnamefont {T.}~\bibnamefont
  {Multamaki}}\ and\ \bibinfo {author} {\bibfnamefont {I.}~\bibnamefont
  {Vilja}},\ }\href {https://doi.org/10.1016/S0370-2693(00)00679-1} {\bibfield
  {journal} {\bibinfo  {journal} {Phys. Lett. B}\ }\textbf {\bibinfo {volume}
  {484}},\ \bibinfo {pages} {283} (\bibinfo {year} {2000}{\natexlab{b}})},\
  \Eprint {https://arxiv.org/abs/hep-ph/0005162} {arXiv:hep-ph/0005162}
  \BibitemShut {NoStop}%
\bibitem [{\citenamefont {Kasuya}\ and\ \citenamefont
  {Kawasaki}(2001)}]{Kasuya:2001hg}%
  \BibitemOpen
  \bibfield  {author} {\bibinfo {author} {\bibfnamefont {S.}~\bibnamefont
  {Kasuya}}\ and\ \bibinfo {author} {\bibfnamefont {M.}~\bibnamefont
  {Kawasaki}},\ }\href {https://doi.org/10.1103/PhysRevD.64.123515} {\bibfield
  {journal} {\bibinfo  {journal} {Phys. Rev. D}\ }\textbf {\bibinfo {volume}
  {64}},\ \bibinfo {pages} {123515} (\bibinfo {year} {2001})},\ \Eprint
  {https://arxiv.org/abs/hep-ph/0106119} {arXiv:hep-ph/0106119} \BibitemShut
  {NoStop}%
\bibitem [{\citenamefont {Multamaki}\ and\ \citenamefont
  {Vilja}(2002{\natexlab{a}})}]{Multamaki:2002hv}%
  \BibitemOpen
  \bibfield  {author} {\bibinfo {author} {\bibfnamefont {T.}~\bibnamefont
  {Multamaki}}\ and\ \bibinfo {author} {\bibfnamefont {I.}~\bibnamefont
  {Vilja}},\ }\href {https://doi.org/10.1016/S0370-2693(02)01730-6} {\bibfield
  {journal} {\bibinfo  {journal} {Phys. Lett. B}\ }\textbf {\bibinfo {volume}
  {535}},\ \bibinfo {pages} {170} (\bibinfo {year} {2002}{\natexlab{a}})},\
  \Eprint {https://arxiv.org/abs/hep-ph/0203195} {arXiv:hep-ph/0203195}
  \BibitemShut {NoStop}%
\bibitem [{\citenamefont {Tsumagari}(2009)}]{Tsumagari:2009na}%
  \BibitemOpen
  \bibfield  {author} {\bibinfo {author} {\bibfnamefont {M.~I.}\ \bibnamefont
  {Tsumagari}},\ }\href {https://doi.org/10.1103/PhysRevD.80.085010} {\bibfield
   {journal} {\bibinfo  {journal} {Phys. Rev. D}\ }\textbf {\bibinfo {volume}
  {80}},\ \bibinfo {pages} {085010} (\bibinfo {year} {2009})},\ \Eprint
  {https://arxiv.org/abs/0907.4197} {arXiv:0907.4197 [hep-th]} \BibitemShut
  {NoStop}%
\bibitem [{\citenamefont {Hiramatsu}\ \emph {et~al.}(2010)\citenamefont
  {Hiramatsu}, \citenamefont {Kawasaki},\ and\ \citenamefont
  {Takahashi}}]{Hiramatsu:2010dx}%
  \BibitemOpen
  \bibfield  {author} {\bibinfo {author} {\bibfnamefont {T.}~\bibnamefont
  {Hiramatsu}}, \bibinfo {author} {\bibfnamefont {M.}~\bibnamefont
  {Kawasaki}},\ and\ \bibinfo {author} {\bibfnamefont {F.}~\bibnamefont
  {Takahashi}},\ }\href {https://doi.org/10.1088/1475-7516/2010/06/008}
  {\bibfield  {journal} {\bibinfo  {journal} {JCAP}\ }\textbf {\bibinfo
  {volume} {06}},\ \bibinfo {pages} {008}},\ \Eprint
  {https://arxiv.org/abs/1003.1779} {arXiv:1003.1779 [hep-ph]} \BibitemShut
  {NoStop}%
\bibitem [{\citenamefont {Chiba}\ \emph
  {et~al.}(2010{\natexlab{a}})\citenamefont {Chiba}, \citenamefont {Kamada},
  \citenamefont {Kasuya},\ and\ \citenamefont {Yamaguchi}}]{Chiba:2010ff}%
  \BibitemOpen
  \bibfield  {author} {\bibinfo {author} {\bibfnamefont {T.}~\bibnamefont
  {Chiba}}, \bibinfo {author} {\bibfnamefont {K.}~\bibnamefont {Kamada}},
  \bibinfo {author} {\bibfnamefont {S.}~\bibnamefont {Kasuya}},\ and\ \bibinfo
  {author} {\bibfnamefont {M.}~\bibnamefont {Yamaguchi}},\ }\href
  {https://doi.org/10.1103/PhysRevD.82.103534} {\bibfield  {journal} {\bibinfo
  {journal} {Phys. Rev. D}\ }\textbf {\bibinfo {volume} {82}},\ \bibinfo
  {pages} {103534} (\bibinfo {year} {2010}{\natexlab{a}})},\ \Eprint
  {https://arxiv.org/abs/1007.4235} {arXiv:1007.4235 [hep-ph]} \BibitemShut
  {NoStop}%
\bibitem [{\citenamefont {Zhou}(2015)}]{Zhou:2015yfa}%
  \BibitemOpen
  \bibfield  {author} {\bibinfo {author} {\bibfnamefont {S.-Y.}\ \bibnamefont
  {Zhou}},\ }\href {https://doi.org/10.1088/1475-7516/2015/06/033} {\bibfield
  {journal} {\bibinfo  {journal} {JCAP}\ }\textbf {\bibinfo {volume} {06}},\
  \bibinfo {pages} {033}},\ \Eprint {https://arxiv.org/abs/1501.01217}
  {arXiv:1501.01217 [astro-ph.CO]} \BibitemShut {NoStop}%
\bibitem [{\citenamefont {Wang}\ and\ \citenamefont
  {Wang}(2022)}]{Wang:2021rfk}%
  \BibitemOpen
  \bibfield  {author} {\bibinfo {author} {\bibfnamefont {F.}~\bibnamefont
  {Wang}}\ and\ \bibinfo {author} {\bibfnamefont {R.}~\bibnamefont {Wang}},\
  }\href {https://doi.org/10.1140/epjc/s10052-022-10291-3} {\bibfield
  {journal} {\bibinfo  {journal} {Eur. Phys. J. C}\ }\textbf {\bibinfo {volume}
  {82}},\ \bibinfo {pages} {325} (\bibinfo {year} {2022})},\ \Eprint
  {https://arxiv.org/abs/2104.04682} {arXiv:2104.04682 [gr-qc]} \BibitemShut
  {NoStop}%
\bibitem [{\citenamefont {Enqvist}\ \emph
  {et~al.}(2002{\natexlab{a}})\citenamefont {Enqvist}, \citenamefont {Kasuya},\
  and\ \citenamefont {Mazumdar}}]{Enqvist:2002si}%
  \BibitemOpen
  \bibfield  {author} {\bibinfo {author} {\bibfnamefont {K.}~\bibnamefont
  {Enqvist}}, \bibinfo {author} {\bibfnamefont {S.}~\bibnamefont {Kasuya}},\
  and\ \bibinfo {author} {\bibfnamefont {A.}~\bibnamefont {Mazumdar}},\ }\href
  {https://doi.org/10.1103/PhysRevD.66.043505} {\bibfield  {journal} {\bibinfo
  {journal} {Phys. Rev. D}\ }\textbf {\bibinfo {volume} {66}},\ \bibinfo
  {pages} {043505} (\bibinfo {year} {2002}{\natexlab{a}})},\ \Eprint
  {https://arxiv.org/abs/hep-ph/0206272} {arXiv:hep-ph/0206272} \BibitemShut
  {NoStop}%
\bibitem [{\citenamefont {Enqvist}\ \emph
  {et~al.}(2002{\natexlab{b}})\citenamefont {Enqvist}, \citenamefont {Kasuya},\
  and\ \citenamefont {Mazumdar}}]{Enqvist:2002rj}%
  \BibitemOpen
  \bibfield  {author} {\bibinfo {author} {\bibfnamefont {K.}~\bibnamefont
  {Enqvist}}, \bibinfo {author} {\bibfnamefont {S.}~\bibnamefont {Kasuya}},\
  and\ \bibinfo {author} {\bibfnamefont {A.}~\bibnamefont {Mazumdar}},\ }\href
  {https://doi.org/10.1103/PhysRevLett.89.091301} {\bibfield  {journal}
  {\bibinfo  {journal} {Phys. Rev. Lett.}\ }\textbf {\bibinfo {volume} {89}},\
  \bibinfo {pages} {091301} (\bibinfo {year} {2002}{\natexlab{b}})},\ \Eprint
  {https://arxiv.org/abs/hep-ph/0204270} {arXiv:hep-ph/0204270} \BibitemShut
  {NoStop}%
\bibitem [{\citenamefont {Laine}\ and\ \citenamefont
  {Shaposhnikov}(1998)}]{Laine:1998rg}%
  \BibitemOpen
  \bibfield  {author} {\bibinfo {author} {\bibfnamefont {M.}~\bibnamefont
  {Laine}}\ and\ \bibinfo {author} {\bibfnamefont {M.~E.}\ \bibnamefont
  {Shaposhnikov}},\ }\href {https://doi.org/10.1016/S0550-3213(98)00474-X}
  {\bibfield  {journal} {\bibinfo  {journal} {Nucl. Phys. B}\ }\textbf
  {\bibinfo {volume} {532}},\ \bibinfo {pages} {376} (\bibinfo {year}
  {1998})},\ \Eprint {https://arxiv.org/abs/hep-ph/9804237}
  {arXiv:hep-ph/9804237} \BibitemShut {NoStop}%
\bibitem [{\citenamefont {Enqvist}\ and\ \citenamefont
  {McDonald}(1998{\natexlab{b}})}]{Enqvist:1998ds}%
  \BibitemOpen
  \bibfield  {author} {\bibinfo {author} {\bibfnamefont {K.}~\bibnamefont
  {Enqvist}}\ and\ \bibinfo {author} {\bibfnamefont {J.}~\bibnamefont
  {McDonald}},\ }\href {https://doi.org/10.1103/PhysRevLett.81.3071} {\bibfield
   {journal} {\bibinfo  {journal} {Phys. Rev. Lett.}\ }\textbf {\bibinfo
  {volume} {81}},\ \bibinfo {pages} {3071} (\bibinfo {year}
  {1998}{\natexlab{b}})},\ \Eprint {https://arxiv.org/abs/hep-ph/9806213}
  {arXiv:hep-ph/9806213} \BibitemShut {NoStop}%
\bibitem [{\citenamefont {Banerjee}\ and\ \citenamefont
  {Jedamzik}(2000)}]{Banerjee:2000mb}%
  \BibitemOpen
  \bibfield  {author} {\bibinfo {author} {\bibfnamefont {R.}~\bibnamefont
  {Banerjee}}\ and\ \bibinfo {author} {\bibfnamefont {K.}~\bibnamefont
  {Jedamzik}},\ }\href {https://doi.org/10.1016/S0370-2693(00)00688-2}
  {\bibfield  {journal} {\bibinfo  {journal} {Phys. Lett. B}\ }\textbf
  {\bibinfo {volume} {484}},\ \bibinfo {pages} {278} (\bibinfo {year}
  {2000})},\ \Eprint {https://arxiv.org/abs/hep-ph/0005031}
  {arXiv:hep-ph/0005031} \BibitemShut {NoStop}%
\bibitem [{\citenamefont {Fujii}\ and\ \citenamefont
  {Yanagida}(2002)}]{Fujii:2002aj}%
  \BibitemOpen
  \bibfield  {author} {\bibinfo {author} {\bibfnamefont {M.}~\bibnamefont
  {Fujii}}\ and\ \bibinfo {author} {\bibfnamefont {T.}~\bibnamefont
  {Yanagida}},\ }\href {https://doi.org/10.1016/S0370-2693(02)02341-9}
  {\bibfield  {journal} {\bibinfo  {journal} {Phys. Lett. B}\ }\textbf
  {\bibinfo {volume} {542}},\ \bibinfo {pages} {80} (\bibinfo {year} {2002})},\
  \Eprint {https://arxiv.org/abs/hep-ph/0206066} {arXiv:hep-ph/0206066}
  \BibitemShut {NoStop}%
\bibitem [{\citenamefont {Kasuya}\ \emph {et~al.}(2013)\citenamefont {Kasuya},
  \citenamefont {Kawasaki},\ and\ \citenamefont {Yamada}}]{Kasuya:2012mh}%
  \BibitemOpen
  \bibfield  {author} {\bibinfo {author} {\bibfnamefont {S.}~\bibnamefont
  {Kasuya}}, \bibinfo {author} {\bibfnamefont {M.}~\bibnamefont {Kawasaki}},\
  and\ \bibinfo {author} {\bibfnamefont {M.}~\bibnamefont {Yamada}},\ }\href
  {https://doi.org/10.1016/j.physletb.2013.08.008} {\bibfield  {journal}
  {\bibinfo  {journal} {Phys. Lett. B}\ }\textbf {\bibinfo {volume} {726}},\
  \bibinfo {pages} {1} (\bibinfo {year} {2013})},\ \Eprint
  {https://arxiv.org/abs/1211.4743} {arXiv:1211.4743 [hep-ph]} \BibitemShut
  {NoStop}%
\bibitem [{\citenamefont {Arafune}\ \emph {et~al.}(2000)\citenamefont
  {Arafune}, \citenamefont {Yoshida}, \citenamefont {Nakamura},\ and\
  \citenamefont {Ogure}}]{Arafune:2000yv}%
  \BibitemOpen
  \bibfield  {author} {\bibinfo {author} {\bibfnamefont {J.}~\bibnamefont
  {Arafune}}, \bibinfo {author} {\bibfnamefont {T.}~\bibnamefont {Yoshida}},
  \bibinfo {author} {\bibfnamefont {S.}~\bibnamefont {Nakamura}},\ and\
  \bibinfo {author} {\bibfnamefont {K.}~\bibnamefont {Ogure}},\ }\href
  {https://doi.org/10.1103/PhysRevD.62.105013} {\bibfield  {journal} {\bibinfo
  {journal} {Phys. Rev. D}\ }\textbf {\bibinfo {volume} {62}},\ \bibinfo
  {pages} {105013} (\bibinfo {year} {2000})},\ \Eprint
  {https://arxiv.org/abs/hep-ph/0005103} {arXiv:hep-ph/0005103} \BibitemShut
  {NoStop}%
\bibitem [{\citenamefont {Kusenko}\ \emph {et~al.}(2005)\citenamefont
  {Kusenko}, \citenamefont {Loveridge},\ and\ \citenamefont
  {Shaposhnikov}}]{Kusenko:2005du}%
  \BibitemOpen
  \bibfield  {author} {\bibinfo {author} {\bibfnamefont {A.}~\bibnamefont
  {Kusenko}}, \bibinfo {author} {\bibfnamefont {L.~C.}\ \bibnamefont
  {Loveridge}},\ and\ \bibinfo {author} {\bibfnamefont {M.}~\bibnamefont
  {Shaposhnikov}},\ }\href {https://doi.org/10.1088/1475-7516/2005/08/011}
  {\bibfield  {journal} {\bibinfo  {journal} {JCAP}\ }\textbf {\bibinfo
  {volume} {08}},\ \bibinfo {pages} {011}},\ \Eprint
  {https://arxiv.org/abs/astro-ph/0507225} {arXiv:astro-ph/0507225}
  \BibitemShut {NoStop}%
\bibitem [{\citenamefont {Takenaga}\ \emph {et~al.}(2007)\citenamefont
  {Takenaga} \emph {et~al.}}]{Super-Kamiokande:2006sdq}%
  \BibitemOpen
  \bibfield  {author} {\bibinfo {author} {\bibfnamefont {Y.}~\bibnamefont
  {Takenaga}} \emph {et~al.} (\bibinfo {collaboration} {Super-Kamiokande}),\
  }\href {https://doi.org/10.1016/j.physletb.2007.01.047} {\bibfield  {journal}
  {\bibinfo  {journal} {Phys. Lett. B}\ }\textbf {\bibinfo {volume} {647}},\
  \bibinfo {pages} {18} (\bibinfo {year} {2007})},\ \Eprint
  {https://arxiv.org/abs/hep-ex/0608057} {arXiv:hep-ex/0608057} \BibitemShut
  {NoStop}%
\bibitem [{\citenamefont {Jackson~Kimball}\ \emph {et~al.}(2018)\citenamefont
  {Jackson~Kimball}, \citenamefont {Budker}, \citenamefont {Eby}, \citenamefont
  {Pospelov}, \citenamefont {Pustelny}, \citenamefont {Scholtes}, \citenamefont
  {Stadnik}, \citenamefont {Weis},\ and\ \citenamefont
  {Wickenbrock}}]{JacksonKimball:2017qgk}%
  \BibitemOpen
  \bibfield  {author} {\bibinfo {author} {\bibfnamefont {D.~F.}\ \bibnamefont
  {Jackson~Kimball}}, \bibinfo {author} {\bibfnamefont {D.}~\bibnamefont
  {Budker}}, \bibinfo {author} {\bibfnamefont {J.}~\bibnamefont {Eby}},
  \bibinfo {author} {\bibfnamefont {M.}~\bibnamefont {Pospelov}}, \bibinfo
  {author} {\bibfnamefont {S.}~\bibnamefont {Pustelny}}, \bibinfo {author}
  {\bibfnamefont {T.}~\bibnamefont {Scholtes}}, \bibinfo {author}
  {\bibfnamefont {Y.~V.}\ \bibnamefont {Stadnik}}, \bibinfo {author}
  {\bibfnamefont {A.}~\bibnamefont {Weis}},\ and\ \bibinfo {author}
  {\bibfnamefont {A.}~\bibnamefont {Wickenbrock}},\ }\href
  {https://doi.org/10.1103/PhysRevD.97.043002} {\bibfield  {journal} {\bibinfo
  {journal} {Phys. Rev. D}\ }\textbf {\bibinfo {volume} {97}},\ \bibinfo
  {pages} {043002} (\bibinfo {year} {2018})},\ \Eprint
  {https://arxiv.org/abs/1710.04323} {arXiv:1710.04323 [physics.atom-ph]}
  \BibitemShut {NoStop}%
\bibitem [{\citenamefont {Hong}\ \emph {et~al.}(2016)\citenamefont {Hong},
  \citenamefont {Kawasaki},\ and\ \citenamefont {Yamada}}]{Hong:2016ict}%
  \BibitemOpen
  \bibfield  {author} {\bibinfo {author} {\bibfnamefont {J.-P.}\ \bibnamefont
  {Hong}}, \bibinfo {author} {\bibfnamefont {M.}~\bibnamefont {Kawasaki}},\
  and\ \bibinfo {author} {\bibfnamefont {M.}~\bibnamefont {Yamada}},\ }\href
  {https://doi.org/10.1088/1475-7516/2016/08/053} {\bibfield  {journal}
  {\bibinfo  {journal} {JCAP}\ }\textbf {\bibinfo {volume} {08}},\ \bibinfo
  {pages} {053}},\ \Eprint {https://arxiv.org/abs/1604.04352} {arXiv:1604.04352
  [hep-ph]} \BibitemShut {NoStop}%
\bibitem [{\citenamefont {Hong}\ and\ \citenamefont
  {Kawasaki}(2017{\natexlab{b}})}]{Hong:2017qvx}%
  \BibitemOpen
  \bibfield  {author} {\bibinfo {author} {\bibfnamefont {J.-P.}\ \bibnamefont
  {Hong}}\ and\ \bibinfo {author} {\bibfnamefont {M.}~\bibnamefont
  {Kawasaki}},\ }\href {https://doi.org/10.1103/PhysRevD.95.123532} {\bibfield
  {journal} {\bibinfo  {journal} {Phys. Rev. D}\ }\textbf {\bibinfo {volume}
  {95}},\ \bibinfo {pages} {123532} (\bibinfo {year} {2017}{\natexlab{b}})},\
  \Eprint {https://arxiv.org/abs/1702.00889} {arXiv:1702.00889 [hep-ph]}
  \BibitemShut {NoStop}%
\bibitem [{\citenamefont {Bakari}\ \emph {et~al.}(2001)\citenamefont {Bakari},
  \citenamefont {Dekhissi}, \citenamefont {Derkaoui}, \citenamefont
  {Giacomelli}, \citenamefont {Mandrioli}, \citenamefont {Ouchrif},
  \citenamefont {Patrizii},\ and\ \citenamefont {Popa}}]{Bakari:2000dq}%
  \BibitemOpen
  \bibfield  {author} {\bibinfo {author} {\bibfnamefont {D.}~\bibnamefont
  {Bakari}}, \bibinfo {author} {\bibfnamefont {H.}~\bibnamefont {Dekhissi}},
  \bibinfo {author} {\bibfnamefont {J.}~\bibnamefont {Derkaoui}}, \bibinfo
  {author} {\bibfnamefont {G.}~\bibnamefont {Giacomelli}}, \bibinfo {author}
  {\bibfnamefont {G.}~\bibnamefont {Mandrioli}}, \bibinfo {author}
  {\bibfnamefont {M.}~\bibnamefont {Ouchrif}}, \bibinfo {author} {\bibfnamefont
  {L.}~\bibnamefont {Patrizii}},\ and\ \bibinfo {author} {\bibfnamefont
  {V.}~\bibnamefont {Popa}},\ }\href
  {https://doi.org/10.1016/S0927-6505(00)00140-7} {\bibfield  {journal}
  {\bibinfo  {journal} {Astropart. Phys.}\ }\textbf {\bibinfo {volume} {15}},\
  \bibinfo {pages} {137} (\bibinfo {year} {2001})},\ \Eprint
  {https://arxiv.org/abs/hep-ex/0003003} {arXiv:hep-ex/0003003} \BibitemShut
  {NoStop}%
\bibitem [{\citenamefont {Belolaptikov}\ \emph {et~al.}(1998)\citenamefont
  {Belolaptikov} \emph {et~al.}}]{Belolaptikov:1998mn}%
  \BibitemOpen
  \bibfield  {author} {\bibinfo {author} {\bibfnamefont {I.~A.}\ \bibnamefont
  {Belolaptikov}} \emph {et~al.},\ }\href@noop {} {\  (\bibinfo {year}
  {1998})},\ \Eprint {https://arxiv.org/abs/astro-ph/9802223}
  {arXiv:astro-ph/9802223} \BibitemShut {NoStop}%
\bibitem [{\citenamefont {Kusenko}\ \emph
  {et~al.}(1998{\natexlab{b}})\citenamefont {Kusenko}, \citenamefont {Kuzmin},
  \citenamefont {Shaposhnikov},\ and\ \citenamefont
  {Tinyakov}}]{Kusenko:1997vp}%
  \BibitemOpen
  \bibfield  {author} {\bibinfo {author} {\bibfnamefont {A.}~\bibnamefont
  {Kusenko}}, \bibinfo {author} {\bibfnamefont {V.}~\bibnamefont {Kuzmin}},
  \bibinfo {author} {\bibfnamefont {M.~E.}\ \bibnamefont {Shaposhnikov}},\ and\
  \bibinfo {author} {\bibfnamefont {P.~G.}\ \bibnamefont {Tinyakov}},\ }\href
  {https://doi.org/10.1103/PhysRevLett.80.3185} {\bibfield  {journal} {\bibinfo
   {journal} {Phys. Rev. Lett.}\ }\textbf {\bibinfo {volume} {80}},\ \bibinfo
  {pages} {3185} (\bibinfo {year} {1998}{\natexlab{b}})},\ \Eprint
  {https://arxiv.org/abs/hep-ph/9712212} {arXiv:hep-ph/9712212} \BibitemShut
  {NoStop}%
\bibitem [{\citenamefont {Attanasio}\ \emph {et~al.}(2022)\citenamefont
  {Attanasio} \emph {et~al.}}]{Windchime:2022whs}%
  \BibitemOpen
  \bibfield  {author} {\bibinfo {author} {\bibfnamefont {A.}~\bibnamefont
  {Attanasio}} \emph {et~al.} (\bibinfo {collaboration} {Windchime}),\ }in\
  \href@noop {} {\emph {\bibinfo {booktitle} {{Snowmass 2021}}}}\ (\bibinfo
  {year} {2022})\ \Eprint {https://arxiv.org/abs/2203.07242} {arXiv:2203.07242
  [hep-ex]} \BibitemShut {NoStop}%
\bibitem [{\citenamefont {Cotner}\ and\ \citenamefont
  {Kusenko}(2016)}]{Cotner:2016dhw}%
  \BibitemOpen
  \bibfield  {author} {\bibinfo {author} {\bibfnamefont {E.}~\bibnamefont
  {Cotner}}\ and\ \bibinfo {author} {\bibfnamefont {A.}~\bibnamefont
  {Kusenko}},\ }\href {https://doi.org/10.1103/PhysRevD.94.123006} {\bibfield
  {journal} {\bibinfo  {journal} {Phys. Rev. D}\ }\textbf {\bibinfo {volume}
  {94}},\ \bibinfo {pages} {123006} (\bibinfo {year} {2016})},\ \Eprint
  {https://arxiv.org/abs/1609.00970} {arXiv:1609.00970 [hep-ph]} \BibitemShut
  {NoStop}%
\bibitem [{\citenamefont {Kusenko}\ \emph
  {et~al.}(1998{\natexlab{c}})\citenamefont {Kusenko}, \citenamefont
  {Shaposhnikov}, \citenamefont {Tinyakov},\ and\ \citenamefont
  {Tkachev}}]{Kusenko:1997it}%
  \BibitemOpen
  \bibfield  {author} {\bibinfo {author} {\bibfnamefont {A.}~\bibnamefont
  {Kusenko}}, \bibinfo {author} {\bibfnamefont {M.~E.}\ \bibnamefont
  {Shaposhnikov}}, \bibinfo {author} {\bibfnamefont {P.~G.}\ \bibnamefont
  {Tinyakov}},\ and\ \bibinfo {author} {\bibfnamefont {I.~I.}\ \bibnamefont
  {Tkachev}},\ }\href {https://doi.org/10.1016/S0370-2693(98)00133-6}
  {\bibfield  {journal} {\bibinfo  {journal} {Phys. Lett. B}\ }\textbf
  {\bibinfo {volume} {423}},\ \bibinfo {pages} {104} (\bibinfo {year}
  {1998}{\natexlab{c}})},\ \Eprint {https://arxiv.org/abs/hep-ph/9801212}
  {arXiv:hep-ph/9801212} \BibitemShut {NoStop}%
\bibitem [{\citenamefont {Hisano}\ \emph {et~al.}(2001)\citenamefont {Hisano},
  \citenamefont {Nojiri},\ and\ \citenamefont {Okada}}]{Hisano:2001dr}%
  \BibitemOpen
  \bibfield  {author} {\bibinfo {author} {\bibfnamefont {J.}~\bibnamefont
  {Hisano}}, \bibinfo {author} {\bibfnamefont {M.~M.}\ \bibnamefont {Nojiri}},\
  and\ \bibinfo {author} {\bibfnamefont {N.}~\bibnamefont {Okada}},\ }\href
  {https://doi.org/10.1103/PhysRevD.64.023511} {\bibfield  {journal} {\bibinfo
  {journal} {Phys. Rev. D}\ }\textbf {\bibinfo {volume} {64}},\ \bibinfo
  {pages} {023511} (\bibinfo {year} {2001})},\ \Eprint
  {https://arxiv.org/abs/hep-ph/0102045} {arXiv:hep-ph/0102045} \BibitemShut
  {NoStop}%
\bibitem [{\citenamefont {Kawasaki}\ \emph {et~al.}(2005)\citenamefont
  {Kawasaki}, \citenamefont {Konya},\ and\ \citenamefont
  {Takahashi}}]{Kawasaki:2005xc}%
  \BibitemOpen
  \bibfield  {author} {\bibinfo {author} {\bibfnamefont {M.}~\bibnamefont
  {Kawasaki}}, \bibinfo {author} {\bibfnamefont {K.}~\bibnamefont {Konya}},\
  and\ \bibinfo {author} {\bibfnamefont {F.}~\bibnamefont {Takahashi}},\ }\href
  {https://doi.org/10.1016/j.physletb.2005.05.082} {\bibfield  {journal}
  {\bibinfo  {journal} {Phys. Lett. B}\ }\textbf {\bibinfo {volume} {619}},\
  \bibinfo {pages} {233} (\bibinfo {year} {2005})},\ \Eprint
  {https://arxiv.org/abs/hep-ph/0504105} {arXiv:hep-ph/0504105} \BibitemShut
  {NoStop}%
\bibitem [{\citenamefont {Kawasaki}\ and\ \citenamefont
  {Nakatsuka}(2020)}]{Kawasaki:2019ywz}%
  \BibitemOpen
  \bibfield  {author} {\bibinfo {author} {\bibfnamefont {M.}~\bibnamefont
  {Kawasaki}}\ and\ \bibinfo {author} {\bibfnamefont {H.}~\bibnamefont
  {Nakatsuka}},\ }\href {https://doi.org/10.1088/1475-7516/2020/04/017}
  {\bibfield  {journal} {\bibinfo  {journal} {JCAP}\ }\textbf {\bibinfo
  {volume} {04}},\ \bibinfo {pages} {017}},\ \Eprint
  {https://arxiv.org/abs/1912.06993} {arXiv:1912.06993 [hep-ph]} \BibitemShut
  {NoStop}%
\bibitem [{\citenamefont {Fujii}\ and\ \citenamefont
  {Hamaguchi}(2002{\natexlab{a}})}]{Fujii:2001xp}%
  \BibitemOpen
  \bibfield  {author} {\bibinfo {author} {\bibfnamefont {M.}~\bibnamefont
  {Fujii}}\ and\ \bibinfo {author} {\bibfnamefont {K.}~\bibnamefont
  {Hamaguchi}},\ }\href {https://doi.org/10.1016/S0370-2693(01)01412-5}
  {\bibfield  {journal} {\bibinfo  {journal} {Phys. Lett. B}\ }\textbf
  {\bibinfo {volume} {525}},\ \bibinfo {pages} {143} (\bibinfo {year}
  {2002}{\natexlab{a}})},\ \Eprint {https://arxiv.org/abs/hep-ph/0110072}
  {arXiv:hep-ph/0110072} \BibitemShut {NoStop}%
\bibitem [{\citenamefont {Fujii}\ and\ \citenamefont
  {Hamaguchi}(2002{\natexlab{b}})}]{Fujii:2002kr}%
  \BibitemOpen
  \bibfield  {author} {\bibinfo {author} {\bibfnamefont {M.}~\bibnamefont
  {Fujii}}\ and\ \bibinfo {author} {\bibfnamefont {K.}~\bibnamefont
  {Hamaguchi}},\ }\href {https://doi.org/10.1103/PhysRevD.66.083501} {\bibfield
   {journal} {\bibinfo  {journal} {Phys. Rev. D}\ }\textbf {\bibinfo {volume}
  {66}},\ \bibinfo {pages} {083501} (\bibinfo {year} {2002}{\natexlab{b}})},\
  \Eprint {https://arxiv.org/abs/hep-ph/0205044} {arXiv:hep-ph/0205044}
  \BibitemShut {NoStop}%
\bibitem [{\citenamefont {Kusenko}\ and\ \citenamefont
  {Steinhardt}(2001)}]{Kusenko:2001vu}%
  \BibitemOpen
  \bibfield  {author} {\bibinfo {author} {\bibfnamefont {A.}~\bibnamefont
  {Kusenko}}\ and\ \bibinfo {author} {\bibfnamefont {P.~J.}\ \bibnamefont
  {Steinhardt}},\ }\href {https://doi.org/10.1103/PhysRevLett.87.141301}
  {\bibfield  {journal} {\bibinfo  {journal} {Phys. Rev. Lett.}\ }\textbf
  {\bibinfo {volume} {87}},\ \bibinfo {pages} {141301} (\bibinfo {year}
  {2001})},\ \Eprint {https://arxiv.org/abs/astro-ph/0106008}
  {arXiv:astro-ph/0106008} \BibitemShut {NoStop}%
\bibitem [{\citenamefont {Troitsky}(2016)}]{Troitsky:2015mda}%
  \BibitemOpen
  \bibfield  {author} {\bibinfo {author} {\bibfnamefont {S.}~\bibnamefont
  {Troitsky}},\ }\href {https://doi.org/10.1088/1475-7516/2016/11/027}
  {\bibfield  {journal} {\bibinfo  {journal} {JCAP}\ }\textbf {\bibinfo
  {volume} {11}},\ \bibinfo {pages} {027}},\ \Eprint
  {https://arxiv.org/abs/1510.07132} {arXiv:1510.07132 [hep-ph]} \BibitemShut
  {NoStop}%
\bibitem [{\citenamefont {Ansari}\ \emph {et~al.}(2024)\citenamefont {Ansari},
  \citenamefont {Singh~Bhandari},\ and\ \citenamefont
  {Thalapillil}}]{Ansari:2023cay}%
  \BibitemOpen
  \bibfield  {author} {\bibinfo {author} {\bibfnamefont {A.}~\bibnamefont
  {Ansari}}, \bibinfo {author} {\bibfnamefont {L.}~\bibnamefont
  {Singh~Bhandari}},\ and\ \bibinfo {author} {\bibfnamefont {A.~M.}\
  \bibnamefont {Thalapillil}},\ }\href
  {https://doi.org/10.1103/PhysRevD.109.023003} {\bibfield  {journal} {\bibinfo
   {journal} {Phys. Rev. D}\ }\textbf {\bibinfo {volume} {109}},\ \bibinfo
  {pages} {023003} (\bibinfo {year} {2024})},\ \Eprint
  {https://arxiv.org/abs/2302.11590} {arXiv:2302.11590 [hep-ph]} \BibitemShut
  {NoStop}%
\bibitem [{\citenamefont {Kusenko}\ and\ \citenamefont
  {Mazumdar}(2008)}]{Kusenko:2008zm}%
  \BibitemOpen
  \bibfield  {author} {\bibinfo {author} {\bibfnamefont {A.}~\bibnamefont
  {Kusenko}}\ and\ \bibinfo {author} {\bibfnamefont {A.}~\bibnamefont
  {Mazumdar}},\ }\href {https://doi.org/10.1103/PhysRevLett.101.211301}
  {\bibfield  {journal} {\bibinfo  {journal} {Phys. Rev. Lett.}\ }\textbf
  {\bibinfo {volume} {101}},\ \bibinfo {pages} {211301} (\bibinfo {year}
  {2008})},\ \Eprint {https://arxiv.org/abs/0807.4554} {arXiv:0807.4554
  [astro-ph]} \BibitemShut {NoStop}%
\bibitem [{\citenamefont {Kusenko}\ \emph {et~al.}(2009)\citenamefont
  {Kusenko}, \citenamefont {Mazumdar},\ and\ \citenamefont
  {Multamaki}}]{Kusenko:2009cv}%
  \BibitemOpen
  \bibfield  {author} {\bibinfo {author} {\bibfnamefont {A.}~\bibnamefont
  {Kusenko}}, \bibinfo {author} {\bibfnamefont {A.}~\bibnamefont {Mazumdar}},\
  and\ \bibinfo {author} {\bibfnamefont {T.}~\bibnamefont {Multamaki}},\ }\href
  {https://doi.org/10.1103/PhysRevD.79.124034} {\bibfield  {journal} {\bibinfo
  {journal} {Phys. Rev. D}\ }\textbf {\bibinfo {volume} {79}},\ \bibinfo
  {pages} {124034} (\bibinfo {year} {2009})},\ \Eprint
  {https://arxiv.org/abs/0902.2197} {arXiv:0902.2197 [astro-ph.CO]}
  \BibitemShut {NoStop}%
\bibitem [{\citenamefont {Kawasaki}\ and\ \citenamefont
  {Yamada}(2013)}]{Kawasaki:2012gk}%
  \BibitemOpen
  \bibfield  {author} {\bibinfo {author} {\bibfnamefont {M.}~\bibnamefont
  {Kawasaki}}\ and\ \bibinfo {author} {\bibfnamefont {M.}~\bibnamefont
  {Yamada}},\ }\href {https://doi.org/10.1103/PhysRevD.87.023517} {\bibfield
  {journal} {\bibinfo  {journal} {Phys. Rev. D}\ }\textbf {\bibinfo {volume}
  {87}},\ \bibinfo {pages} {023517} (\bibinfo {year} {2013})},\ \Eprint
  {https://arxiv.org/abs/1209.5781} {arXiv:1209.5781 [hep-ph]} \BibitemShut
  {NoStop}%
\bibitem [{\citenamefont {Chiba}\ \emph
  {et~al.}(2010{\natexlab{b}})\citenamefont {Chiba}, \citenamefont {Kamada},\
  and\ \citenamefont {Yamaguchi}}]{Chiba:2009zu}%
  \BibitemOpen
  \bibfield  {author} {\bibinfo {author} {\bibfnamefont {T.}~\bibnamefont
  {Chiba}}, \bibinfo {author} {\bibfnamefont {K.}~\bibnamefont {Kamada}},\ and\
  \bibinfo {author} {\bibfnamefont {M.}~\bibnamefont {Yamaguchi}},\ }\href
  {https://doi.org/10.1103/PhysRevD.81.083503} {\bibfield  {journal} {\bibinfo
  {journal} {Phys. Rev. D}\ }\textbf {\bibinfo {volume} {81}},\ \bibinfo
  {pages} {083503} (\bibinfo {year} {2010}{\natexlab{b}})},\ \Eprint
  {https://arxiv.org/abs/0912.3585} {arXiv:0912.3585 [astro-ph.CO]}
  \BibitemShut {NoStop}%
\bibitem [{\citenamefont {White}\ \emph {et~al.}(2021)\citenamefont {White},
  \citenamefont {Pearce}, \citenamefont {Vagie},\ and\ \citenamefont
  {Kusenko}}]{White:2021hwi}%
  \BibitemOpen
  \bibfield  {author} {\bibinfo {author} {\bibfnamefont {G.}~\bibnamefont
  {White}}, \bibinfo {author} {\bibfnamefont {L.}~\bibnamefont {Pearce}},
  \bibinfo {author} {\bibfnamefont {D.}~\bibnamefont {Vagie}},\ and\ \bibinfo
  {author} {\bibfnamefont {A.}~\bibnamefont {Kusenko}},\ }\href
  {https://doi.org/10.1103/PhysRevLett.127.181601} {\bibfield  {journal}
  {\bibinfo  {journal} {Phys. Rev. Lett.}\ }\textbf {\bibinfo {volume} {127}},\
  \bibinfo {pages} {181601} (\bibinfo {year} {2021})},\ \Eprint
  {https://arxiv.org/abs/2105.11655} {arXiv:2105.11655 [hep-ph]} \BibitemShut
  {NoStop}%
\bibitem [{\citenamefont {Inomata}\ \emph
  {et~al.}(2019{\natexlab{a}})\citenamefont {Inomata}, \citenamefont {Kohri},
  \citenamefont {Nakama},\ and\ \citenamefont {Terada}}]{Inomata:2019zqy}%
  \BibitemOpen
  \bibfield  {author} {\bibinfo {author} {\bibfnamefont {K.}~\bibnamefont
  {Inomata}}, \bibinfo {author} {\bibfnamefont {K.}~\bibnamefont {Kohri}},
  \bibinfo {author} {\bibfnamefont {T.}~\bibnamefont {Nakama}},\ and\ \bibinfo
  {author} {\bibfnamefont {T.}~\bibnamefont {Terada}},\ }\href
  {https://doi.org/10.1088/1475-7516/2019/10/071} {\bibfield  {journal}
  {\bibinfo  {journal} {JCAP}\ }\textbf {\bibinfo {volume} {10}},\ \bibinfo
  {pages} {071}},\ \bibinfo {note} {[Erratum: JCAP 08, E01 (2023)]},\ \Eprint
  {https://arxiv.org/abs/1904.12878} {arXiv:1904.12878 [astro-ph.CO]}
  \BibitemShut {NoStop}%
\bibitem [{\citenamefont {Kawasaki}\ and\ \citenamefont
  {Murai}(2024)}]{Kawasaki:2023rfx}%
  \BibitemOpen
  \bibfield  {author} {\bibinfo {author} {\bibfnamefont {M.}~\bibnamefont
  {Kawasaki}}\ and\ \bibinfo {author} {\bibfnamefont {K.}~\bibnamefont
  {Murai}},\ }\href {https://doi.org/10.1088/1475-7516/2024/01/050} {\bibfield
  {journal} {\bibinfo  {journal} {JCAP}\ }\textbf {\bibinfo {volume} {01}},\
  \bibinfo {pages} {050}},\ \Eprint {https://arxiv.org/abs/2308.13134}
  {arXiv:2308.13134 [astro-ph.CO]} \BibitemShut {NoStop}%
\bibitem [{\citenamefont {Sakai}\ and\ \citenamefont
  {Tamaki}(2012)}]{Sakai:2011wn}%
  \BibitemOpen
  \bibfield  {author} {\bibinfo {author} {\bibfnamefont {N.}~\bibnamefont
  {Sakai}}\ and\ \bibinfo {author} {\bibfnamefont {T.}~\bibnamefont {Tamaki}},\
  }\href {https://doi.org/10.1103/PhysRevD.85.104008} {\bibfield  {journal}
  {\bibinfo  {journal} {Phys. Rev. D}\ }\textbf {\bibinfo {volume} {85}},\
  \bibinfo {pages} {104008} (\bibinfo {year} {2012})},\ \Eprint
  {https://arxiv.org/abs/1112.5559} {arXiv:1112.5559 [gr-qc]} \BibitemShut
  {NoStop}%
\bibitem [{\citenamefont {Cotner}\ and\ \citenamefont
  {Kusenko}(2017)}]{Cotner:2016cvr}%
  \BibitemOpen
  \bibfield  {author} {\bibinfo {author} {\bibfnamefont {E.}~\bibnamefont
  {Cotner}}\ and\ \bibinfo {author} {\bibfnamefont {A.}~\bibnamefont
  {Kusenko}},\ }\href {https://doi.org/10.1103/PhysRevLett.119.031103}
  {\bibfield  {journal} {\bibinfo  {journal} {Phys. Rev. Lett.}\ }\textbf
  {\bibinfo {volume} {119}},\ \bibinfo {pages} {031103} (\bibinfo {year}
  {2017})},\ \Eprint {https://arxiv.org/abs/1612.02529} {arXiv:1612.02529
  [astro-ph.CO]} \BibitemShut {NoStop}%
\bibitem [{\citenamefont {Cotner}\ \emph {et~al.}(2019)\citenamefont {Cotner},
  \citenamefont {Kusenko}, \citenamefont {Sasaki},\ and\ \citenamefont
  {Takhistov}}]{Cotner:2019ykd}%
  \BibitemOpen
  \bibfield  {author} {\bibinfo {author} {\bibfnamefont {E.}~\bibnamefont
  {Cotner}}, \bibinfo {author} {\bibfnamefont {A.}~\bibnamefont {Kusenko}},
  \bibinfo {author} {\bibfnamefont {M.}~\bibnamefont {Sasaki}},\ and\ \bibinfo
  {author} {\bibfnamefont {V.}~\bibnamefont {Takhistov}},\ }\href
  {https://doi.org/10.1088/1475-7516/2019/10/077} {\bibfield  {journal}
  {\bibinfo  {journal} {JCAP}\ }\textbf {\bibinfo {volume} {10}},\ \bibinfo
  {pages} {077}},\ \Eprint {https://arxiv.org/abs/1907.10613} {arXiv:1907.10613
  [astro-ph.CO]} \BibitemShut {NoStop}%
\bibitem [{\citenamefont {Multamaki}\ and\ \citenamefont
  {Vilja}(2002{\natexlab{b}})}]{Multamaki:2002wk}%
  \BibitemOpen
  \bibfield  {author} {\bibinfo {author} {\bibfnamefont {T.}~\bibnamefont
  {Multamaki}}\ and\ \bibinfo {author} {\bibfnamefont {I.}~\bibnamefont
  {Vilja}},\ }\href {https://doi.org/10.1016/S0370-2693(02)02274-8} {\bibfield
  {journal} {\bibinfo  {journal} {Phys. Lett. B}\ }\textbf {\bibinfo {volume}
  {542}},\ \bibinfo {pages} {137} (\bibinfo {year} {2002}{\natexlab{b}})},\
  \Eprint {https://arxiv.org/abs/hep-ph/0205302} {arXiv:hep-ph/0205302}
  \BibitemShut {NoStop}%
\bibitem [{\citenamefont {Chodos}\ \emph {et~al.}(1974)\citenamefont {Chodos},
  \citenamefont {Jaffe}, \citenamefont {Johnson}, \citenamefont {Thorn},\ and\
  \citenamefont {Weisskopf}}]{Chodos:1974je}%
  \BibitemOpen
  \bibfield  {author} {\bibinfo {author} {\bibfnamefont {A.}~\bibnamefont
  {Chodos}}, \bibinfo {author} {\bibfnamefont {R.~L.}\ \bibnamefont {Jaffe}},
  \bibinfo {author} {\bibfnamefont {K.}~\bibnamefont {Johnson}}, \bibinfo
  {author} {\bibfnamefont {C.~B.}\ \bibnamefont {Thorn}},\ and\ \bibinfo
  {author} {\bibfnamefont {V.~F.}\ \bibnamefont {Weisskopf}},\ }\href
  {https://doi.org/10.1103/PhysRevD.9.3471} {\bibfield  {journal} {\bibinfo
  {journal} {Phys. Rev. D}\ }\textbf {\bibinfo {volume} {9}},\ \bibinfo {pages}
  {3471} (\bibinfo {year} {1974})}\BibitemShut {NoStop}%
\bibitem [{\citenamefont {Witten}(1984)}]{Witten:1984rs}%
  \BibitemOpen
  \bibfield  {author} {\bibinfo {author} {\bibfnamefont {E.}~\bibnamefont
  {Witten}},\ }\href {https://doi.org/10.1103/PhysRevD.30.272} {\bibfield
  {journal} {\bibinfo  {journal} {Phys. Rev. D}\ }\textbf {\bibinfo {volume}
  {30}},\ \bibinfo {pages} {272} (\bibinfo {year} {1984})}\BibitemShut
  {NoStop}%
\bibitem [{\citenamefont {Terazawa}(1989)}]{Terazawa:1989iw}%
  \BibitemOpen
  \bibfield  {author} {\bibinfo {author} {\bibfnamefont {H.}~\bibnamefont
  {Terazawa}},\ }\href {https://doi.org/10.1143/JPSJ.58.3555} {\bibfield
  {journal} {\bibinfo  {journal} {J. Phys. Soc. Jap.}\ }\textbf {\bibinfo
  {volume} {58}},\ \bibinfo {pages} {3555} (\bibinfo {year}
  {1989})}\BibitemShut {NoStop}%
\bibitem [{\citenamefont {Farhi}\ and\ \citenamefont
  {Jaffe}(1984)}]{Farhi:1984qu}%
  \BibitemOpen
  \bibfield  {author} {\bibinfo {author} {\bibfnamefont {E.}~\bibnamefont
  {Farhi}}\ and\ \bibinfo {author} {\bibfnamefont {R.~L.}\ \bibnamefont
  {Jaffe}},\ }\href {https://doi.org/10.1103/PhysRevD.30.2379} {\bibfield
  {journal} {\bibinfo  {journal} {Phys. Rev. D}\ }\textbf {\bibinfo {volume}
  {30}},\ \bibinfo {pages} {2379} (\bibinfo {year} {1984})}\BibitemShut
  {NoStop}%
\bibitem [{\citenamefont {Buballa}\ and\ \citenamefont
  {Oertel}(1999)}]{Buballa:1998pr}%
  \BibitemOpen
  \bibfield  {author} {\bibinfo {author} {\bibfnamefont {M.}~\bibnamefont
  {Buballa}}\ and\ \bibinfo {author} {\bibfnamefont {M.}~\bibnamefont
  {Oertel}},\ }\href {https://doi.org/10.1016/S0370-2693(99)00533-X} {\bibfield
   {journal} {\bibinfo  {journal} {Phys. Lett. B}\ }\textbf {\bibinfo {volume}
  {457}},\ \bibinfo {pages} {261} (\bibinfo {year} {1999})},\ \Eprint
  {https://arxiv.org/abs/hep-ph/9810529} {arXiv:hep-ph/9810529} \BibitemShut
  {NoStop}%
\bibitem [{\citenamefont {Fukushima}\ \emph {et~al.}(2005)\citenamefont
  {Fukushima}, \citenamefont {Kouvaris},\ and\ \citenamefont
  {Rajagopal}}]{Fukushima:2004zq}%
  \BibitemOpen
  \bibfield  {author} {\bibinfo {author} {\bibfnamefont {K.}~\bibnamefont
  {Fukushima}}, \bibinfo {author} {\bibfnamefont {C.}~\bibnamefont
  {Kouvaris}},\ and\ \bibinfo {author} {\bibfnamefont {K.}~\bibnamefont
  {Rajagopal}},\ }\href {https://doi.org/10.1103/PhysRevD.71.034002} {\bibfield
   {journal} {\bibinfo  {journal} {Phys. Rev. D}\ }\textbf {\bibinfo {volume}
  {71}},\ \bibinfo {pages} {034002} (\bibinfo {year} {2005})},\ \Eprint
  {https://arxiv.org/abs/hep-ph/0408322} {arXiv:hep-ph/0408322} \BibitemShut
  {NoStop}%
\bibitem [{\citenamefont {Liang}\ and\ \citenamefont
  {Zhitnitsky}(2016)}]{Liang:2016tqc}%
  \BibitemOpen
  \bibfield  {author} {\bibinfo {author} {\bibfnamefont {X.}~\bibnamefont
  {Liang}}\ and\ \bibinfo {author} {\bibfnamefont {A.}~\bibnamefont
  {Zhitnitsky}},\ }\href {https://doi.org/10.1103/PhysRevD.94.083502}
  {\bibfield  {journal} {\bibinfo  {journal} {Phys. Rev. D}\ }\textbf {\bibinfo
  {volume} {94}},\ \bibinfo {pages} {083502} (\bibinfo {year} {2016})},\
  \Eprint {https://arxiv.org/abs/1606.00435} {arXiv:1606.00435 [hep-ph]}
  \BibitemShut {NoStop}%
\bibitem [{\citenamefont {Holdom}\ \emph {et~al.}(2018)\citenamefont {Holdom},
  \citenamefont {Ren},\ and\ \citenamefont {Zhang}}]{Holdom:2017gdc}%
  \BibitemOpen
  \bibfield  {author} {\bibinfo {author} {\bibfnamefont {B.}~\bibnamefont
  {Holdom}}, \bibinfo {author} {\bibfnamefont {J.}~\bibnamefont {Ren}},\ and\
  \bibinfo {author} {\bibfnamefont {C.}~\bibnamefont {Zhang}},\ }\href
  {https://doi.org/10.1103/PhysRevLett.120.222001} {\bibfield  {journal}
  {\bibinfo  {journal} {Phys. Rev. Lett.}\ }\textbf {\bibinfo {volume} {120}},\
  \bibinfo {pages} {222001} (\bibinfo {year} {2018})},\ \Eprint
  {https://arxiv.org/abs/1707.06610} {arXiv:1707.06610 [hep-ph]} \BibitemShut
  {NoStop}%
\bibitem [{\citenamefont {Bai}\ \emph {et~al.}(2019)\citenamefont {Bai},
  \citenamefont {Long},\ and\ \citenamefont {Lu}}]{Bai:2018dxf}%
  \BibitemOpen
  \bibfield  {author} {\bibinfo {author} {\bibfnamefont {Y.}~\bibnamefont
  {Bai}}, \bibinfo {author} {\bibfnamefont {A.~J.}\ \bibnamefont {Long}},\ and\
  \bibinfo {author} {\bibfnamefont {S.}~\bibnamefont {Lu}},\ }\href
  {https://doi.org/10.1103/PhysRevD.99.055047} {\bibfield  {journal} {\bibinfo
  {journal} {Phys. Rev. D}\ }\textbf {\bibinfo {volume} {99}},\ \bibinfo
  {pages} {055047} (\bibinfo {year} {2019})},\ \Eprint
  {https://arxiv.org/abs/1810.04360} {arXiv:1810.04360 [hep-ph]} \BibitemShut
  {NoStop}%
\bibitem [{\citenamefont {Itoh}(1970)}]{Itoh:1970uw}%
  \BibitemOpen
  \bibfield  {author} {\bibinfo {author} {\bibfnamefont {N.}~\bibnamefont
  {Itoh}},\ }\href {https://doi.org/10.1143/PTP.44.291} {\bibfield  {journal}
  {\bibinfo  {journal} {Prog. Theor. Phys.}\ }\textbf {\bibinfo {volume}
  {44}},\ \bibinfo {pages} {291} (\bibinfo {year} {1970})}\BibitemShut
  {NoStop}%
\bibitem [{\citenamefont {Alcock}\ \emph {et~al.}(1986)\citenamefont {Alcock},
  \citenamefont {Farhi},\ and\ \citenamefont {Olinto}}]{Alcock:1986hz}%
  \BibitemOpen
  \bibfield  {author} {\bibinfo {author} {\bibfnamefont {C.}~\bibnamefont
  {Alcock}}, \bibinfo {author} {\bibfnamefont {E.}~\bibnamefont {Farhi}},\ and\
  \bibinfo {author} {\bibfnamefont {A.}~\bibnamefont {Olinto}},\ }\href
  {https://doi.org/10.1086/164679} {\bibfield  {journal} {\bibinfo  {journal}
  {Astrophys. J.}\ }\textbf {\bibinfo {volume} {310}},\ \bibinfo {pages} {261}
  (\bibinfo {year} {1986})}\BibitemShut {NoStop}%
\bibitem [{\citenamefont {Haensel}\ \emph {et~al.}(1986)\citenamefont
  {Haensel}, \citenamefont {Zdunik},\ and\ \citenamefont
  {Schaeffer}}]{Haensel:1986qb}%
  \BibitemOpen
  \bibfield  {author} {\bibinfo {author} {\bibfnamefont {P.}~\bibnamefont
  {Haensel}}, \bibinfo {author} {\bibfnamefont {J.~L.}\ \bibnamefont
  {Zdunik}},\ and\ \bibinfo {author} {\bibfnamefont {R.}~\bibnamefont
  {Schaeffer}},\ }\href@noop {} {\bibfield  {journal} {\bibinfo  {journal}
  {Astron. Astrophys.}\ }\textbf {\bibinfo {volume} {160}},\ \bibinfo {pages}
  {121} (\bibinfo {year} {1986})}\BibitemShut {NoStop}%
\bibitem [{\citenamefont {Holdom}(1987)}]{Holdom:1987ep}%
  \BibitemOpen
  \bibfield  {author} {\bibinfo {author} {\bibfnamefont {B.}~\bibnamefont
  {Holdom}},\ }\href {https://doi.org/10.1103/PhysRevD.36.1000} {\bibfield
  {journal} {\bibinfo  {journal} {Phys. Rev. D}\ }\textbf {\bibinfo {volume}
  {36}},\ \bibinfo {pages} {1000} (\bibinfo {year} {1987})}\BibitemShut
  {NoStop}%
\bibitem [{\citenamefont {Macpherson}\ and\ \citenamefont
  {Campbell}(1995)}]{Macpherson:1994wf}%
  \BibitemOpen
  \bibfield  {author} {\bibinfo {author} {\bibfnamefont {A.~L.}\ \bibnamefont
  {Macpherson}}\ and\ \bibinfo {author} {\bibfnamefont {B.~A.}\ \bibnamefont
  {Campbell}},\ }\href {https://doi.org/10.1016/0370-2693(95)00080-5}
  {\bibfield  {journal} {\bibinfo  {journal} {Phys. Lett. B}\ }\textbf
  {\bibinfo {volume} {347}},\ \bibinfo {pages} {205} (\bibinfo {year}
  {1995})},\ \Eprint {https://arxiv.org/abs/hep-ph/9408387}
  {arXiv:hep-ph/9408387} \BibitemShut {NoStop}%
\bibitem [{\citenamefont {Hong}\ \emph {et~al.}(2020)\citenamefont {Hong},
  \citenamefont {Jung},\ and\ \citenamefont {Xie}}]{Hong:2020est}%
  \BibitemOpen
  \bibfield  {author} {\bibinfo {author} {\bibfnamefont {J.-P.}\ \bibnamefont
  {Hong}}, \bibinfo {author} {\bibfnamefont {S.}~\bibnamefont {Jung}},\ and\
  \bibinfo {author} {\bibfnamefont {K.-P.}\ \bibnamefont {Xie}},\ }\href
  {https://doi.org/10.1103/PhysRevD.102.075028} {\bibfield  {journal} {\bibinfo
   {journal} {Phys. Rev. D}\ }\textbf {\bibinfo {volume} {102}},\ \bibinfo
  {pages} {075028} (\bibinfo {year} {2020})},\ \Eprint
  {https://arxiv.org/abs/2008.04430} {arXiv:2008.04430 [hep-ph]} \BibitemShut
  {NoStop}%
\bibitem [{\citenamefont {Kawana}\ and\ \citenamefont
  {Xie}(2022)}]{Kawana:2021tde}%
  \BibitemOpen
  \bibfield  {author} {\bibinfo {author} {\bibfnamefont {K.}~\bibnamefont
  {Kawana}}\ and\ \bibinfo {author} {\bibfnamefont {K.-P.}\ \bibnamefont
  {Xie}},\ }\href {https://doi.org/10.1016/j.physletb.2021.136791} {\bibfield
  {journal} {\bibinfo  {journal} {Phys. Lett. B}\ }\textbf {\bibinfo {volume}
  {824}},\ \bibinfo {pages} {136791} (\bibinfo {year} {2022})},\ \Eprint
  {https://arxiv.org/abs/2106.00111} {arXiv:2106.00111 [astro-ph.CO]}
  \BibitemShut {NoStop}%
\bibitem [{\citenamefont {Del~Grosso}\ \emph {et~al.}(2023)\citenamefont
  {Del~Grosso}, \citenamefont {Franciolini}, \citenamefont {Pani},\ and\
  \citenamefont {Urbano}}]{DelGrosso:2023trq}%
  \BibitemOpen
  \bibfield  {author} {\bibinfo {author} {\bibfnamefont {L.}~\bibnamefont
  {Del~Grosso}}, \bibinfo {author} {\bibfnamefont {G.}~\bibnamefont
  {Franciolini}}, \bibinfo {author} {\bibfnamefont {P.}~\bibnamefont {Pani}},\
  and\ \bibinfo {author} {\bibfnamefont {A.}~\bibnamefont {Urbano}},\ }\href
  {https://doi.org/10.1103/PhysRevD.108.044024} {\bibfield  {journal} {\bibinfo
   {journal} {Phys. Rev. D}\ }\textbf {\bibinfo {volume} {108}},\ \bibinfo
  {pages} {044024} (\bibinfo {year} {2023})},\ \Eprint
  {https://arxiv.org/abs/2301.08709} {arXiv:2301.08709 [gr-qc]} \BibitemShut
  {NoStop}%
\bibitem [{\citenamefont {Saly}(1983)}]{Saly:1983fbs}%
  \BibitemOpen
  \bibfield  {author} {\bibinfo {author} {\bibfnamefont {R.}~\bibnamefont
  {Saly}},\ }\href {https://doi.org/10.1016/0010-4655(83)90082-6} {\bibfield
  {journal} {\bibinfo  {journal} {Comput. Phys. Commun.}\ }\textbf {\bibinfo
  {volume} {30}},\ \bibinfo {pages} {411} (\bibinfo {year} {1983})}\BibitemShut
  {NoStop}%
\bibitem [{\citenamefont {Koeppel}\ and\ \citenamefont
  {Harvey}(1985)}]{Koeppel:1985tt}%
  \BibitemOpen
  \bibfield  {author} {\bibinfo {author} {\bibfnamefont {T.}~\bibnamefont
  {Koeppel}}\ and\ \bibinfo {author} {\bibfnamefont {M.}~\bibnamefont
  {Harvey}},\ }\href {https://doi.org/10.1103/PhysRevD.31.171} {\bibfield
  {journal} {\bibinfo  {journal} {Phys. Rev. D}\ }\textbf {\bibinfo {volume}
  {31}},\ \bibinfo {pages} {171} (\bibinfo {year} {1985})}\BibitemShut
  {NoStop}%
\bibitem [{\citenamefont {Horn}\ \emph {et~al.}(1986)\citenamefont {Horn},
  \citenamefont {Wilets},\ and\ \citenamefont {Goldflam}}]{Horn:1986qj}%
  \BibitemOpen
  \bibfield  {author} {\bibinfo {author} {\bibfnamefont {R.}~\bibnamefont
  {Horn}}, \bibinfo {author} {\bibfnamefont {L.}~\bibnamefont {Wilets}},\ and\
  \bibinfo {author} {\bibfnamefont {R.}~\bibnamefont {Goldflam}},\ }\href
  {https://doi.org/10.1016/0010-4655(86)90235-3} {\bibfield  {journal}
  {\bibinfo  {journal} {Comput. Phys. Commun.}\ }\textbf {\bibinfo {volume}
  {42}},\ \bibinfo {pages} {105} (\bibinfo {year} {1986})}\BibitemShut
  {NoStop}%
\bibitem [{\citenamefont {Lee}(1979)}]{Lee:1978mf}%
  \BibitemOpen
  \bibfield  {author} {\bibinfo {author} {\bibfnamefont {T.~D.}\ \bibnamefont
  {Lee}},\ }\href {https://doi.org/10.1103/PhysRevD.19.1802} {\bibfield
  {journal} {\bibinfo  {journal} {Phys. Rev. D}\ }\textbf {\bibinfo {volume}
  {19}},\ \bibinfo {pages} {1802} (\bibinfo {year} {1979})}\BibitemShut
  {NoStop}%
\bibitem [{\citenamefont {Tang}\ and\ \citenamefont
  {Wilets}(1990)}]{Tang:1990hd}%
  \BibitemOpen
  \bibfield  {author} {\bibinfo {author} {\bibfnamefont {P.}~\bibnamefont
  {Tang}}\ and\ \bibinfo {author} {\bibfnamefont {L.}~\bibnamefont {Wilets}},\
  }\href {https://doi.org/10.1063/1.528712} {\bibfield  {journal} {\bibinfo
  {journal} {J. Math. Phys.}\ }\textbf {\bibinfo {volume} {31}},\ \bibinfo
  {pages} {1661} (\bibinfo {year} {1990})}\BibitemShut {NoStop}%
\bibitem [{\citenamefont {Dethier}\ \emph {et~al.}(1983)\citenamefont
  {Dethier}, \citenamefont {Goldflam}, \citenamefont {Henley},\ and\
  \citenamefont {Wilets}}]{Dethier:1982ax}%
  \BibitemOpen
  \bibfield  {author} {\bibinfo {author} {\bibfnamefont {J.~L.}\ \bibnamefont
  {Dethier}}, \bibinfo {author} {\bibfnamefont {R.}~\bibnamefont {Goldflam}},
  \bibinfo {author} {\bibfnamefont {E.~M.}\ \bibnamefont {Henley}},\ and\
  \bibinfo {author} {\bibfnamefont {L.}~\bibnamefont {Wilets}},\ }\href
  {https://doi.org/10.1103/PhysRevD.27.2191} {\bibfield  {journal} {\bibinfo
  {journal} {Phys. Rev. D}\ }\textbf {\bibinfo {volume} {27}},\ \bibinfo
  {pages} {2191} (\bibinfo {year} {1983})}\BibitemShut {NoStop}%
\bibitem [{\citenamefont {Peierls}\ and\ \citenamefont
  {Yoccoz}(1957)}]{Peierls:1957er}%
  \BibitemOpen
  \bibfield  {author} {\bibinfo {author} {\bibfnamefont {R.~E.}\ \bibnamefont
  {Peierls}}\ and\ \bibinfo {author} {\bibfnamefont {J.}~\bibnamefont
  {Yoccoz}},\ }\href {https://doi.org/10.1088/0370-1298/70/5/309} {\bibfield
  {journal} {\bibinfo  {journal} {Proc. Phys. Soc. A}\ }\textbf {\bibinfo
  {volume} {70}},\ \bibinfo {pages} {381} (\bibinfo {year} {1957})}\BibitemShut
  {NoStop}%
\bibitem [{\citenamefont {Lubeck}\ \emph {et~al.}(1986)\citenamefont {Lubeck},
  \citenamefont {Birse}, \citenamefont {Henley},\ and\ \citenamefont
  {Wilets}}]{Lubeck:1986if}%
  \BibitemOpen
  \bibfield  {author} {\bibinfo {author} {\bibfnamefont {E.~G.}\ \bibnamefont
  {Lubeck}}, \bibinfo {author} {\bibfnamefont {M.~C.}\ \bibnamefont {Birse}},
  \bibinfo {author} {\bibfnamefont {E.~M.}\ \bibnamefont {Henley}},\ and\
  \bibinfo {author} {\bibfnamefont {L.}~\bibnamefont {Wilets}},\ }\href
  {https://doi.org/10.1103/PhysRevD.33.234} {\bibfield  {journal} {\bibinfo
  {journal} {Phys. Rev. D}\ }\textbf {\bibinfo {volume} {33}},\ \bibinfo
  {pages} {234} (\bibinfo {year} {1986})}\BibitemShut {NoStop}%
\bibitem [{\citenamefont {Lubeck}\ \emph {et~al.}(1987)\citenamefont {Lubeck},
  \citenamefont {Henley},\ and\ \citenamefont {Wilets}}]{Lubeck:1987pj}%
  \BibitemOpen
  \bibfield  {author} {\bibinfo {author} {\bibfnamefont {E.~G.}\ \bibnamefont
  {Lubeck}}, \bibinfo {author} {\bibfnamefont {E.~M.}\ \bibnamefont {Henley}},\
  and\ \bibinfo {author} {\bibfnamefont {L.}~\bibnamefont {Wilets}},\ }\href
  {https://doi.org/10.1103/PhysRevD.35.2809} {\bibfield  {journal} {\bibinfo
  {journal} {Phys. Rev. D}\ }\textbf {\bibinfo {volume} {35}},\ \bibinfo
  {pages} {2809} (\bibinfo {year} {1987})}\BibitemShut {NoStop}%
\bibitem [{\citenamefont {Nielsen}\ and\ \citenamefont
  {Patkos}(1982)}]{Nielsen:1981fi}%
  \BibitemOpen
  \bibfield  {author} {\bibinfo {author} {\bibfnamefont {H.~B.}\ \bibnamefont
  {Nielsen}}\ and\ \bibinfo {author} {\bibfnamefont {A.}~\bibnamefont
  {Patkos}},\ }\href {https://doi.org/10.1016/0550-3213(82)90051-7} {\bibfield
  {journal} {\bibinfo  {journal} {Nucl. Phys. B}\ }\textbf {\bibinfo {volume}
  {195}},\ \bibinfo {pages} {137} (\bibinfo {year} {1982})}\BibitemShut
  {NoStop}%
\bibitem [{\citenamefont {Kahana}\ \emph {et~al.}(1984)\citenamefont {Kahana},
  \citenamefont {Ripka},\ and\ \citenamefont {Soni}}]{Kahana:1984dx}%
  \BibitemOpen
  \bibfield  {author} {\bibinfo {author} {\bibfnamefont {S.}~\bibnamefont
  {Kahana}}, \bibinfo {author} {\bibfnamefont {G.}~\bibnamefont {Ripka}},\ and\
  \bibinfo {author} {\bibfnamefont {V.}~\bibnamefont {Soni}},\ }\href
  {https://doi.org/10.1016/0375-9474(84)90306-3} {\bibfield  {journal}
  {\bibinfo  {journal} {Nucl. Phys. A}\ }\textbf {\bibinfo {volume} {415}},\
  \bibinfo {pages} {351} (\bibinfo {year} {1984})}\BibitemShut {NoStop}%
\bibitem [{\citenamefont {Kahana}\ and\ \citenamefont
  {Ripka}(1984)}]{Kahana:1984be}%
  \BibitemOpen
  \bibfield  {author} {\bibinfo {author} {\bibfnamefont {S.}~\bibnamefont
  {Kahana}}\ and\ \bibinfo {author} {\bibfnamefont {G.}~\bibnamefont {Ripka}},\
  }\href {https://doi.org/10.1016/0375-9474(84)90692-4} {\bibfield  {journal}
  {\bibinfo  {journal} {Nucl. Phys. A}\ }\textbf {\bibinfo {volume} {429}},\
  \bibinfo {pages} {462} (\bibinfo {year} {1984})}\BibitemShut {NoStop}%
\bibitem [{\citenamefont {Birse}\ and\ \citenamefont
  {Banerjee}(1985)}]{Birse:1984js}%
  \BibitemOpen
  \bibfield  {author} {\bibinfo {author} {\bibfnamefont {M.~C.}\ \bibnamefont
  {Birse}}\ and\ \bibinfo {author} {\bibfnamefont {M.~K.}\ \bibnamefont
  {Banerjee}},\ }\href {https://doi.org/10.1103/PhysRevD.31.118} {\bibfield
  {journal} {\bibinfo  {journal} {Phys. Rev. D}\ }\textbf {\bibinfo {volume}
  {31}},\ \bibinfo {pages} {118} (\bibinfo {year} {1985})}\BibitemShut
  {NoStop}%
\bibitem [{\citenamefont {Williams}\ and\ \citenamefont
  {Thomas}(1986)}]{Williams:1985ku}%
  \BibitemOpen
  \bibfield  {author} {\bibinfo {author} {\bibfnamefont {A.~G.}\ \bibnamefont
  {Williams}}\ and\ \bibinfo {author} {\bibfnamefont {A.~W.}\ \bibnamefont
  {Thomas}},\ }\href {https://doi.org/10.1103/PhysRevC.33.1070} {\bibfield
  {journal} {\bibinfo  {journal} {Phys. Rev. C}\ }\textbf {\bibinfo {volume}
  {33}},\ \bibinfo {pages} {1070} (\bibinfo {year} {1986})}\BibitemShut
  {NoStop}%
\bibitem [{\citenamefont {Dodd}\ \emph {et~al.}(1987)\citenamefont {Dodd},
  \citenamefont {Williams},\ and\ \citenamefont {Thomas}}]{Dodd:1987pw}%
  \BibitemOpen
  \bibfield  {author} {\bibinfo {author} {\bibfnamefont {L.~R.}\ \bibnamefont
  {Dodd}}, \bibinfo {author} {\bibfnamefont {A.~G.}\ \bibnamefont {Williams}},\
  and\ \bibinfo {author} {\bibfnamefont {A.~W.}\ \bibnamefont {Thomas}},\
  }\href {https://doi.org/10.1103/PhysRevD.35.1040} {\bibfield  {journal}
  {\bibinfo  {journal} {Phys. Rev. D}\ }\textbf {\bibinfo {volume} {35}},\
  \bibinfo {pages} {1040} (\bibinfo {year} {1987})}\BibitemShut {NoStop}%
\bibitem [{\citenamefont {Witten}(1977)}]{Witten:1976ck}%
  \BibitemOpen
  \bibfield  {author} {\bibinfo {author} {\bibfnamefont {E.}~\bibnamefont
  {Witten}},\ }\href {https://doi.org/10.1103/PhysRevLett.38.121} {\bibfield
  {journal} {\bibinfo  {journal} {Phys. Rev. Lett.}\ }\textbf {\bibinfo
  {volume} {38}},\ \bibinfo {pages} {121} (\bibinfo {year} {1977})}\BibitemShut
  {NoStop}%
\bibitem [{\citenamefont {Gleiser}\ and\ \citenamefont
  {Thorarinson}(2007)}]{Gleiser:2007te}%
  \BibitemOpen
  \bibfield  {author} {\bibinfo {author} {\bibfnamefont {M.}~\bibnamefont
  {Gleiser}}\ and\ \bibinfo {author} {\bibfnamefont {J.}~\bibnamefont
  {Thorarinson}},\ }\href {https://doi.org/10.1103/PhysRevD.76.041701}
  {\bibfield  {journal} {\bibinfo  {journal} {Phys. Rev. D}\ }\textbf {\bibinfo
  {volume} {76}},\ \bibinfo {pages} {041701} (\bibinfo {year} {2007})},\
  \Eprint {https://arxiv.org/abs/hep-th/0701294} {arXiv:hep-th/0701294}
  \BibitemShut {NoStop}%
\bibitem [{\citenamefont {Kolb}\ and\ \citenamefont
  {Tkachev}(1994)}]{Kolb:1993hw}%
  \BibitemOpen
  \bibfield  {author} {\bibinfo {author} {\bibfnamefont {E.~W.}\ \bibnamefont
  {Kolb}}\ and\ \bibinfo {author} {\bibfnamefont {I.~I.}\ \bibnamefont
  {Tkachev}},\ }\href {https://doi.org/10.1103/PhysRevD.49.5040} {\bibfield
  {journal} {\bibinfo  {journal} {Phys. Rev. D}\ }\textbf {\bibinfo {volume}
  {49}},\ \bibinfo {pages} {5040} (\bibinfo {year} {1994})},\ \Eprint
  {https://arxiv.org/abs/astro-ph/9311037} {arXiv:astro-ph/9311037}
  \BibitemShut {NoStop}%
\bibitem [{\citenamefont {Oll\'e}\ \emph {et~al.}(2020)\citenamefont {Oll\'e},
  \citenamefont {Pujol\`as},\ and\ \citenamefont {Rompineve}}]{Olle:2019kbo}%
  \BibitemOpen
  \bibfield  {author} {\bibinfo {author} {\bibfnamefont {J.}~\bibnamefont
  {Oll\'e}}, \bibinfo {author} {\bibfnamefont {O.}~\bibnamefont {Pujol\`as}},\
  and\ \bibinfo {author} {\bibfnamefont {F.}~\bibnamefont {Rompineve}},\ }\href
  {https://doi.org/10.1088/1475-7516/2020/02/006} {\bibfield  {journal}
  {\bibinfo  {journal} {JCAP}\ }\textbf {\bibinfo {volume} {02}},\ \bibinfo
  {pages} {006}},\ \Eprint {https://arxiv.org/abs/1906.06352} {arXiv:1906.06352
  [hep-ph]} \BibitemShut {NoStop}%
\bibitem [{\citenamefont {Amin}\ and\ \citenamefont
  {Mou}(2021)}]{Amin:2020vja}%
  \BibitemOpen
  \bibfield  {author} {\bibinfo {author} {\bibfnamefont {M.~A.}\ \bibnamefont
  {Amin}}\ and\ \bibinfo {author} {\bibfnamefont {Z.-G.}\ \bibnamefont {Mou}},\
  }\href {https://doi.org/10.1088/1475-7516/2021/02/024} {\bibfield  {journal}
  {\bibinfo  {journal} {JCAP}\ }\textbf {\bibinfo {volume} {02}},\ \bibinfo
  {pages} {024}},\ \Eprint {https://arxiv.org/abs/2009.11337} {arXiv:2009.11337
  [astro-ph.CO]} \BibitemShut {NoStop}%
\bibitem [{\citenamefont {Amin}\ \emph {et~al.}(2021)\citenamefont {Amin},
  \citenamefont {Long}, \citenamefont {Mou},\ and\ \citenamefont
  {Saffin}}]{Amin:2021tnq}%
  \BibitemOpen
  \bibfield  {author} {\bibinfo {author} {\bibfnamefont {M.~A.}\ \bibnamefont
  {Amin}}, \bibinfo {author} {\bibfnamefont {A.~J.}\ \bibnamefont {Long}},
  \bibinfo {author} {\bibfnamefont {Z.-G.}\ \bibnamefont {Mou}},\ and\ \bibinfo
  {author} {\bibfnamefont {P.}~\bibnamefont {Saffin}},\ }\href
  {https://doi.org/10.1007/JHEP06(2021)182} {\bibfield  {journal} {\bibinfo
  {journal} {JHEP}\ }\textbf {\bibinfo {volume} {06}},\ \bibinfo {pages}
  {182}},\ \Eprint {https://arxiv.org/abs/2103.12082} {arXiv:2103.12082
  [hep-ph]} \BibitemShut {NoStop}%
\bibitem [{\citenamefont {Amin}\ \emph {et~al.}(2023)\citenamefont {Amin},
  \citenamefont {Long},\ and\ \citenamefont {Schiappacasse}}]{Amin:2023imi}%
  \BibitemOpen
  \bibfield  {author} {\bibinfo {author} {\bibfnamefont {M.~A.}\ \bibnamefont
  {Amin}}, \bibinfo {author} {\bibfnamefont {A.~J.}\ \bibnamefont {Long}},\
  and\ \bibinfo {author} {\bibfnamefont {E.~D.}\ \bibnamefont
  {Schiappacasse}},\ }\href {https://doi.org/10.1088/1475-7516/2023/05/015}
  {\bibfield  {journal} {\bibinfo  {journal} {JCAP}\ }\textbf {\bibinfo
  {volume} {05}},\ \bibinfo {pages} {015}},\ \Eprint
  {https://arxiv.org/abs/2301.11470} {arXiv:2301.11470 [hep-ph]} \BibitemShut
  {NoStop}%
\bibitem [{\citenamefont {Graham}(2007{\natexlab{a}})}]{Graham:2006vy}%
  \BibitemOpen
  \bibfield  {author} {\bibinfo {author} {\bibfnamefont {N.}~\bibnamefont
  {Graham}},\ }\href {https://doi.org/10.1103/PhysRevLett.98.101801} {\bibfield
   {journal} {\bibinfo  {journal} {Phys. Rev. Lett.}\ }\textbf {\bibinfo
  {volume} {98}},\ \bibinfo {pages} {101801} (\bibinfo {year}
  {2007}{\natexlab{a}})},\ \bibinfo {note} {[Erratum: Phys.Rev.Lett. 98, 189904
  (2007)]},\ \Eprint {https://arxiv.org/abs/hep-th/0610267}
  {arXiv:hep-th/0610267} \BibitemShut {NoStop}%
\bibitem [{\citenamefont {Graham}(2007{\natexlab{b}})}]{Graham:2007ds}%
  \BibitemOpen
  \bibfield  {author} {\bibinfo {author} {\bibfnamefont {N.}~\bibnamefont
  {Graham}},\ }\href {https://doi.org/10.1103/PhysRevD.76.085017} {\bibfield
  {journal} {\bibinfo  {journal} {Phys. Rev. D}\ }\textbf {\bibinfo {volume}
  {76}},\ \bibinfo {pages} {085017} (\bibinfo {year} {2007}{\natexlab{b}})},\
  \Eprint {https://arxiv.org/abs/0706.4125} {arXiv:0706.4125 [hep-th]}
  \BibitemShut {NoStop}%
\bibitem [{\citenamefont {Farhi}\ \emph {et~al.}(2005)\citenamefont {Farhi},
  \citenamefont {Graham}, \citenamefont {Khemani}, \citenamefont {Markov},\
  and\ \citenamefont {Rosales}}]{Farhi:2005rz}%
  \BibitemOpen
  \bibfield  {author} {\bibinfo {author} {\bibfnamefont {E.}~\bibnamefont
  {Farhi}}, \bibinfo {author} {\bibfnamefont {N.}~\bibnamefont {Graham}},
  \bibinfo {author} {\bibfnamefont {V.}~\bibnamefont {Khemani}}, \bibinfo
  {author} {\bibfnamefont {R.}~\bibnamefont {Markov}},\ and\ \bibinfo {author}
  {\bibfnamefont {R.}~\bibnamefont {Rosales}},\ }\href
  {https://doi.org/10.1103/PhysRevD.72.101701} {\bibfield  {journal} {\bibinfo
  {journal} {Phys. Rev. D}\ }\textbf {\bibinfo {volume} {72}},\ \bibinfo
  {pages} {101701} (\bibinfo {year} {2005})},\ \Eprint
  {https://arxiv.org/abs/hep-th/0505273} {arXiv:hep-th/0505273} \BibitemShut
  {NoStop}%
\bibitem [{\citenamefont {Correa}\ \emph {et~al.}(2015)\citenamefont {Correa},
  \citenamefont {da~Rocha},\ and\ \citenamefont
  {de~Souza~Dutra}}]{Correa:2015kha}%
  \BibitemOpen
  \bibfield  {author} {\bibinfo {author} {\bibfnamefont {R.~A.~C.}\
  \bibnamefont {Correa}}, \bibinfo {author} {\bibfnamefont {R.}~\bibnamefont
  {da~Rocha}},\ and\ \bibinfo {author} {\bibfnamefont {A.}~\bibnamefont
  {de~Souza~Dutra}},\ }\href {https://doi.org/10.1103/PhysRevD.91.125021}
  {\bibfield  {journal} {\bibinfo  {journal} {Phys. Rev. D}\ }\textbf {\bibinfo
  {volume} {91}},\ \bibinfo {pages} {125021} (\bibinfo {year} {2015})},\
  \Eprint {https://arxiv.org/abs/1504.04038} {arXiv:1504.04038 [hep-th]}
  \BibitemShut {NoStop}%
\bibitem [{\citenamefont {Farhi}\ \emph {et~al.}(2008)\citenamefont {Farhi},
  \citenamefont {Graham}, \citenamefont {Guth}, \citenamefont {Iqbal},
  \citenamefont {Rosales},\ and\ \citenamefont {Stamatopoulos}}]{Farhi:2007wj}%
  \BibitemOpen
  \bibfield  {author} {\bibinfo {author} {\bibfnamefont {E.}~\bibnamefont
  {Farhi}}, \bibinfo {author} {\bibfnamefont {N.}~\bibnamefont {Graham}},
  \bibinfo {author} {\bibfnamefont {A.~H.}\ \bibnamefont {Guth}}, \bibinfo
  {author} {\bibfnamefont {N.}~\bibnamefont {Iqbal}}, \bibinfo {author}
  {\bibfnamefont {R.~R.}\ \bibnamefont {Rosales}},\ and\ \bibinfo {author}
  {\bibfnamefont {N.}~\bibnamefont {Stamatopoulos}},\ }\href
  {https://doi.org/10.1103/PhysRevD.77.085019} {\bibfield  {journal} {\bibinfo
  {journal} {Phys. Rev. D}\ }\textbf {\bibinfo {volume} {77}},\ \bibinfo
  {pages} {085019} (\bibinfo {year} {2008})},\ \Eprint
  {https://arxiv.org/abs/0712.3034} {arXiv:0712.3034 [hep-th]} \BibitemShut
  {NoStop}%
\bibitem [{\citenamefont {Riotto}(1996)}]{Riotto:1995yy}%
  \BibitemOpen
  \bibfield  {author} {\bibinfo {author} {\bibfnamefont {A.}~\bibnamefont
  {Riotto}},\ }\href {https://doi.org/10.1016/0370-2693(95)01239-7} {\bibfield
  {journal} {\bibinfo  {journal} {Phys. Lett. B}\ }\textbf {\bibinfo {volume}
  {365}},\ \bibinfo {pages} {64} (\bibinfo {year} {1996})},\ \Eprint
  {https://arxiv.org/abs/hep-ph/9507201} {arXiv:hep-ph/9507201} \BibitemShut
  {NoStop}%
\bibitem [{\citenamefont {Gleiser}\ \emph {et~al.}(2010)\citenamefont
  {Gleiser}, \citenamefont {Graham},\ and\ \citenamefont
  {Stamatopoulos}}]{Gleiser:2010qt}%
  \BibitemOpen
  \bibfield  {author} {\bibinfo {author} {\bibfnamefont {M.}~\bibnamefont
  {Gleiser}}, \bibinfo {author} {\bibfnamefont {N.}~\bibnamefont {Graham}},\
  and\ \bibinfo {author} {\bibfnamefont {N.}~\bibnamefont {Stamatopoulos}},\
  }\href {https://doi.org/10.1103/PhysRevD.82.043517} {\bibfield  {journal}
  {\bibinfo  {journal} {Phys. Rev. D}\ }\textbf {\bibinfo {volume} {82}},\
  \bibinfo {pages} {043517} (\bibinfo {year} {2010})},\ \Eprint
  {https://arxiv.org/abs/1004.4658} {arXiv:1004.4658 [astro-ph.CO]}
  \BibitemShut {NoStop}%
\bibitem [{\citenamefont {P\^\i{}rvu}\ \emph {et~al.}(2024)\citenamefont
  {P\^\i{}rvu}, \citenamefont {Johnson},\ and\ \citenamefont
  {Sibiryakov}}]{Pirvu:2023plk}%
  \BibitemOpen
  \bibfield  {author} {\bibinfo {author} {\bibfnamefont {D.}~\bibnamefont
  {P\^\i{}rvu}}, \bibinfo {author} {\bibfnamefont {M.~C.}\ \bibnamefont
  {Johnson}},\ and\ \bibinfo {author} {\bibfnamefont {S.}~\bibnamefont
  {Sibiryakov}},\ }\href {https://doi.org/10.1007/JHEP11(2024)064} {\bibfield
  {journal} {\bibinfo  {journal} {JHEP}\ }\textbf {\bibinfo {volume} {11}},\
  \bibinfo {pages} {064}},\ \Eprint {https://arxiv.org/abs/2312.13364}
  {arXiv:2312.13364 [hep-th]} \BibitemShut {NoStop}%
\bibitem [{\citenamefont {Gleiser}\ and\ \citenamefont
  {Howell}(2002)}]{Gleiser:2002pw}%
  \BibitemOpen
  \bibfield  {author} {\bibinfo {author} {\bibfnamefont {M.}~\bibnamefont
  {Gleiser}}\ and\ \bibinfo {author} {\bibfnamefont {R.~C.}\ \bibnamefont
  {Howell}},\ }\href@noop {} {\  (\bibinfo {year} {2002})},\ \Eprint
  {https://arxiv.org/abs/hep-ph/0209176} {arXiv:hep-ph/0209176} \BibitemShut
  {NoStop}%
\bibitem [{\citenamefont {Gleiser}\ and\ \citenamefont
  {Howell}(2005)}]{Gleiser:2004iy}%
  \BibitemOpen
  \bibfield  {author} {\bibinfo {author} {\bibfnamefont {M.}~\bibnamefont
  {Gleiser}}\ and\ \bibinfo {author} {\bibfnamefont {R.~C.}\ \bibnamefont
  {Howell}},\ }\href {https://doi.org/10.1103/PhysRevLett.94.151601} {\bibfield
   {journal} {\bibinfo  {journal} {Phys. Rev. Lett.}\ }\textbf {\bibinfo
  {volume} {94}},\ \bibinfo {pages} {151601} (\bibinfo {year} {2005})},\
  \Eprint {https://arxiv.org/abs/hep-ph/0409179} {arXiv:hep-ph/0409179}
  \BibitemShut {NoStop}%
\bibitem [{\citenamefont {Hindmarsh}\ and\ \citenamefont
  {Salmi}(2008)}]{Hindmarsh:2007jb}%
  \BibitemOpen
  \bibfield  {author} {\bibinfo {author} {\bibfnamefont {M.}~\bibnamefont
  {Hindmarsh}}\ and\ \bibinfo {author} {\bibfnamefont {P.}~\bibnamefont
  {Salmi}},\ }\href {https://doi.org/10.1103/PhysRevD.77.105025} {\bibfield
  {journal} {\bibinfo  {journal} {Phys. Rev. D}\ }\textbf {\bibinfo {volume}
  {77}},\ \bibinfo {pages} {105025} (\bibinfo {year} {2008})},\ \Eprint
  {https://arxiv.org/abs/0712.0614} {arXiv:0712.0614 [hep-th]} \BibitemShut
  {NoStop}%
\bibitem [{\citenamefont {Adib}\ \emph {et~al.}(2002)\citenamefont {Adib},
  \citenamefont {Gleiser},\ and\ \citenamefont {Almeida}}]{Adib:2002ff}%
  \BibitemOpen
  \bibfield  {author} {\bibinfo {author} {\bibfnamefont {A.~B.}\ \bibnamefont
  {Adib}}, \bibinfo {author} {\bibfnamefont {M.}~\bibnamefont {Gleiser}},\ and\
  \bibinfo {author} {\bibfnamefont {C.~A.~S.}\ \bibnamefont {Almeida}},\ }\href
  {https://doi.org/10.1103/PhysRevD.66.085011} {\bibfield  {journal} {\bibinfo
  {journal} {Phys. Rev. D}\ }\textbf {\bibinfo {volume} {66}},\ \bibinfo
  {pages} {085011} (\bibinfo {year} {2002})},\ \Eprint
  {https://arxiv.org/abs/hep-th/0203072} {arXiv:hep-th/0203072} \BibitemShut
  {NoStop}%
\bibitem [{\citenamefont {Jain}\ \emph {et~al.}(2024)\citenamefont {Jain},
  \citenamefont {Wanichwecharungruang},\ and\ \citenamefont
  {Thomas}}]{Jain:2023tsr}%
  \BibitemOpen
  \bibfield  {author} {\bibinfo {author} {\bibfnamefont {M.}~\bibnamefont
  {Jain}}, \bibinfo {author} {\bibfnamefont {W.}~\bibnamefont
  {Wanichwecharungruang}},\ and\ \bibinfo {author} {\bibfnamefont
  {J.}~\bibnamefont {Thomas}},\ }\href
  {https://doi.org/10.1103/PhysRevD.109.016002} {\bibfield  {journal} {\bibinfo
   {journal} {Phys. Rev. D}\ }\textbf {\bibinfo {volume} {109}},\ \bibinfo
  {pages} {016002} (\bibinfo {year} {2024})},\ \Eprint
  {https://arxiv.org/abs/2310.00058} {arXiv:2310.00058 [astro-ph.CO]}
  \BibitemShut {NoStop}%
\bibitem [{\citenamefont {Jain}\ \emph {et~al.}(2023)\citenamefont {Jain},
  \citenamefont {Amin}, \citenamefont {Thomas},\ and\ \citenamefont
  {Wanichwecharungruang}}]{Jain:2023ojg}%
  \BibitemOpen
  \bibfield  {author} {\bibinfo {author} {\bibfnamefont {M.}~\bibnamefont
  {Jain}}, \bibinfo {author} {\bibfnamefont {M.~A.}\ \bibnamefont {Amin}},
  \bibinfo {author} {\bibfnamefont {J.}~\bibnamefont {Thomas}},\ and\ \bibinfo
  {author} {\bibfnamefont {W.}~\bibnamefont {Wanichwecharungruang}},\ }\href
  {https://doi.org/10.1103/PhysRevD.108.043535} {\bibfield  {journal} {\bibinfo
   {journal} {Phys. Rev. D}\ }\textbf {\bibinfo {volume} {108}},\ \bibinfo
  {pages} {043535} (\bibinfo {year} {2023})},\ \Eprint
  {https://arxiv.org/abs/2304.01985} {arXiv:2304.01985 [astro-ph.CO]}
  \BibitemShut {NoStop}%
\bibitem [{\citenamefont {Gleiser}(2007)}]{Gleiser:2006te}%
  \BibitemOpen
  \bibfield  {author} {\bibinfo {author} {\bibfnamefont {M.}~\bibnamefont
  {Gleiser}},\ }\href {https://doi.org/10.1142/S0218271807009954} {\bibfield
  {journal} {\bibinfo  {journal} {Int. J. Mod. Phys. D}\ }\textbf {\bibinfo
  {volume} {16}},\ \bibinfo {pages} {219} (\bibinfo {year} {2007})},\ \Eprint
  {https://arxiv.org/abs/hep-th/0602187} {arXiv:hep-th/0602187} \BibitemShut
  {NoStop}%
\bibitem [{\citenamefont {McDonald}(2002)}]{McDonald:2001iv}%
  \BibitemOpen
  \bibfield  {author} {\bibinfo {author} {\bibfnamefont {J.}~\bibnamefont
  {McDonald}},\ }\href {https://doi.org/10.1103/PhysRevD.66.043525} {\bibfield
  {journal} {\bibinfo  {journal} {Phys. Rev. D}\ }\textbf {\bibinfo {volume}
  {66}},\ \bibinfo {pages} {043525} (\bibinfo {year} {2002})},\ \Eprint
  {https://arxiv.org/abs/hep-ph/0105235} {arXiv:hep-ph/0105235} \BibitemShut
  {NoStop}%
\bibitem [{\citenamefont {Gleiser}\ \emph {et~al.}(2011)\citenamefont
  {Gleiser}, \citenamefont {Graham},\ and\ \citenamefont
  {Stamatopoulos}}]{Gleiser:2011xj}%
  \BibitemOpen
  \bibfield  {author} {\bibinfo {author} {\bibfnamefont {M.}~\bibnamefont
  {Gleiser}}, \bibinfo {author} {\bibfnamefont {N.}~\bibnamefont {Graham}},\
  and\ \bibinfo {author} {\bibfnamefont {N.}~\bibnamefont {Stamatopoulos}},\
  }\href {https://doi.org/10.1103/PhysRevD.83.096010} {\bibfield  {journal}
  {\bibinfo  {journal} {Phys. Rev. D}\ }\textbf {\bibinfo {volume} {83}},\
  \bibinfo {pages} {096010} (\bibinfo {year} {2011})},\ \Eprint
  {https://arxiv.org/abs/1103.1911} {arXiv:1103.1911 [hep-th]} \BibitemShut
  {NoStop}%
\bibitem [{\citenamefont {Lozanov}\ and\ \citenamefont
  {Amin}(2014)}]{Lozanov:2014zfa}%
  \BibitemOpen
  \bibfield  {author} {\bibinfo {author} {\bibfnamefont {K.~D.}\ \bibnamefont
  {Lozanov}}\ and\ \bibinfo {author} {\bibfnamefont {M.~A.}\ \bibnamefont
  {Amin}},\ }\href {https://doi.org/10.1103/PhysRevD.90.083528} {\bibfield
  {journal} {\bibinfo  {journal} {Phys. Rev. D}\ }\textbf {\bibinfo {volume}
  {90}},\ \bibinfo {pages} {083528} (\bibinfo {year} {2014})},\ \Eprint
  {https://arxiv.org/abs/1408.1811} {arXiv:1408.1811 [hep-ph]} \BibitemShut
  {NoStop}%
\bibitem [{\citenamefont {Gleiser}\ and\ \citenamefont
  {Graham}(2014)}]{Gleiser:2014ipa}%
  \BibitemOpen
  \bibfield  {author} {\bibinfo {author} {\bibfnamefont {M.}~\bibnamefont
  {Gleiser}}\ and\ \bibinfo {author} {\bibfnamefont {N.}~\bibnamefont
  {Graham}},\ }\href {https://doi.org/10.1103/PhysRevD.89.083502} {\bibfield
  {journal} {\bibinfo  {journal} {Phys. Rev. D}\ }\textbf {\bibinfo {volume}
  {89}},\ \bibinfo {pages} {083502} (\bibinfo {year} {2014})},\ \Eprint
  {https://arxiv.org/abs/1401.6225} {arXiv:1401.6225 [astro-ph.CO]}
  \BibitemShut {NoStop}%
\bibitem [{\citenamefont {Adshead}\ \emph {et~al.}(2015)\citenamefont
  {Adshead}, \citenamefont {Giblin}, \citenamefont {Scully},\ and\
  \citenamefont {Sfakianakis}}]{Adshead:2015pva}%
  \BibitemOpen
  \bibfield  {author} {\bibinfo {author} {\bibfnamefont {P.}~\bibnamefont
  {Adshead}}, \bibinfo {author} {\bibfnamefont {J.~T.}\ \bibnamefont {Giblin}},
  \bibinfo {author} {\bibfnamefont {T.~R.}\ \bibnamefont {Scully}},\ and\
  \bibinfo {author} {\bibfnamefont {E.~I.}\ \bibnamefont {Sfakianakis}},\
  }\href {https://doi.org/10.1088/1475-7516/2015/12/034} {\bibfield  {journal}
  {\bibinfo  {journal} {JCAP}\ }\textbf {\bibinfo {volume} {12}},\ \bibinfo
  {pages} {034}},\ \Eprint {https://arxiv.org/abs/1502.06506} {arXiv:1502.06506
  [astro-ph.CO]} \BibitemShut {NoStop}%
\bibitem [{\citenamefont {Antusch}\ \emph
  {et~al.}(2018{\natexlab{a}})\citenamefont {Antusch}, \citenamefont {Cefala},
  \citenamefont {Krippendorf}, \citenamefont {Muia}, \citenamefont {Orani},\
  and\ \citenamefont {Quevedo}}]{Antusch:2017flz}%
  \BibitemOpen
  \bibfield  {author} {\bibinfo {author} {\bibfnamefont {S.}~\bibnamefont
  {Antusch}}, \bibinfo {author} {\bibfnamefont {F.}~\bibnamefont {Cefala}},
  \bibinfo {author} {\bibfnamefont {S.}~\bibnamefont {Krippendorf}}, \bibinfo
  {author} {\bibfnamefont {F.}~\bibnamefont {Muia}}, \bibinfo {author}
  {\bibfnamefont {S.}~\bibnamefont {Orani}},\ and\ \bibinfo {author}
  {\bibfnamefont {F.}~\bibnamefont {Quevedo}},\ }\href
  {https://doi.org/10.1007/JHEP01(2018)083} {\bibfield  {journal} {\bibinfo
  {journal} {JHEP}\ }\textbf {\bibinfo {volume} {01}},\ \bibinfo {pages}
  {083}},\ \Eprint {https://arxiv.org/abs/1708.08922} {arXiv:1708.08922
  [hep-th]} \BibitemShut {NoStop}%
\bibitem [{\citenamefont {Lozanov}\ and\ \citenamefont
  {Amin}(2018)}]{Lozanov:2017hjm}%
  \BibitemOpen
  \bibfield  {author} {\bibinfo {author} {\bibfnamefont {K.~D.}\ \bibnamefont
  {Lozanov}}\ and\ \bibinfo {author} {\bibfnamefont {M.~A.}\ \bibnamefont
  {Amin}},\ }\href {https://doi.org/10.1103/PhysRevD.97.023533} {\bibfield
  {journal} {\bibinfo  {journal} {Phys. Rev. D}\ }\textbf {\bibinfo {volume}
  {97}},\ \bibinfo {pages} {023533} (\bibinfo {year} {2018})},\ \Eprint
  {https://arxiv.org/abs/1710.06851} {arXiv:1710.06851 [astro-ph.CO]}
  \BibitemShut {NoStop}%
\bibitem [{\citenamefont {Sang}\ and\ \citenamefont
  {Huang}(2021)}]{Sang:2020kpd}%
  \BibitemOpen
  \bibfield  {author} {\bibinfo {author} {\bibfnamefont {Y.}~\bibnamefont
  {Sang}}\ and\ \bibinfo {author} {\bibfnamefont {Q.-G.}\ \bibnamefont
  {Huang}},\ }\href {https://doi.org/10.1016/j.physletb.2021.136781} {\bibfield
   {journal} {\bibinfo  {journal} {Phys. Lett. B}\ }\textbf {\bibinfo {volume}
  {823}},\ \bibinfo {pages} {136781} (\bibinfo {year} {2021})},\ \Eprint
  {https://arxiv.org/abs/2012.14697} {arXiv:2012.14697 [hep-th]} \BibitemShut
  {NoStop}%
\bibitem [{\citenamefont {Mahbub}\ and\ \citenamefont
  {Mishra}(2023)}]{Mahbub:2023faw}%
  \BibitemOpen
  \bibfield  {author} {\bibinfo {author} {\bibfnamefont {R.}~\bibnamefont
  {Mahbub}}\ and\ \bibinfo {author} {\bibfnamefont {S.~S.}\ \bibnamefont
  {Mishra}},\ }\href {https://doi.org/10.1103/PhysRevD.108.063524} {\bibfield
  {journal} {\bibinfo  {journal} {Phys. Rev. D}\ }\textbf {\bibinfo {volume}
  {108}},\ \bibinfo {pages} {063524} (\bibinfo {year} {2023})},\ \Eprint
  {https://arxiv.org/abs/2303.07503} {arXiv:2303.07503 [astro-ph.CO]}
  \BibitemShut {NoStop}%
\bibitem [{\citenamefont {Jia}\ \emph {et~al.}(2024)\citenamefont {Jia},
  \citenamefont {Sang},\ and\ \citenamefont {Zhang}}]{Jia:2024fmo}%
  \BibitemOpen
  \bibfield  {author} {\bibinfo {author} {\bibfnamefont {T.}~\bibnamefont
  {Jia}}, \bibinfo {author} {\bibfnamefont {Y.}~\bibnamefont {Sang}},\ and\
  \bibinfo {author} {\bibfnamefont {X.}~\bibnamefont {Zhang}},\ }\href@noop {}
  {\  (\bibinfo {year} {2024})},\ \Eprint {https://arxiv.org/abs/2409.04046}
  {arXiv:2409.04046 [astro-ph.CO]} \BibitemShut {NoStop}%
\bibitem [{\citenamefont {Shafi}\ \emph {et~al.}(2024)\citenamefont {Shafi},
  \citenamefont {Copeland}, \citenamefont {Mahbub}, \citenamefont {Mishra},\
  and\ \citenamefont {Basak}}]{Shafi:2024jig}%
  \BibitemOpen
  \bibfield  {author} {\bibinfo {author} {\bibfnamefont {M.}~\bibnamefont
  {Shafi}}, \bibinfo {author} {\bibfnamefont {E.~J.}\ \bibnamefont {Copeland}},
  \bibinfo {author} {\bibfnamefont {R.}~\bibnamefont {Mahbub}}, \bibinfo
  {author} {\bibfnamefont {S.~S.}\ \bibnamefont {Mishra}},\ and\ \bibinfo
  {author} {\bibfnamefont {S.}~\bibnamefont {Basak}},\ }\href
  {https://doi.org/10.1088/1475-7516/2024/10/082} {\bibfield  {journal}
  {\bibinfo  {journal} {JCAP}\ }\textbf {\bibinfo {volume} {10}},\ \bibinfo
  {pages} {082}},\ \Eprint {https://arxiv.org/abs/2406.00108} {arXiv:2406.00108
  [hep-ph]} \BibitemShut {NoStop}%
\bibitem [{\citenamefont {Amin}\ \emph
  {et~al.}(2014{\natexlab{b}})\citenamefont {Amin}, \citenamefont {Hertzberg},
  \citenamefont {Kaiser},\ and\ \citenamefont {Karouby}}]{Amin:2014eta}%
  \BibitemOpen
  \bibfield  {author} {\bibinfo {author} {\bibfnamefont {M.~A.}\ \bibnamefont
  {Amin}}, \bibinfo {author} {\bibfnamefont {M.~P.}\ \bibnamefont {Hertzberg}},
  \bibinfo {author} {\bibfnamefont {D.~I.}\ \bibnamefont {Kaiser}},\ and\
  \bibinfo {author} {\bibfnamefont {J.}~\bibnamefont {Karouby}},\ }\href
  {https://doi.org/10.1142/S0218271815300037} {\bibfield  {journal} {\bibinfo
  {journal} {Int. J. Mod. Phys. D}\ }\textbf {\bibinfo {volume} {24}},\
  \bibinfo {pages} {1530003} (\bibinfo {year} {2014}{\natexlab{b}})},\ \Eprint
  {https://arxiv.org/abs/1410.3808} {arXiv:1410.3808 [hep-ph]} \BibitemShut
  {NoStop}%
\bibitem [{\citenamefont {Kawasaki}\ and\ \citenamefont
  {Takeda}(2014)}]{Kawasaki:2013hka}%
  \BibitemOpen
  \bibfield  {author} {\bibinfo {author} {\bibfnamefont {M.}~\bibnamefont
  {Kawasaki}}\ and\ \bibinfo {author} {\bibfnamefont {N.}~\bibnamefont
  {Takeda}},\ }\href {https://doi.org/10.1088/1475-7516/2014/07/038} {\bibfield
   {journal} {\bibinfo  {journal} {JCAP}\ }\textbf {\bibinfo {volume} {07}},\
  \bibinfo {pages} {038}},\ \Eprint {https://arxiv.org/abs/1310.4615}
  {arXiv:1310.4615 [astro-ph.CO]} \BibitemShut {NoStop}%
\bibitem [{\citenamefont {Saffin}(2017)}]{Saffin:2016kof}%
  \BibitemOpen
  \bibfield  {author} {\bibinfo {author} {\bibfnamefont {P.~M.}\ \bibnamefont
  {Saffin}},\ }\href {https://doi.org/10.1007/JHEP07(2017)126} {\bibfield
  {journal} {\bibinfo  {journal} {JHEP}\ }\textbf {\bibinfo {volume} {07}},\
  \bibinfo {pages} {126}},\ \Eprint {https://arxiv.org/abs/1612.02014}
  {arXiv:1612.02014 [hep-th]} \BibitemShut {NoStop}%
\bibitem [{\citenamefont {Zhou}\ \emph {et~al.}(2013)\citenamefont {Zhou},
  \citenamefont {Copeland}, \citenamefont {Easther}, \citenamefont {Finkel},
  \citenamefont {Mou},\ and\ \citenamefont {Saffin}}]{Zhou:2013tsa}%
  \BibitemOpen
  \bibfield  {author} {\bibinfo {author} {\bibfnamefont {S.-Y.}\ \bibnamefont
  {Zhou}}, \bibinfo {author} {\bibfnamefont {E.~J.}\ \bibnamefont {Copeland}},
  \bibinfo {author} {\bibfnamefont {R.}~\bibnamefont {Easther}}, \bibinfo
  {author} {\bibfnamefont {H.}~\bibnamefont {Finkel}}, \bibinfo {author}
  {\bibfnamefont {Z.-G.}\ \bibnamefont {Mou}},\ and\ \bibinfo {author}
  {\bibfnamefont {P.~M.}\ \bibnamefont {Saffin}},\ }\href
  {https://doi.org/10.1007/JHEP10(2013)026} {\bibfield  {journal} {\bibinfo
  {journal} {JHEP}\ }\textbf {\bibinfo {volume} {10}},\ \bibinfo {pages}
  {026}},\ \Eprint {https://arxiv.org/abs/1304.6094} {arXiv:1304.6094
  [astro-ph.CO]} \BibitemShut {NoStop}%
\bibitem [{\citenamefont {Antusch}\ \emph
  {et~al.}(2018{\natexlab{b}})\citenamefont {Antusch}, \citenamefont {Cefala},\
  and\ \citenamefont {Orani}}]{Antusch:2017vga}%
  \BibitemOpen
  \bibfield  {author} {\bibinfo {author} {\bibfnamefont {S.}~\bibnamefont
  {Antusch}}, \bibinfo {author} {\bibfnamefont {F.}~\bibnamefont {Cefala}},\
  and\ \bibinfo {author} {\bibfnamefont {S.}~\bibnamefont {Orani}},\ }\href
  {https://doi.org/10.1088/1475-7516/2018/03/032} {\bibfield  {journal}
  {\bibinfo  {journal} {JCAP}\ }\textbf {\bibinfo {volume} {03}},\ \bibinfo
  {pages} {032}},\ \Eprint {https://arxiv.org/abs/1712.03231} {arXiv:1712.03231
  [astro-ph.CO]} \BibitemShut {NoStop}%
\bibitem [{\citenamefont {Amin}\ \emph {et~al.}(2018)\citenamefont {Amin},
  \citenamefont {Braden}, \citenamefont {Copeland}, \citenamefont {Giblin},
  \citenamefont {Solorio}, \citenamefont {Weiner},\ and\ \citenamefont
  {Zhou}}]{Amin:2018xfe}%
  \BibitemOpen
  \bibfield  {author} {\bibinfo {author} {\bibfnamefont {M.~A.}\ \bibnamefont
  {Amin}}, \bibinfo {author} {\bibfnamefont {J.}~\bibnamefont {Braden}},
  \bibinfo {author} {\bibfnamefont {E.~J.}\ \bibnamefont {Copeland}}, \bibinfo
  {author} {\bibfnamefont {J.~T.}\ \bibnamefont {Giblin}}, \bibinfo {author}
  {\bibfnamefont {C.}~\bibnamefont {Solorio}}, \bibinfo {author} {\bibfnamefont
  {Z.~J.}\ \bibnamefont {Weiner}},\ and\ \bibinfo {author} {\bibfnamefont
  {S.-Y.}\ \bibnamefont {Zhou}},\ }\href
  {https://doi.org/10.1103/PhysRevD.98.024040} {\bibfield  {journal} {\bibinfo
  {journal} {Phys. Rev. D}\ }\textbf {\bibinfo {volume} {98}},\ \bibinfo
  {pages} {024040} (\bibinfo {year} {2018})},\ \Eprint
  {https://arxiv.org/abs/1803.08047} {arXiv:1803.08047 [astro-ph.CO]}
  \BibitemShut {NoStop}%
\bibitem [{\citenamefont {Lozanov}\ and\ \citenamefont
  {Amin}(2019)}]{Lozanov:2019ylm}%
  \BibitemOpen
  \bibfield  {author} {\bibinfo {author} {\bibfnamefont {K.~D.}\ \bibnamefont
  {Lozanov}}\ and\ \bibinfo {author} {\bibfnamefont {M.~A.}\ \bibnamefont
  {Amin}},\ }\href {https://doi.org/10.1103/PhysRevD.99.123504} {\bibfield
  {journal} {\bibinfo  {journal} {Phys. Rev. D}\ }\textbf {\bibinfo {volume}
  {99}},\ \bibinfo {pages} {123504} (\bibinfo {year} {2019})},\ \Eprint
  {https://arxiv.org/abs/1902.06736} {arXiv:1902.06736 [astro-ph.CO]}
  \BibitemShut {NoStop}%
\bibitem [{\citenamefont {Hiramatsu}\ \emph {et~al.}(2021)\citenamefont
  {Hiramatsu}, \citenamefont {Sfakianakis},\ and\ \citenamefont
  {Yamaguchi}}]{Hiramatsu:2020obh}%
  \BibitemOpen
  \bibfield  {author} {\bibinfo {author} {\bibfnamefont {T.}~\bibnamefont
  {Hiramatsu}}, \bibinfo {author} {\bibfnamefont {E.~I.}\ \bibnamefont
  {Sfakianakis}},\ and\ \bibinfo {author} {\bibfnamefont {M.}~\bibnamefont
  {Yamaguchi}},\ }\href {https://doi.org/10.1007/JHEP03(2021)021} {\bibfield
  {journal} {\bibinfo  {journal} {JHEP}\ }\textbf {\bibinfo {volume} {03}},\
  \bibinfo {pages} {021}},\ \Eprint {https://arxiv.org/abs/2011.12201}
  {arXiv:2011.12201 [hep-ph]} \BibitemShut {NoStop}%
\bibitem [{\citenamefont {Antusch}\ \emph {et~al.}(2017)\citenamefont
  {Antusch}, \citenamefont {Cefala},\ and\ \citenamefont
  {Orani}}]{Antusch:2016con}%
  \BibitemOpen
  \bibfield  {author} {\bibinfo {author} {\bibfnamefont {S.}~\bibnamefont
  {Antusch}}, \bibinfo {author} {\bibfnamefont {F.}~\bibnamefont {Cefala}},\
  and\ \bibinfo {author} {\bibfnamefont {S.}~\bibnamefont {Orani}},\ }\href
  {https://doi.org/10.1103/PhysRevLett.118.011303} {\bibfield  {journal}
  {\bibinfo  {journal} {Phys. Rev. Lett.}\ }\textbf {\bibinfo {volume} {118}},\
  \bibinfo {pages} {011303} (\bibinfo {year} {2017})},\ \bibinfo {note}
  {[Erratum: Phys.Rev.Lett. 120, 219901 (2018)]},\ \Eprint
  {https://arxiv.org/abs/1607.01314} {arXiv:1607.01314 [astro-ph.CO]}
  \BibitemShut {NoStop}%
\bibitem [{\citenamefont {Lozanov}\ \emph {et~al.}(2023)\citenamefont
  {Lozanov}, \citenamefont {Sasaki},\ and\ \citenamefont
  {Takhistov}}]{Lozanov:2023aez}%
  \BibitemOpen
  \bibfield  {author} {\bibinfo {author} {\bibfnamefont {K.~D.}\ \bibnamefont
  {Lozanov}}, \bibinfo {author} {\bibfnamefont {M.}~\bibnamefont {Sasaki}},\
  and\ \bibinfo {author} {\bibfnamefont {V.}~\bibnamefont {Takhistov}},\
  }\href@noop {} {\  (\bibinfo {year} {2023})},\ \Eprint
  {https://arxiv.org/abs/2304.06709} {arXiv:2304.06709 [astro-ph.CO]}
  \BibitemShut {NoStop}%
\bibitem [{\citenamefont {Dufaux}\ \emph {et~al.}(2007)\citenamefont {Dufaux},
  \citenamefont {Bergman}, \citenamefont {Felder}, \citenamefont {Kofman},\
  and\ \citenamefont {Uzan}}]{Dufaux:2007pt}%
  \BibitemOpen
  \bibfield  {author} {\bibinfo {author} {\bibfnamefont {J.~F.}\ \bibnamefont
  {Dufaux}}, \bibinfo {author} {\bibfnamefont {A.}~\bibnamefont {Bergman}},
  \bibinfo {author} {\bibfnamefont {G.~N.}\ \bibnamefont {Felder}}, \bibinfo
  {author} {\bibfnamefont {L.}~\bibnamefont {Kofman}},\ and\ \bibinfo {author}
  {\bibfnamefont {J.-P.}\ \bibnamefont {Uzan}},\ }\href
  {https://doi.org/10.1103/PhysRevD.76.123517} {\bibfield  {journal} {\bibinfo
  {journal} {Phys. Rev. D}\ }\textbf {\bibinfo {volume} {76}},\ \bibinfo
  {pages} {123517} (\bibinfo {year} {2007})},\ \Eprint
  {https://arxiv.org/abs/0707.0875} {arXiv:0707.0875 [astro-ph]} \BibitemShut
  {NoStop}%
\bibitem [{\citenamefont {Liu}\ \emph {et~al.}(2018)\citenamefont {Liu},
  \citenamefont {Guo}, \citenamefont {Cai},\ and\ \citenamefont
  {Shiu}}]{Liu:2017hua}%
  \BibitemOpen
  \bibfield  {author} {\bibinfo {author} {\bibfnamefont {J.}~\bibnamefont
  {Liu}}, \bibinfo {author} {\bibfnamefont {Z.-K.}\ \bibnamefont {Guo}},
  \bibinfo {author} {\bibfnamefont {R.-G.}\ \bibnamefont {Cai}},\ and\ \bibinfo
  {author} {\bibfnamefont {G.}~\bibnamefont {Shiu}},\ }\href
  {https://doi.org/10.1103/PhysRevLett.120.031301} {\bibfield  {journal}
  {\bibinfo  {journal} {Phys. Rev. Lett.}\ }\textbf {\bibinfo {volume} {120}},\
  \bibinfo {pages} {031301} (\bibinfo {year} {2018})},\ \Eprint
  {https://arxiv.org/abs/1707.09841} {arXiv:1707.09841 [astro-ph.CO]}
  \BibitemShut {NoStop}%
\bibitem [{\citenamefont {Liu}\ \emph {et~al.}(2019)\citenamefont {Liu},
  \citenamefont {Guo}, \citenamefont {Cai},\ and\ \citenamefont
  {Shiu}}]{Liu:2018rrt}%
  \BibitemOpen
  \bibfield  {author} {\bibinfo {author} {\bibfnamefont {J.}~\bibnamefont
  {Liu}}, \bibinfo {author} {\bibfnamefont {Z.-K.}\ \bibnamefont {Guo}},
  \bibinfo {author} {\bibfnamefont {R.-G.}\ \bibnamefont {Cai}},\ and\ \bibinfo
  {author} {\bibfnamefont {G.}~\bibnamefont {Shiu}},\ }\href
  {https://doi.org/10.1103/PhysRevD.99.103506} {\bibfield  {journal} {\bibinfo
  {journal} {Phys. Rev. D}\ }\textbf {\bibinfo {volume} {99}},\ \bibinfo
  {pages} {103506} (\bibinfo {year} {2019})},\ \Eprint
  {https://arxiv.org/abs/1812.09235} {arXiv:1812.09235 [astro-ph.CO]}
  \BibitemShut {NoStop}%
\bibitem [{\citenamefont {Sang}\ and\ \citenamefont
  {Huang}(2019)}]{Sang:2019ndv}%
  \BibitemOpen
  \bibfield  {author} {\bibinfo {author} {\bibfnamefont {Y.}~\bibnamefont
  {Sang}}\ and\ \bibinfo {author} {\bibfnamefont {Q.-G.}\ \bibnamefont
  {Huang}},\ }\href {https://doi.org/10.1103/PhysRevD.100.063516} {\bibfield
  {journal} {\bibinfo  {journal} {Phys. Rev. D}\ }\textbf {\bibinfo {volume}
  {100}},\ \bibinfo {pages} {063516} (\bibinfo {year} {2019})},\ \Eprint
  {https://arxiv.org/abs/1905.00371} {arXiv:1905.00371 [astro-ph.CO]}
  \BibitemShut {NoStop}%
\bibitem [{\citenamefont {Kou}\ \emph {et~al.}(2022)\citenamefont {Kou},
  \citenamefont {Mertens}, \citenamefont {Tian},\ and\ \citenamefont
  {Zhou}}]{Kou:2021bij}%
  \BibitemOpen
  \bibfield  {author} {\bibinfo {author} {\bibfnamefont {X.-X.}\ \bibnamefont
  {Kou}}, \bibinfo {author} {\bibfnamefont {J.~B.}\ \bibnamefont {Mertens}},
  \bibinfo {author} {\bibfnamefont {C.}~\bibnamefont {Tian}},\ and\ \bibinfo
  {author} {\bibfnamefont {S.-Y.}\ \bibnamefont {Zhou}},\ }\href
  {https://doi.org/10.1103/PhysRevD.105.123505} {\bibfield  {journal} {\bibinfo
   {journal} {Phys. Rev. D}\ }\textbf {\bibinfo {volume} {105}},\ \bibinfo
  {pages} {123505} (\bibinfo {year} {2022})},\ \Eprint
  {https://arxiv.org/abs/2112.07626} {arXiv:2112.07626 [gr-qc]} \BibitemShut
  {NoStop}%
\bibitem [{\citenamefont {Kou}\ \emph {et~al.}(2021)\citenamefont {Kou},
  \citenamefont {Tian},\ and\ \citenamefont {Zhou}}]{Kou:2019bbc}%
  \BibitemOpen
  \bibfield  {author} {\bibinfo {author} {\bibfnamefont {X.-X.}\ \bibnamefont
  {Kou}}, \bibinfo {author} {\bibfnamefont {C.}~\bibnamefont {Tian}},\ and\
  \bibinfo {author} {\bibfnamefont {S.-Y.}\ \bibnamefont {Zhou}},\ }\href
  {https://doi.org/10.1088/1361-6382/abd09f} {\bibfield  {journal} {\bibinfo
  {journal} {Class. Quant. Grav.}\ }\textbf {\bibinfo {volume} {38}},\ \bibinfo
  {pages} {045005} (\bibinfo {year} {2021})},\ \Eprint
  {https://arxiv.org/abs/1912.09658} {arXiv:1912.09658 [gr-qc]} \BibitemShut
  {NoStop}%
\bibitem [{\citenamefont {Aurrekoetxea}\ \emph {et~al.}(2023)\citenamefont
  {Aurrekoetxea}, \citenamefont {Clough},\ and\ \citenamefont
  {Muia}}]{Aurrekoetxea:2023jwd}%
  \BibitemOpen
  \bibfield  {author} {\bibinfo {author} {\bibfnamefont {J.~C.}\ \bibnamefont
  {Aurrekoetxea}}, \bibinfo {author} {\bibfnamefont {K.}~\bibnamefont
  {Clough}},\ and\ \bibinfo {author} {\bibfnamefont {F.}~\bibnamefont {Muia}},\
  }\href {https://doi.org/10.1103/PhysRevD.108.023501} {\bibfield  {journal}
  {\bibinfo  {journal} {Phys. Rev. D}\ }\textbf {\bibinfo {volume} {108}},\
  \bibinfo {pages} {023501} (\bibinfo {year} {2023})},\ \Eprint
  {https://arxiv.org/abs/2304.01673} {arXiv:2304.01673 [gr-qc]} \BibitemShut
  {NoStop}%
\bibitem [{\citenamefont {Baumgarte}\ and\ \citenamefont
  {Shapiro}(1998)}]{Baumgarte:1998te}%
  \BibitemOpen
  \bibfield  {author} {\bibinfo {author} {\bibfnamefont {T.~W.}\ \bibnamefont
  {Baumgarte}}\ and\ \bibinfo {author} {\bibfnamefont {S.~L.}\ \bibnamefont
  {Shapiro}},\ }\href {https://doi.org/10.1103/PhysRevD.59.024007} {\bibfield
  {journal} {\bibinfo  {journal} {Phys. Rev. D}\ }\textbf {\bibinfo {volume}
  {59}},\ \bibinfo {pages} {024007} (\bibinfo {year} {1998})},\ \Eprint
  {https://arxiv.org/abs/gr-qc/9810065} {arXiv:gr-qc/9810065} \BibitemShut
  {NoStop}%
\bibitem [{\citenamefont {Shibata}\ and\ \citenamefont
  {Nakamura}(1995)}]{Shibata:1995we}%
  \BibitemOpen
  \bibfield  {author} {\bibinfo {author} {\bibfnamefont {M.}~\bibnamefont
  {Shibata}}\ and\ \bibinfo {author} {\bibfnamefont {T.}~\bibnamefont
  {Nakamura}},\ }\href {https://doi.org/10.1103/PhysRevD.52.5428} {\bibfield
  {journal} {\bibinfo  {journal} {Phys. Rev. D}\ }\textbf {\bibinfo {volume}
  {52}},\ \bibinfo {pages} {5428} (\bibinfo {year} {1995})}\BibitemShut
  {NoStop}%
\bibitem [{\citenamefont {Aurrekoetxea}\ \emph {et~al.}(2024)\citenamefont
  {Aurrekoetxea}, \citenamefont {Clough},\ and\ \citenamefont
  {Lim}}]{Aurrekoetxea:2024mdy}%
  \BibitemOpen
  \bibfield  {author} {\bibinfo {author} {\bibfnamefont {J.~C.}\ \bibnamefont
  {Aurrekoetxea}}, \bibinfo {author} {\bibfnamefont {K.}~\bibnamefont
  {Clough}},\ and\ \bibinfo {author} {\bibfnamefont {E.~A.}\ \bibnamefont
  {Lim}},\ }\href@noop {} {\  (\bibinfo {year} {2024})},\ \Eprint
  {https://arxiv.org/abs/2409.01939} {arXiv:2409.01939 [gr-qc]} \BibitemShut
  {NoStop}%
\bibitem [{\citenamefont {Nazari}\ \emph {et~al.}(2021)\citenamefont {Nazari},
  \citenamefont {Cicoli}, \citenamefont {Clough},\ and\ \citenamefont
  {Muia}}]{Nazari:2020fmk}%
  \BibitemOpen
  \bibfield  {author} {\bibinfo {author} {\bibfnamefont {Z.}~\bibnamefont
  {Nazari}}, \bibinfo {author} {\bibfnamefont {M.}~\bibnamefont {Cicoli}},
  \bibinfo {author} {\bibfnamefont {K.}~\bibnamefont {Clough}},\ and\ \bibinfo
  {author} {\bibfnamefont {F.}~\bibnamefont {Muia}},\ }\href
  {https://doi.org/10.1088/1475-7516/2021/05/027} {\bibfield  {journal}
  {\bibinfo  {journal} {JCAP}\ }\textbf {\bibinfo {volume} {05}},\ \bibinfo
  {pages} {027}},\ \Eprint {https://arxiv.org/abs/2010.05933} {arXiv:2010.05933
  [gr-qc]} \BibitemShut {NoStop}%
\bibitem [{\citenamefont {Cotner}\ \emph {et~al.}(2018)\citenamefont {Cotner},
  \citenamefont {Kusenko},\ and\ \citenamefont {Takhistov}}]{Cotner:2018vug}%
  \BibitemOpen
  \bibfield  {author} {\bibinfo {author} {\bibfnamefont {E.}~\bibnamefont
  {Cotner}}, \bibinfo {author} {\bibfnamefont {A.}~\bibnamefont {Kusenko}},\
  and\ \bibinfo {author} {\bibfnamefont {V.}~\bibnamefont {Takhistov}},\ }\href
  {https://doi.org/10.1103/PhysRevD.98.083513} {\bibfield  {journal} {\bibinfo
  {journal} {Phys. Rev. D}\ }\textbf {\bibinfo {volume} {98}},\ \bibinfo
  {pages} {083513} (\bibinfo {year} {2018})},\ \Eprint
  {https://arxiv.org/abs/1801.03321} {arXiv:1801.03321 [astro-ph.CO]}
  \BibitemShut {NoStop}%
\bibitem [{\citenamefont {Lozanov}\ and\ \citenamefont
  {Takhistov}(2023)}]{Lozanov:2022yoy}%
  \BibitemOpen
  \bibfield  {author} {\bibinfo {author} {\bibfnamefont {K.~D.}\ \bibnamefont
  {Lozanov}}\ and\ \bibinfo {author} {\bibfnamefont {V.}~\bibnamefont
  {Takhistov}},\ }\href {https://doi.org/10.1103/PhysRevLett.130.181002}
  {\bibfield  {journal} {\bibinfo  {journal} {Phys. Rev. Lett.}\ }\textbf
  {\bibinfo {volume} {130}},\ \bibinfo {pages} {181002} (\bibinfo {year}
  {2023})},\ \Eprint {https://arxiv.org/abs/2204.07152} {arXiv:2204.07152
  [astro-ph.CO]} \BibitemShut {NoStop}%
\bibitem [{\citenamefont {Lozanov}\ \emph {et~al.}(2024)\citenamefont
  {Lozanov}, \citenamefont {Sasaki},\ and\ \citenamefont
  {Takhistov}}]{Lozanov:2023knf}%
  \BibitemOpen
  \bibfield  {author} {\bibinfo {author} {\bibfnamefont {K.~D.}\ \bibnamefont
  {Lozanov}}, \bibinfo {author} {\bibfnamefont {M.}~\bibnamefont {Sasaki}},\
  and\ \bibinfo {author} {\bibfnamefont {V.}~\bibnamefont {Takhistov}},\ }\href
  {https://doi.org/10.1016/j.physletb.2023.138392} {\bibfield  {journal}
  {\bibinfo  {journal} {Phys. Lett. B}\ }\textbf {\bibinfo {volume} {848}},\
  \bibinfo {pages} {138392} (\bibinfo {year} {2024})},\ \Eprint
  {https://arxiv.org/abs/2309.14193} {arXiv:2309.14193 [astro-ph.CO]}
  \BibitemShut {NoStop}%
\bibitem [{\citenamefont {Ananda}\ \emph {et~al.}(2007)\citenamefont {Ananda},
  \citenamefont {Clarkson},\ and\ \citenamefont {Wands}}]{Ananda:2006af}%
  \BibitemOpen
  \bibfield  {author} {\bibinfo {author} {\bibfnamefont {K.~N.}\ \bibnamefont
  {Ananda}}, \bibinfo {author} {\bibfnamefont {C.}~\bibnamefont {Clarkson}},\
  and\ \bibinfo {author} {\bibfnamefont {D.}~\bibnamefont {Wands}},\ }\href
  {https://doi.org/10.1103/PhysRevD.75.123518} {\bibfield  {journal} {\bibinfo
  {journal} {Phys. Rev. D}\ }\textbf {\bibinfo {volume} {75}},\ \bibinfo
  {pages} {123518} (\bibinfo {year} {2007})},\ \Eprint
  {https://arxiv.org/abs/gr-qc/0612013} {arXiv:gr-qc/0612013} \BibitemShut
  {NoStop}%
\bibitem [{\citenamefont {Inomata}\ \emph
  {et~al.}(2019{\natexlab{b}})\citenamefont {Inomata}, \citenamefont {Kohri},
  \citenamefont {Nakama},\ and\ \citenamefont {Terada}}]{Inomata:2019ivs}%
  \BibitemOpen
  \bibfield  {author} {\bibinfo {author} {\bibfnamefont {K.}~\bibnamefont
  {Inomata}}, \bibinfo {author} {\bibfnamefont {K.}~\bibnamefont {Kohri}},
  \bibinfo {author} {\bibfnamefont {T.}~\bibnamefont {Nakama}},\ and\ \bibinfo
  {author} {\bibfnamefont {T.}~\bibnamefont {Terada}},\ }\href
  {https://doi.org/10.1103/PhysRevD.108.049901} {\bibfield  {journal} {\bibinfo
   {journal} {Phys. Rev. D}\ }\textbf {\bibinfo {volume} {100}},\ \bibinfo
  {pages} {043532} (\bibinfo {year} {2019}{\natexlab{b}})},\ \bibinfo {note}
  {[Erratum: Phys.Rev.D 108, 049901 (2023)]},\ \Eprint
  {https://arxiv.org/abs/1904.12879} {arXiv:1904.12879 [astro-ph.CO]}
  \BibitemShut {NoStop}%
\end{thebibliography}%

\end{document}